%% file: template.tex
\documentclass{article}

\usepackage{arxiv}

\usepackage[utf8]{inputenc} 
\usepackage[T1]{fontenc}    
\usepackage{hyperref}       
\usepackage{url}            
\usepackage{booktabs}       
\usepackage{amsfonts}       
\usepackage{nicefrac}       
\usepackage{microtype}      
\usepackage{lipsum}
\usepackage{graphicx}
\usepackage{subfigure}
\usepackage{amssymb}
\usepackage{amsmath}
\usepackage[ruled,vlined]{algorithm2e}
\graphicspath{ {./images/} }

\title{Full waveform inversion by model extension: practical applications}

\author{
    Guillaume Barnier \\
    Geophysics Department\\
    Stanford University\\
    Stanford, CA 94305 \\
    \texttt{gbarnier@sep.stanford.edu} \\
\And
    Ettore Biondi \\
    Seismology Laboratory\\
    California Institute of Technology\\
    Pasadena, CA 91125\\
    \texttt{ettore88@sep.stanford.edu} \\
\And
    Robert G. Clapp \\
    Geophysics Department\\
    Stanford University\\
    Stanford, CA 94305 \\
    \texttt{bob@sep.stanford.edu} \\
\And
 Biondo Biondi \\
  Geophysics Department\\
  Stanford University\\
  Stanford, CA 94305 \\
  \texttt{biondo@sep.stanford.edu} \\  
}

\begin{document}
\maketitle

\begin{abstract}
\input{abstract}
\end{abstract}

\section{Introduction}
\input{introduction}

\section{FWIME theory}
\input{theory}

\section{Marmousi2}
\input{Marmousi}

\section{BP 2004 model: the North Sea region}
\input{bpCaspian}

\section{Seiscope's syncline model}
\input{syncline}

\section{Salt model}
\input{bpSalt}
\section{Towards a more flexible model parametrization}
\input{bpSaltFeeder}

\section{Computational aspects}
\input{computational_aspects}

\section{Embedding FWIME in a production workflow}
\input{discussions}

\section{Conclusions}
\input{conclusions}

\bibliographystyle{unsrt}  
\bibliography{references}  

\end{document}

%% file: abstract.tex
Producing reliable acoustic subsurface velocity models still remains the main bottleneck of the oil and gas industry's traditional imaging sequence. In complex geological settings, the output of conventional ray-based or wave-equation-based tomographic methods may not be accurate enough for full waveform inversion (FWI) to converge to a geologically satisfactory Earth model. We create a new method referred to as full waveform inversion by model extension (FWIME) in which a wave-equation migration velocity analysis (WEMVA) technique is efficiently paired with a modified version of FWI. We show that our method is more powerful than applying WEMVA and FWI sequentially, and that it is able to converge to accurate solutions without the use of a good initial guess or low-frequency energy. We demonstrate FWIME's potential on five realistic and challenging numerical examples that simulate complex geological scenarios often encountered in hydrocarbon exploration. We guide the reader step by step throughout the optimization process. We show that our method can simultaneously invert all wave types with the same simple mechanism and without the need for a user-intensive hyper-parameter tuning process. In an online repository, we provide a fully-reproducible open-source software solution implemented with general-purpose graphics processing units (GPU) and with a user-friendly Python interface. 


%% file: introduction.tex
Seismic imaging is the most ubiquitous and effective method for subsurface hydrocarbon exploration. Energy companies rely on good-quality images to ensure safety and efficiency during exploration, drilling and production. However, reliable images of the subsurface can be quite challenging to generate in certain geological settings where complex overburdens are encountered. The main difficulty within this process is to obtain accurate seismic velocity models of the subsurface. Small errors rapidly degrade the image quality and may lead to erroneous interpretations. Due to the high-dimensional nature of the unknown model space and the large amount of seismic data to process, the velocity-estimation step is typically recast as a nonlinear optimization problem such as acoustic full waveform inversion (FWI), and numerically solved using gradient-descent algorithms \cite[]{lailly1983seismic,tarantola1984inversion}. In the last two decades, the increasing availability of high-performance computing (HPC) resources combined with the advent of general-purpose graphics processing units (GPU) enabled the industry to routinely conduct successful FWI production workflows on 3D field datasets, thereby obtaining substantial uplifts in structural image quality and higher-resolution Earth models \cite[]{sirgue2010thematic,baeten2013use,shen2018full}.

The first drawback hampering FWI is the need to start the inversion scheme with an accurate model. Failing to satisfy this condition may lead gradient-descent optimization algorithms to converge to local minima present in the FWI objective function \cite[]{virieux2009overview}. This effect can be mitigated by employing long-offset low-frequency data (with suitable signal-to-noise ratio) in conjunction with a data-space multi-scale approach \cite[]{bunks1995multiscale,brenders2007waveform,fichtner2010full}. However, acquiring such type of signal is often too impractical or costly, even with the recent advancements in the design of low-frequency sources and ultra-long offset ocean-bottom node (OBN) surveys \cite[]{dellinger2016wolfspar,brenders2018wolfspar,brittan2019fwi,bate2021ultra}. To solve this issue, the conventional imaging process usually begins by applying a tomographic algorithm such as migration velocity analysis (MVA), with which an accurate low-resolution velocity model is produced and used as an initial guess for FWI \cite[]{biondi1999wave,biondi2004angle,sava2004wave1,yang2013wavefield,estebanThesis}. Unfortunately, in certain geological scenarios, the output of the tomographic step is not precise enough for FWI to converge to a useful solution, which leaves an information gap between the two families of techniques \cite[]{claerbout1985imaging}. Moreover, even if the initial model is accurate enough, conventional FWI must often be modified or adapted to the type of waves used for the inversion. This tedious tuning and data-selection process gave rise to a multitude of ad hoc FWI variants that require intensive human inputs, thereby making their application challenging for non-expert geophysicists. Finally, FWI's computational burden may still limit the use of costly numerical schemes capable of modeling more realistic physics such as anisotropic or visco-elastic effects. Therefore, the acoustic approximation (commonly made for production workflows) can prevent the inversion from recovering useful elastic and petrophysical properties of hydrocarbon reservoirs. 

Promising approaches have been developed to improve the convergence properties of conventional FWI by designing more convex objective functions \cite[]{pladys2021cycle}. This goal is achieved by either extending the unknown model search space \cite[]{symes2008migration,fleury2012bi,biondi2014simultaneous,huang2015born,huang2017full,guo2017velocity,barnier2020full,metivier2021receiver}, by relaxing certain constraints related to the physics of the problem \cite[]{van2013mitigating,van2014new,warner2014adaptive,warner2016adaptive,guasch2019adaptive,aghamiry2019improving,li2021extended}, or by measuring the data misfit with a more appropriate norm \cite[]{metivier2016measuring,fortini2018robust,metivier2018optimal}. 

We build upon the idea introduced by \cite{symes2008migration} and \cite{biondi2014simultaneous} which consists in using the concept of extended modeling as a tool to pair a wave-equation migration velocity analysis (WEMVA) approach with conventional FWI into a single workflow. We propose a novel method, full waveform inversion by model extension (FWIME). Our goal is to bridge the information gap previously discussed and produce consistent acoustic seismic velocity models by inverting any type of surface (or borehole) seismic data without requiring accurate initial models or low-frequency long-offset recordings. We make our workflow automatic by reducing the necessity for human inputs and by limiting the number of hyper-parameters to adjust in our cost function. In addition, we simultaneously invert the full dataset (which may include all wave types and all frequencies), thereby mitigating the need for tedious data selection or filtering. Hence, our method is consistently applied with the exact same mechanism, regardless of the geological scenario. 

In a companion paper \cite[]{barnierFwime1}, we provide an exhaustive description of the theory, design, and optimization scheme for FWIME. In this paper, we focus on the method's applications and we give practical implementation details on how to efficiently conduct FWIME and select the optimal hyper-parameter values. We thoroughly guide the reader step by step through a sequence of five realistic 2D tests, which simulate some of the most challenging geological scenarios encountered in field exploration. We demonstrate FWIME's ability to automatically and simultaneously invert complex datasets composed of all types of waves. In each scenario, the recorded signal lacks low-frequency energy and the initial model is inaccurate. Such conditions are quite common in field-data applications, and lead standard methods to fail at recovering useful Earth models. 

We begin with a brief summary of the theory and design of FWIME. Then, we present our results on five numerical tests. In the first example, we invert cycle-skipped data dominated by reflected events generated by the Marmousi2 model \cite[]{martin2006marmousi2}. Additionally, we assess the effect of coherent noise by inverting an analogous dataset containing free-surface multiples. We show that for a shallow water layer, the presence of free-surface multiples can hamper the quality of the FWIME solution. The second example focuses on the use of refracted energy (diving waves) to simultaneously retrieve high- and low-resolution features from the North Sea region of the 2004 BP model \cite[]{billette20052004}. The third example is conducted on a model created and shared by the Seiscope consortium, which is designed to assess the robustness of FWIME against the ill-posedness coming from wrong association of predicted and observed waveforms. In the fourth test, we leverage recorded signal from ultra-long offset surveys (which contains a mix of reflected and refracted energy) to successfully recover a salt body without the use of low-frequency data and by starting from a pure sediment model. In the last example, we highlight some current limitations of our method at dealing with complex overburdens and we propose potential research directions to address these limitations. Finally, we discuss the computational aspects of FWIME (numerical implementation and computational cost). In addition, all numerical examples proposed in this paper are fully reproducible and can be freely accessed (along with our open-source software solution) on our online repository.

%% file: theory.tex
We summarize of the theory of full waveform inversion by model extension (FWIME). We describe the design of our objective function, we discuss the concept of extended modeling, and we introduce the model-space multi-scale approach employed in our inversion scheme. A more thorough explanation with detailed mathematical derivations can be found in our complementary paper \cite[]{barnierFwimeTheory}. 

\subsection{Formulation}
We minimize the FWIME objective function defined by

\begin{eqnarray}
    \label{eqn:fwime.obj}
    \Phi_{\epsilon}(\mathbf{m}) &=& \frac{1}{2} \left\| \mathbf{f}(\mathbf{Sm}) + \tilde{\mathbf{B}} (\mathbf{Sm}) \mathbf{\tilde{p}}_{\epsilon}^{opt}(\mathbf{Sm}) - \mathbf{d}^{obs} \right\|^2_2 + \frac{\epsilon^2}{2} \left\| \mathbf{D}{\mathbf{\tilde p}}_{\epsilon}^{opt}(\mathbf{Sm}) \right\|^2_2,
\end{eqnarray}

where $\mathbf{m}$ is the (non-extended) acoustic velocity model parametrized on a spline grid, and $\mathbf{S}$ is a linear spline interpolation operator that maps models from a coarse grid onto a finite-difference grid \cite[]{barnier2019waveform}. $\mathbf{f}$ is the wavefield generated by an acoustic isotropic constant-density two-way wave-equation operator (extracted at some predefined receivers' locations), and $\tilde{\mathbf{B}}$ denotes the Born modeling operator extended with either time lags or horizontal subsurface offsets \cite[]{biondi2014simultaneous,barnier2018full}. In this paper, all extended operations and operators are denoted by the $\sim$ symbol. $\mathbf{d}^{obs}$ is the observed data, and the diagonal matrix $\mathbf{D}$ is an invertible modified version of the differential semblance optimization (DSO) operator that penalizes defocused energy within extended images \cite[]{symes1994inversion}. $\epsilon$ is the fixed trade-off parameter that balances the two components of the objective function. The variable $\tilde{\mathbf{p}}_{\epsilon}^{opt}$ is always parametrized on the finite-difference grid, and is defined as the minimizer of the quadratic objective function $\Phi_{\epsilon,\mathbf{m}}$ (for a constant $\mathbf{m}$),

\begin{eqnarray}
    \label{eqn:vp.obj}
    \Phi_{\epsilon,\mathbf{m}}(\mathbf{\mathbf{\tilde{p}}}) &=& \frac{1}{2} \left\| \tilde{\mathbf{B}}(\mathbf{Sm})\mathbf{\tilde{p}}  - \left [ \mathbf{d}^{obs} - \mathbf{f}(\mathbf{Sm}) \right ] \right\|^2_2 + \frac{\epsilon^2}{2} \left\| \mathbf{D} \tilde{\mathbf{p}} \right\|^2_2.
\end{eqnarray}

The Hessian matrix of $\Phi_{\epsilon,\mathbf{m}}$ is given by the following expression,

\begin{eqnarray}
    \label{eqn:pert.opt}
    \mathbf{H}_{{\Phi}_{\epsilon,\mathbf{m}}} &=& \tilde{\mathbf{B}}^*(\mathbf{Sm}) \; \tilde{\mathbf{B}}(\mathbf{Sm}) + \epsilon^2 \mathbf{D}^* \mathbf{D},
\end{eqnarray}

where $*$ symbolizes adjoint operations. For $\epsilon > 0$, $\Phi_{\epsilon,\mathbf{m}}$ is the sum of a real symmetric positive semi-definite matrix $\tilde{\mathbf{B}}^*(\mathbf{Sm}) \; \tilde{\mathbf{B}}(\mathbf{Sm})$ and a real symmetric positive definite matrix, ${\mathbf{D}^*} {\mathbf{D}}$, so $\Phi_{\epsilon,\mathbf{m}}$ is a real symmetric positive definite matrix. Therefore, there exists a unique minimizer of $\Phi_{\epsilon,\mathbf{m}}$, referred to as \emph{the} optimal extended perturbation, denoted by $\mathbf{\tilde{p}}_{\epsilon}^{opt}$. Its expression is given by the normal equation,

\begin{eqnarray}
    \label{eqn:pert.opt}
    \tilde{\mathbf{p}}_{\epsilon}^{opt}(\mathbf{\mathbf{S}m}) &=& \tilde{\mathbf{B}}_{\epsilon, \mathbf{D}}^{\dagger}(\mathbf{S}\mathbf{m}) \left [ \mathbf{d}^{obs} - \mathbf{f}(\mathbf{S}\mathbf{m}) \right ],
\end{eqnarray}

where $\tilde{\mathbf{B}}_{\epsilon, \mathbf{D}}^{\dagger}$ is the pseudo-inverse of $\tilde{\mathbf{B}}$ in equation~\ref{eqn:vp.obj}:

\begin{eqnarray}
    \label{eqn:vp.pseudo.inv}
    \tilde{\mathbf{B}}_{\epsilon, \mathbf{D}}^{\dagger} (\mathbf{S}\mathbf{m})&=& \left [ \tilde{\mathbf{B}}^*(\mathbf{S}\mathbf{m}) \; \tilde{\mathbf{B}}(\mathbf{S}\mathbf{m}) + \epsilon^2 {\mathbf{D}^*} {\mathbf{D}} \right ]^{-1} \tilde{\mathbf{B}}^*(\mathbf{S}\mathbf{m}). 
\end{eqnarray}

The minimization of equation~\ref{eqn:vp.obj} corresponds to the variable projection step of FWIME and is conducted with a linear conjugate-gradient scheme \cite[]{golub1973differentiation,aster2018parameter}. The number of iterations needed for convergence is problem-dependent, but in practice, we observe that approximately 50 iterations are usually necessary, for both 2D and 3D applications. The estimation of $\mathbf{\tilde{p}}_{\epsilon}^{opt}$ is equivalent to conducting an extended least-square reverse-time migration \cite[]{leader2015separation}, and it is the main computational bottleneck of our algorithm. Finally, the nonlinear objective function defined in equation~\ref{eqn:fwime.obj} is minimized with L-BFGS. 

\subsection{Preventing ``cycle-skipping" with extended modeling}
The main idea driving FWIME is that in equation~\ref{eqn:fwime.obj}, we modify the conventional FWI objective function by adding a data-correcting term $\tilde{\mathbf{B}} (\mathbf{Sm}) \mathbf{\tilde{p}}_{\epsilon}^{opt}(\mathbf{Sm})$ to ensure phase alignment between modeled and observed data. This terms corresponds to the extended Born modeling (also referred to as extended demigration) of the extended perturbation $\mathbf{\tilde{p}}_{\epsilon}^{opt}$. It has been shown that for an appropriate extension and extended Born modeling operator $\tilde{\mathbf{B}}$, there exists such a $\mathbf{\tilde{p}}_{\epsilon}^{opt}$ that satisfies

\begin{eqnarray}
    \label{eqn:data_fitting}
    \tilde{\mathbf{B}}(\mathbf{S}\mathbf{m})\tilde{\mathbf{p}}_{\epsilon}^{opt}(\mathbf{S}\mathbf{m}) \approx \mathbf{d}^{obs} - \mathbf{f}(\mathbf{S}\mathbf{m}).
\end{eqnarray}

For extremely inaccurate velocity models $\mathbf{m}$, this property is only valid if $\mathbf{\tilde{p}}_{\epsilon}^{opt}$ is extended \cite[]{symes2008migration}. With this powerful tool, we are able to construct a new forward modeling operator such that $\mathbf{f}(\mathbf{{S}\mathbf{m}}) + \tilde{\mathbf{B}}(\mathbf{S}\mathbf{m})\tilde{\mathbf{p}}_{\epsilon}^{opt}(\mathbf{S}\mathbf{m}) \approx \mathbf{d}^{obs}$ at any given precision, thereby mitigating the cycle-skipping phenomenon. For a fixed velocity model $\mathbf{m}$, the level of data-matching is controlled by the value of $\epsilon$: lower $\epsilon$-values are such that equation~\ref{eqn:data_fitting} is satisfied with more accuracy, and vice-versa. However, preventing cycle-skipping (a data-space phenomenon) is not sufficient to guarantee that the objective function defined in equation~\ref{eqn:fwime.obj} is free of local minima. Therefore, an annihilating component is added to our objective function (second term on the right side of equation~\ref{eqn:fwime.obj}), whose goal is to gradually reduce the contributions of the correcting term with iterations. This task can be achieved by penalizing the energy within $\tilde{\mathbf{p}}_{\epsilon}^{opt}$. Eventually, when $\mathbf{f}(\mathbf{S} \mathbf{m})$ is able to accurately predict $\mathbf{d}^{obs}$ without the help of the data-correcting term (and with $\mathbf{S}=\mathbf{I}_d$), the inversion has converged to the global solution. In our complementary paper, we illustrate how this annihilating term is similar to a wave-equation migration velocity analysis (WEMVA) objective function. As a result, FWIME can be interpreted as a combination of a modified FWI (first component of equation~\ref{eqn:fwime.obj}) and WEMVA (second component of equation~\ref{eqn:fwime.obj}) into a robust workflow that is more powerful than applying each technique separately or sequentially. In this paper, however, we do not prove mathematically that our new formulation is free of local minima but we show numerical evidence. 

\subsection{Optimization}
Equation~\ref{eqn:fwime.obj} is minimized using a local optimization method and its gradient is composed of two terms,

\begin{eqnarray}
    \label{eqn:fwime.gradient.split}
    \nabla_{\mathbf{m}} \Phi_{\epsilon}(\mathbf{m}) &=& \nabla_{\mathbf{m}} \Phi_{\epsilon}^{Born}(\mathbf{m}) + \nabla_{\mathbf{m}} \Phi_{\epsilon}^{Tomo}(\mathbf{m}),
\end{eqnarray}

with
\begin{eqnarray}
    \label{eqn:fwime.gradient.born}
    \nabla_{\mathbf{m}} \Phi_{\epsilon}^{Born}(\mathbf{m}) &=& \mathbf{S}^* \mathbf{B}^*(\mathbf{Sm}) \; \mathbf{r}^{\epsilon}_d (\mathbf{Sm}), \\
    \label{eqn:fwime.gradient.tomo}
    \nabla_{\mathbf{m}} \Phi_{\epsilon}^{Tomo}(\mathbf{m}) &=& \mathbf{S}^* \mathbf{T}^*(\mathbf{Sm},\mathbf{\tilde{p}}_{\epsilon}^{opt}) \; \mathbf{r}^{\epsilon}_d (\mathbf{Sm}),
\end{eqnarray}

where $\mathbf{r}^{\epsilon}_d (\mathbf{Sm}) = \mathbf{f}(\mathbf{Sm}) + \tilde{\mathbf{B}} (\mathbf{Sm}) \mathbf{\tilde{p}}_{\epsilon}^{opt}(\mathbf{Sm}) - \mathbf{d}^{obs}$ is the adjoint source, $\mathbf{B}^*$ is the adjoint of the non-extended Born modeling operator, $\mathbf{T}^*$ is the adjoint of the data-space tomographic operator \cite[]{barnier2018full}, and $\mathbf{S}^*$ is the adjoint of the spline interpolator, which maps a model from the finite-difference grid onto a coarser spline grid. The Born gradient $\nabla_{\mathbf{m}} \Phi_{\epsilon}^{Born}$ is similar to the conventional FWI gradient (and tends to guide the inversion to local minima at early stages when the initial model is poor). The tomographic component $\nabla_{\mathbf{m}} \Phi_{\epsilon}^{Tomo}$ is responsible for recovering the missing long-wavelength components of the velocity model, thereby mitigating the presence of local minima. 

Our new FWIME formulation is paired with the model-space multi-scale workflow proposed by \cite{barnier2019waveform}. The full data-bandwidth is simultaneously inverted and the resolution of the model updates is controlled by the spatial sampling of the spline grid on which the velocity model $\mathbf{m}$ is parametrized. We begin with a coarse grid to enforce smooth updates and we progressively refine the spline sampling with iterations. The inverted model on a spline grid is taken as an initial guess for the inversion on a finer grid, and the last grid usually coincides with the finite-difference grid. Among many benefits, this multi-scale strategy prevents the inversion scheme from embedding spurious high-resolution artifacts at early stages, such as migration isochrones \cite[]{brossier2015velocity,zhou2015full}, and greatly improves the convergence properties of FWIME.

%% file: Marmousi.tex
The Marmousi2 synthetic model simulates a geologically complex subsurface structure and is well-suited for calibrating velocity-model building algorithms \cite[]{martin2006marmousi2}. We use this benchmark test to show that FWIME can successfully invert reflection-dominated datasets and recover accurate solutions without the need for low-frequency signal or good initial guesses. Compared to the original Marmousi \cite[]{versteeg1990practical}, this model is wider and more representative of current long-offset offshore acquisition geometries. The sediments are placed under a thicker layer of water of approximately 475 m, which reduces the presence of refracted energy recorded in the data. Finally, additional hydrocarbon reservoirs and stratigraphic features have been added to increase the overall geological complexity of the model \cite[]{martin2002marmousi}.   

\subsection{Presentation and challenges}
\begin{figure}[t]
    \centering
    \subfigure[]{\label{fig:Marmousi_init_mod}\includegraphics[width=0.45\columnwidth]{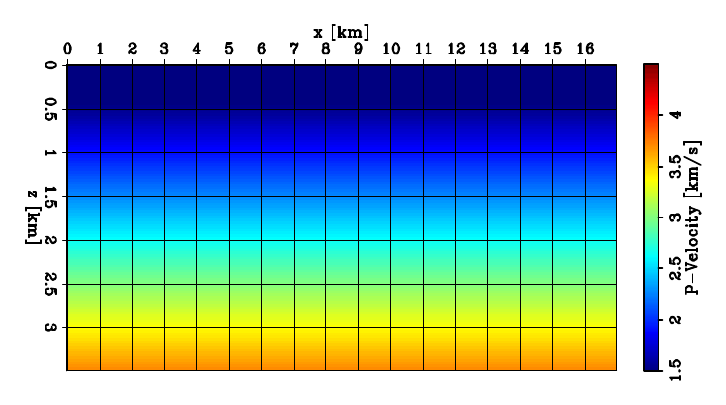}}
    \subfigure[]{\label{fig:Marmousi_fwi_mod}\includegraphics[width=0.45\columnwidth]{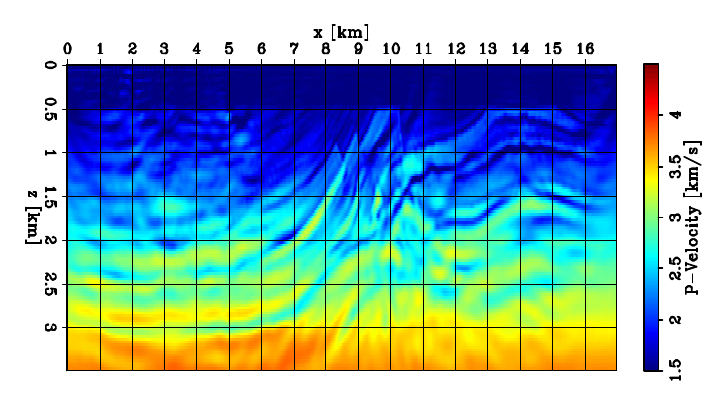}} \\
    \subfigure[]{\label{fig:Marmousi_fwime_mod}\includegraphics[width=0.45\columnwidth]{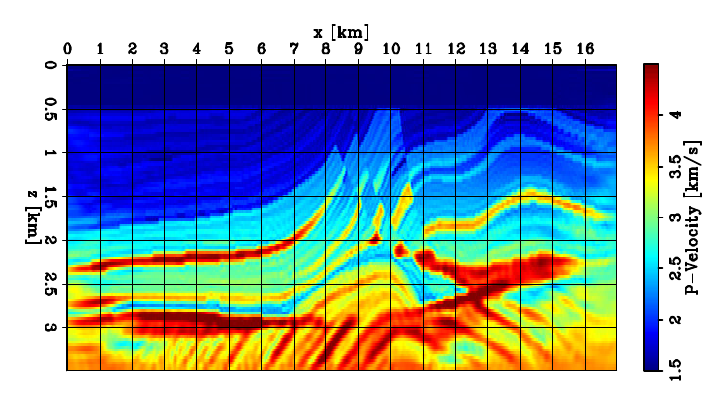}}
    \subfigure[]{\label{fig:Marmousi_true_mod}\includegraphics[width=0.45\columnwidth]{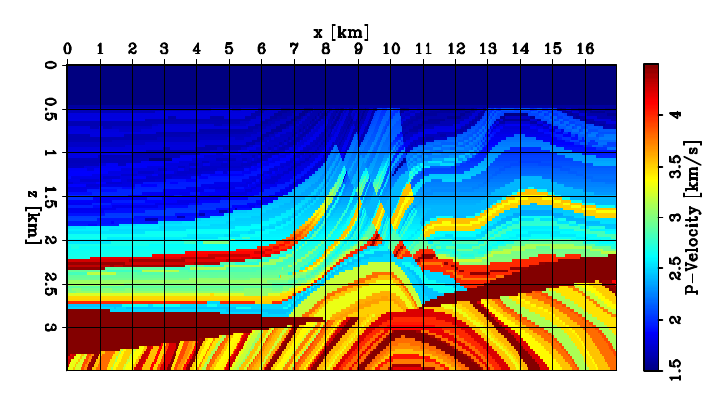}}    
    \caption{2D panels of velocity models. (a) Initial model. (b) Inverted model after conventional data-space multi-scale FWI using five frequency bands. (c) Final FWIME inverted model. (d) True model.}
    \label{fig:Marmousi_mod}
\end{figure}

\begin{figure}[t]
    \centering
    \subfigure[]{\label{fig:Marmousi_mod_1d_km6}\includegraphics[width=0.2\columnwidth]{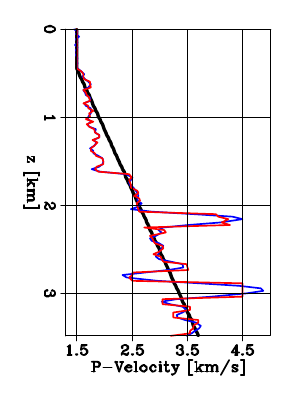}}
    \subfigure[]{\label{fig:Marmousi_mod_1d_km9}\includegraphics[width=0.2\columnwidth]{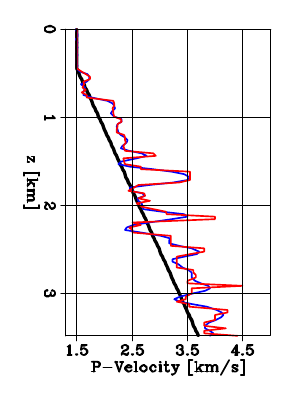}} 
    \subfigure[]{\label{fig:Marmousi_mod_1d_km11}\includegraphics[width=0.2\columnwidth]{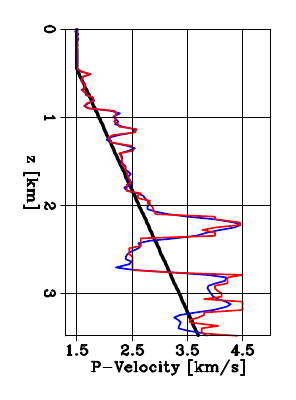}}
    \subfigure[]{\label{fig:Marmousi_mod_1d_km13}\includegraphics[width=0.2\columnwidth]{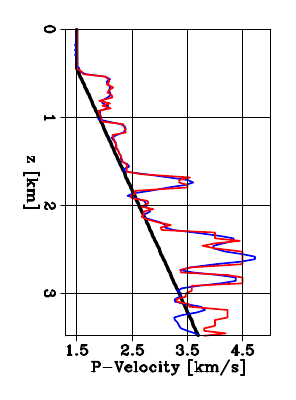}}        
    \caption{Depth velocity profiles extracted at (a) $x=6$ km, (b) $x=9$ km, and (c) $x=11$ km, and (d) $x=13$ km. The black curve represents the initial model, the red curve is the true model, and the blue curve is the final FWIME inverted model.}
    \label{fig:Marmousi_mod_1d}
\end{figure}

\begin{figure}[t]
    \centering
    \subfigure[]{\label{fig:Marmousi_wav_full}\includegraphics[width=0.45\columnwidth]{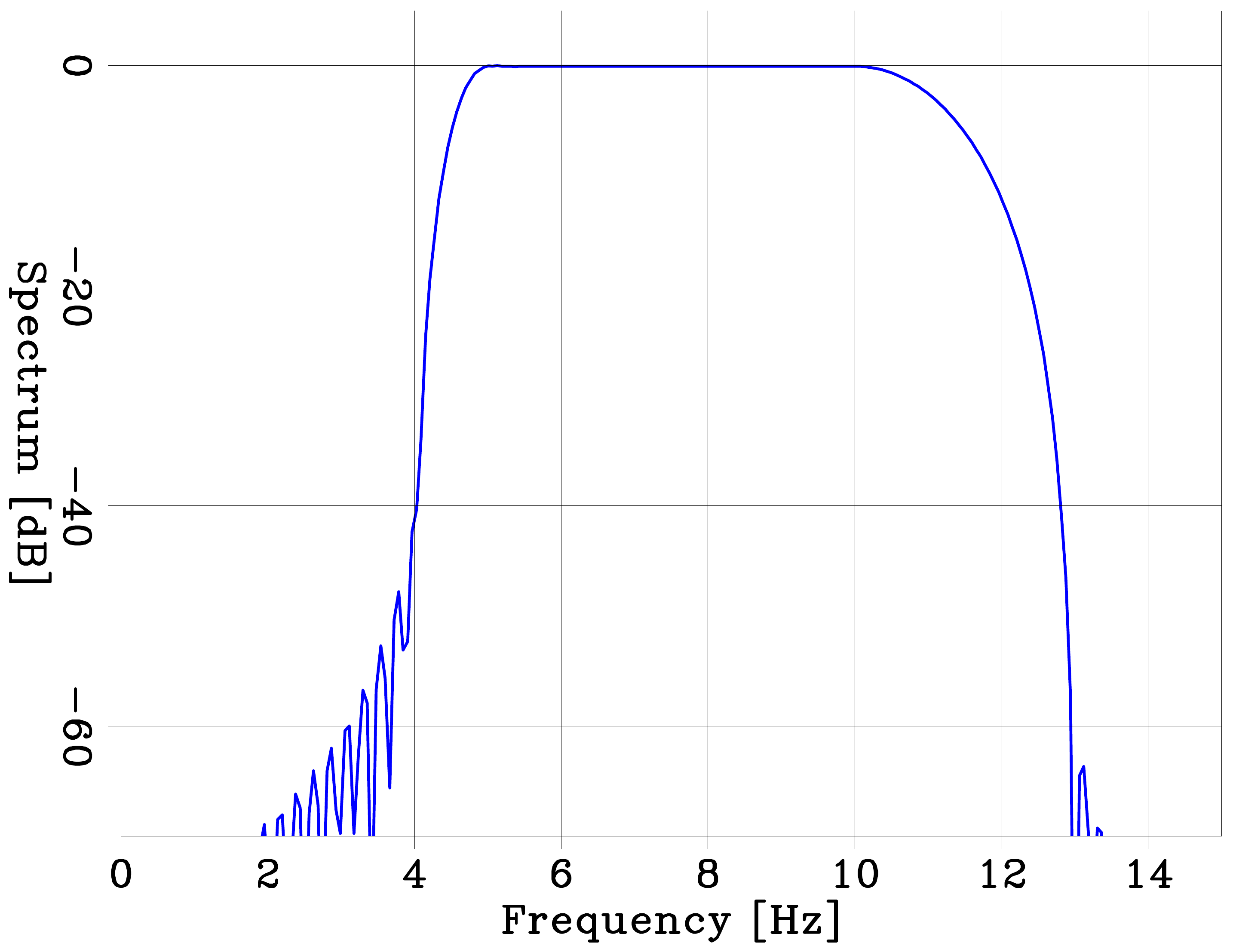}} \hspace{5mm}
    \subfigure[]{\label{fig:Marmousi_wav_multi_scale}\includegraphics[width=0.45\columnwidth]{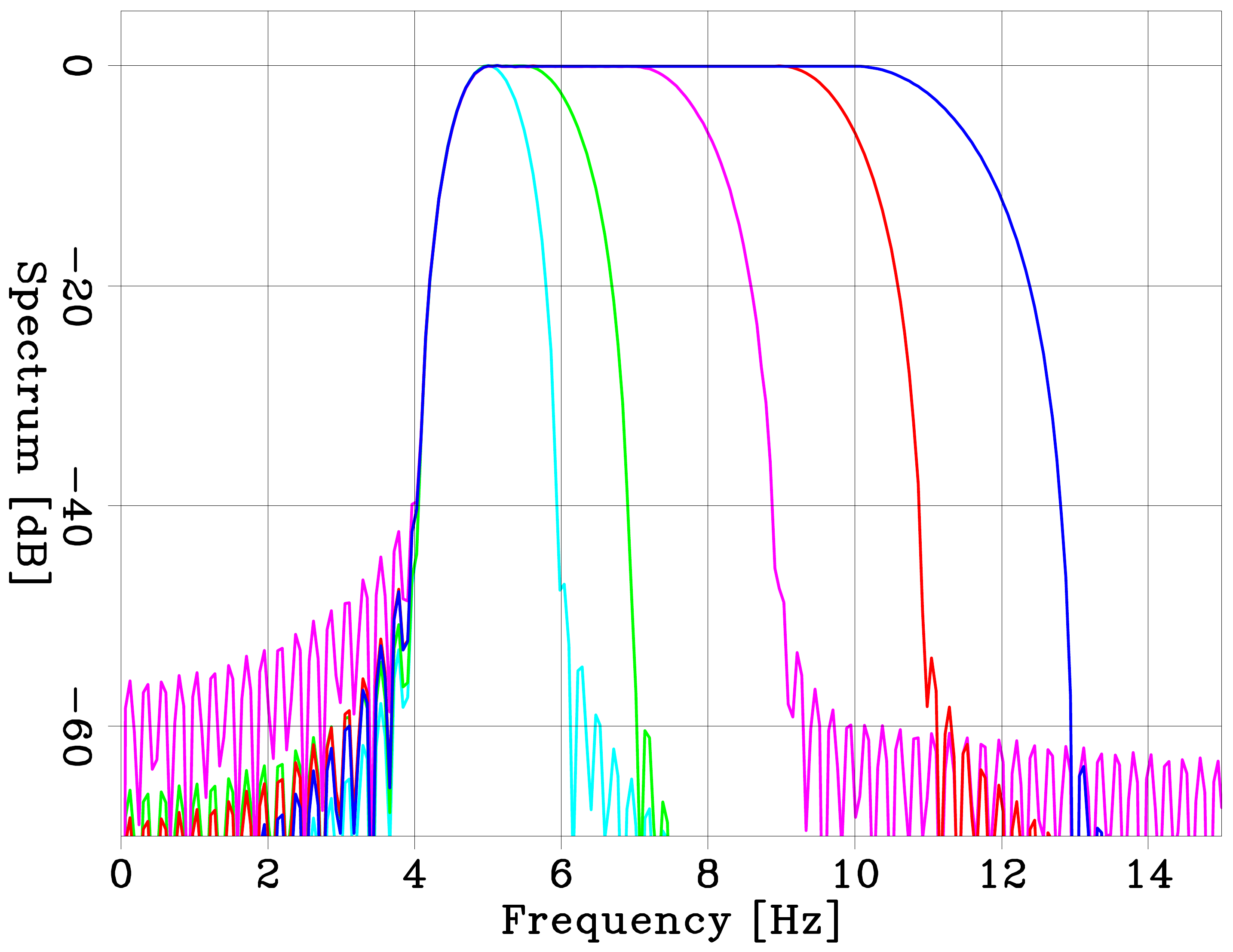}} 
    \caption{Amplitude spectra of the seismic sources used to generate the various datasets for this numerical example. (a) Source used for the FWIME workflow. (b) Sequence of sources used for the data-space multi-scale FWI workflow.}
    \label{fig:Marmousi_wav}
\end{figure}

\begin{figure}[t]
    \centering
    \subfigure[]{\label{fig:Marmousi_data_true_s10}\includegraphics[width=0.30\columnwidth]{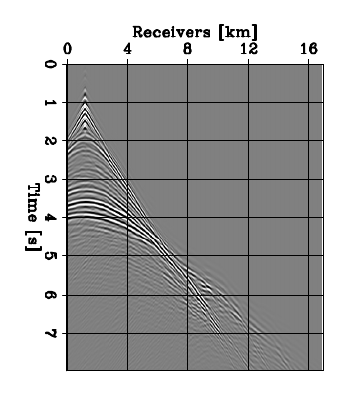}}
    \subfigure[]{\label{fig:Marmousi_data_init_s10}\includegraphics[width=0.30\columnwidth]{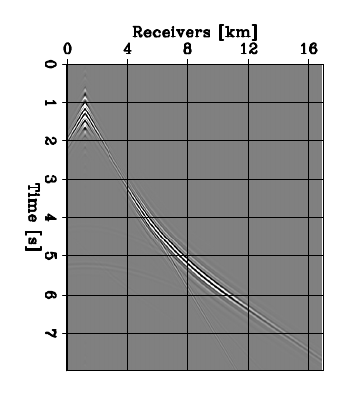}}
    \subfigure[]{\label{fig:Marmousi_data_initDiff_s10}\includegraphics[width=0.30\columnwidth]{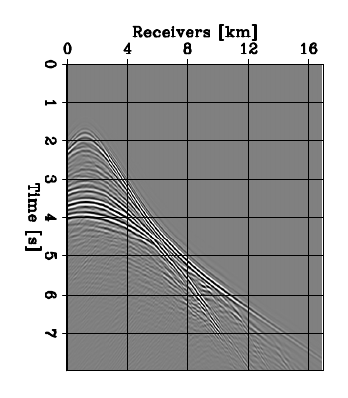}} \\
    \subfigure[]{\label{fig:Marmousi_data_true_s70}\includegraphics[width=0.30\columnwidth]{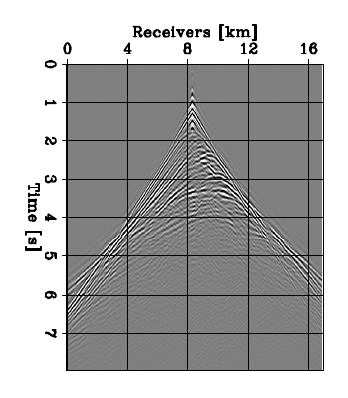}}
    \subfigure[]{\label{fig:Marmousi_data_init_s70}\includegraphics[width=0.30\columnwidth]{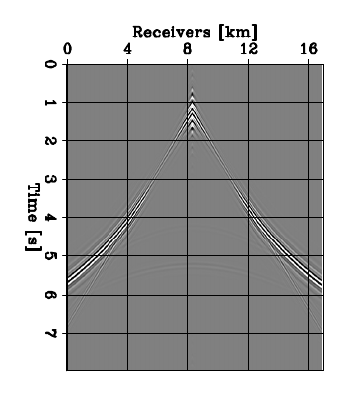}}
    \subfigure[]{\label{fig:Marmousi_data_initDiff_s70}\includegraphics[width=0.30\columnwidth]{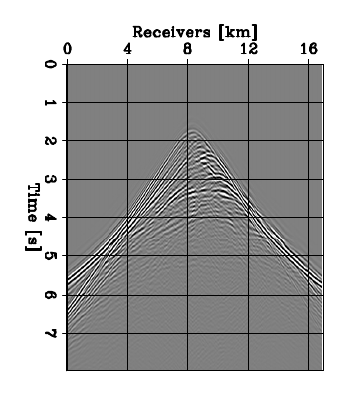}}      
    \caption{Representative shot gathers for sources placed at $x = 1.2$ km (first row) and $x=8.4$ km (second row). Observed data, $\mathbf{d}^{obs}$ (first column), predicted data with the initial model, $\mathbf{f}(\mathbf{m}_{init})$ (second column), and initial data-difference, $\Delta \mathbf{d}(\mathbf{m}_{init})=\mathbf{d}^{obs} - \mathbf{f}(\mathbf{m}_{init})$ (third column). All panels are displayed with the same grayscale.}
    \label{fig:Marmousi_data_obs}
\end{figure}

The true model $\mathbf{m}_{true}$ is 17 km wide and 3.5 km deep (Figure~\ref{fig:Marmousi_true_mod}). The initial model $\mathbf{m}_{init}$ is purposely designed to be inaccurate (Figure~\ref{fig:Marmousi_init_mod}). Below the water layer (the bathymetry is assumed to be known for this problem), the velocity is laterally invariant and linearly increasing with depth. Velocity profiles of the true and initial models extracted at four horizontal positions are displayed in Figure~\ref{fig:Marmousi_mod_1d} (red and and black curves). We place 140 sources every 120 m, and 567 fixed receivers every 30 m. All acquisition devices are located 30 m below the water surface. We generate noise-free pressure data with a source wavelet containing energy restricted to the 4-13 Hz range (Figure~\ref{fig:Marmousi_wav_full}), a uniform finite-difference grid spacing of 30 m, and we use absorbing-boundary conditions in all directions \cite[]{cerjan1985nonreflecting}. The data are modeled and inverted with an acoustic isotropic constant-density two-way wave-equation operator. Figures~\ref{fig:Marmousi_data_obs}a and \ref{fig:Marmousi_data_obs}d show two representative shot gathers for sources placed at $x = 1.2$ km and $x=8.4$ km. Even though some refracted energy is recorded at offsets larger than 7 km, the dataset is strongly dominated by reflected events. Figures~\ref{fig:Marmousi_data_obs}b and \ref{fig:Marmousi_data_obs}e show that the initial prediction $\mathbf{f}(\mathbf{m}_{init})$ fails to generate any reflection. The inaccuracy of $\mathbf{m}_{init}$ coupled with the lack of low-frequency and refracted energy (i.e., diving waves) make this test challenging for conventional velocity-model building techniques. To illustrate this claim, we apply data-space multi-scale FWI using five frequency bands (Figure~\ref{fig:Marmousi_wav_multi_scale}), which fails to retrieve a useful solution, especially in the deeper section of the model (Figure~\ref{fig:Marmousi_fwi_mod}). 

\subsection{FWIME}
We guide the reader through the initial step of FWIME which includes a hyper-parameter-tuning process and the computation of the initial search direction. Then, we present and carefully analyze the inversion results.

\subsubsection{Selection of the extension}
The first step consists in selecting the optimal extension type and the length of the extended axis for $\tilde{\mathbf{p}}_{\epsilon}^{opt}$ such that the initial data-correcting term $\tilde{\mathbf{B}}(\mathbf{m}_{init})\tilde{\mathbf{p}}_{0}^{opt}(\mathbf{m}_{init})$ is able to match the initial data difference $\mathbf{d}^{obs} - \mathbf{f}(\mathbf{m}_{init})$ (i.e., satisfy the condition expressed in equation~\ref{eqn:data_fitting}). This condition ensures that without penalizing defocused energy within $\mathbf{\tilde{p}}_{\epsilon}^{opt}$ (i.e., when $\epsilon=0$), the initial prediction error can be fully explained by the mapping of $\mathbf{\tilde{p}}_{0}^{opt}(\mathbf{m}_{init})$ into the data space. Indeed, failing to select an adequate extension length may result in cycle-skipping, which in turn may lead FWIME to converge to a local minimum. For computational efficiency, we wish to find the optimal extension type that satisfies this condition but with the smallest number of points on the extended axis. For that, we minimize the objective function defined in equation~\ref{eqn:vp.obj} with three different types of extension: using time lags, using horizontal subsurface offsets, and without any extension (we conduct three separate inversions). For all three inversions, we set $\mathbf{m}=\mathbf{m}_{init}$, $\mathbf{S}= \mathbf{I}_d$, and $\epsilon = 0$. 

\begin{figure}[t]
    \centering
    \subfigure[]{\label{fig:Marmousi_vp_obj_eps0}\includegraphics[width=0.45\columnwidth]{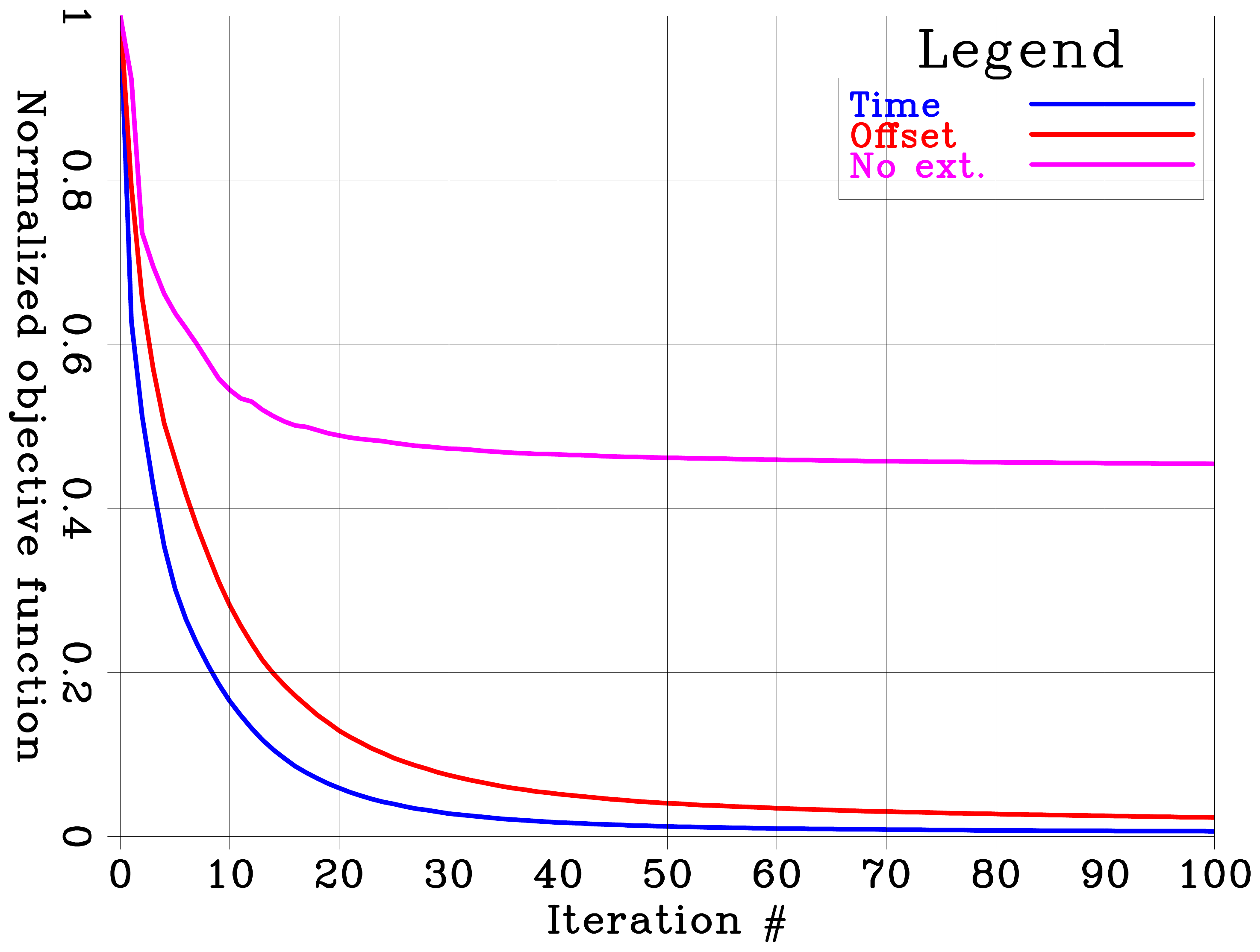}} \hspace{5mm}
    \subfigure[]{\label{fig:Marmousi_vp_obj_eps}\includegraphics[width=0.45\columnwidth]{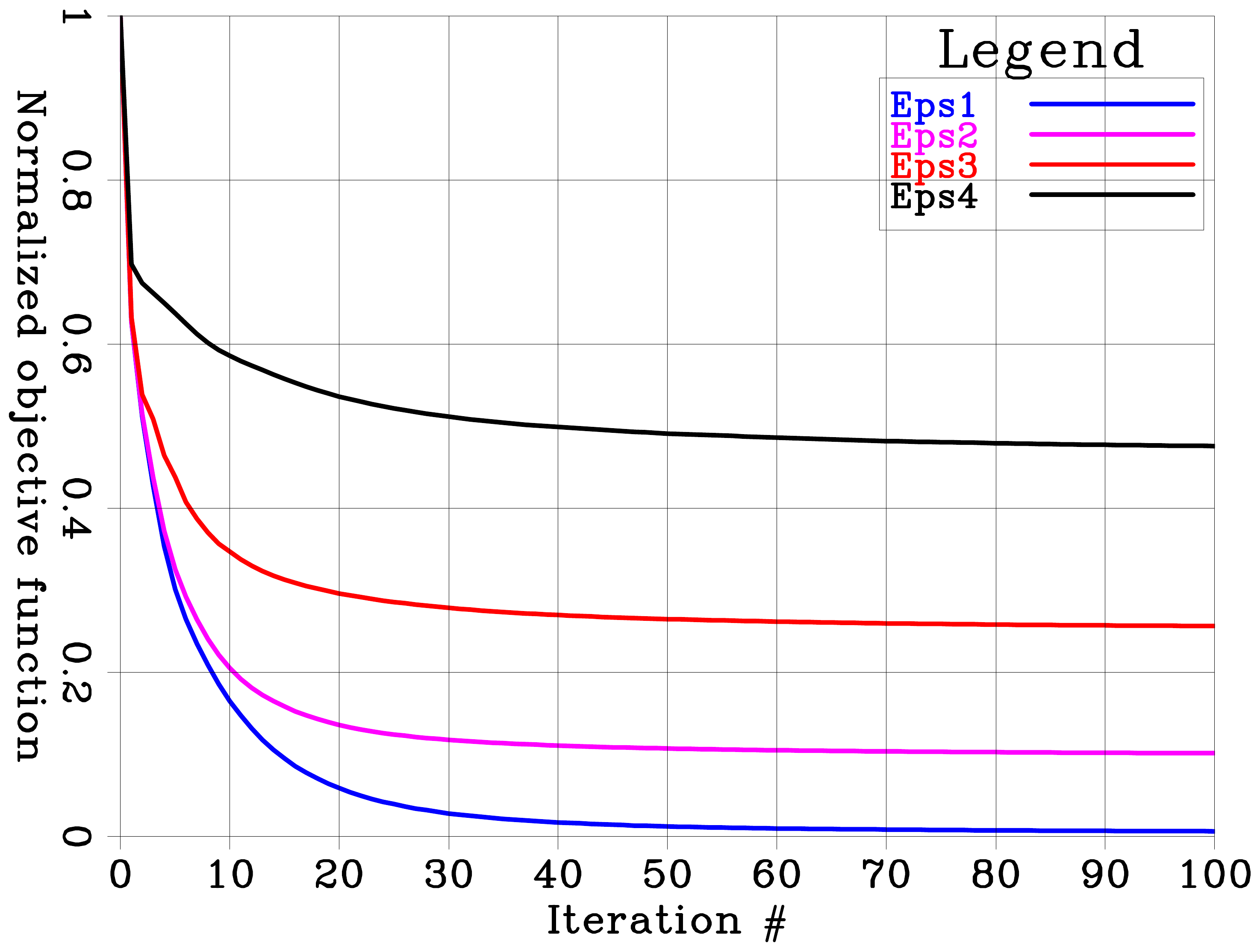}} \hspace{5mm}
    \caption{Normalized objective functions corresponding to the minimization of equation~\ref{eqn:vp.obj} (initial variable projection step) with different extensions and $\epsilon$-values. (a) Time-lag extension (blue curve), horizontal subsurface-offset extension (red curve), and non-extended (pink curve).  For (a), we set $\epsilon=0$. (b) Time-lag extension with different $\epsilon$-values: $\epsilon_1=0$ (blue curve), $\epsilon_2=1.5 \times 10^{-5}$ (red curve), $\epsilon_3=5.0 \times 10^{-5}$ (pink curve), and $\epsilon_4=5.0 \times 10^{-4}$ (black curve). In both panels, all curves are normalized by the same value. The blue curves in panels (a) and (b) are identical. }
    \label{fig:Marmousi_vp_obj}
\end{figure}

We test with an extended axis of 101 points for both time lags and subsurface offsets with a sampling of $\Delta \tau=16$ ms and $\Delta h_x=30$ m, respectively. Hence, $\tau$ ranges from -0.8 s to 0.8 s, whereas $h_x$ ranges from -1.5 km to 1.5 km. Figure~\ref{fig:Marmousi_vp_obj_eps0} shows the normalized convergence curves for the minimization of objective function in equation~\ref{eqn:vp.obj} (the initial variable projection step) using time lags (blue curve), horizontal subsurface offsets (red curve), and with no extension (pink curve). As expected, the non-extended inversion fails to reduce the misfit to zero. In addition, the time-lag extension seems to perform better than subsurface offsets as it manages to decrease the objective function value by approximately $99.4 \%$ after 60 iterations of linear conjugate gradient, and by $99.9 \%$ after 100 iterations (compared to $96\%$ and $97\%$ for subsurface offsets), thereby satisfying the condition expressed in equation~\ref{eqn:data_fitting} more efficiently. Indeed, a more accurate matching (up to numerical precision) would require more iterations. For this numerical example, we choose a time-lag extension with a length of 101 points (a very conservative number) sampled at 16 ms, and we limit the number of linear conjugate gradient iterations to 60 for the variable projection step. In practice, to further reduce the computational cost for 3D field applications, a smaller number of points on the extended axis could have been chosen. 

\subsubsection{Selection of the trade-off parameter $\epsilon$}
\begin{figure}[t]
    \centering
    \subfigure[]{\label{fig:Marmousi_vp_cig_dso0}\includegraphics[width=0.22\columnwidth]{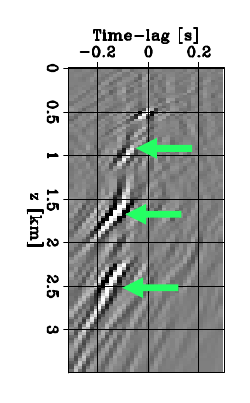}}
    \subfigure[]{\label{fig:Marmousi_vp_cig_dso1}\includegraphics[width=0.22\columnwidth]{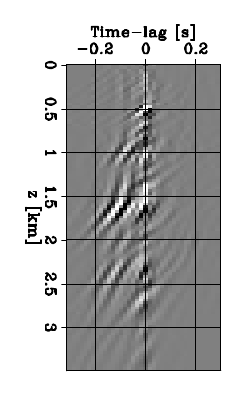}}
    \subfigure[]{\label{fig:Marmousi_vp_cig_dso2a}\includegraphics[width=0.22\columnwidth]{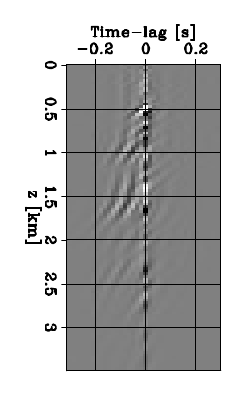}}    
    \subfigure[]{\label{fig:Marmousi_vp_cig_dso3b}\includegraphics[width=0.22\columnwidth]{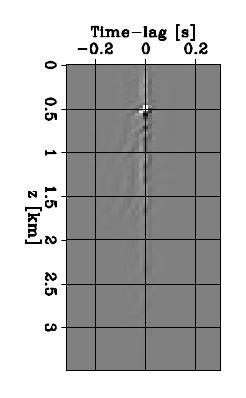}}        
    \caption{Time-lag common image gathers (TLCIG) extracted at $x = 14$ km from $\mathbf{\tilde{p}}_{\epsilon}^{opt}(\mathbf{m}_{init})$ computed with four different $\epsilon$-values. (a) $\epsilon = 0$, (b) $\epsilon=1.5 \times 10^{-5}$, (c) $\epsilon=5.0 \times 10^{-5}$, and (d) $\epsilon=5.0 \times 10^{-4}$. All panels are plotted for $\tau \in [-0.3 \; \rm{s}, \; 0.3 \; \rm{s}]$ for display purposes. However, all the computations are conducted for the full time-lag range with $\tau \in [-0.8 \; \rm{s}, \; 0.8 \; \rm{s}]$. All panels are displayed with the same grayscale.}
    \label{fig:Marmousi_vp_cig}
\end{figure}

\begin{figure}[t]
    \centering
    \subfigure[]{\label{fig:Marmousi_data_vp_dso0_s10}\includegraphics[width=0.22\columnwidth]{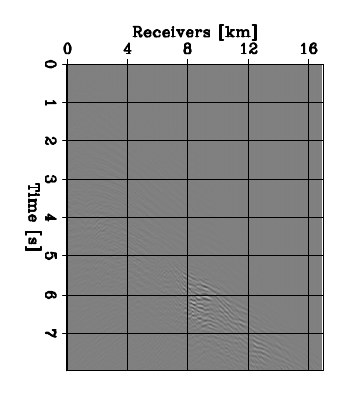}}
    \subfigure[]{\label{fig:Marmousi_data_vp_dso1_s10}\includegraphics[width=0.22\columnwidth]{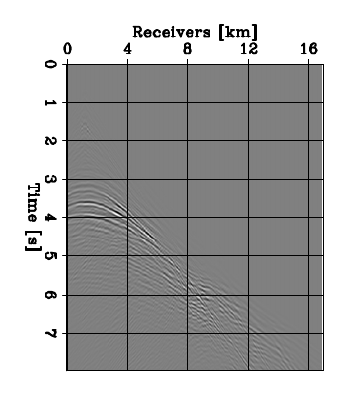}}    
    \subfigure[]{\label{fig:Marmousi_data_vp_dso2a_s10}\includegraphics[width=0.22\columnwidth]{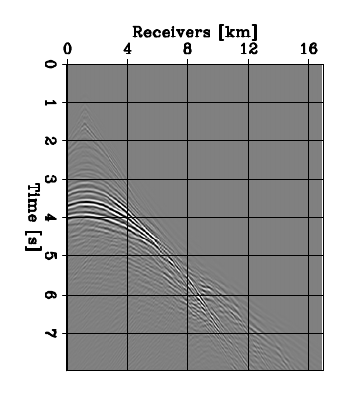}}            
    \subfigure[]{\label{fig:Marmousi_data_vp_dso3b_s10}\includegraphics[width=0.22\columnwidth]{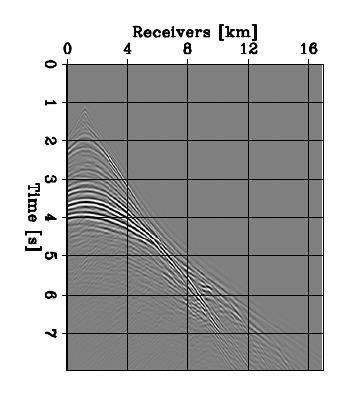}}    
    \caption{Representative shot gathers generated by a source located at $x = 1.2$ km displaying the FWIME adjoint source, $\mathbf{r}^{\epsilon}_d (\mathbf{m}_{init}) =\tilde{\mathbf{B}}(\mathbf{m}_{init})\tilde{\mathbf{p}}_{0}^{opt}(\mathbf{m}_{init})- \left ( \mathbf{d}^{obs} - \mathbf{f}(\mathbf{m}_{init}) \right )$, computed for four different $\epsilon$-values ($\mathbf{m}_{init}$ is fixed). (a) $\epsilon = 0$, (b) $\epsilon=1.5 \times 10^{-5}$, (c) $\epsilon=5.0 \times 10^{-5}$, and (d) $\epsilon=5.0 \times 10^{-4}$. All panels are displayed with the same grayscale.}
    \label{fig:Marmousi_data_vp_s10}
\end{figure}

The second step consists in selecting the optimal $\epsilon$-value. This quantity serves as a trade-off parameter that controls the level of data-fitting by penalizing the presence of energy within $\mathbf{\tilde{p}}_{\epsilon}^{opt}$ during the minimization of the objective functions defined in equations~\ref{eqn:fwime.obj} and \ref{eqn:vp.obj}. To illustrate its effect, we compute $\mathbf{\tilde{p}}_{\epsilon}^{opt}$ (using the initial velocity model $\mathbf{m}_{init}$) for four $\epsilon$-values by minimizing the objective function in equation~\ref{eqn:vp.obj}. The corresponding convergence curves are shown in Figure~\ref{fig:Marmousi_vp_obj_eps}. As the $\epsilon$-value increases, the ability of the data-correcting term $\tilde{\mathbf{B}} (\mathbf{m}_{init}) \mathbf{\tilde{p}}_{\epsilon}^{opt}(\mathbf{m}_{init})$ to predict the initial data misfit $\mathbf{d}^{obs}-\mathbf{f}(\mathbf{m}_{init})$ is reduced. Figure~\ref{fig:Marmousi_vp_cig} shows a time-lag common image gather (TLCIG) extracted at $x=14$ km from the four inverted $\mathbf{\tilde{p}}_{\epsilon}^{opt}$ (corresponding to the four convergence curves shown in Figure~\ref{fig:Marmousi_vp_obj_eps}). Figure~\ref{fig:Marmousi_data_vp_s10} represents the difference between the corresponding data-correcting term and the initial FWIME data-residual (the adjoint source): $\mathbf{r}^{\epsilon}_d (\mathbf{m}_{init}) = \tilde{\mathbf{B}}(\mathbf{m}_{init})\tilde{\mathbf{p}}_{\epsilon}^{opt}(\mathbf{m}_{init})- \left ( \mathbf{d}^{obs} - \mathbf{f}(\mathbf{m}_{init}) \right )$. 

For $\epsilon = 0$, no constraint is applied to $\tilde{\mathbf{p}}_{\epsilon}^{opt}$, and therefore a substantial amount of energy is mapped away from the physical plane, as shown by the green arrows in Figure~\ref{fig:Marmousi_vp_cig_dso0}. Consequently, the data-correcting term is able to match the initial data residual with high accuracy, and almost no coherent signal is present within the adjoint source (Figure~\ref{fig:Marmousi_data_vp_dso0_s10}). Increasing the $\epsilon$-value penalizes defocused events within $\mathbf{\tilde{p}}_{\epsilon}^{opt}$ and reduces the amount of energy spread away from the physical plane (Figures~\ref{fig:Marmousi_vp_cig}b and c). The data-correcting term is then unable to accurately match the initial data-misfit (Figures~\ref{fig:Marmousi_data_vp_s10}b and c). Eventually, for an extremely high $\epsilon$-value, both $\mathbf{\tilde{p}}_{\epsilon}^{opt}$ and the data-correcting term vanish (Figures~\ref{fig:Marmousi_vp_cig_dso3b} and \ref{fig:Marmousi_data_vp_dso3b_s10}), and $\mathbf{r}^{\epsilon}_d (\mathbf{m}_{init}) \approx \mathbf{f}(\mathbf{m}_{init}) - \mathbf{d}^{obs}$. Mathematically, setting $\epsilon$ to a high value corresponds to conducting a non-extended FWIME, or equivalently, conventional FWI.

During the variable projection step, the background velocity model $\mathbf{m}$ is fixed, and thus the $\epsilon$-value only affects the amplitude of the events within $\mathbf{\tilde{p}}_{\epsilon}^{opt}$ rather than the kinematics: the reflectors' positions are not affected by $\epsilon$. In Figure~\ref{fig:Marmousi_vp_cig_dso0} (corresponding to $\epsilon=0$), there are three distinct events located in the extended space at negative time-lag values (green arrows), which provide valuable information on velocity errors present in the initial model. In this particular case, they indicate that the initial velocity is too low \cite[]{barnier2018full}. 

On one hand, selecting a too small $\epsilon$-value is sub-optimal because the amplitude of the corresponding adjoint source $\mathbf{r}^{\epsilon}_d (\mathbf{m}_{init})$ (employed in the gradient computation) would not contain any useful information, as its amplitude would be numerically close to zero (Figure~\ref{fig:Marmousi_data_vp_dso0_s10}). From a physical point of view, a small $\epsilon$-value would not penalize defocused energy within $\mathbf{\tilde{p}}_{\epsilon}^{opt}$. On the other hand, a too-high $\epsilon$-value would force the amplitude of the events in the extended space to vanish, and the information they carry about velocity errors would be lost (Figure~\ref{fig:Marmousi_vp_cig_dso3b}). In addition, the data-correcting term would not be able to match the initial data residual Figures~\ref{fig:Marmousi_data_vp_dso3b_s10}, resulting in cycle-skipping. For this numerical example, we observe that for $\epsilon \in [1.5 \times 10^{-5}, \; 5.0 \times 10^{-5} ]$, FWIME converges to similar accurate solutions. In this paper, we propose to select the $\epsilon$-value by examining a subset of the TLCIGs extracted from the initial $\mathbf{\tilde{p}}_{\epsilon}^{opt}$ computed with a few $\epsilon$-values. We acknowledge the need for a more automatic approach for the selection of $\epsilon$, but we leave such investigation for future work. Fortunately, we observe that our results are relatively insensitive to the choice of $\epsilon$ as long as the proper order of magnitude is determined. 

\subsubsection{Initial search direction}
\begin{figure}[t]
    \centering
    \subfigure[]{\label{fig:Marmousi_born_grad_dso2}\includegraphics[width=0.45\columnwidth]{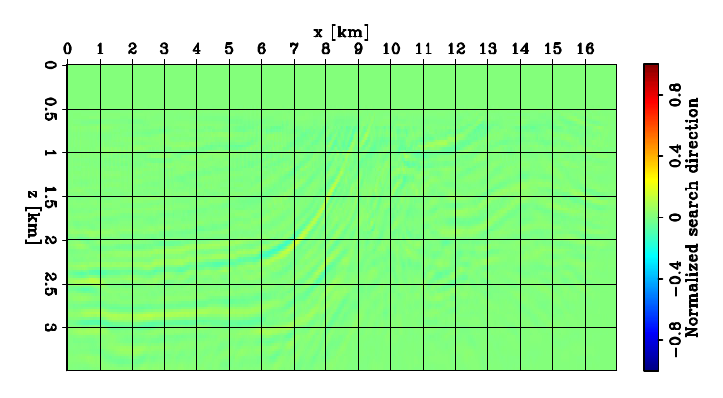}}
    \subfigure[]{\label{fig:Marmousi_born_grad_dso2}\includegraphics[width=0.45\columnwidth]{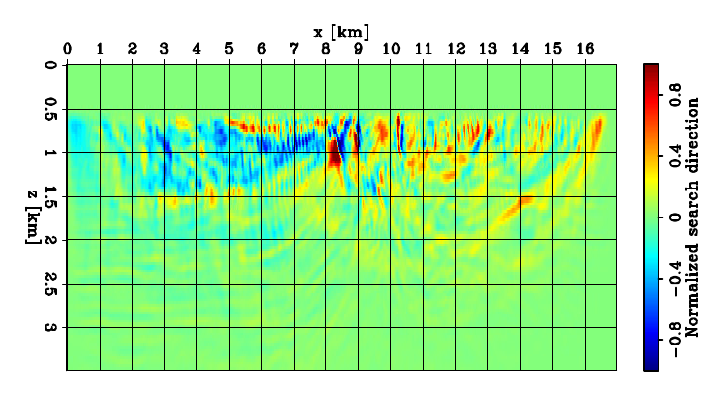}} \\
    \subfigure[]{\label{fig:Marmousi_born_grad_dso2}\includegraphics[width=0.45\columnwidth]{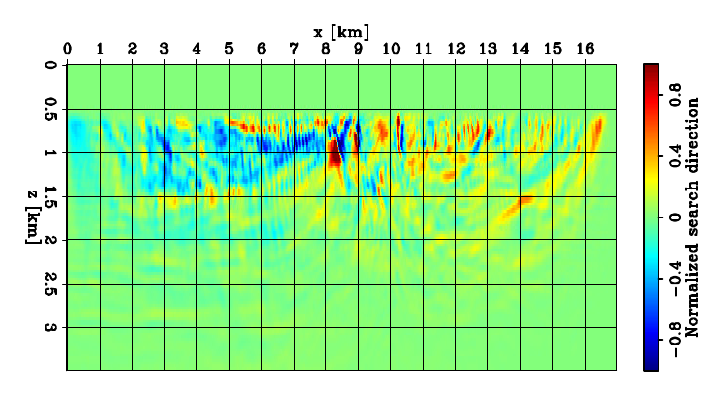}} 
    \subfigure[]{\label{fig:Marmousi_grad_true}\includegraphics[width=0.45\columnwidth]{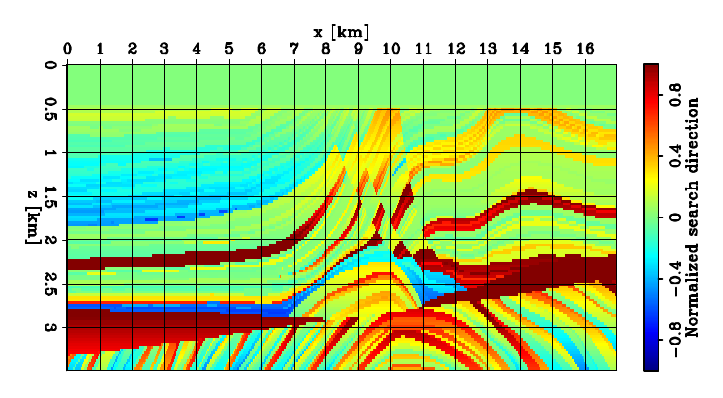}}    
    \caption{Initial FWIME search directions on the finite-difference grid (before spline re-parametrization), computed for $\epsilon= 1.5 \times 10^{-5}$. (a) Born component, (b) tomographic component, (c) total search direction, and (d) true search direction. Panels (a)-(c) are normalized with the same value.}
    \label{fig:Marmousi_vp_grad_split}
\end{figure}

\begin{figure}[t]
    \centering
    \subfigure[]{\label{fig:Marmousi_born_grad_dso2_spline}\includegraphics[width=0.45\columnwidth]{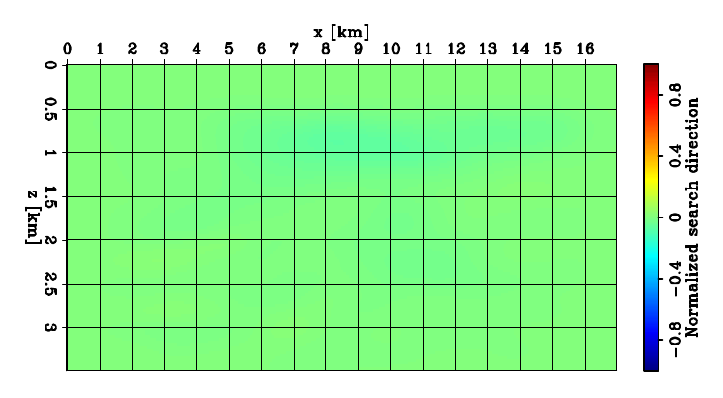}}
    \subfigure[]{\label{fig:Marmousi_born_grad_dso2_spline}\includegraphics[width=0.45\columnwidth]{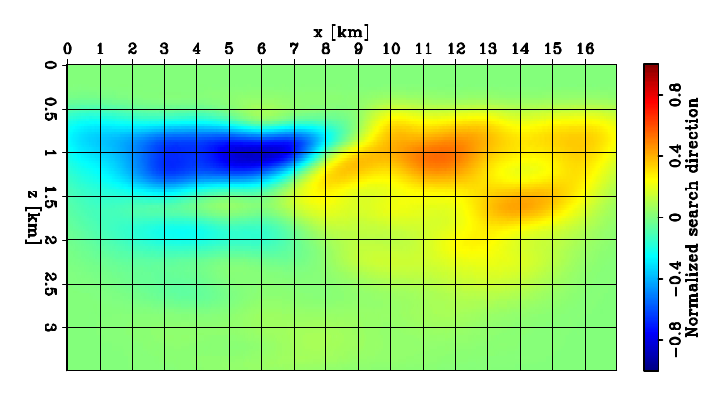}} \\
    \subfigure[]{\label{fig:Marmousi_born_grad_dso2_spline}\includegraphics[width=0.45\columnwidth]{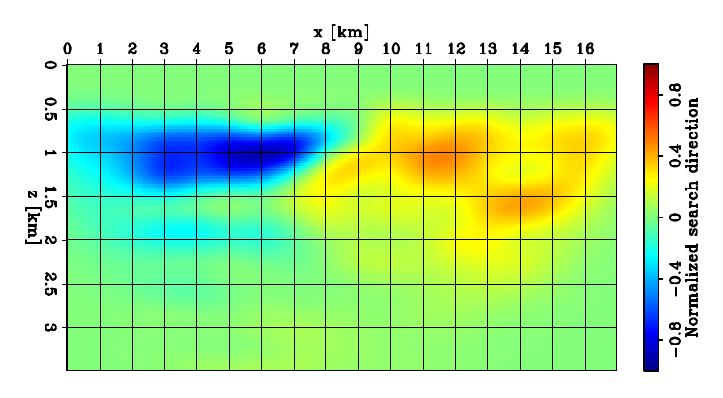}} 
    \subfigure[]{\label{fig:Marmousi_grad_true_spline}\includegraphics[width=0.45\columnwidth]{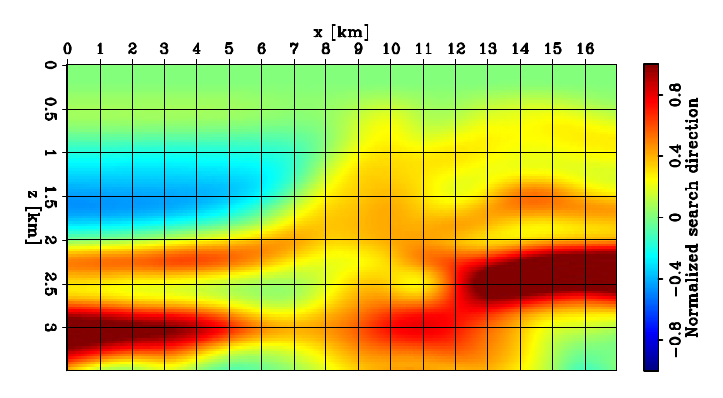}}    
    \caption{Initial FWIME search directions after applying $\mathbf{S}_0 \mathbf{S}_0^*$ to the panels in Figure~\ref{fig:Marmousi_vp_grad_split}, computed for $\epsilon= 1.5 \times 10^{-5}$. (a) Born component, (b) tomographic component, (c) total search direction, and (d) true search direction. Panels (a)-(c) are normalized with the same value.}
    \label{fig:Marmousi_vp_grad_spline_split}
\end{figure}

Figure~\ref{fig:Marmousi_vp_grad_split} shows the Born, tomographic, and total FWIME initial search directions on the finite-difference grid computed with $\epsilon = 1.5 \times 10^{-5}$ (equations~\ref{eqn:fwime.gradient.split}-\ref{eqn:fwime.gradient.tomo}). The total search direction is mainly guided by the tomographic component and seems to accurately capture some of the low-wavenumber features present in the ideal update in the shallow region (by comparing Figures~\ref{fig:Marmousi_vp_grad_split}c and \ref{fig:Marmousi_vp_grad_split}d). However, in panel~\ref{fig:Marmousi_vp_grad_split}c, the update direction is overwhelmed by spurious high-wavenumber artifacts. To mitigate this effect, we use an initial coarse spline grid $\mathbf{S}_0$ with a sampling of 0.2 km and 1 km in the vertical and horizontal directions, respectively. Figure~\ref{fig:Marmousi_vp_grad_spline_split} shows the analogous panels from Figure~\ref{fig:Marmousi_vp_grad_split} after their mapping onto the first spline grid (i.e., after applying operator $\mathbf{S}_0\mathbf{S}_0^*$). As expected, the total FWIME search direction is improved.

\subsubsection{Inversion results}
For the FWIME scheme, we use a sequence of four spline grids and we keep the same $\epsilon$-value throughout the entire process ($\epsilon = 1.5 \times 10^{-5}$). Each spline grid refinement is automatically triggered when the stepper is unable to find an appropriate step length for that particular grid. The spacing in the second and third grids are obtained by halving the spacing from the previous ones. The final spline grid coincides with the finite-difference grid ($\Delta z = \Delta x = 30$ m). 

\begin{table}[h!]
\centering
\begin{tabular}{ |c|c|c|c|  } 
\hline
\textbf{Grid number} & $\Delta z$ [km] & $\Delta x$ [km]\\
\hline
 0 & 0.2 & 0.9 \\
 1 & 0.1 & 0.45 \\
 2 & 0.05 & 0.22 \\
 3 & 0.03 & 0.03 \\
\hline
\end{tabular}
\caption{Parameters of the spline grid sequence used for the model-space multi-scale FWIME scheme. Spline 3 coincides with the finite-difference grid.}
\label{table:Marmousi_spline}
\end{table}

\begin{figure}[t]
    \centering
    \subfigure[]{\label{fig:Marmousi_fwime_init_mod}\includegraphics[width=0.45\columnwidth]{Fig/Marmousi/Marmoush-init-mod.pdf}}
    \subfigure[]{\label{fig:Marmousi_fwime_s1_mod}\includegraphics[width=0.45\columnwidth]{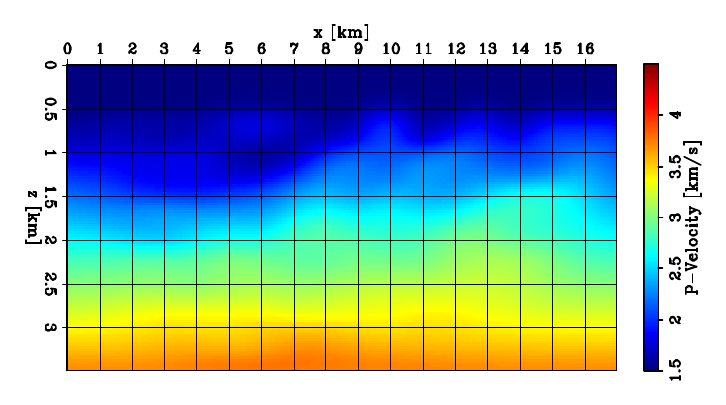}} \\
    \subfigure[]{\label{fig:Marmousi_fwime_s2_mod}\includegraphics[width=0.45\columnwidth]{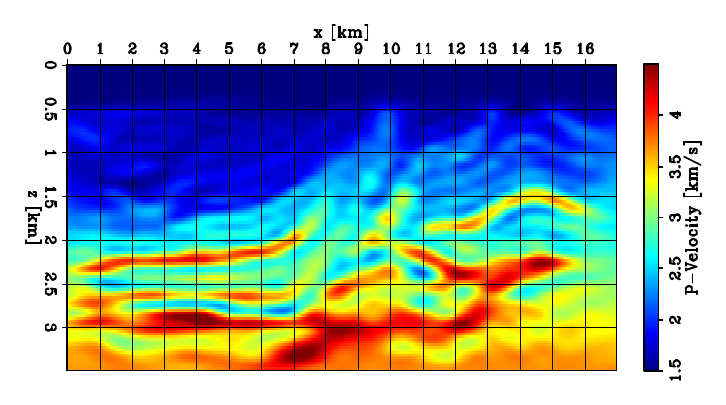}}     
    \subfigure[]{\label{fig:Marmousi_fwime_s3_mod}\includegraphics[width=0.45\columnwidth]{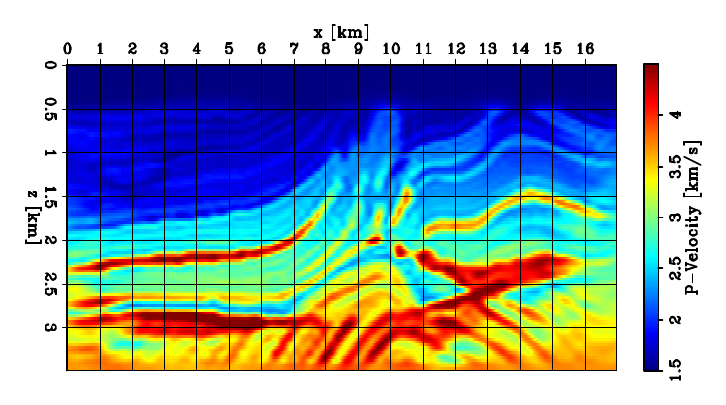}} \\
    \subfigure[]{\label{fig:Marmousi_fwime_noSpline_mod}\includegraphics[width=0.45\columnwidth]{Fig/Marmousi/Marmoush-fwime-noSpline-e3-mod.pdf}} 
    \subfigure[]{\label{fig:Marmousi_fwime_true_mod}\includegraphics[width=0.45\columnwidth]{Fig/Marmousi/Marmoush-true-mod.pdf}}
    \caption{Inverted models at different stages of the FWIME workflow conducted with $\epsilon = 1.5 \times 10^{-5}$. (a) Initial model. (b) Inverted model after 25 iterations on the first spline grid, $\mathbf{S}_0$. (c) Inverted model after 55 iterations on the second spline grid, $\mathbf{S}_1$. (d) Inverted model after 110 iterations on the third spline grid, $\mathbf{S}_2$. (e) Inverted model after 50 iterations on the fourth (finite-difference) grid. (f) True model. The FWIME models in panels (b), (c), and (d) are inverted on their respective spline grids $\mathbf{S}_i$ but are shown on the finite-difference grid.}
    \label{fig:Marmousi_fwime_grid_mod}
\end{figure}

Figures~\ref{fig:Marmousi_fwime_grid_mod}b-d show the FWIME inverted model after spline 0, 1, and 2, respectively. The final FWIME inverted model after 240 iterations of L-BFGS is shown in Figure~\ref{fig:Marmousi_fwime_noSpline_mod}. Even though it suffers from minor edge effects due to the limited acquisition aperture, it is very accurate, as confirmed by the velocity profiles shown in Figure~\ref{fig:Marmousi_mod_1d} (blue curves).

\begin{figure}[t]
    \centering
    \subfigure[]{\label{fig:Marmousi_fwime_obs_data}\includegraphics[width=0.26\columnwidth]{Fig/Marmousi/Marmousi-thesis-data-true-s10.pdf}}
    \subfigure[]{\label{fig:Marmousi_fwime_predData_s1}\includegraphics[width=0.26\columnwidth]{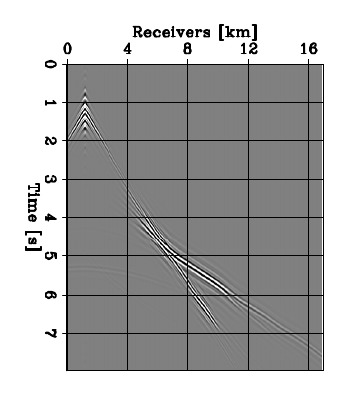}} 
    \subfigure[]{\label{fig:Marmousi_fwime_dataDiff_s1}\includegraphics[width=0.26\columnwidth]{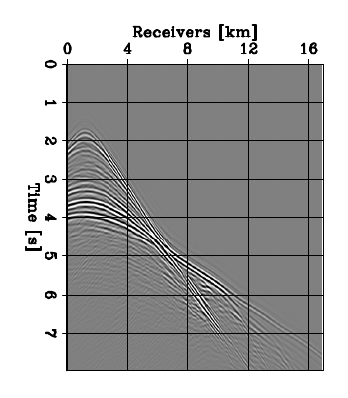}} \\
    \subfigure[]{\label{fig:Marmousi_fwime_obs_data}\includegraphics[width=0.26\columnwidth]{Fig/Marmousi/Marmousi-thesis-data-true-s10.pdf}}
    \subfigure[]{\label{fig:Marmousi_fwime_predData_s2}\includegraphics[width=0.26\columnwidth]{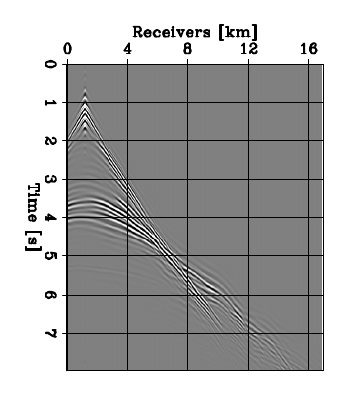}} 
    \subfigure[]{\label{fig:Marmousi_fwime_dataDiff_s2}\includegraphics[width=0.26\columnwidth]{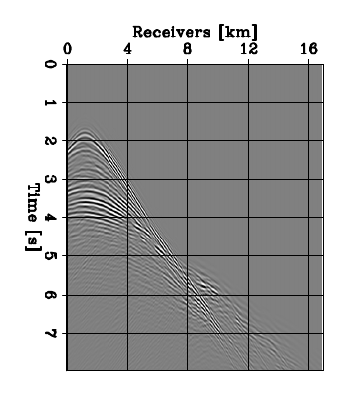}}   \\  
    \subfigure[]{\label{fig:Marmousi_fwime_obs_data}\includegraphics[width=0.26\columnwidth]{Fig/Marmousi/Marmousi-thesis-data-true-s10.pdf}}
    \subfigure[]{\label{fig:Marmousi_fwime_predData_noSpline}\includegraphics[width=0.26\columnwidth]{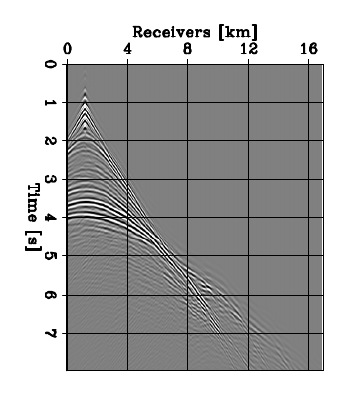}} 
    \subfigure[]{\label{fig:Marmousi_fwime_dataDiff_noSpline}\includegraphics[width=0.26\columnwidth]{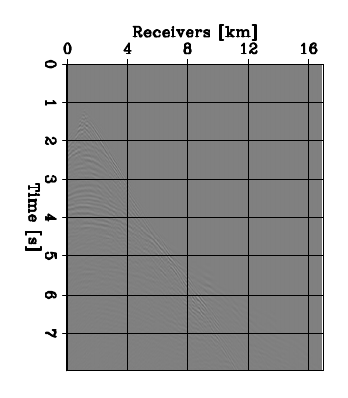}} \\            
    \caption{Representative shot gathers for a source placed at $x = 1.2$ km. Observed data $\mathbf{d}^{obs}$ (left column), predicted data $\mathbf{f}(\mathbf{S}_i \mathbf{m}_i)$ (middle column), and data difference $\mathbf{d}^{obs}-\mathbf{f}(\mathbf{S}_i \mathbf{m}_i)$ computed with the inverted model on the first spline grid (first row), second spline grid (second row), and final inverted model (third row). All panels are displayed with the same grayscale.}
    \label{fig:Marmousi_fwime_data}
\end{figure}

Figure~\ref{fig:Marmousi_fwime_data} shows the observed (left column), predicted (middle column), and data-difference (right column) for a source located at $x=1.2$ km, computed with the FWIME inverted models obtained after the first, second, and final grids (the panels for the third grid are not shown). The inversion workflow begins by allowing exclusively low-wavenumber updates into the model (due to the coarse spline grid parametrization), which enforces a better matching of the observed data at larger offsets (i.e., diving waves), as shown in Figures~\ref{fig:Marmousi_fwime_data}b and \ref{fig:Marmousi_fwime_data}c (compared to Figures~\ref{fig:Marmousi_data_obs}b and \ref{fig:Marmousi_data_obs}c). As we refine the spline grid and allow for higher-wavenumber updates, reflections are progressively matched (second row of Figure~\ref{fig:Marmousi_fwime_data}), and the data-residuals eventually vanish (third row in Figure~\ref{fig:Marmousi_fwime_data}).

\begin{figure}[t]
    \centering
    \subfigure[]{\label{fig:Marmousi_obj_fwime_total}\includegraphics[width=0.45\columnwidth]{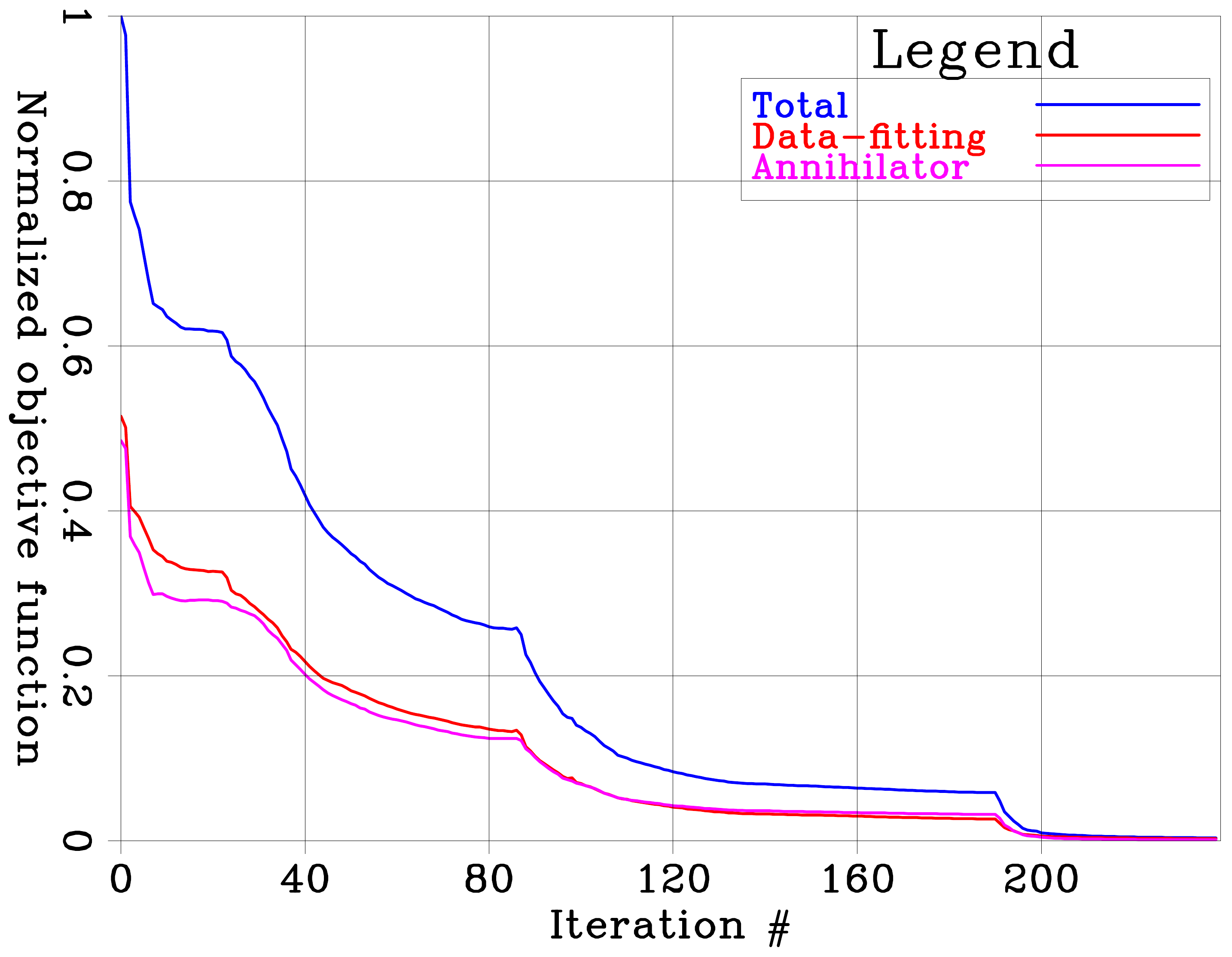}} \hspace{5mm}
    \subfigure[]{\label{fig:Marmousi_obj_fwime_fwi}\includegraphics[width=0.45\columnwidth]{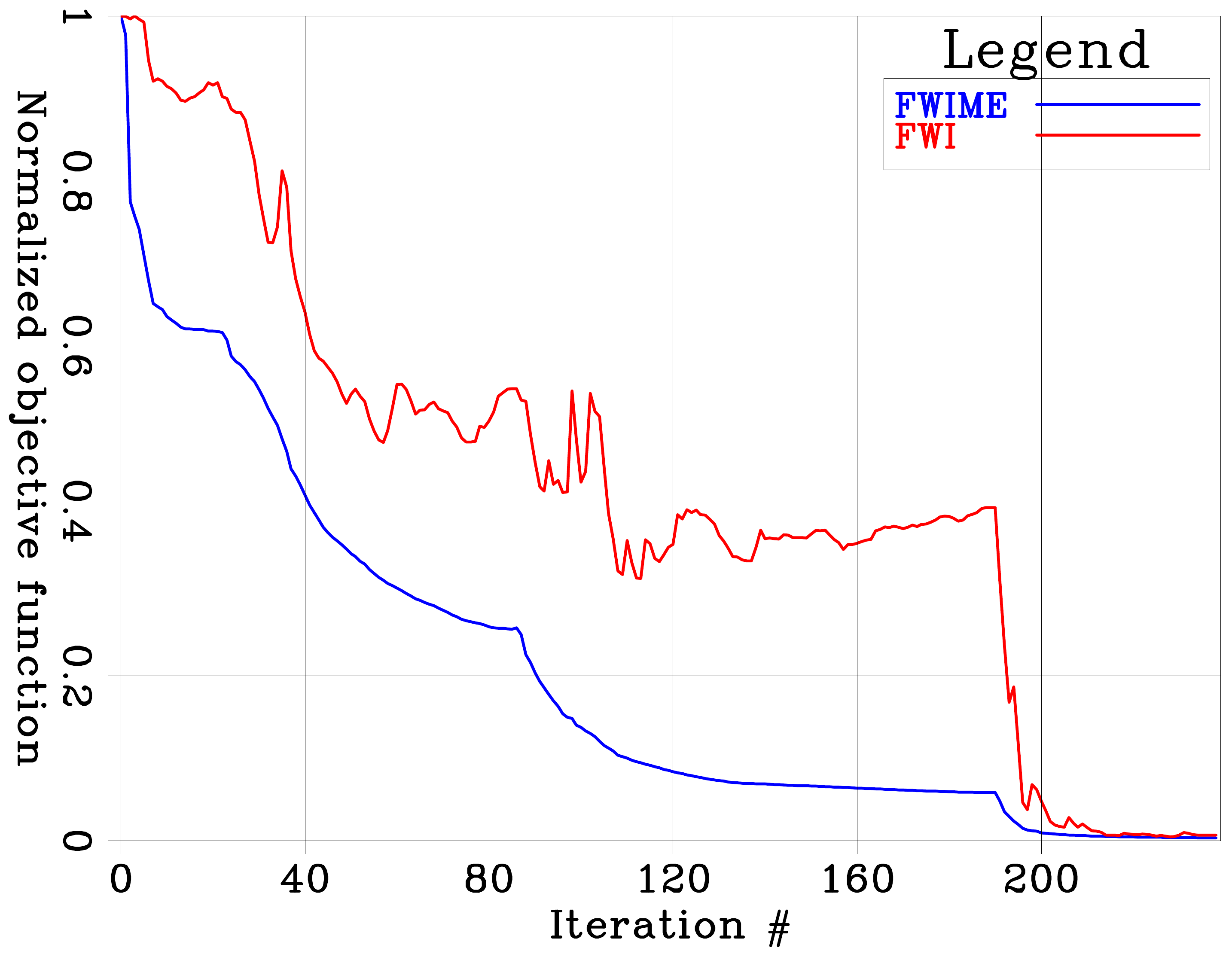}}
    \caption{Normalized objective functions. (a) Total  FWIME  objective  function (blue curve), FWIME data-fitting component (red curve), and FWIME annihilating component (pink curve). (b) Total FWIME objective function (blue curve), and FWI objective function evaluated at each FWIME inverted model (red curve).}
    \label{fig:Marmousi_obj}
\end{figure}

Figure~\ref{fig:Marmousi_obj_fwime_total} shows the FWIME convergence curves as a function of iterations throughout the four stages of the inversion process (the four spline grids). The blue curve corresponds to the total objective function, the red curve corresponds to the data-fitting component, and the pink curve displays the annihilating component. The three major discontinuities in the rate of convergence occurring at iterations 25, 80, and 190 indicate a spline grid refinement. Figure~\ref{fig:Marmousi_obj_fwime_fwi} displays the normalized total FWIME objective function (blue curve) along with the FWI objective function evaluated at each inverted model during the optimization sequence (red curve). The red curve is not the result of an inversion process, but simply an evaluation of the FWI objective function at each FWIME inverted model. It can also be interpreted as a measure of how well the FWIME estimated model explains all the events in the recorded data. Moreover, it is not monotonically decreasing, which illustrates that the FWIME algorithm has successfully created an alternate descent-path towards the optimal solution that could not have been taken by FWI. Eventually, both curves converge to zero (up to numerical precision) which means that FWIME has managed to find an inverted model that matches all the events on the observed data without the need for the additional data-correcting term.

\begin{figure}[t]
    \centering
    \subfigure[]{\label{fig:Marmousi_fwime_pOpt_zero_offset_init}\includegraphics[width=0.45\columnwidth]{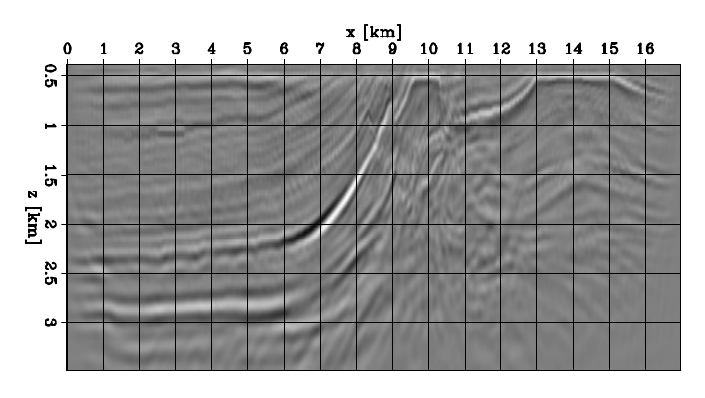}}
    \subfigure[]{\label{fig:Marmousi_fwime_pOpt_zero_offset_s1}\includegraphics[width=0.45\columnwidth]{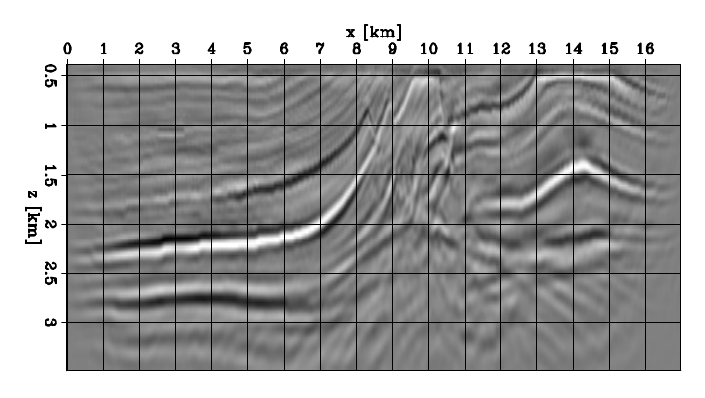}} \\
    \subfigure[]{\label{fig:Marmousi_fwime_pOpt_zero_offset_s2}\includegraphics[width=0.45\columnwidth]{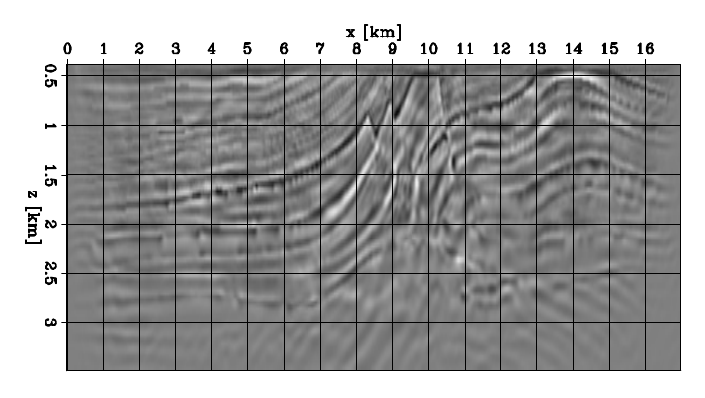}}    
    \subfigure[]{\label{fig:Marmousi_fwime_pOpt_zero_offset_s3}\includegraphics[width=0.45\columnwidth]{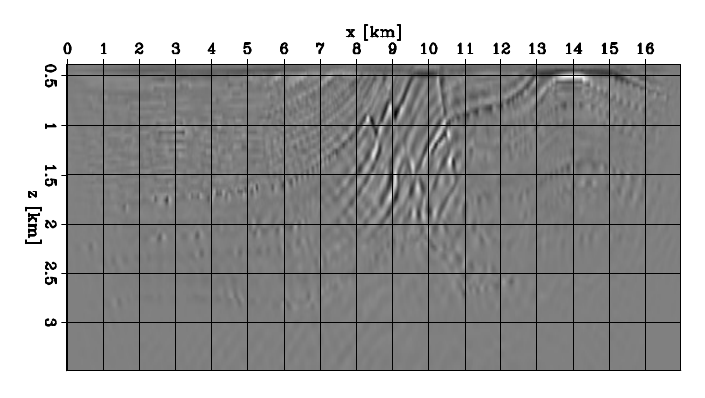}} \\        
    \subfigure[]{\label{fig:Marmousi_fwime_pOpt_zero_offset_final}\includegraphics[width=0.45\columnwidth]{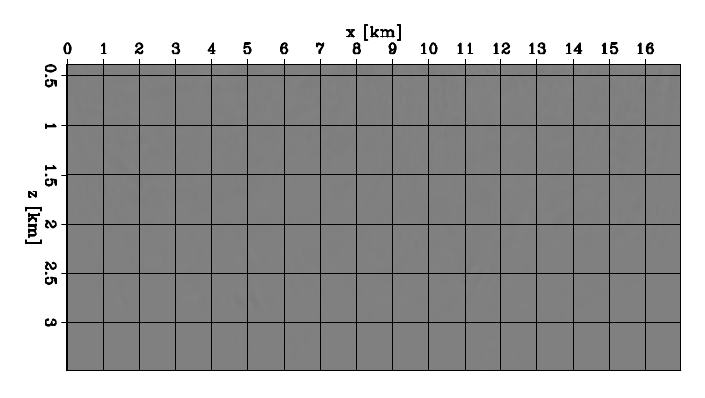}}            
    \caption{Zero time-lag sections of $\mathbf{p}_{\epsilon}^{opt}$ computed at four stages of the FWIME workflow. (a) Initial step. (b) After inversion on spline 1. (c) After inversion on spline 2. (d) After inversion on spline 3. (e) Final step. All panels are displayed with the same grayscale.}
    \label{fig:Marmousi_fwime_pOpt_zero_offset}
\end{figure}

\begin{figure}[t]
    \centering
    \subfigure[]{\label{fig:Marmousi_pOpt_cig_init}\includegraphics[width=0.18\columnwidth]{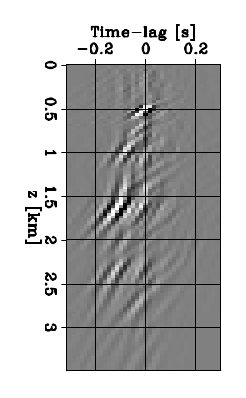}}
    \subfigure[]{\label{fig:Marmousi_pOpt_cig_s1}\includegraphics[width=0.18\columnwidth]{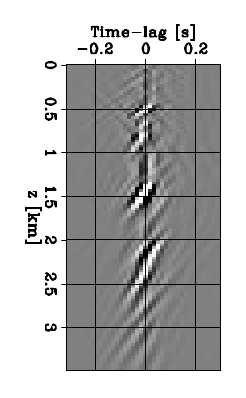}}    
    \subfigure[]{\label{fig:Marmousi_pOpt_cig_s2}\includegraphics[width=0.18\columnwidth]{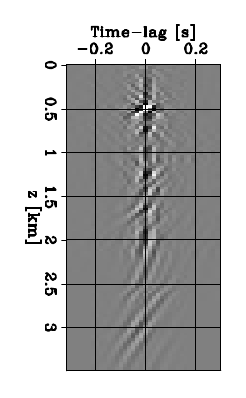}} 
    \subfigure[]{\label{fig:Marmousi_pOpt_cig_s3}\includegraphics[width=0.18\columnwidth]{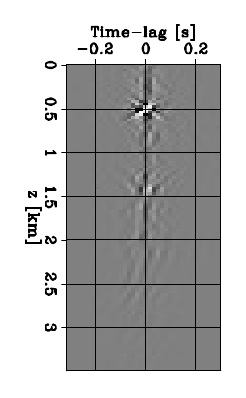}}     
    \subfigure[]{\label{fig:Marmousi_pOpt_cig_final}\includegraphics[width=0.18\columnwidth]{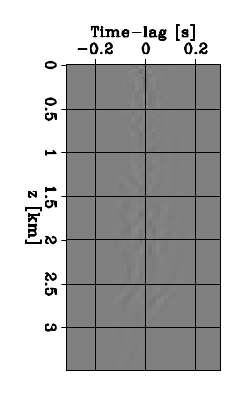}}     
    \caption{TLCIG extracted at $x=14$ km from $\mathbf{p}_{\epsilon}^{opt}$ computed at four stages of the FWIME workflow. (a) Initial step. (b) After inversion on spline 1. (c) After inversion on spline 2. (d) Final step.  All panels are plotted for $\tau \in [-0.3 \; \rm{s}, \; 0.3 \; \rm{s}]$ for display purposes. However, all the computations are conducted for the full time-lag range with $\tau \in [-0.8 \; \rm{s}, \; 0.8 \; \rm{s}]$. All panels are displayed with the same grayscale.}
    \label{fig:Marmousi_pOpt_cig}
\end{figure}

We analyze $\mathbf{\tilde{p}}_{\epsilon}^{opt}$ by examining the evolution of its zero-lag cross section (Figure~\ref{fig:Marmousi_fwime_pOpt_zero_offset}) and a representative TLCIG extracted at $x=14$ km (Figure~\ref{fig:Marmousi_pOpt_cig}) at four stages of the inversion process. Initially, the energy is clustered away from the physical plane (Figure~\ref{fig:Marmousi_pOpt_cig_init}) and the zero-lag cross-section lacks coherency in the region where the reservoirs are located (Figure~\ref{fig:Marmousi_fwime_pOpt_zero_offset_init}). As the velocity model becomes more accurate, the energy gradually focuses towards the physical plane (Figures~\ref{fig:Marmousi_pOpt_cig}b-d), and the coherency of the zero-lag section is simultaneously enhanced (Figures~\ref{fig:Marmousi_fwime_pOpt_zero_offset}b-d). As expected, when the algorithm converges to the optimal solution, $\mathbf{\tilde{p}}_{\epsilon}^{opt}$ vanishes (Figures~\ref{fig:Marmousi_pOpt_cig_final} and \ref{fig:Marmousi_fwime_pOpt_zero_offset_final}). 

\subsection{Reducing FWIME's computational cost}
\label{reducing_cost_marmousi2}
\begin{figure}[t]
    \centering
    \subfigure[]{\label{fig:Marmousi_fwime_s1_fwi_mod}\includegraphics[width=0.45\columnwidth]{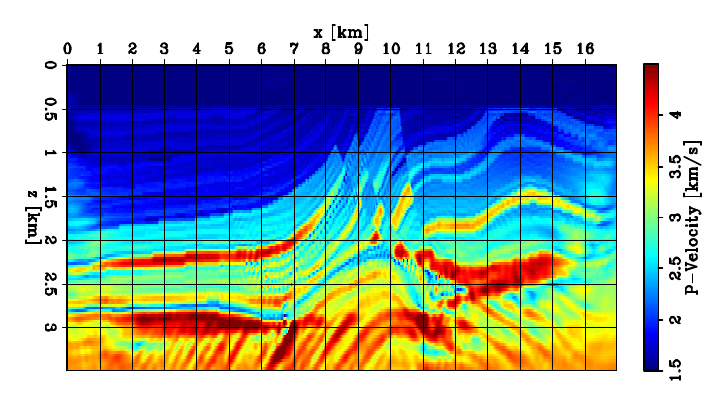}}
    \subfigure[]{\label{fig:Marmousi_fwime_s2_fwi_mod}\includegraphics[width=0.45\columnwidth]{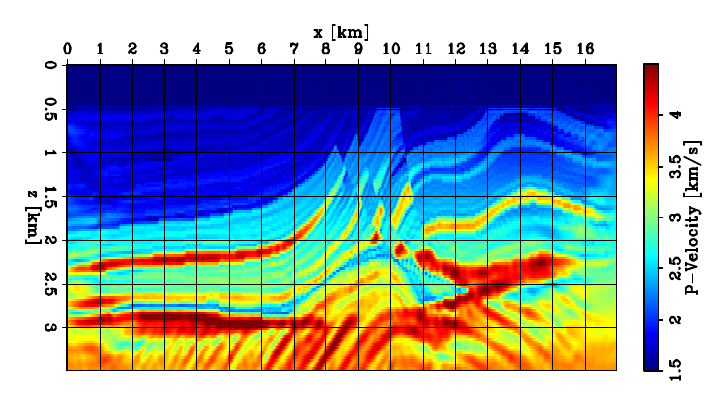}} \\
    \subfigure[]{\label{fig:Marmousi_fwime_s3_fwi_mod}\includegraphics[width=0.45\columnwidth]{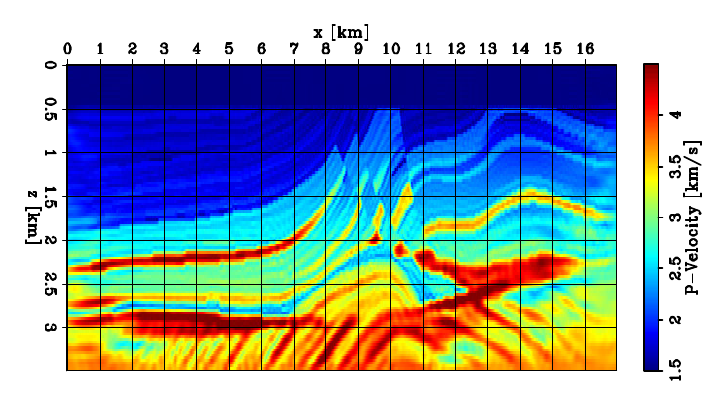}}     
    \subfigure[]{\label{fig:Marmousi_fwime_mod}\includegraphics[width=0.45\columnwidth]{Fig/Marmousi/Marmoush-fwime-noSpline-e3-mod.pdf}} \\
    \caption{2D panels showing the result of applying FWI using various FWIME inverted models as initial guesses. (a) FWIME inverted model on the first spline grid $\mathbf{S}_0$ (Figure~\ref{fig:Marmousi_fwime_s1_mod}). (b) FWIME inverted model on the second spline grid, $\mathbf{S}_1$ (Figure~\ref{fig:Marmousi_fwime_s2_mod}). (c) FWIME inverted model on the third grid $\mathbf{S}_2$ (Figure~\ref{fig:Marmousi_fwime_s3_mod}). (d) Final FWIME inverted model (Figure~\ref{fig:Marmousi_fwime_noSpline_mod}).}
    \label{fig:Marmousi_fwime_fwi_mod}
\end{figure}

To mitigate the computational cost of FWIME, our algorithm could typically be used to recover a good-enough starting model for FWI. In this numerical example, we conduct FWIME until completion in order to illustrate the potential of the method. However, for 3D field applications, the optimization could have been stopped at earlier stages. Figure~\ref{fig:Marmousi_fwime_s1_fwi_mod} shows the result of applying data-space multi-scale FWI using the FWIME inverted model on the first spline grid, and indicates that in this case, 25 iterations of FWIME would have been sufficient for FWI to converge to an accurate solution. Beyond that point, conducting FWI using the initial model from FWIME would perform as well as conducting FWIME until completion (Figure~\ref{fig:Marmousi_fwime_fwi_mod}b-d). 

The most computationally intensive component of FWIME is the variable projection step, which consists in iteratively minimizing the quadratic objective function defined in equation~\ref{eqn:vp.obj} with a linear conjugate-gradient scheme. From a numerical aspect, this condition implies that equation~\ref{eqn:vp.obj} should be minimized until full convergence. That is, by conducting ``enough" linear conjugate-gradient iterations. In order to further reduce the computational cost of our method, we assess how solving the variable projection step less accurately (i.e., with less linear conjugate-gradient iterations) impacts the quality of the FWIME solution. 

\begin{figure}[t]
    \centering
    \subfigure[]{\label{fig:Marmousi_fwimeLinIter7}\includegraphics[width=0.45\columnwidth]{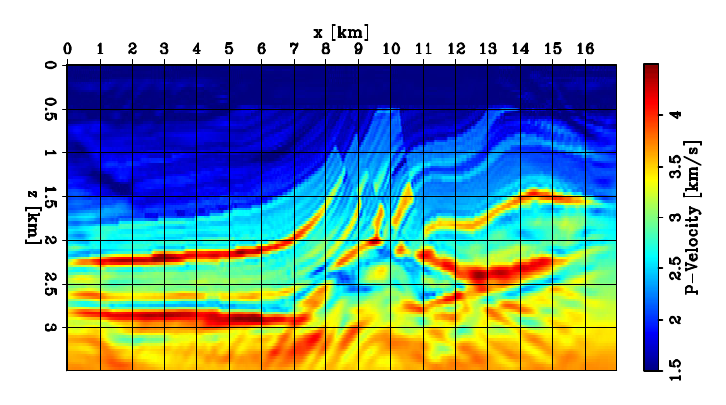}}
    \subfigure[]{\label{fig:Marmousi_fwimeLinIter10}\includegraphics[width=0.45\columnwidth]{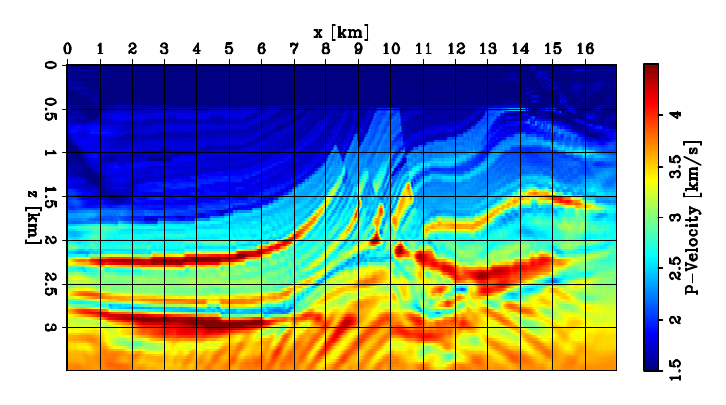}}     
    \subfigure[]{\label{fig:Marmousi_fwimeLinIter15}\includegraphics[width=0.45\columnwidth]{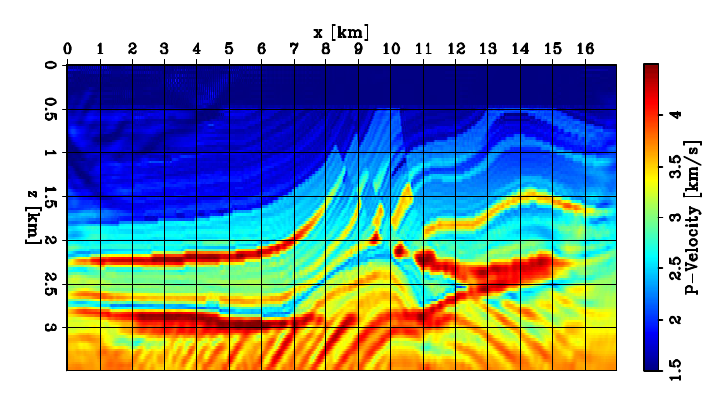}}         
    \subfigure[]{\label{fig:Marmousi_fwimeLinIter20}\includegraphics[width=0.45\columnwidth]{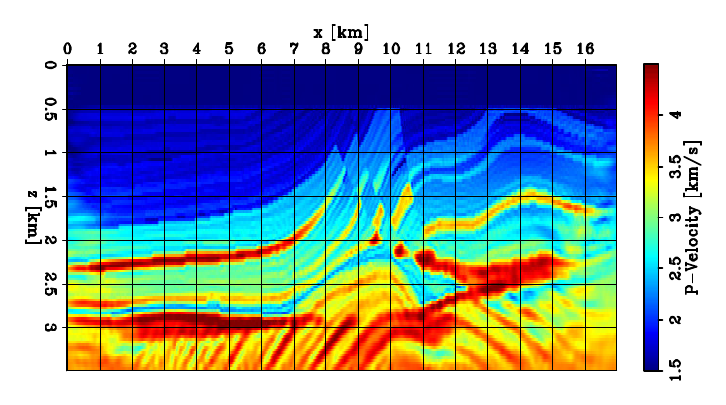}} 
    \caption{FWIME inverted models computed with the same hyper-parameter selection as for the result shown in Figure~\ref{fig:Marmousi_fwime_noSpline_mod} but with less linear conjugate-gradient iterations for the variable projection step. (a) 7 iterations, (b) 10 iterations, (c) 15 iterations, and (d) 20 iterations. }
    \label{fig:Marmousi_fwime_linIter}
\end{figure}

The inverted model shown in Figure~\ref{fig:Marmousi_fwime_noSpline_mod} is obtained by solving the variable projection problem with 60 iterations, which accounts for 97$\%$ of the total FWIME cost. We conduct analogous FWIME schemes with the same hyper-parameter selection but by reducing the number of linear iterations during the variable projection step. Figures~\ref{fig:Marmousi_fwime_linIter}a-d show the FWIME inverted models obtained by minimizing equation~\ref{eqn:vp.obj} with 7, 10, 15, and 20 linear iterations, respectively. For a fair cost comparison, the inverted models in Figures~\ref{fig:Marmousi_fwime_linIter} were obtained with at most 240 nonlinear iterations of L-BFGS. The inverted results become less accurate for 10 linear iterations or less (Figures~\ref{fig:Marmousi_fwimeLinIter7} and \ref{fig:Marmousi_fwimeLinIter10}), but seem unaffected when the number of iterations is set to 15 or higher (Figures~\ref{fig:Marmousi_fwimeLinIter15} and \ref{fig:Marmousi_fwimeLinIter20}). In this numerical test, reducing the number of linear iteration to 15 corresponds to a computational cost decrease of 73$\%$. Even though this behavior may be case dependent, it shows potential value in trying to reduce the number of linear iterations for the variable projection step in FWIME, especially for 3D applications. 

\subsection{Effect of coherent source of noise}
\begin{figure}[t]
    \centering
    \subfigure[]{\label{fig:MoushNoFs_data_s10}\includegraphics[width=0.40\columnwidth]{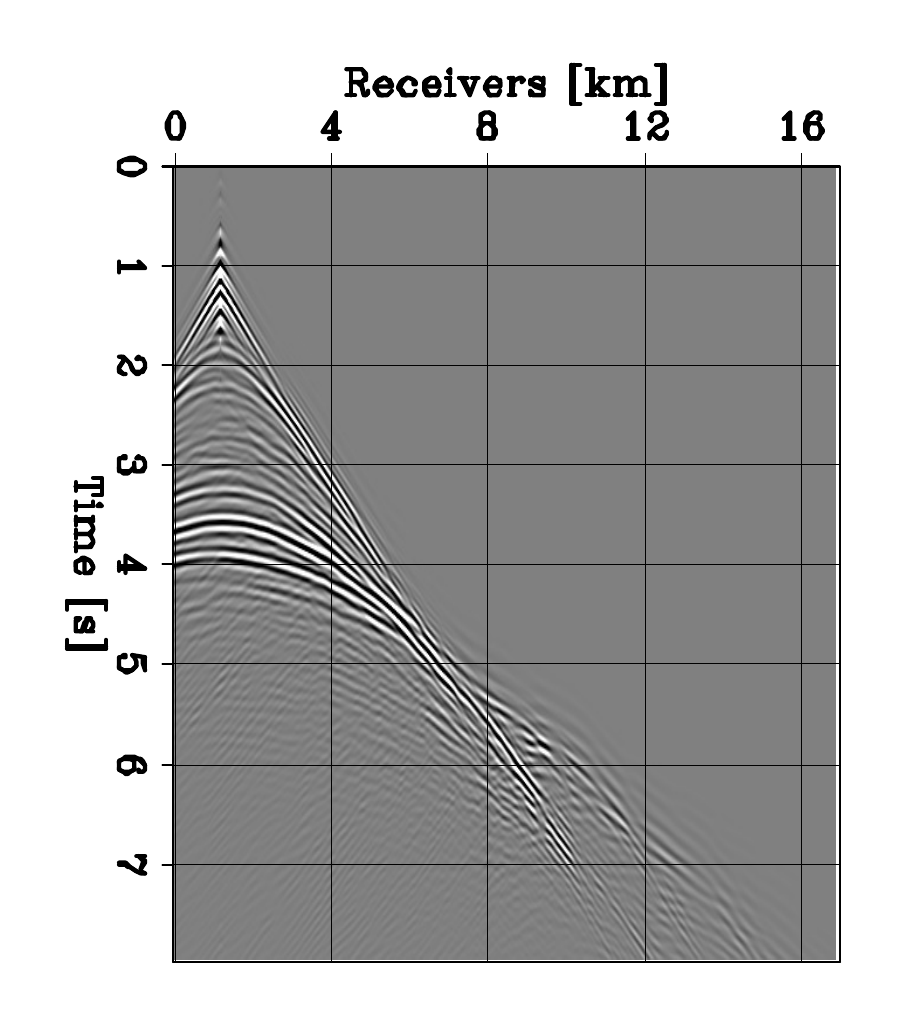}} 
    \subfigure[]{\label{fig:MoushNoFs_data_s120}\includegraphics[width=0.40\columnwidth]{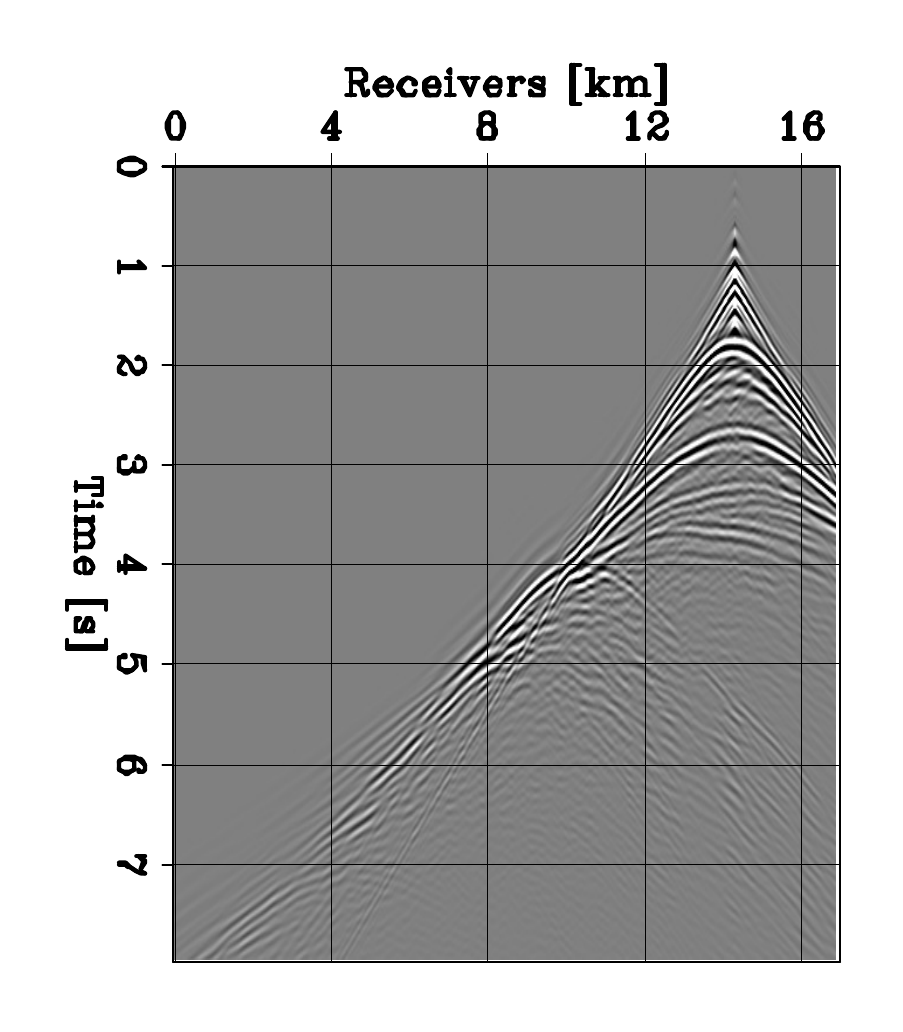}} \\
    \subfigure[]{\label{fig:MoushFs_data_s10}\includegraphics[width=0.40\columnwidth]{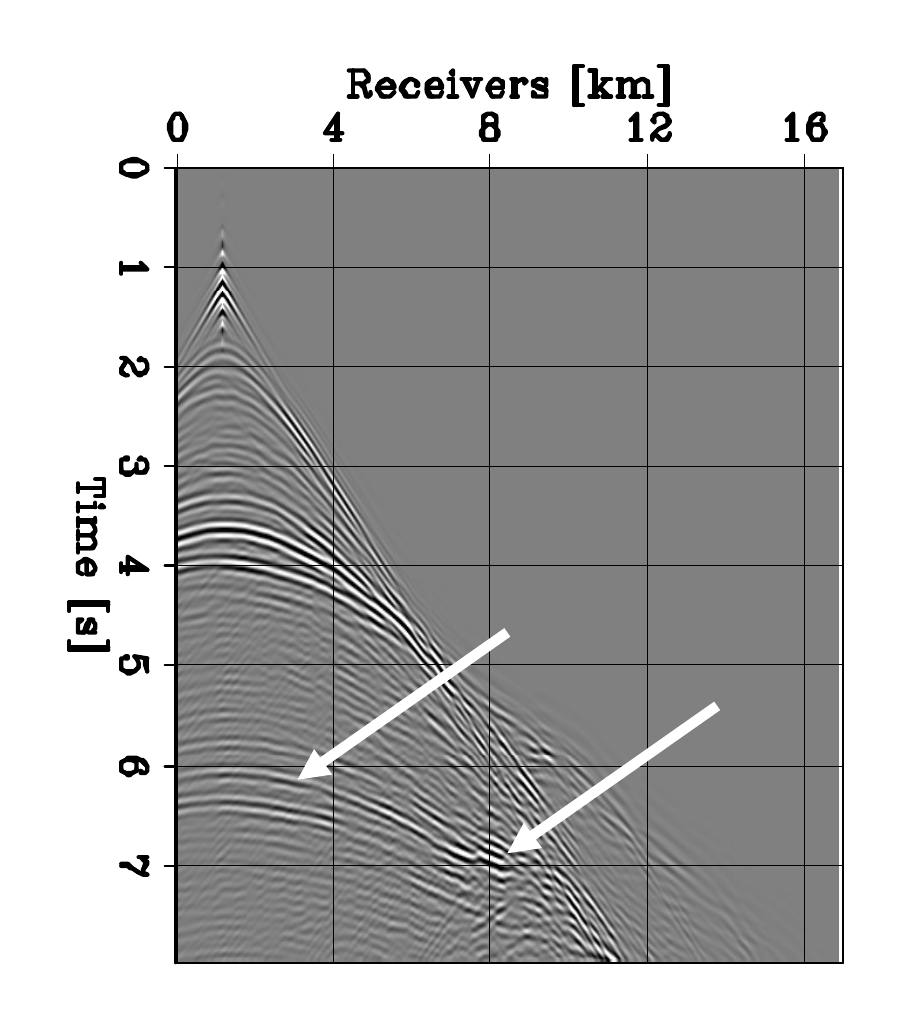}} 
    \subfigure[]{\label{fig:MoushFs_data_s120}\includegraphics[width=0.40\columnwidth]{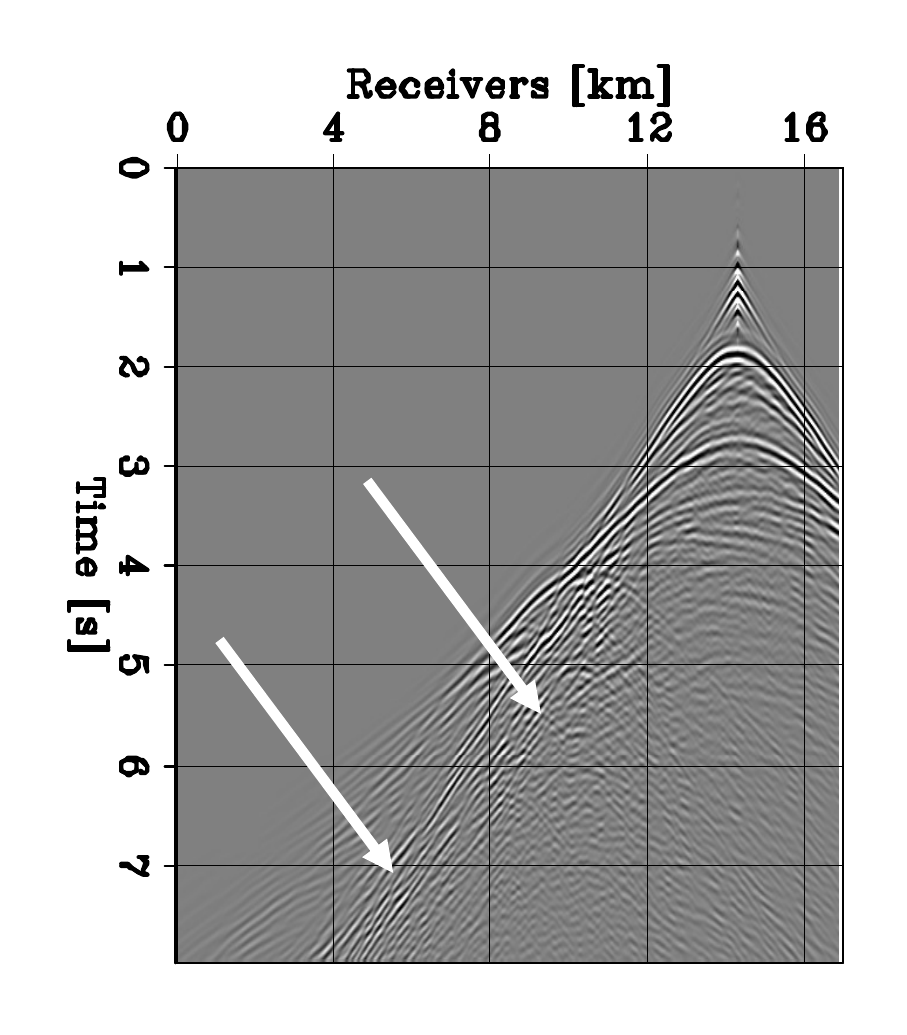}} 
    \caption{Representative shot gathers for sources placed at $x = 1.2$ km (first column) and $x=14.4$ km (second column). The top row shows the data modeled with absorbing-boundary conditions in all directions. The bottom row shows the data modeled with a free-surface boundary condition at the water surface. All panels are displayed with the same grayscale.}
    \label{fig:MoushFs_data}
\end{figure}

We investigate how the presence of free-surface multiples recorded with marine acquisition geometries affects the performance of FWIME. We generate a new dataset with a free-surface boundary condition at the water surface, and absorbing-boundary conditions in all other directions \cite[]{robertsson1996numerical}. Both sources and receivers are placed at a depth of 30 m below the surface. All other acquisition parameters are identical to the ones used for our previous analysis. Figure~\ref{fig:MoushFs_data} shows two shot gathers generated by sources positioned at $x=1.2$ km (first column) and $x=14.4$ km (second column) using an absorbing-boundary condition (first row) and a free-surface boundary condition at the water/air interface (second row). The free-surface multiples are clearly observable (white arrows) and their amplitude is strong due to the relatively small thickness of the water layer. In the central part of the section (corresponding to the original Marmousi model), the multiples overlap with the various reflected events generated by the complex geological structures, as shown in Figure~\ref{fig:MoushFs_data_s120} \cite[]{guo2020data}. 

\begin{figure}[t]
    \centering
    \subfigure[]{\label{fig:MoushFs_mod_fwi}\includegraphics[width=0.45\columnwidth]{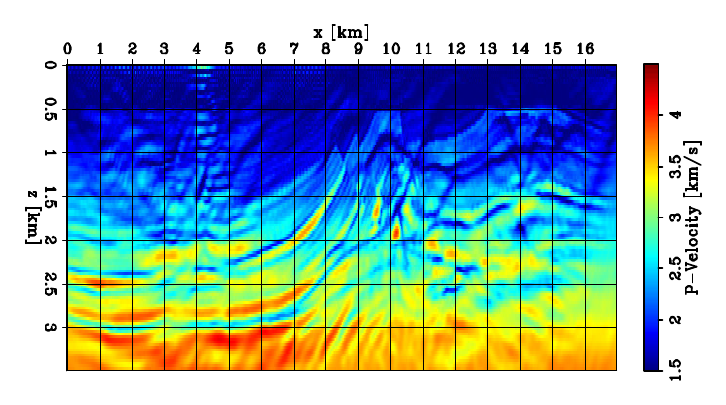}} 
    \subfigure[]{\label{fig:MoushFs_mod_fwime}\includegraphics[width=0.45\columnwidth]{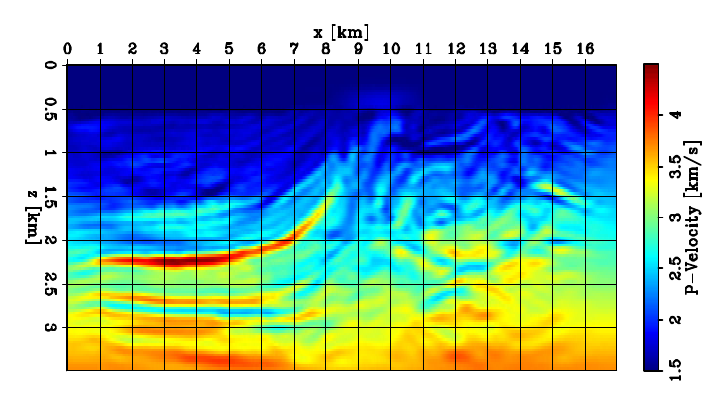}} \\
    \subfigure[]{\label{fig:MoushNoFs_mod_fwime}\includegraphics[width=0.45\columnwidth]{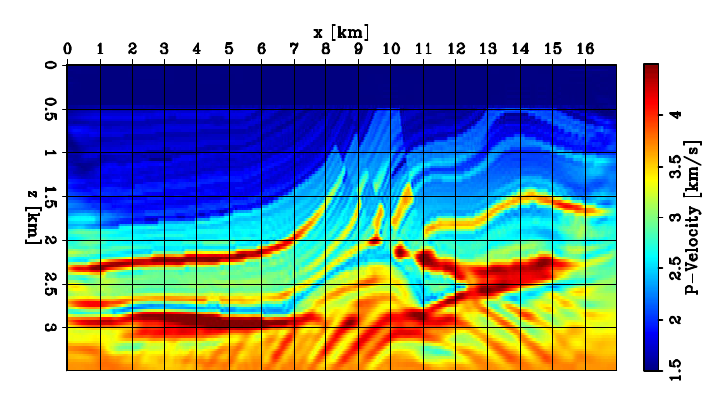}} 
    \subfigure[]{\label{fig:MoushFs_mod_true}\includegraphics[width=0.45\columnwidth]{Fig/Marmousi/Marmoush-true-mod.pdf}} 
    \caption{2D panels of velocity models. (a) FWI model inverted using the dataset modeled with a free-surface. (c) Final FWIME model inverted using the dataset modeled with a free-surface. (c) FWIME model inverted with the original dataset. (d) True model.}
    \label{fig:MoushFs_mod}
\end{figure}

\begin{figure}[t]
    \centering
    \subfigure[]{\label{fig:}\includegraphics[width=0.20\columnwidth]{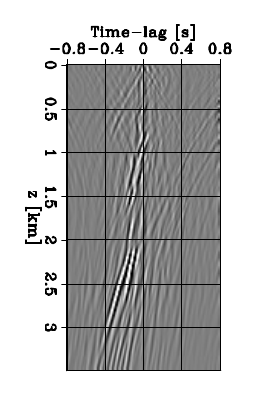}} 
    \subfigure[]{\label{fig:}\includegraphics[width=0.20\columnwidth]{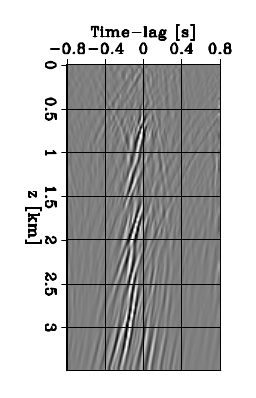}}     
    \subfigure[]{\label{fig:}\includegraphics[width=0.20\columnwidth]{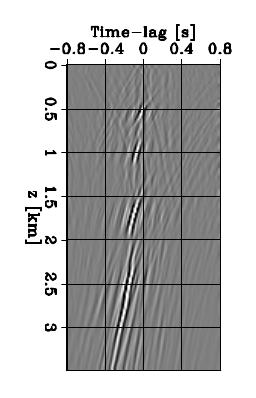}}    
    \subfigure[]{\label{fig:}\includegraphics[width=0.20\columnwidth]{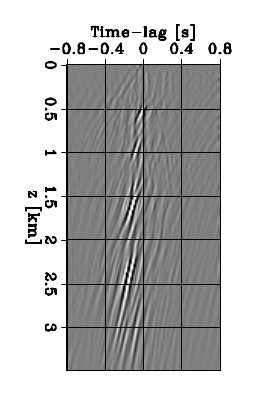}} \\
    \subfigure[]{\label{fig:}\includegraphics[width=0.20\columnwidth]{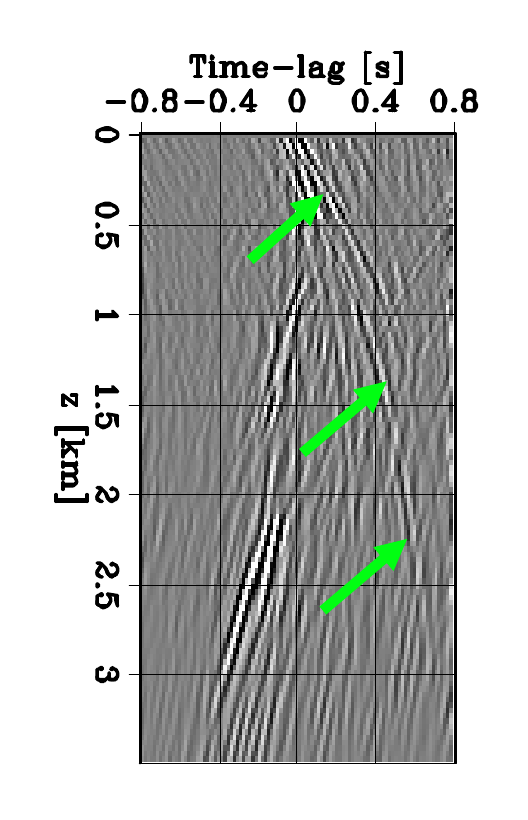}} 
    \subfigure[]{\label{fig:}\includegraphics[width=0.20\columnwidth]{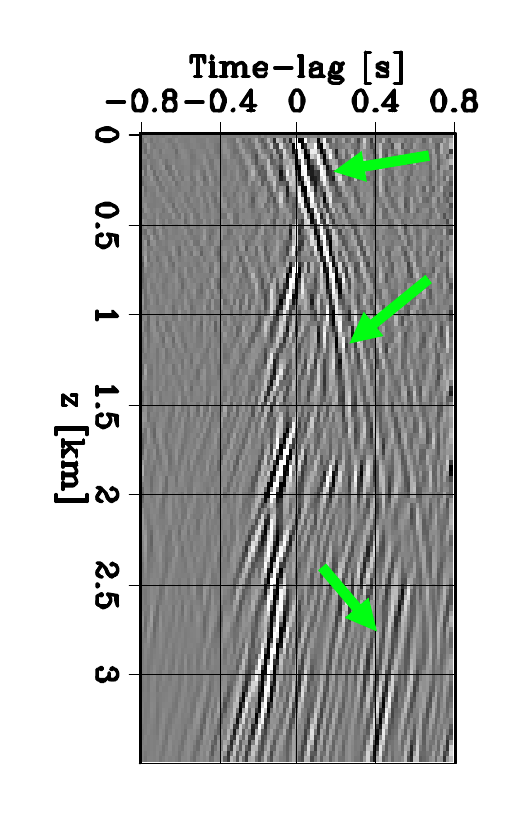}}     
    \subfigure[]{\label{fig:}\includegraphics[width=0.20\columnwidth]{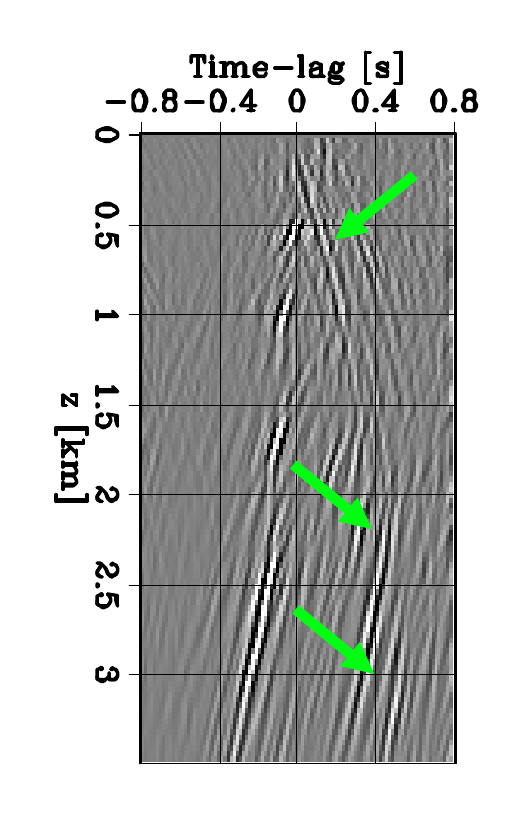}}    
    \subfigure[]{\label{fig:}\includegraphics[width=0.20\columnwidth]{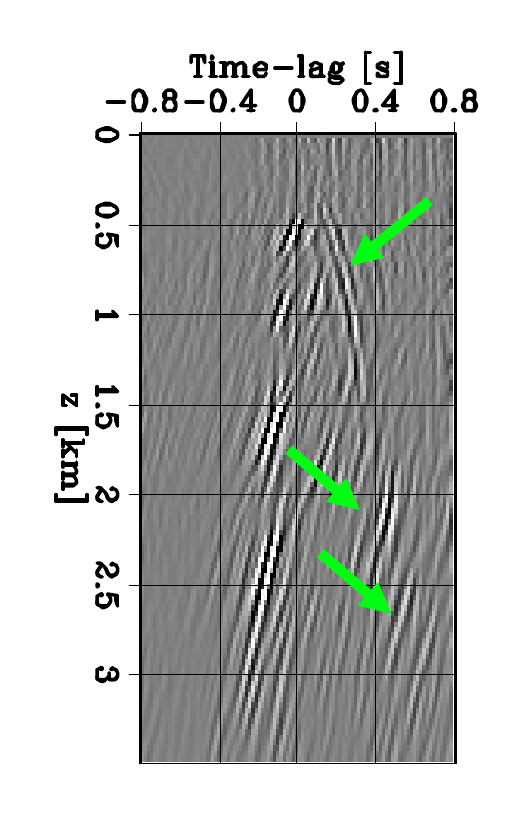}} 
    \caption{TLCIGs extracted at $x=11$ km, $x=12.5$ km, $x=13.5$ km, and $x=14.5$ km from $\mathbf{\tilde{p}}_{\epsilon}^{opt}(\mathbf{m}_{init})$ computed with the dataset using absorbing boundaries (top row), and with a free-surface boundary condition (bottom row). All panels are displayed with the same grayscale.}
    \label{fig:MousFs_cig}
\end{figure}

\begin{figure}[t]
    \centering
    \subfigure[]{\label{fig:MoushFs_grad_init}\includegraphics[width=0.45\columnwidth]{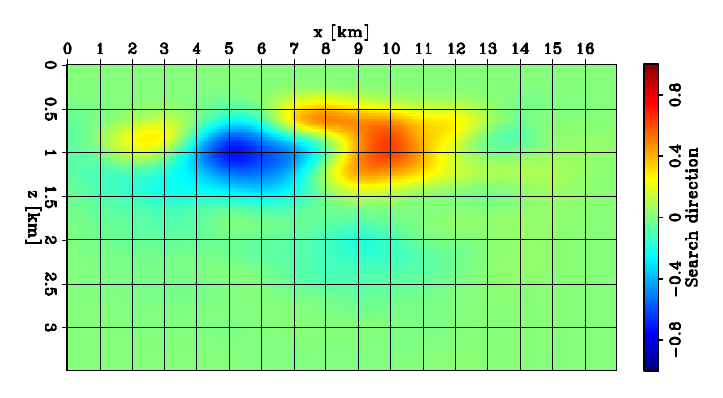}} 
    \subfigure[]{\label{fig:MoushNoFs_grad_init}\includegraphics[width=0.45\columnwidth]{Fig/Marmousi/Marmousi-thesis-vp-gradient-dso2-spline1.pdf}} \\
    \subfigure[]{\label{fig:MoushFs_grad_true}\includegraphics[width=0.45\columnwidth]{Fig/Marmousi/Marmousi-thesis-vp-gradient-true-spline1.pdf}}        
    \caption{Normalized initial search directions computed on the first spline grid. (a) FWIME search direction computed with the new dataset containing free-surface multiples. (b) FWIME search direction computed with the original dataset. (c) True search direction. Panels (a) and (b) are normalized with the same value.}
    \label{fig:MousFs_grad}
\end{figure}

We conduct data-space multi-scale FWI on this new dataset, which converges to an unsatisfactory solution (Figure~\ref{fig:MoushFs_mod_fwi}). We then apply FWIME in a similar fashion as for our previous analysis. For both FWI and FWIME, we invert the data generated with a free surface with the same engine (our forward modeling also uses a free surface). Figure~\ref{fig:MousFs_cig} shows TLCIGs extracted at four horizontal positions (ranging from $x=11$ km to $x=14.5$ km) from $\mathbf{\tilde{p}}_{\epsilon}^{opt}(\mathbf{m}_{init})$ computed at the intial step using the original dataset with absorbing boundaries in all directions (top row), and then using the new dataset with the free-surface boundary condition (bottom row). In both rows of Figures~\ref{fig:MousFs_cig}, we observe clusters of energy located at negative time lags indicating that the initial velocity values are too low for this region of the model (as we saw in the previous analysis). However, for the free-surface case (bottom row), additional clusters of coherent energy are present at positive time lags, which correspond to the coherent mapping of the free-surface multiples into the extended space of $\mathbf{\tilde{p}}_{\epsilon}^{opt}$ (indicated by the green arrows in Figures~\ref{fig:MousFs_cig}e-h). Since free-surface multiples generally propagate with a lower velocity than primary reflections recorded with the same traveltime, their positions on the extended axis wrongfully indicate that the initial velocity $\mathbf{m}_{init}$ is too high, thereby misleading the FWIME search direction for that region of the model. Figures~\ref{fig:MousFs_grad}a and \ref{fig:MousFs_grad}b show the initial FWIME search directions computed on the first spline grid with the original dataset and with the new dataset, respectively. By comparing these panels to the true search direction (Figure~\ref{fig:MoushFs_grad_true}), we can see that the free-surface multiples seem to guide the inversion in the wrong direction, especially in the right side of the model. The final FWIME result obtained after a total of 120 iterations of L-BFGS (using the same sequence of spline grids as for the original test) is shown in Figure~\ref{fig:MoushFs_mod_fwime}. The inversion manages to accurately reconstruct the sharp horizontal reflectors in the left part of the model, but the fails to recover the shallow complex region of the model located between $x=9$ km and $x=15$ km. For this acquisition geometry, the presence of free-surface multiples seems to harm the quality of the inverted solution (compared to the original result shown in Figure~\ref{fig:MoushNoFs_mod_fwime}). To mitigate this effect for field applications, we propose to use surface-related multiple elimination techniques when applying FWIME to offshore data acquired in a similar scenario \cite[]{verschuur1992adaptive,baumstein20063d,dragoset2010perspective,siahkoohi2019surface}. 

%% file: bpCaspian.tex
We invert a diving-wave dominated dataset generated from the North Sea region of the BP 2004 benchmark model \cite[]{billette20052004}. Our goal is to show that FWIME can use diving waves to converge to the global minimum with the same automatic and user-friendly workflow as the one conducted for reflection data in the previous example. The initial velocity model is designed to be extremely inaccurate, and no energy below 3 Hz is present in the recorded data. 

\subsection{Presentation and challenges}
\begin{figure}[t]
    \centering
    \subfigure[]{\label{fig:bpCaspian_init_mod}\includegraphics[width=0.45\columnwidth]{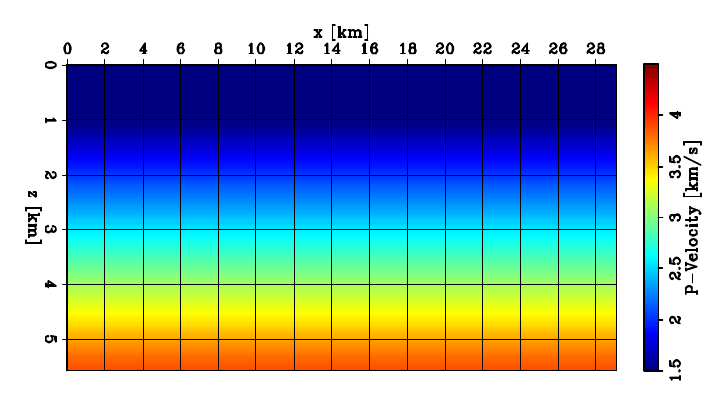}}
    \subfigure[]{\label{fig:bpCaspian_fwi_mod}\includegraphics[width=0.45\columnwidth]{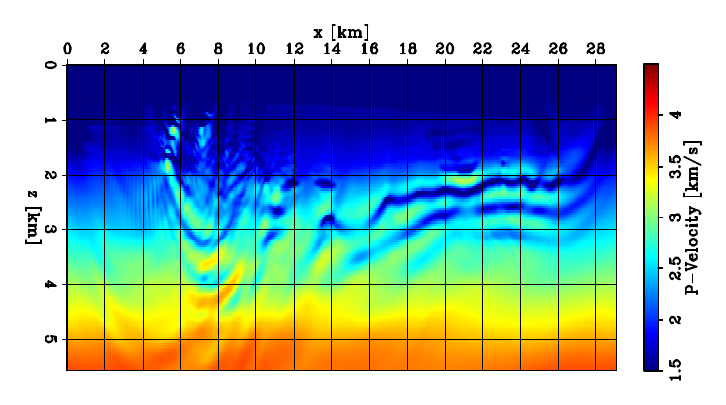}} \\
    \subfigure[]{\label{fig:bpCaspian_fwime_mod}\includegraphics[width=0.45\columnwidth]{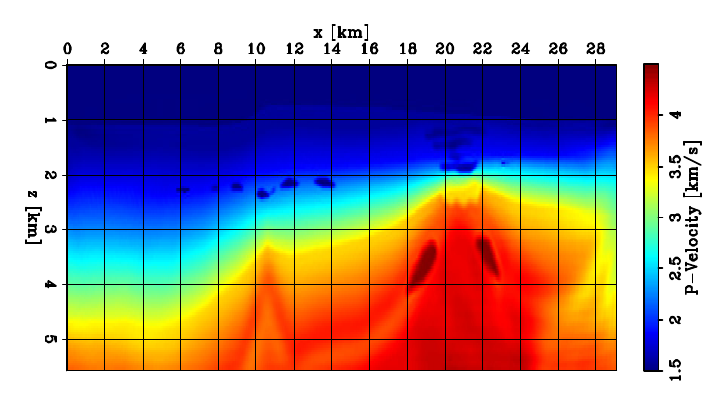}}        
    \subfigure[]{\label{fig:bpCaspian_true_mod}\includegraphics[width=0.45\columnwidth]{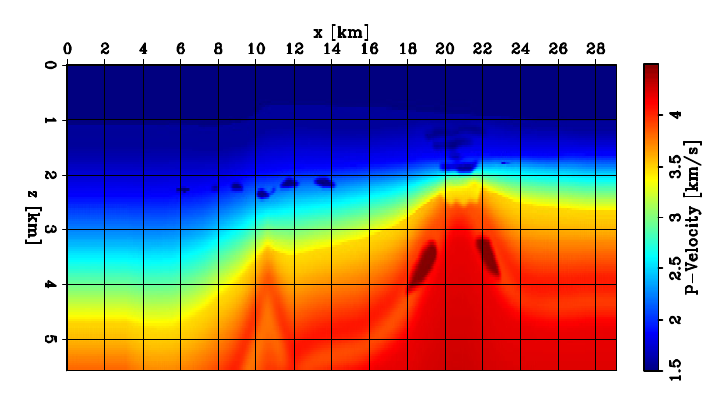}}            
    \caption{2D panels of velocity models. (a) Initial model. (b) Inverted model after conventional data-space multi-scale FWI using four frequency bands. (c) Final FWIME inverted model. (d) True model.}
    \label{fig:bpCaspian_mod}
\end{figure}

\begin{figure}[t]
    \centering
    \subfigure[]{\label{fig:bpCaspian_mod_1d_x10.5}\includegraphics[width=0.22\columnwidth]{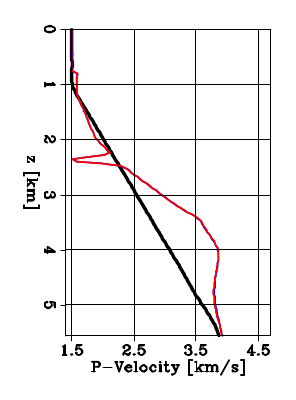}}
    \subfigure[]{\label{fig:bpCaspian_mod_1d_x19}\includegraphics[width=0.22\columnwidth]{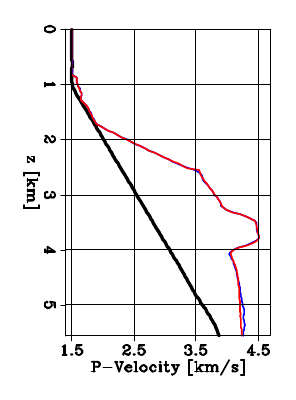}} 
    \subfigure[]{\label{fig:bpCaspian_mod_1d_x21}\includegraphics[width=0.22\columnwidth]{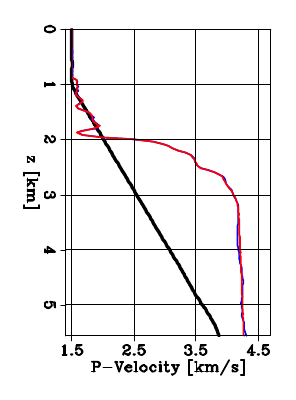}}
    \subfigure[]{\label{fig:bpCaspian_mod_1d_x22.5}\includegraphics[width=0.22\columnwidth]{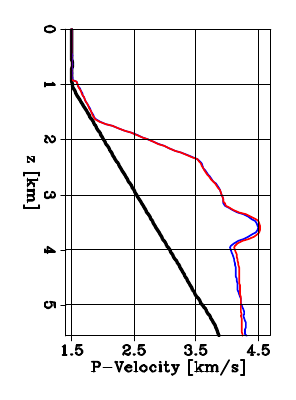}} 
    \caption{Depth velocity profiles extracted at (a) $x=6$ km, (b) $x=9$ km, and (c) $x=11$ km, and (d) $x=13$ km. The black curve represents the initial model, the red curve is the true model, and the blue curve is the FWIME inverted model.}
    \label{fig:bpCaspian_mod_1d_x}
\end{figure}

\begin{figure}[t]
    \centering
    \subfigure[]{\label{fig:bpMod_mod_1d_z2}\includegraphics[width=0.45\columnwidth]{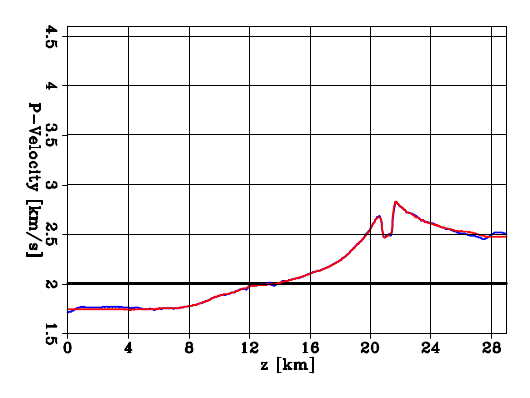}} \hspace{5mm}
    \subfigure[]{\label{fig:bpMod_mod_1d_z3}\includegraphics[width=0.45\columnwidth]{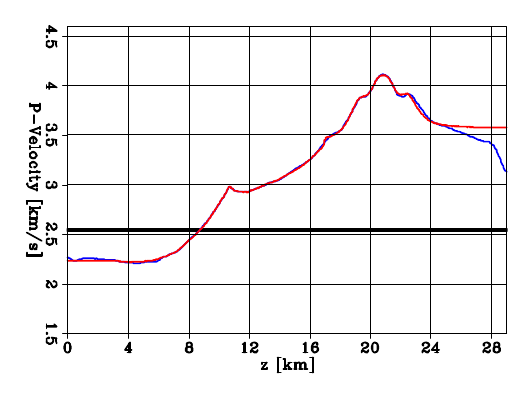}} \\
    \subfigure[]{\label{fig:bpMod_mod_1d_z3.5}\includegraphics[width=0.45\columnwidth]{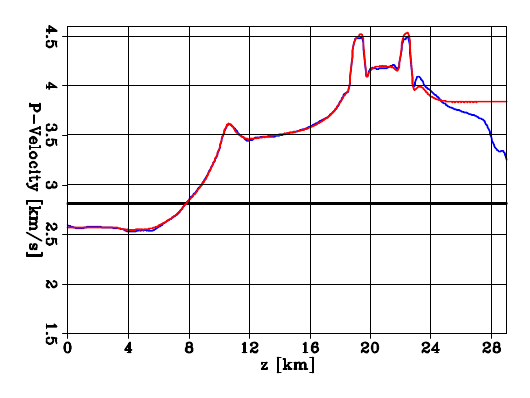}} \hspace{5mm}
    \subfigure[]{\label{fig:bpMod_mod_1d_z4}\includegraphics[width=0.45\columnwidth]{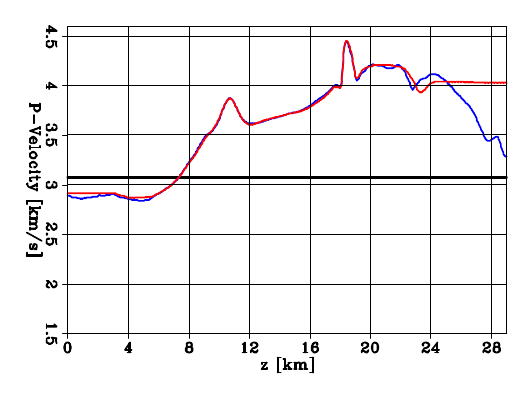}} 
    \caption{Horizontal velocity profiles extracted at (a) $z=2$ km, (b) $z=3$ km, (c) $x=3.5$ km, and (d) $x=4$ km. The black curve represents the initial model, the red curve is the true model, and the blue curve is the FWIME inverted model.}
    \label{fig:bpCaspian_mod_1d_z}
\end{figure}

\begin{figure}[t]
    \centering
    \subfigure[]{\label{fig:bpCaspian_wav_fwime}\includegraphics[width=0.45\columnwidth]{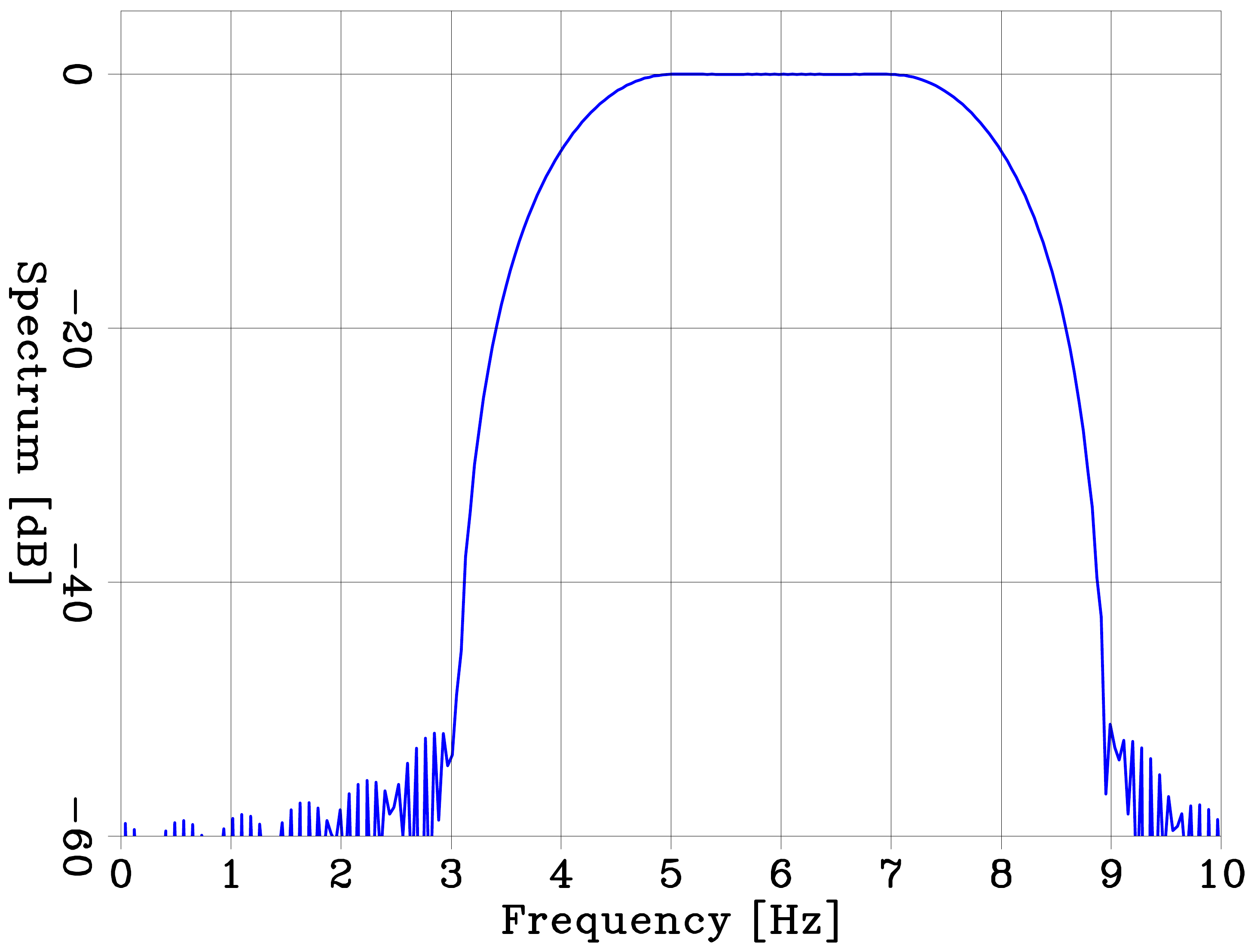}} \hspace{5mm}
    \subfigure[]{\label{fig:bpCaspian_wav_fwi}\includegraphics[width=0.45\columnwidth]{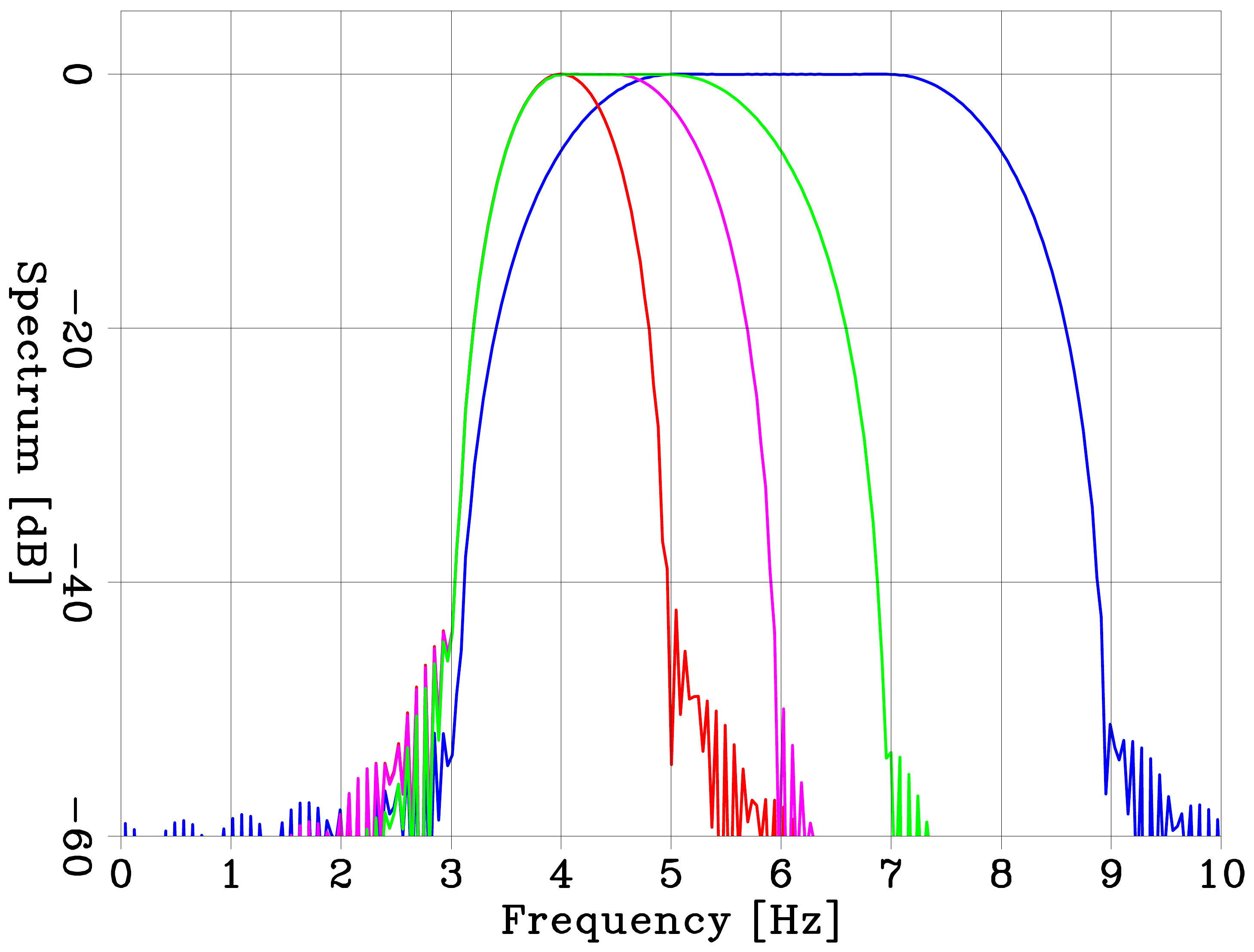}} 
    \caption{Amplitude spectra of the seismic sources employed in this numerical example. (a) Source used for the FWIME workflow. (b) Sequence of sources used for the data-space multi-scale FWI workflow.}
    \label{fig:bpCaspian_wav}
\end{figure}

\begin{figure}[t]
    \centering
    \subfigure[]{\label{fig:bpCaspian_true_data_s0}\includegraphics[width=0.30\columnwidth]{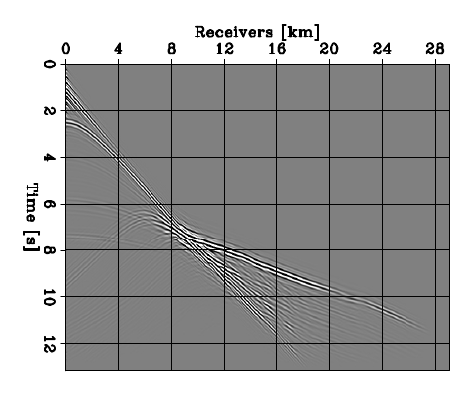}} 
    \subfigure[]{\label{fig:bpCaspian_init_data_s0}\includegraphics[width=0.30\columnwidth]{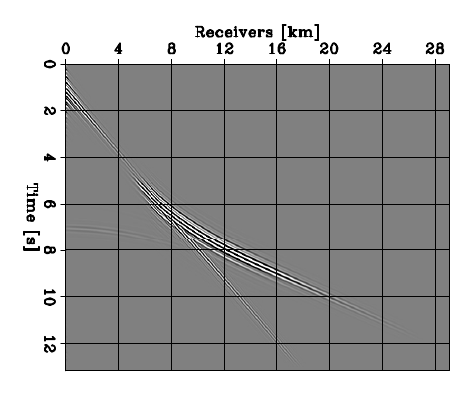}} 
    \subfigure[]{\label{fig:bpCaspian_initDiff_data_s0}\includegraphics[width=0.30\columnwidth]{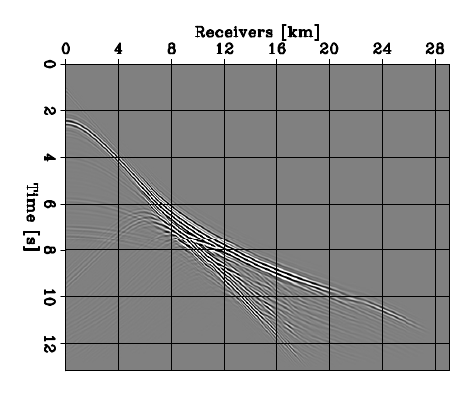}} \\
    \subfigure[]{\label{fig:bpCaspian_true_data_s120}\includegraphics[width=0.30\columnwidth]{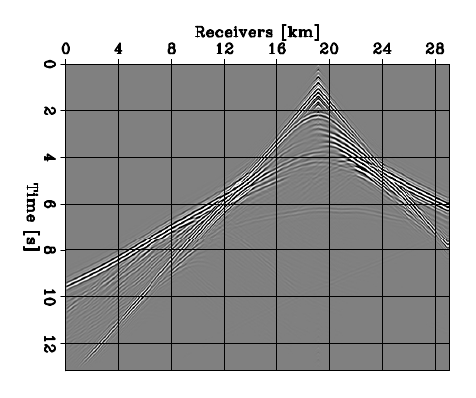}} 
    \subfigure[]{\label{fig:bpCaspian_init_data_s120}\includegraphics[width=0.30\columnwidth]{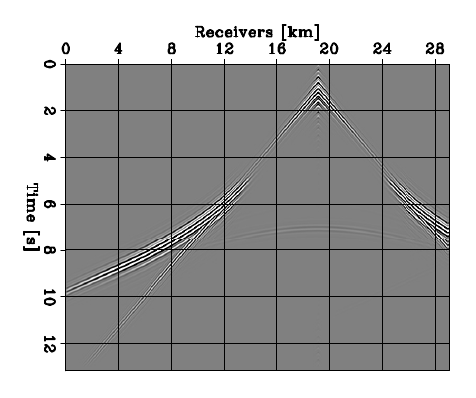}} 
    \subfigure[]{\label{fig:bpCaspian_initDiff_data_s120}\includegraphics[width=0.30\columnwidth]{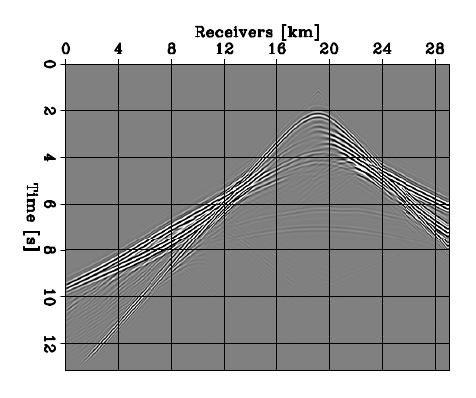}}    
     \caption{Representative shot gathers for sources placed at $x = 0$ km (first row) and $x=19$ km (second row). Observed data, $\mathbf{d}^{obs}$ (first column), predicted data with the initial model, $\mathbf{f}(\mathbf{m}_{init})$ (second column), and initial data-difference, $\Delta \mathbf{d}(\mathbf{m}_{init})=\mathbf{d}^{obs} - \mathbf{f}(\mathbf{m}_{init})$ (third column). All panels are displayed with the same grayscale.}
    \label{fig:bpCaspian_data_init}
\end{figure}

The true model $\mathbf{m}_{true}$ is 29 km-wide and 5.5-km deep, and is shown in Figure~\ref{fig:bpCaspian_true_mod}. The initial velocity model $\mathbf{m}_{init}$ (Figure~\ref{fig:bpCaspian_init_mod}) is horizontally invariant and linearly increasing with depth (below a 1 km-thick water layer), and contains substantial errors as shown by the vertical and horizontal velocity profiles in Figures~\ref{fig:bpCaspian_mod_1d_x} and \ref{fig:bpCaspian_mod_1d_z}. In certain regions, the velocity errors between true and initial models are close to 2.0 km/s. We place 182 sources every 160 m, and 728 fixed receivers every 40 m. All acquisition devices are positioned at a depth of 40 m below the water surface. The noise-free data are generated with a source containing energy strictly limited to the 3-9 Hz frequency range (Figure~\ref{fig:bpCaspian_wav_fwime}), and are recorded for 13 s. As shown in Figures~\ref{fig:bpCaspian_data_init}a and \ref{fig:bpCaspian_data_init}d, the data are dominated by diving waves and the initial prediction generates multiple cycle-skipped events (Figure~\ref{fig:bpCaspian_data_init}c and \ref{fig:bpCaspian_data_init}f).  
This example tests FWIME's ability to simultaneously recover model features of various scales. The first difficulty is to obtain the correct velocity trend (low-resolution component) in the deeper regions of the model. The second challenge consists in accurately delineating the high-resolution features, which include the low-velocity zones in the shallow region and two high-velocity anomalies underneath the mud volcano. As expected, conventional multi-scale FWI (using a sequence of four frequency bands whose spectra are shown in Figure~\ref{fig:bpCaspian_wav_fwi}) fails to retrieve a useful solution (Figure~\ref{fig:bpCaspian_fwi_mod}).

\subsection{FWIME}
For the FWIME process, we follow a similar hyper-parameter-tuning analysis as for the Marmousi2 example, and we set $\epsilon=1.75 \times 10^{-5}$. In order to account for the large kinematic errors in the initial data prediction, we use a time-lag extension spanning the $[-1.2 \; \rm{s}, \; 1.2 \; \rm{s}]$ interval with 101 points sampled at $\Delta \tau =24$ ms. We use a sequence of 4 spline grids (Table~\ref{table:bpCasian_spline}), and each spline grid refinement is automatically triggered when the stepper is unable to find a step length that decreases the total objective function value. The initial grid is chosen to be very coarse with $\Delta z = 1.0$ km and $\Delta x = 2.4$ km. The spacing in the second and third grids are obtained by halving the spacing from the previous ones. The final spline grid coincides with the finite-difference grid ($\Delta z = \Delta x = 40$ m). For the first, second and third grid, we use a finer spatial sampling of 100 m in the vicinity of the water bottom to account for the sharp interface. However, the bathymetry is not assumed to be precisely known. 

\begin{table}[h!]
\centering
\begin{tabular}{ |c|c|c|c|  } 
\hline
\textbf{Grid number} & $\Delta z$ [km] & $\Delta x$ [km]\\
\hline
 0 & 0.9 & 2.4 \\
 1 & 0.6 & 1.2 \\
 2 & 0.2 & 0.28 \\
 3 & 0.04 & 0.04 \\
\hline
\end{tabular}
\caption{Parameters of the spline grid sequence used for the model-space multi-scale FWIME scheme. Spline 3 coincides with the finite-difference grid.}
\label{table:bpCasian_spline}
\end{table}

\begin{figure}[t]
    \centering
    \subfigure[]{\label{fig:bpCaspian_fwime_grad_init}\includegraphics[width=0.45\columnwidth]{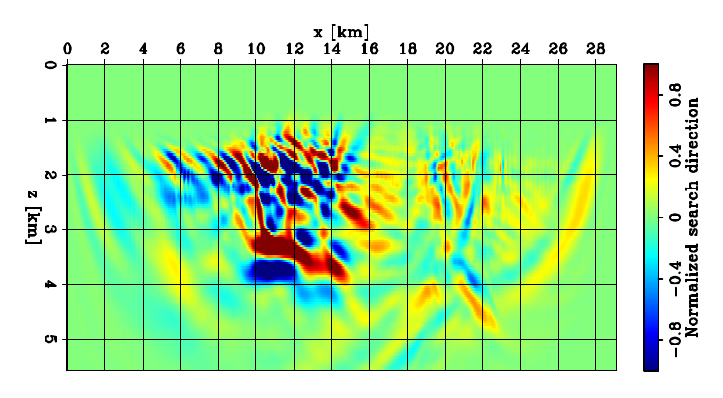}}
    \subfigure[]{\label{fig:bpCaspian_grad_true}\includegraphics[width=0.45\columnwidth]{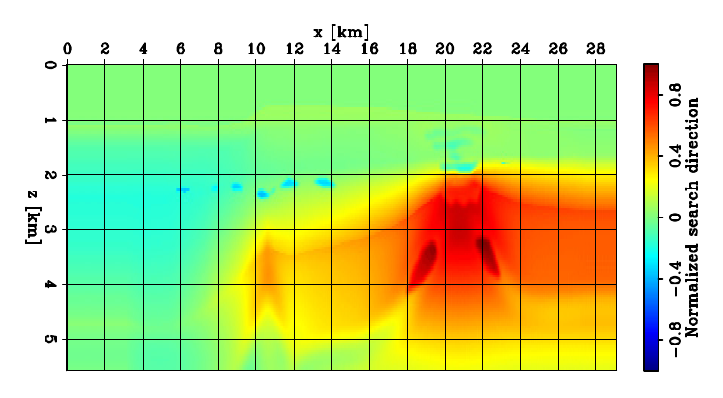}} \\
    \subfigure[]{\label{fig:bpCaspian_fwime_grad_init_spline}\includegraphics[width=0.45\columnwidth]{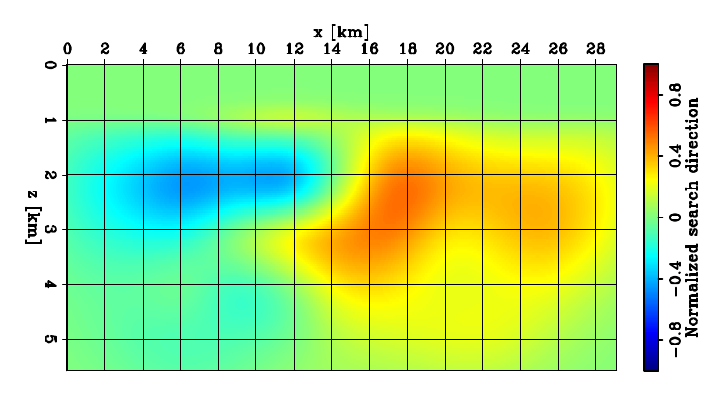}}
    \subfigure[]{\label{fig:bpCaspian_grad_true_spline}\includegraphics[width=0.45\columnwidth]{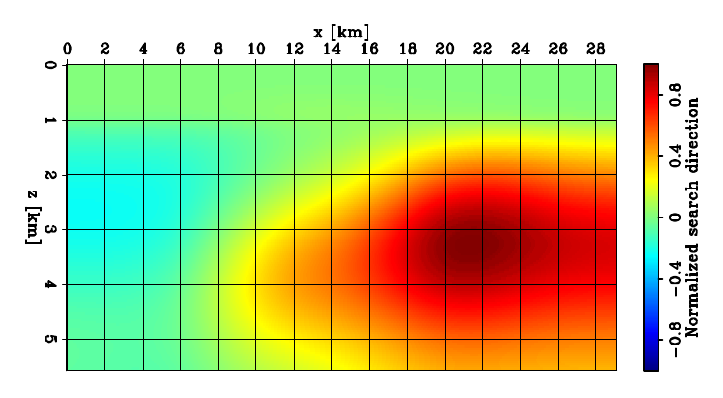}}
    \caption{Normalized initial search directions. (a) FWIME initial search direction $\mathbf{s}_{init}$ before applying any spline parametrization, computed with $\epsilon= 1.75 \times 10^{-5}$. (b) True search direction, $\mathbf{s}_{true}=\mathbf{m}_{true}-\mathbf{m}_{init}$. (c) FWIME initial search direction after its mapping on the initial spline grid (displayed on the finite-difference grid), $\mathbf{s}^{spline}_{init} = \mathbf{S}_0 \mathbf{S}_0^* \mathbf{s}_{init}$. (d) True search direction after its mapping on the initial spline grid (displayed on the finite-difference grid), $\mathbf{s}^{spline}_{true} = \mathbf{S}_0 \mathbf{S}_0^* \mathbf{s}_{true}$.}
    \label{fig:bpCaspian_grad_init}
\end{figure}

The benefit of the spline parametrization can be appreciated by examining the initial search direction (Figure~\ref{fig:bpCaspian_grad_init}). Before its mapping onto the initial spline grid, the FWIME search direction $\mathbf{s}_{init}$ (Figure~\ref{fig:bpCaspian_fwime_grad_init}) contains spurious high-resolution artifacts, which are not present in the true search direction $\mathbf{s}_{true}=\mathbf{m}_{true}-\mathbf{m}_{init}$ (Figure~\ref{fig:bpCaspian_grad_true}). Figures~\ref{fig:bpCaspian_grad_init}c and \ref{fig:bpCaspian_grad_init}d show the FWIME and ideal search directions after their mapping onto the first spline grid (i.e., after applying $\mathbf{S}_0 \mathbf{S}_0^*$ to $\mathbf{s}_{init}$ and to $\mathbf{s}_{true}$, respectively). The high-wavenumber artifacts are removed and the FWIME search direction is improved. 

\begin{figure}[t]
    \centering
    \subfigure[]{\label{fig:bpCaspian_fwime_init_mod}\includegraphics[width=0.45\columnwidth]{Fig/BP_Caspian/bpEttore40-init-mod.pdf}}
    \subfigure[]{\label{fig:bpCaspian_fwime_s0_mod}\includegraphics[width=0.45\columnwidth]{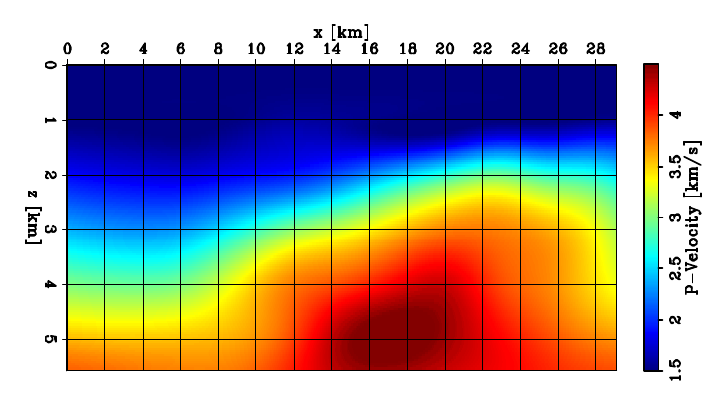}} \\
    \subfigure[]{\label{fig:bpCaspian_fwime_s1_mod}\includegraphics[width=0.45\columnwidth]{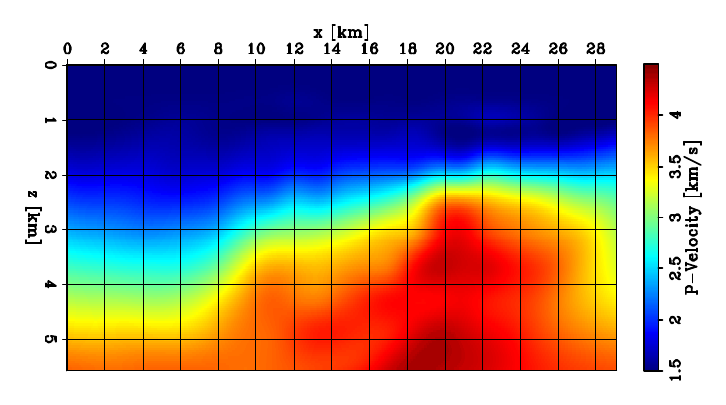}} 
    \subfigure[]{\label{fig:bpCaspian_fwime_s3_mod}\includegraphics[width=0.45\columnwidth]{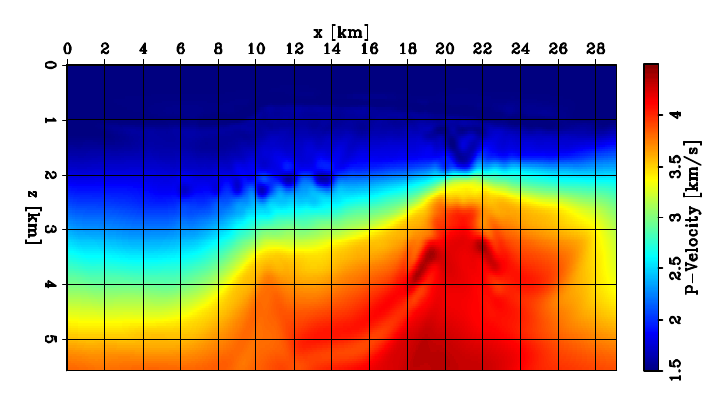}} \\
    \subfigure[]{\label{fig:bpCaspian_fwime_noSpline_mod}\includegraphics[width=0.45\columnwidth]{Fig/BP_Caspian/bpEttore40-noSpline-e5-mod.pdf}}                
    \subfigure[]{\label{fig:bpCaspian_fwime_true_mod}\includegraphics[width=0.45\columnwidth]{Fig/BP_Caspian/bpEttore40-true-mod.pdf}}            
    \caption{Inverted models at different stages of the model-space multi-scale FWIME workflow with $\epsilon = 1.75 \times 10^{-5}$. (a) Initial model. (b) Inverted model after 77 iterations on the first spline grid. (c) Inverted model after 60 iterations on the second spline grid. (d) Inverted model after 33 iterations on the third spline grid. (e) Inverted model after 124 iterations on the fourth (finite-difference) grid. (f) True model. The total number of iterations used to obtained panel \label{fig:bpCaspian_fwime_noSpline_mod} is 217.}
    \label{fig:bpCaspian_fwime_mod_recap}
\end{figure}

Figure~\ref{fig:bpCaspian_fwime_mod_recap} shows the sequence of FWIME inverted models throughout the model-space multi-scale process (for clarity purposes, the models inverted on $\mathbf{S}_0$, $\mathbf{S}_1$, and $\mathbf{S}_2$ are displayed on the finite-difference grid). The solutions after the first and second grids (Figures~\ref{fig:bpCaspian_fwime_mod_recap}b and \ref{fig:bpCaspian_fwime_mod_recap}c) show an accurate recovery of the long-wavelength components (the velocity trend) that were completely missing in the initial model. Moreover, in a test not shown here, we verified that the inverted models from both the second and third grids were accurate enough for conventional FWI to retrieve a solution very similar to the final FWIME inverted model. Here, we only conduct FWIME until full convergence to illustrate its potential. The quality of the final FWIME inverted model (obtained after a total of 217 iterations of L-BFGS) is excellent (Figure~\ref{fig:bpCaspian_fwime_true_mod}). This result is also confirmed by observing vertical and horizontal velocity profiles extracted at various positions in Figures~\ref{fig:bpCaspian_mod_1d_x} and \ref{fig:bpCaspian_mod_1d_z} (blue curves). 

\begin{figure}[t]
    \centering
    \subfigure[]{\label{fig:bpCaspian_pOpt_cig_init}\includegraphics[width=0.18\columnwidth]{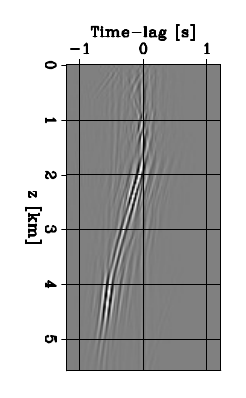}} 
    \subfigure[]{\label{fig:bpCaspian_pOpt_cig_s0}\includegraphics[width=0.18\columnwidth]{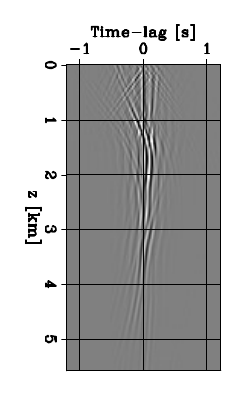}}
    \subfigure[]{\label{fig:bpCaspian_pOpt_cig_s1}\includegraphics[width=0.18\columnwidth]{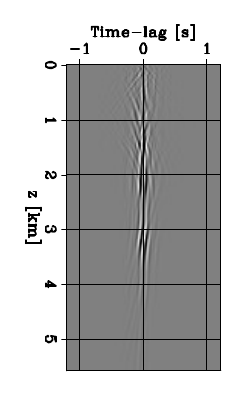}}
    \subfigure[]{\label{fig:bpCaspian_pOpt_cig_s3}\includegraphics[width=0.18\columnwidth]{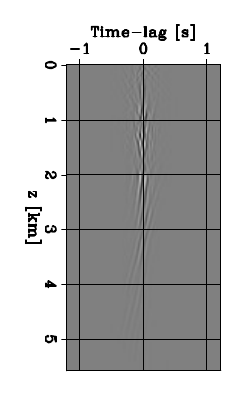}}
    \subfigure[]{\label{fig:bpCaspian_pOpt_cig_final}\includegraphics[width=0.18\columnwidth]{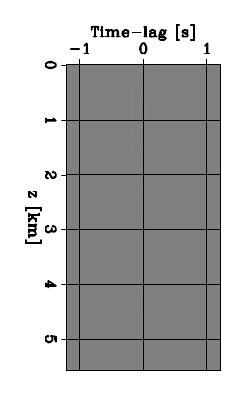}}    
    \caption{Time-lag common image gathers (TLCIG) extracted at $x=22$ km from $\mathbf{p}_{\epsilon}^{opt}$ computed at five stages of the FWIME workflow. (a) Initial step. (b) After inversion on spline 0. (c) After inversion on spline 1. (d)  After inversion on spline 2. (e) Final step. All panels are displayed with the same grayscale.}
    \label{fig:bpCaspian_pOpt_cig}
\end{figure}

\begin{figure}[t]
    \centering
    \subfigure[]{\label{fig:bpCaspian_pOpt_zero_lag_init}\includegraphics[width=0.45\columnwidth]{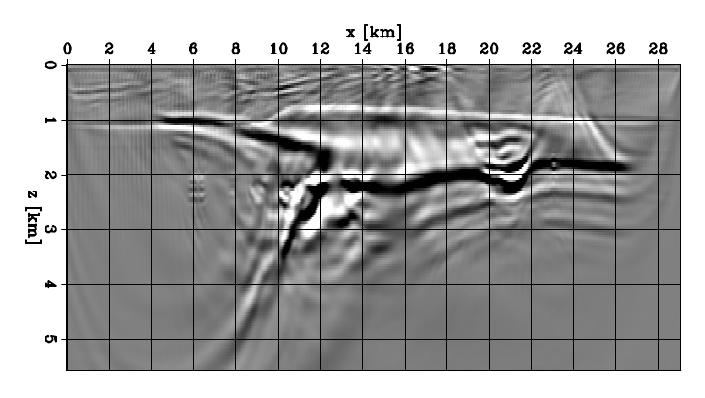}} \hspace{5mm}
    \subfigure[]{\label{fig:bpCaspian_pOpt_zero_lag_s0}\includegraphics[width=0.45\columnwidth]{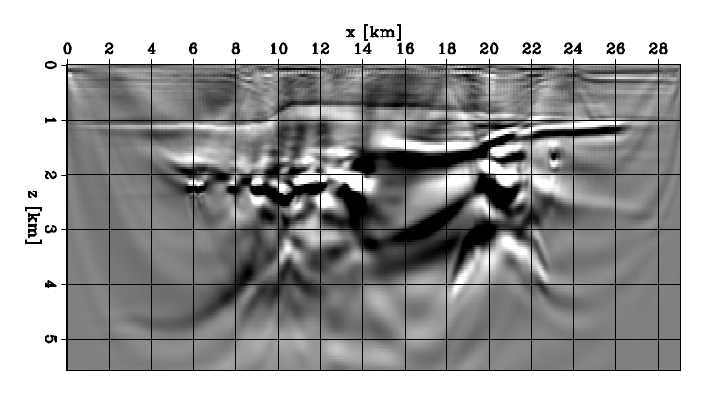}} \\
    \subfigure[]{\label{fig:bpCaspian_pOpt_zero_lag_s1}\includegraphics[width=0.45\columnwidth]{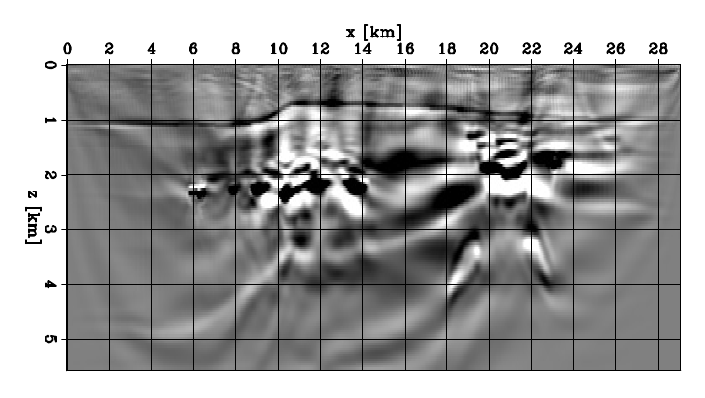}} \hspace{5mm}  
    \subfigure[]{\label{fig:bpCaspian_pOpt_zero_lag_s3}\includegraphics[width=0.45\columnwidth]{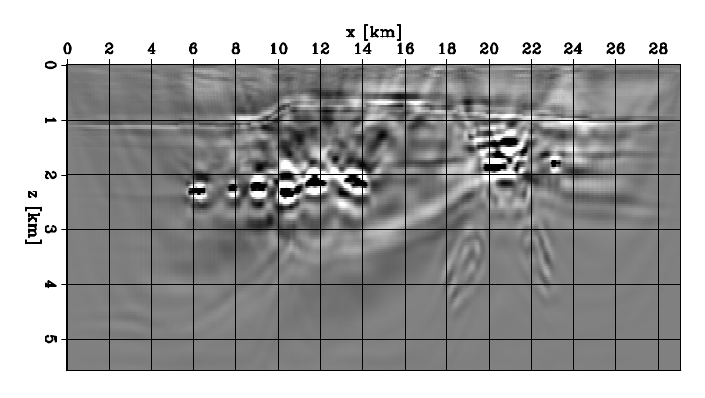}} \\
    \subfigure[]{\label{fig:bpCaspian_pOpt_zero_lag_final}\includegraphics[width=0.45\columnwidth]{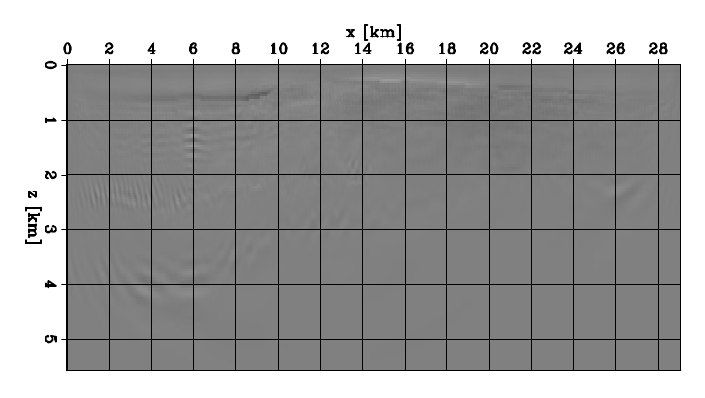}}
    \caption{Zero time-lag sections of $\mathbf{p}_{\epsilon}^{opt}$ computed at five stages of the FWIME workflow. (a) Initial step. (b) After inversion on spline 0. (c) After inversion on spline 1. (d) After inversion on spline 2. (e) Final step. All panels are displayed with the same grayscale.}
    \label{fig:bpCaspian_pOpt_zero_lag}
\end{figure}

We examine the evolution of $\mathbf{\tilde{p}}_{\epsilon}^{opt}$ at five stages of the inversion process. Figure~\ref{fig:bpCaspian_pOpt_cig} shows one TLCIG extracted at $x=22$ km from $\mathbf{\tilde{p}}_{\epsilon}^{opt}$, and Figure~\ref{fig:bpCaspian_pOpt_zero_lag} displays the zero-lag cross-section (the physical plane) of $\mathbf{\tilde{p}}_{\epsilon}^{opt}$. In a similar fashion as for the Marmousi2 example, a considerable amount of energy is initially mapped away from the physical plane and located at negative time lags, confirming that the initial velocity is too low (Figure~\ref{fig:bpCaspian_pOpt_cig_init}). However, notice that the nature of the events (and their moveout) in the extended space is fundamentally different than for the Marmousi2 case (Figure~\ref{fig:Marmousi_pOpt_cig_init}). Here, most of the energy contained within $\mathbf{\tilde{p}}_{\epsilon}^{opt}$ corresponds to the mapping of refracted events (diving waves) present within $\mathbf{d}^{obs}$ that our initial prediction $\mathbf{f}(\mathbf{m}_{init})$ was not able to match. Hence, this example illustrates how FWIME can successfully be applied with the same mechanism regardless of the type of waves present in the data. As the inversion progresses, the FWIME inverted models become more accurate. The energy focuses in the vicinity of the physical plane of $\mathbf{\tilde{p}}_{\epsilon}^{opt}$ (Figures~\ref{fig:bpCaspian_pOpt_cig}b-d) and the events within the physical plane become more coherent (Figures~\ref{fig:bpCaspian_pOpt_zero_lag}b-d). Eventually, $\mathbf{\tilde{p}}_{\epsilon}^{opt}$ completely vanishes (Figures~\ref{fig:bpCaspian_pOpt_cig_final} and \ref{fig:bpCaspian_pOpt_zero_lag_final}). 

%% file: syncline.tex
\begin{figure}[t]
    \centering
    \subfigure[]{\label{fig:Virieux_init_mod}\includegraphics[width=0.45\columnwidth]{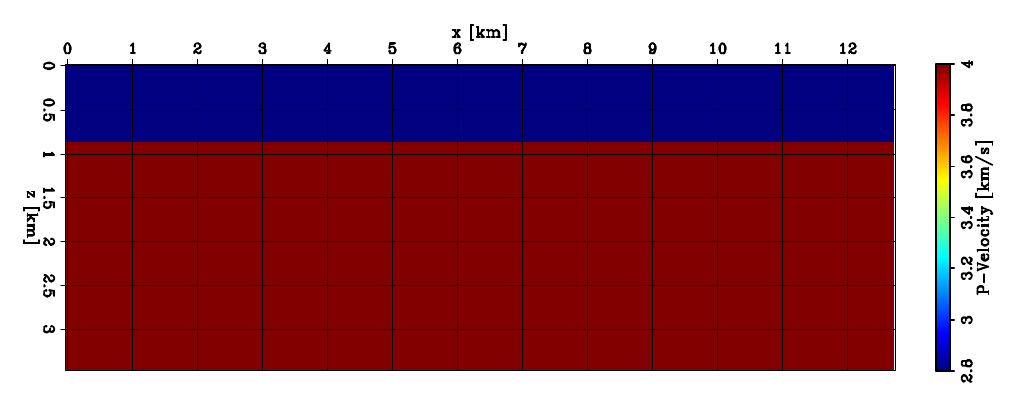}}
    \subfigure[]{\label{fig:Virieux_fwi_mod}\includegraphics[width=0.45\columnwidth]{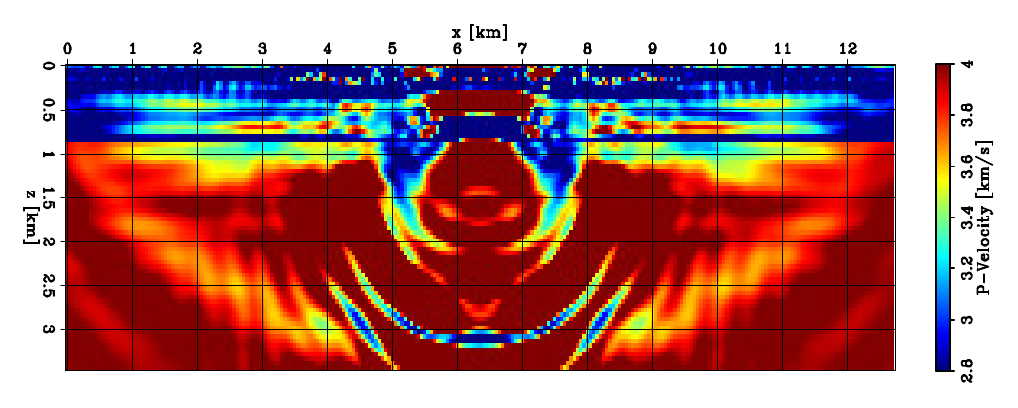}} \\
    \subfigure[]{\label{fig:Virieux_fwime_mod}\includegraphics[width=0.45\columnwidth]{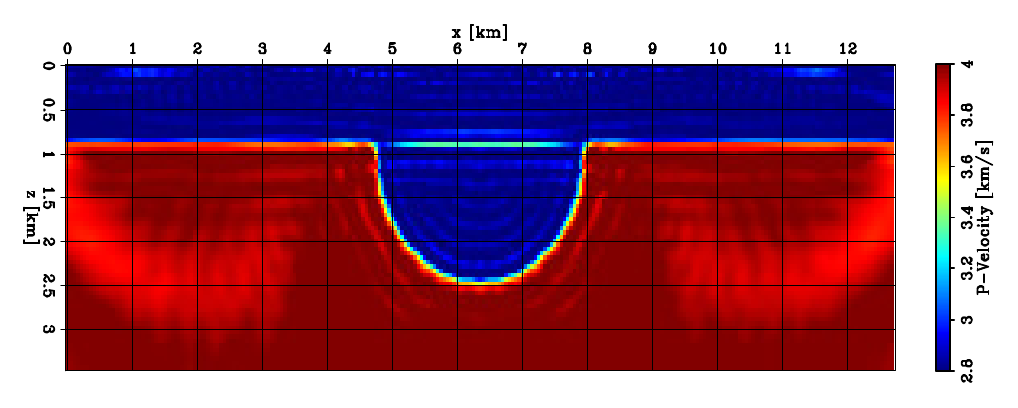}}     
    \subfigure[]{\label{fig:Virieux_true_mod}\includegraphics[width=0.45\columnwidth]{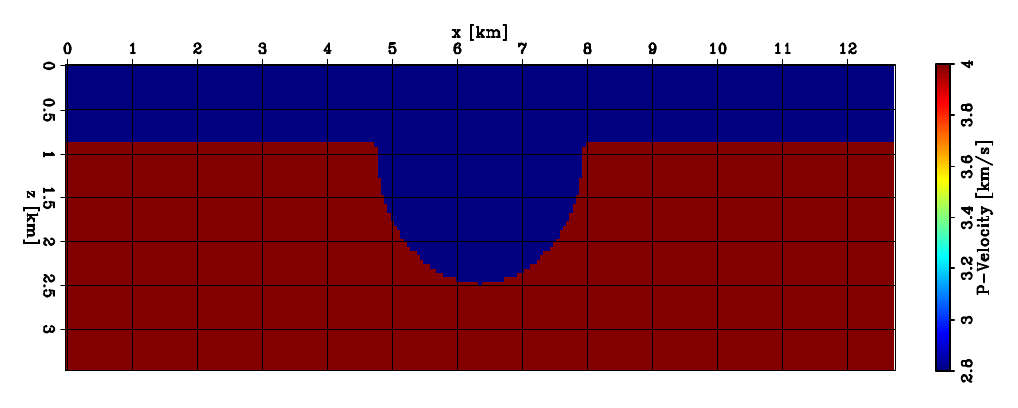}}    
    \caption{2D panels of velocity models. (a) Initial model. (b) Inverted model after conventional data-space multi-scale FWI. (c) Final FWIME inverted model. (d) True model.}
    \label{fig:Virieux_mod}
\end{figure}

We apply FWIME on a synthetic example designed and shared by the Seiscope consortium. The true velocity model is shown in Figure~\ref{fig:Virieux_true_mod}. The model is approximately 12 km wide and 3.5 km deep. It is composed of two horizontal layers with a low-velocity synclinal inclusion (basin) embedded in the second (deeper) layer. The initial velocity (Figure~\ref{fig:Virieux_init_mod}) is identical to the true model but does not contain the synclinal basin. The velocity values are set to $v_s = 2.8$ km/s and $v_d = 4.0$ km/s in the top and bottom layers, respectively. The difference between the two velocity models is shown in Figure~\ref{fig:Virieux_gradient_no_spline_true}.

\begin{figure}[t]
    \centering
    \includegraphics[width=0.45\columnwidth]{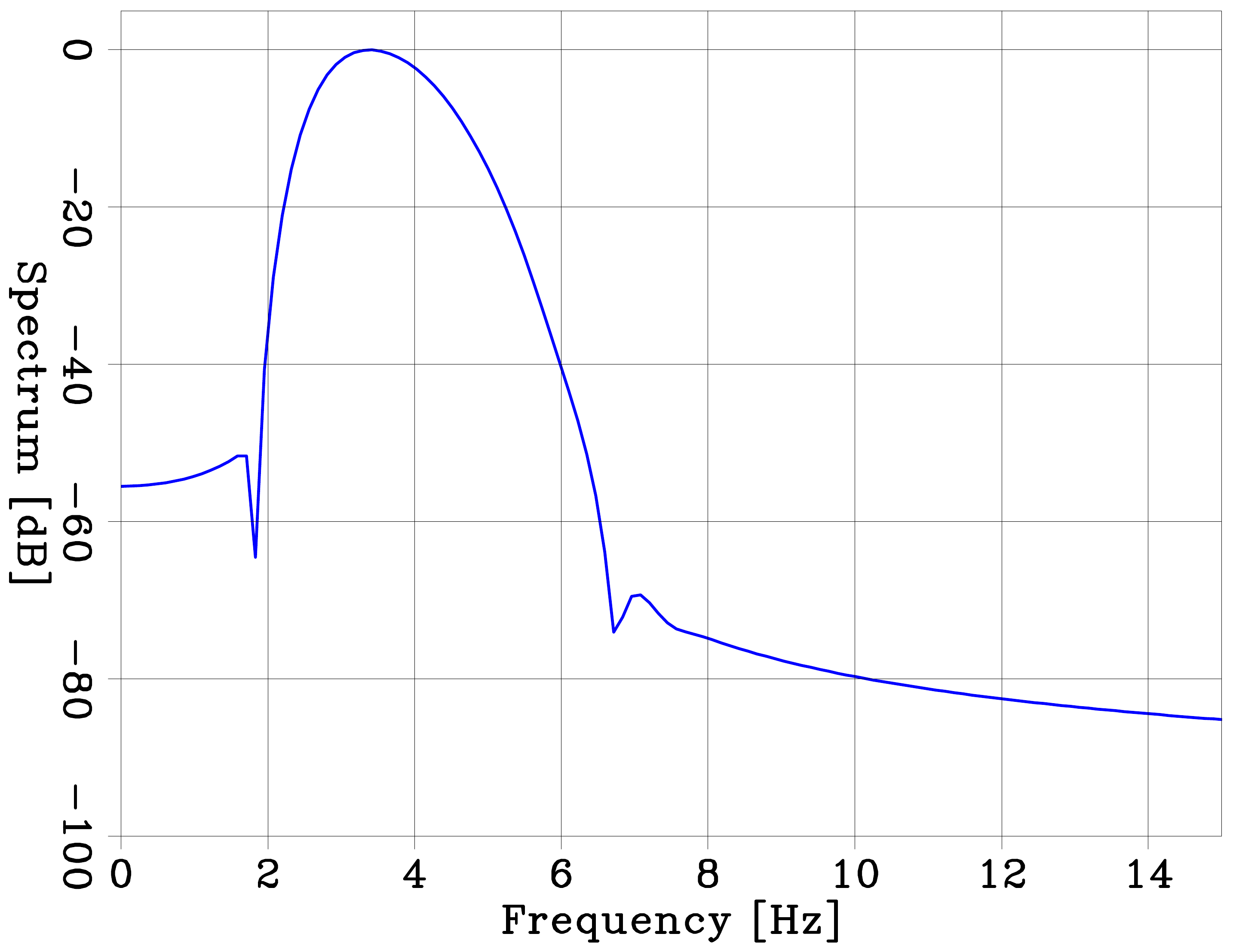}
    \caption{Amplitude spectrum of the seismic source used to generate the dataset for this numerical example.}
    \label{fig:Virieux_wav}
\end{figure}

We generate a noise-free dataset with a finite-difference numerical scheme using a grid spacing of 50 m in both directions and a band-passed Ricker wavelet containing energy restricted to the 1.5-6.5 Hz frequency range (Figure~\ref{fig:Virieux_wav}). We place 48 shots spaced every 250 m and 255 receivers every 50 m. All acquisition devices are placed at a constant depth of 50 m. Figure~\ref{fig:Virieux_shot_gather} shows two representative shot gathers for sources located at $x = 0.3$ km and $x = 6.3$ km.

\subsection{Goal of the experiment}
\begin{figure}[t]
    \centering
    \subfigure[]{\label{fig:Virieux_data_shot1}\includegraphics[width=0.45\columnwidth]{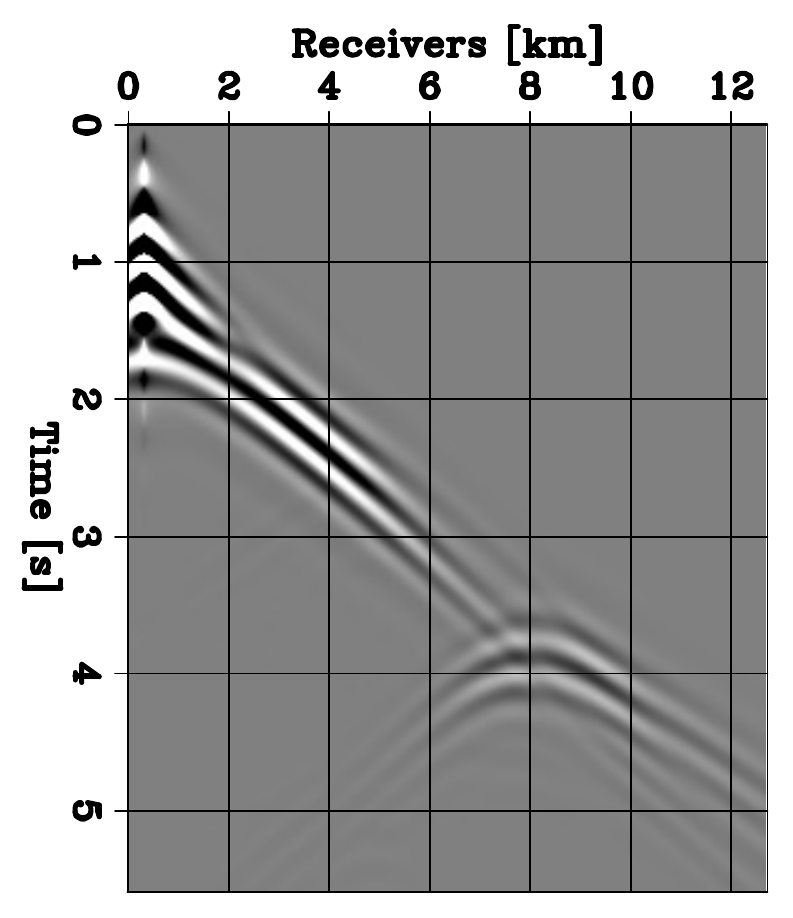}} \hspace{5mm}
    \subfigure[]{\label{fig:Virieux_data_shot2}\includegraphics[width=0.45\columnwidth]{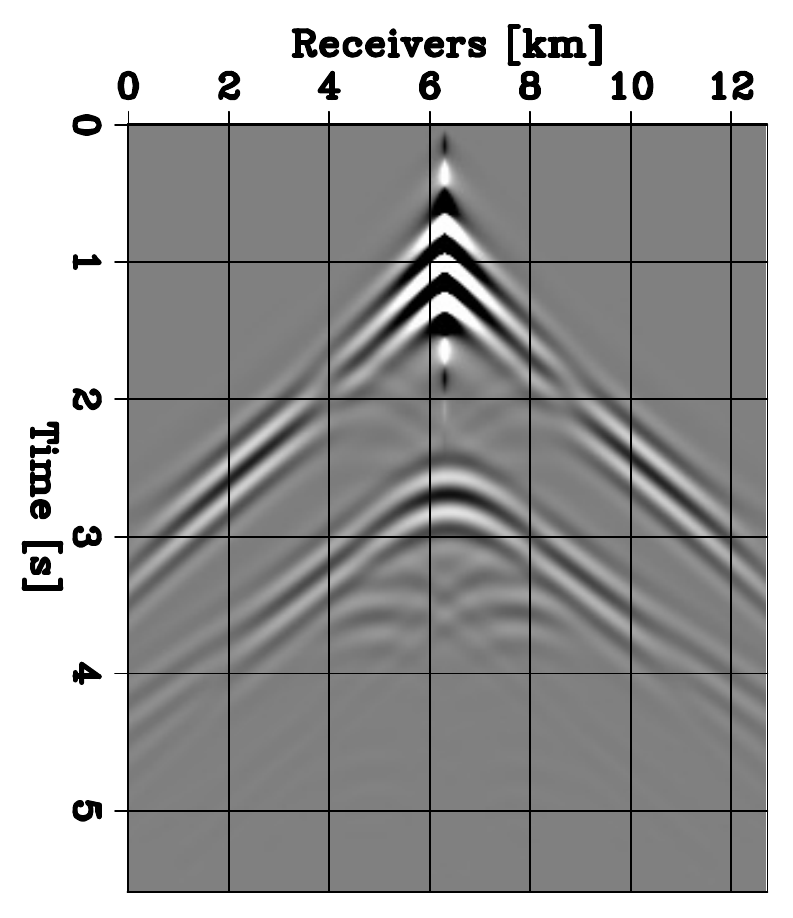}} \hspace{5mm}
    \caption{Representative shot gathers generated by a seismic source containing energy restricted to the 1.5-6 Hz frequency range. (a) Source located at $x = 0.3$ km. (b) Source located at $x = 6.3$ km. All panels are displayed with the same grayscale.}
    \label{fig:Virieux_shot_gather}
\end{figure}

We test the robustness of FWIME against the ill-posedness coming from wrong association of predicted and complex observed waveforms (J. Virieux, personal communication, 2019). In the proposed model, the basin curvature is such that the reflected events from the bottom of the basin generate wavefield triplications in the data, which overlap with the reflections from the shallow interface between the two layers. The complexity of the recorded waveforms can be observed in the shot gathers displayed in Figure~\ref{fig:Virieux_shot_gather}. Moreover, the initial model contains a vast region (relative to the dimensions of the features of interest) with mispositioned sharp interfaces and strong velocity contrasts with its surroundings, which represents a similar (though simpler) scenario as the one encountered when delineating complex overburdens/geobodies, such as salt bodies. 

\subsection{Analyzing the events in the data}
\begin{figure}[t]
    \centering
    \subfigure[]{\label{fig:Virieux_data_shot1_high_frequency}\includegraphics[width=0.45\columnwidth]{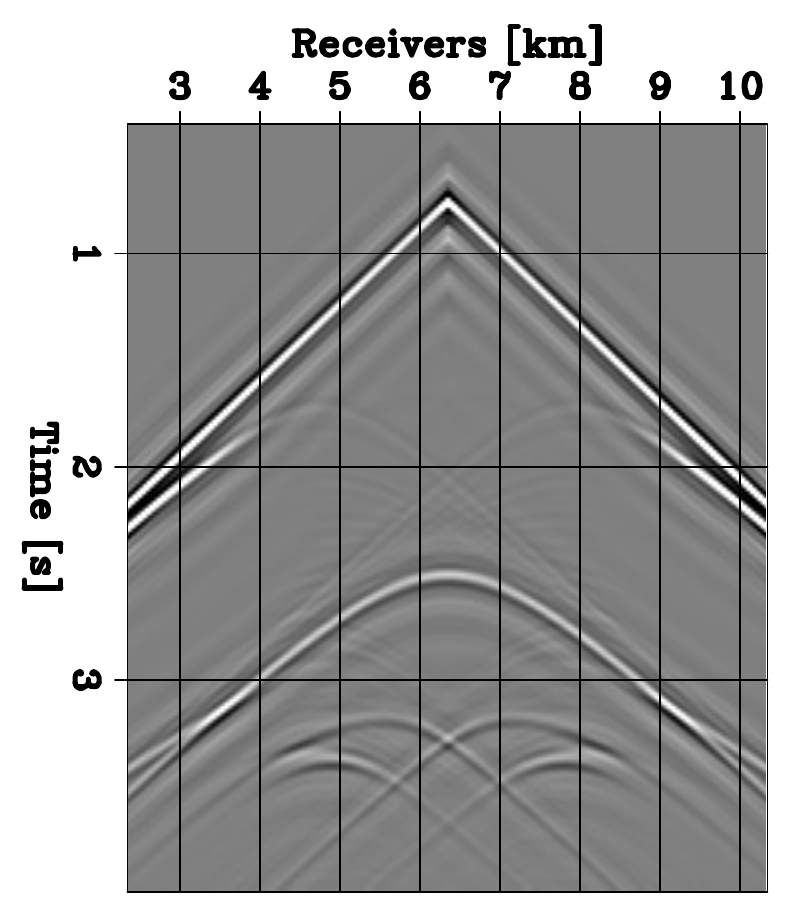}} \hspace{5mm}
    \subfigure[]{\label{fig:Virieux_data_shot1_annotated}\includegraphics[width=0.45\columnwidth]{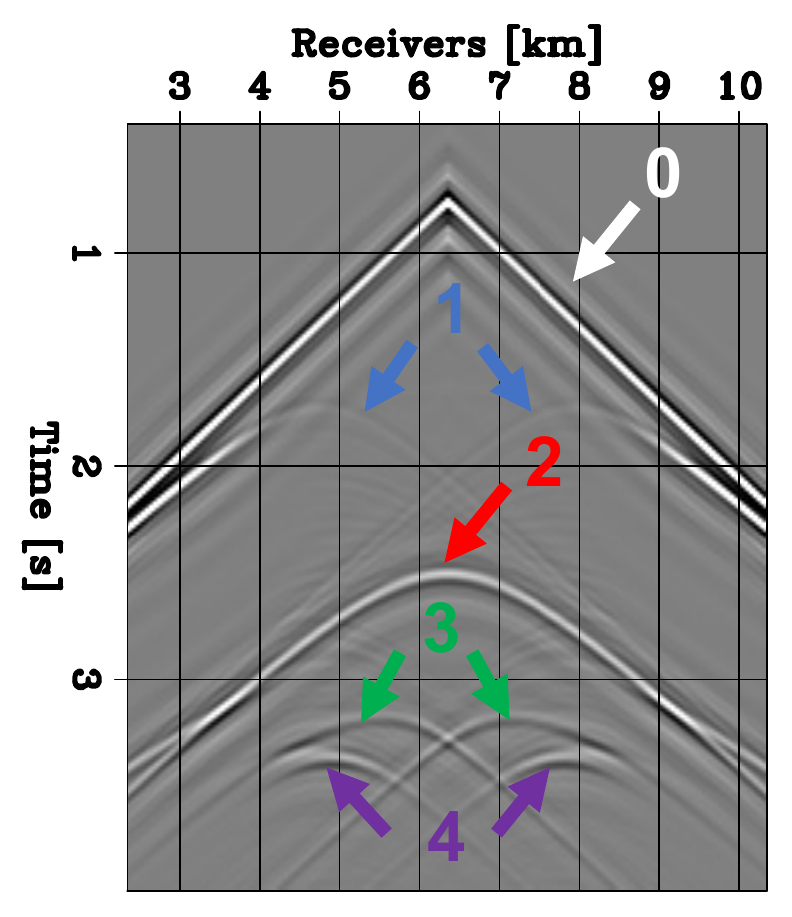}}
    \caption{(a) Shot gather for a source located at x = 6.3 km containing energy within the 2-30 Hz frequency range. (b) Identical shot gather as (a) with labeled seismic events. All panels are displayed with the same grayscale.}
    \label{fig:Virieux_data_high_frequency}
\end{figure}

\begin{figure}[t]
    \centering
    \subfigure[]{\label{fig:Virieux_wave1}\includegraphics[width=0.35\columnwidth]{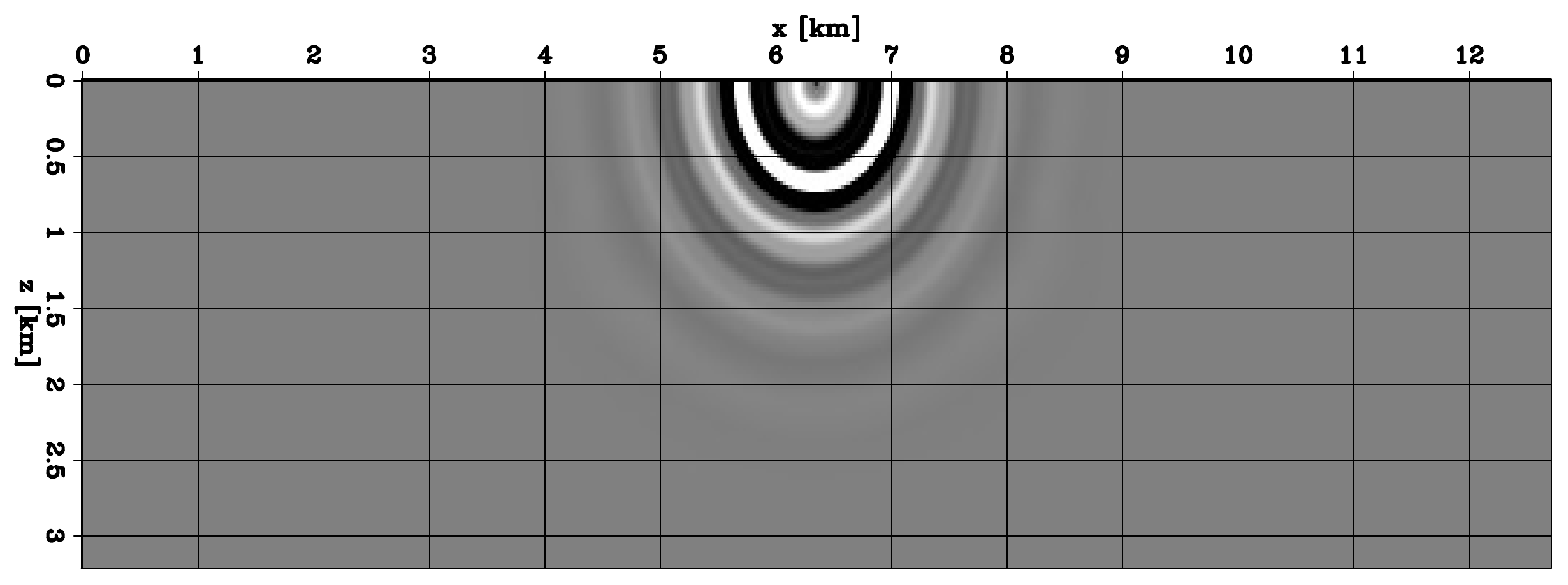}} \hspace{5mm}
    \subfigure[]{\label{fig:Virieux_wave2}\includegraphics[width=0.35\columnwidth]{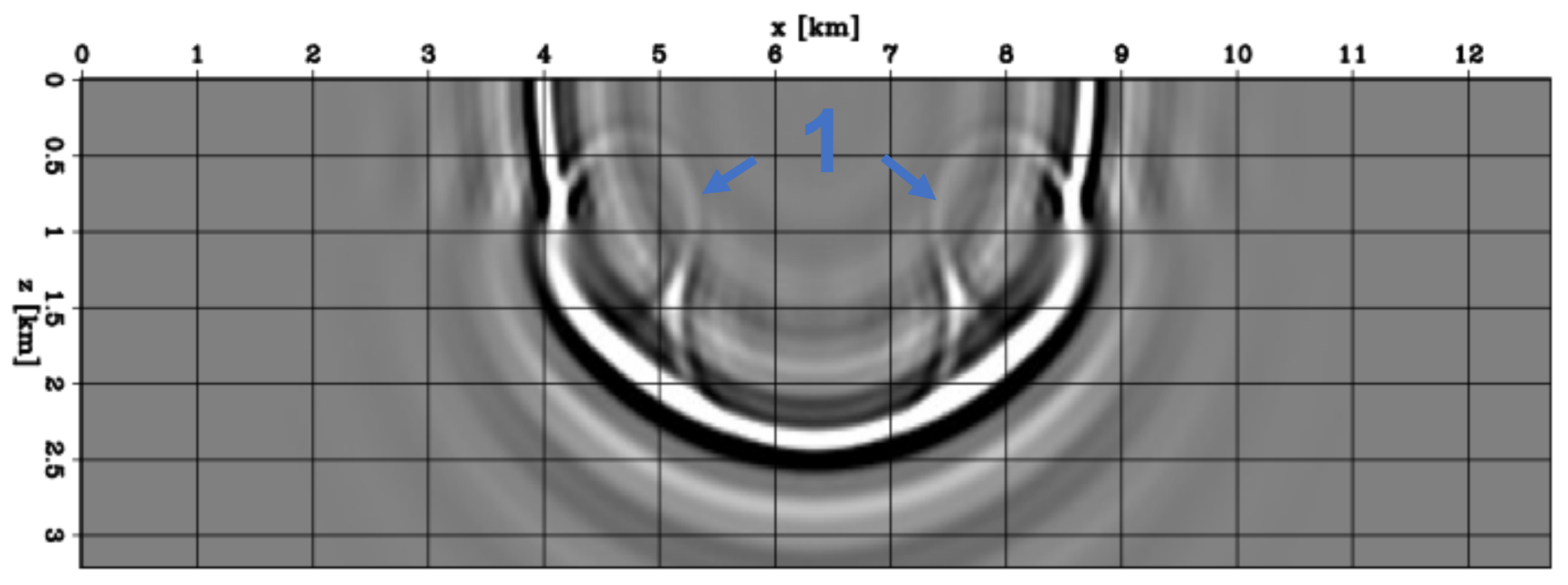}}\\
    \subfigure[]{\label{fig:Virieux_wave3}\includegraphics[width=0.35\columnwidth]{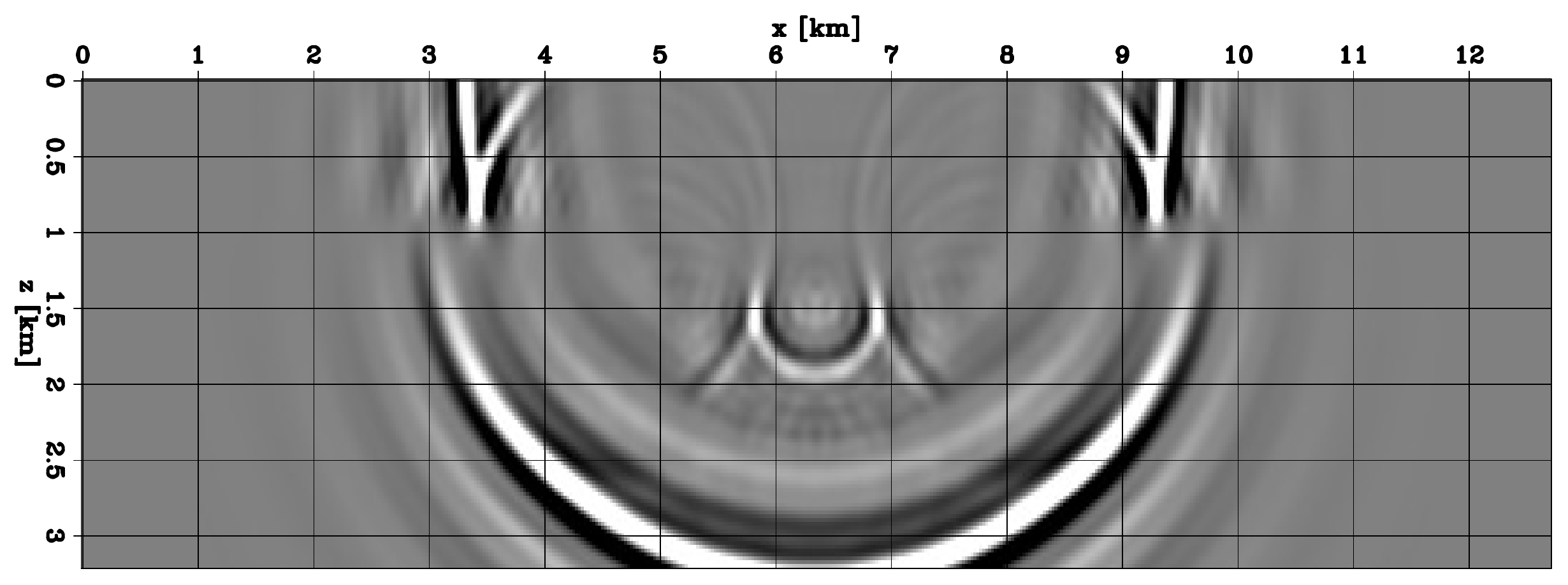}} \hspace{5mm}
    \subfigure[]{\label{fig:Virieux_wave4}\includegraphics[width=0.35\columnwidth]{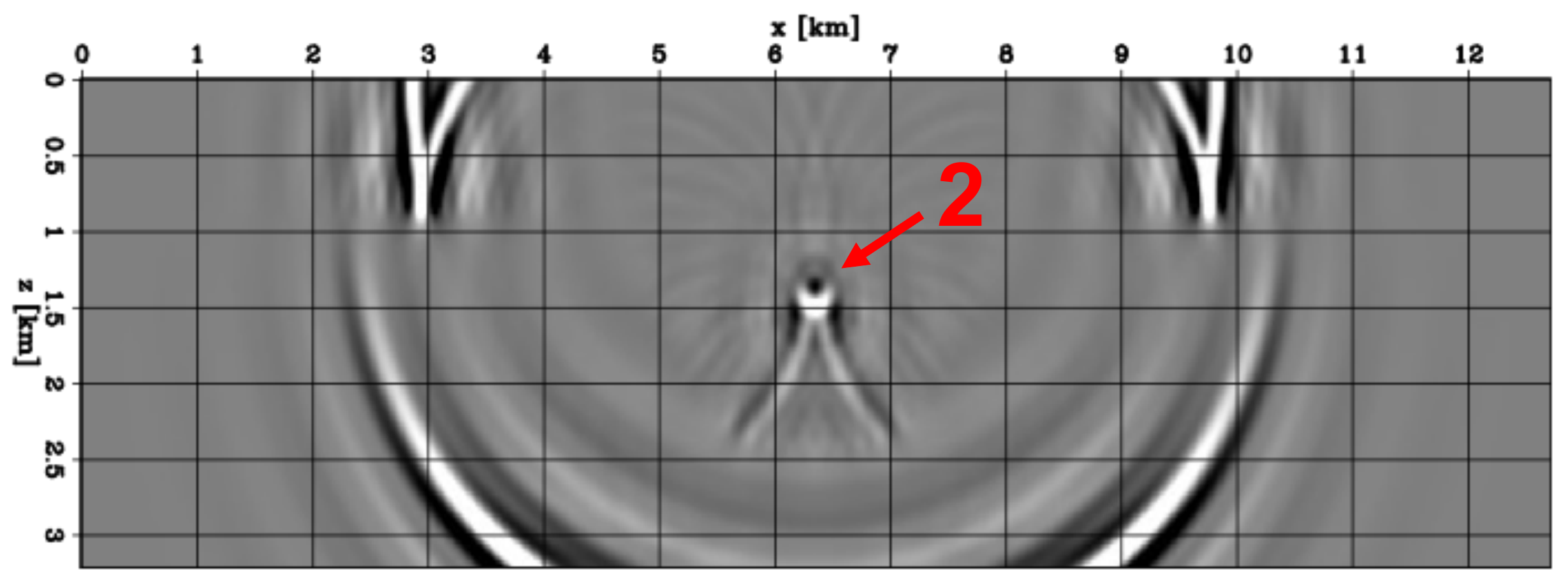}}\\
    \subfigure[]{\label{fig:Virieux_wave5}\includegraphics[width=0.35\columnwidth]{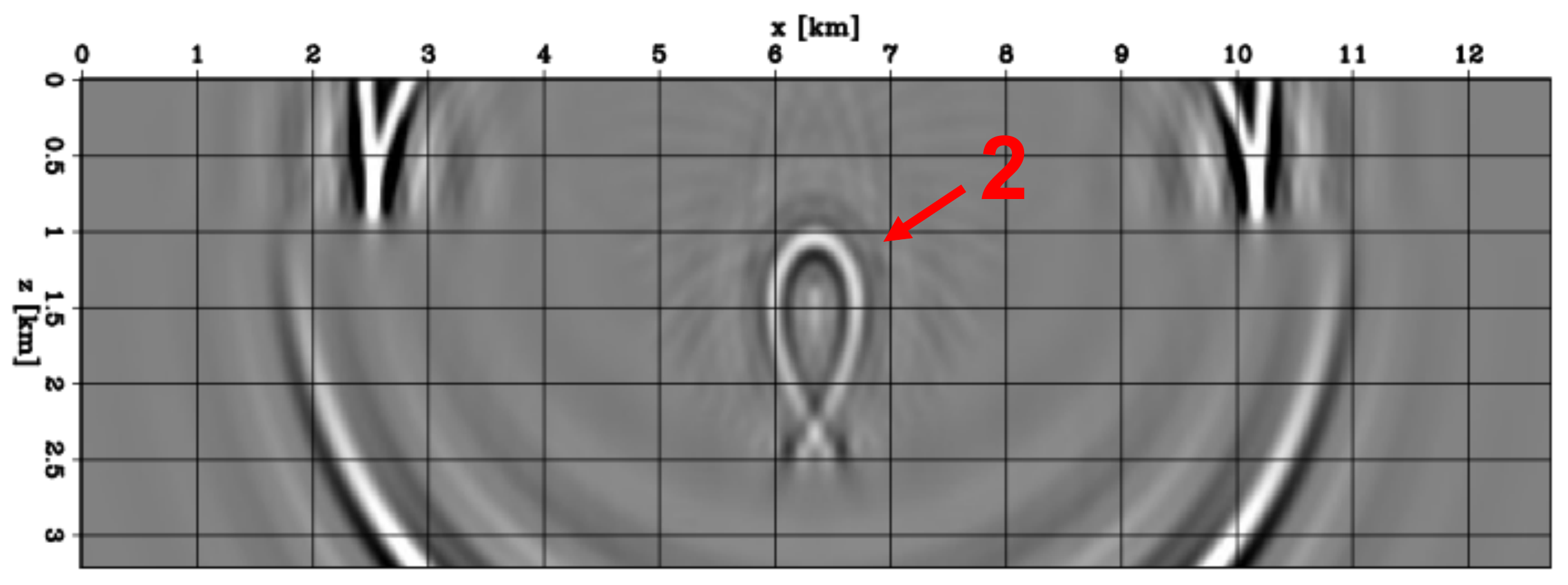}} \hspace{5mm}
    \subfigure[]{\label{fig:Virieux_wave6}\includegraphics[width=0.35\columnwidth]{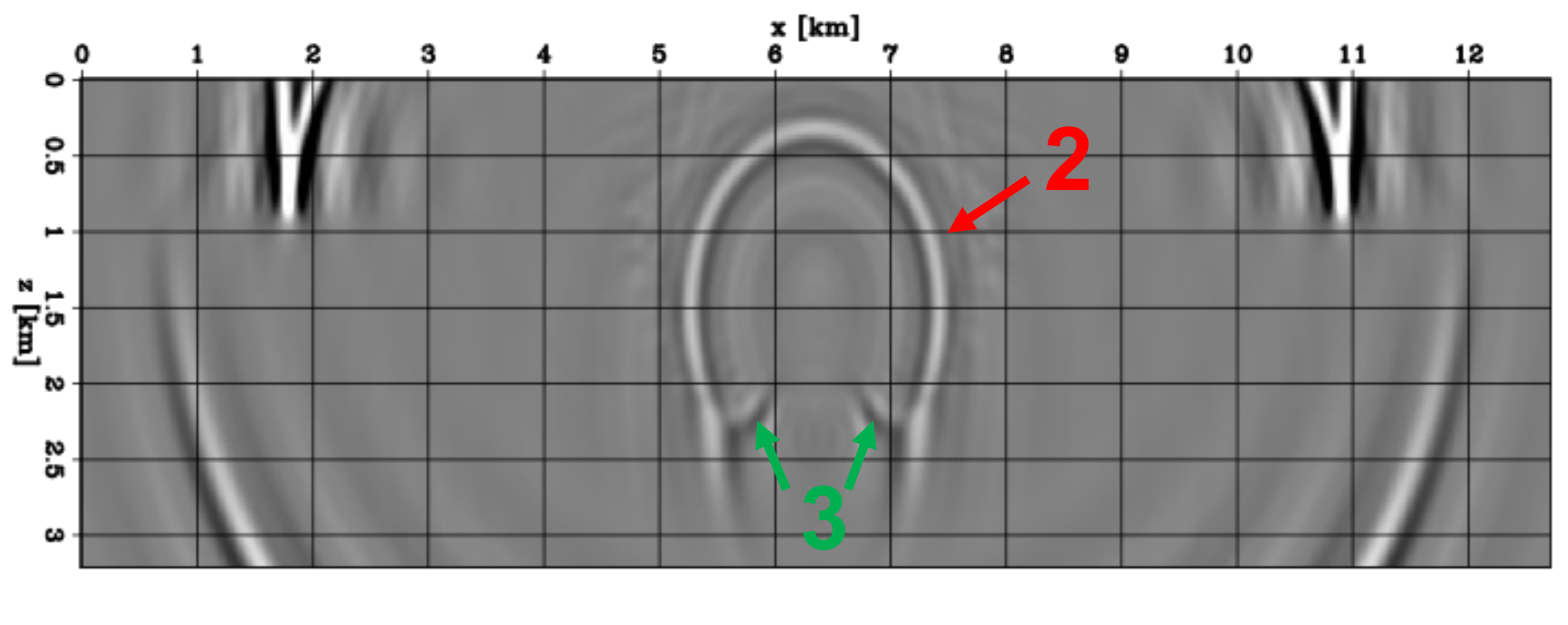}}\\
    \subfigure[]{\label{fig:Virieux_wave7}\includegraphics[width=0.35\columnwidth]{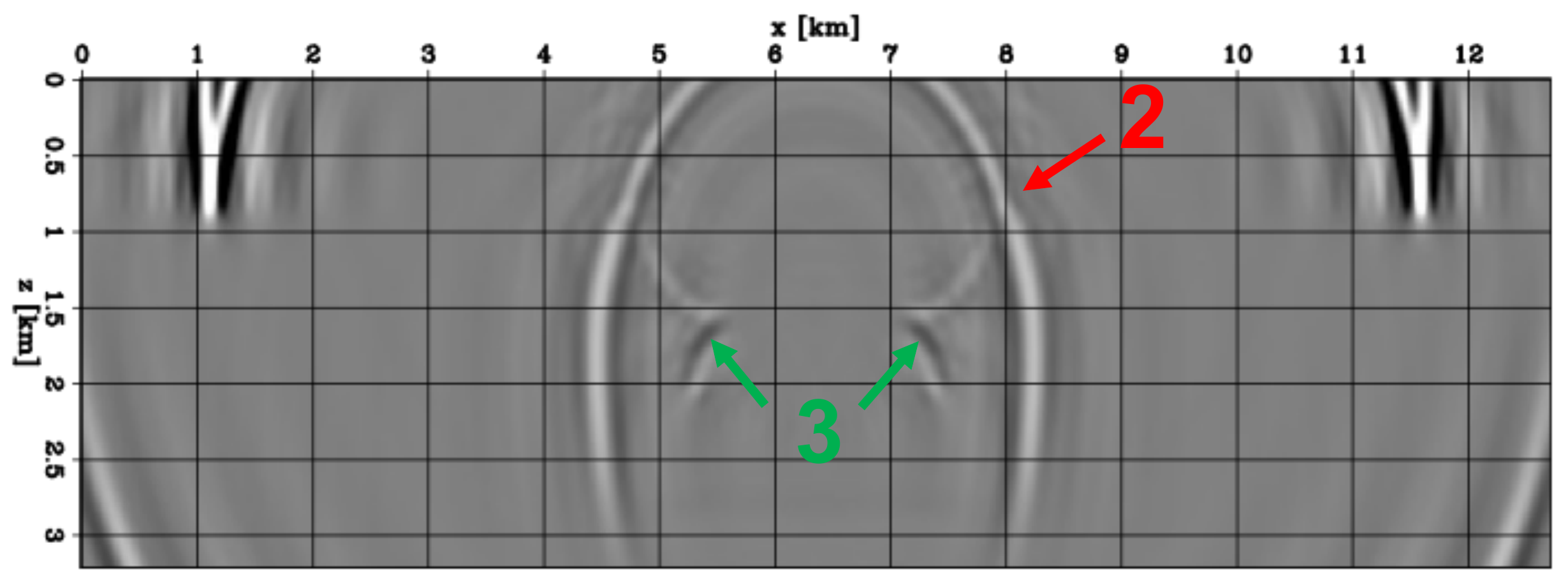}} \hspace{5mm}
    \subfigure[]{\label{fig:Virieux_wave8}\includegraphics[width=0.35\columnwidth]{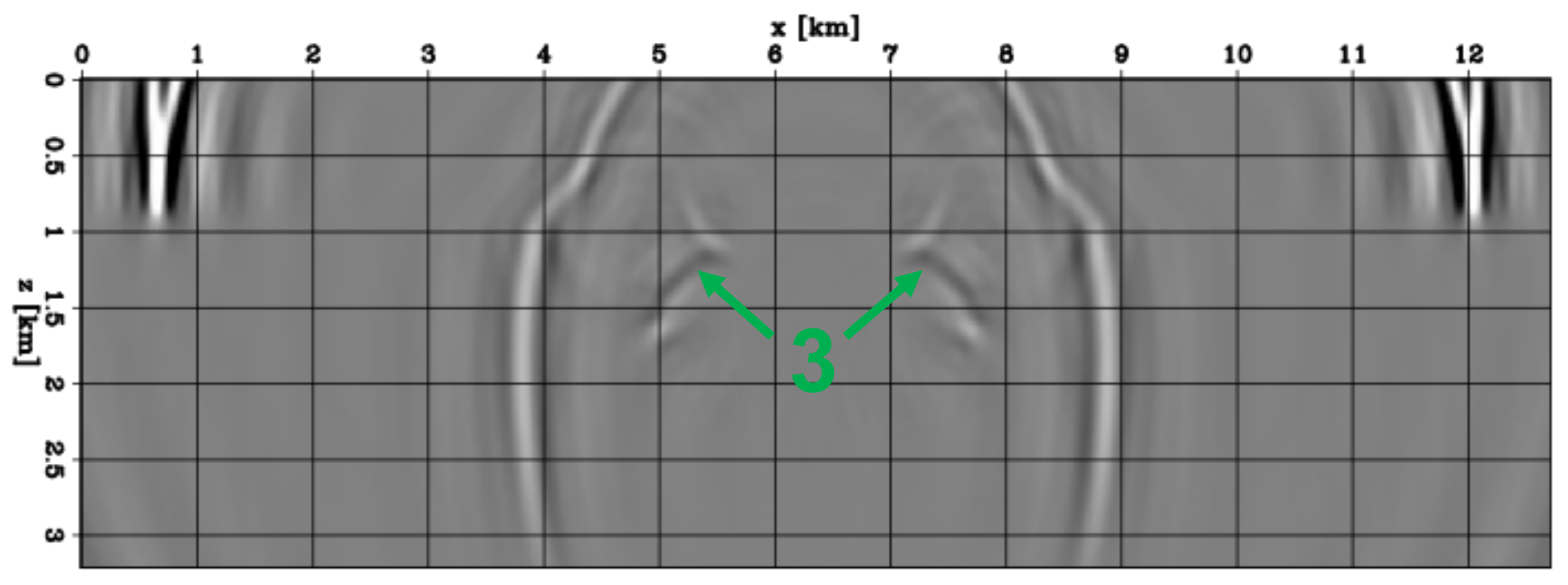}}\\    
    \subfigure[]{\label{fig:Virieux_wave9}\includegraphics[width=0.35\columnwidth]{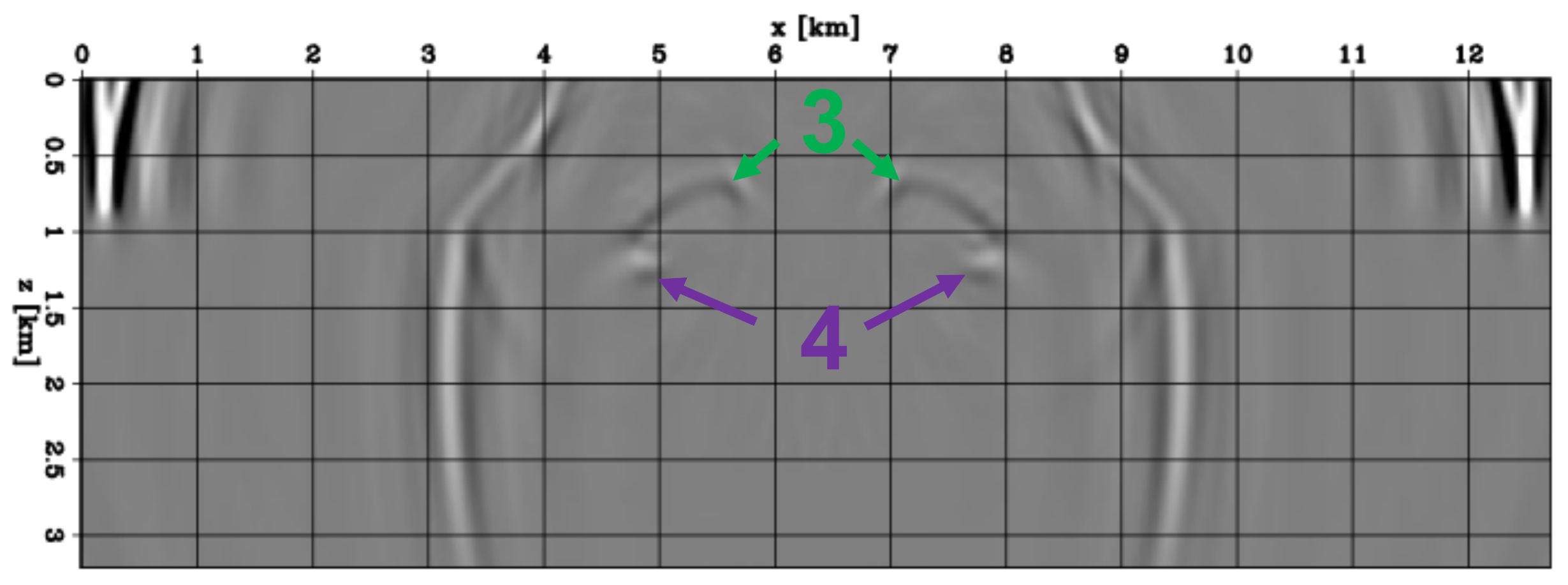}} \hspace{5mm}
    \subfigure[]{\label{fig:Virieux_wave10}\includegraphics[width=0.35\columnwidth]{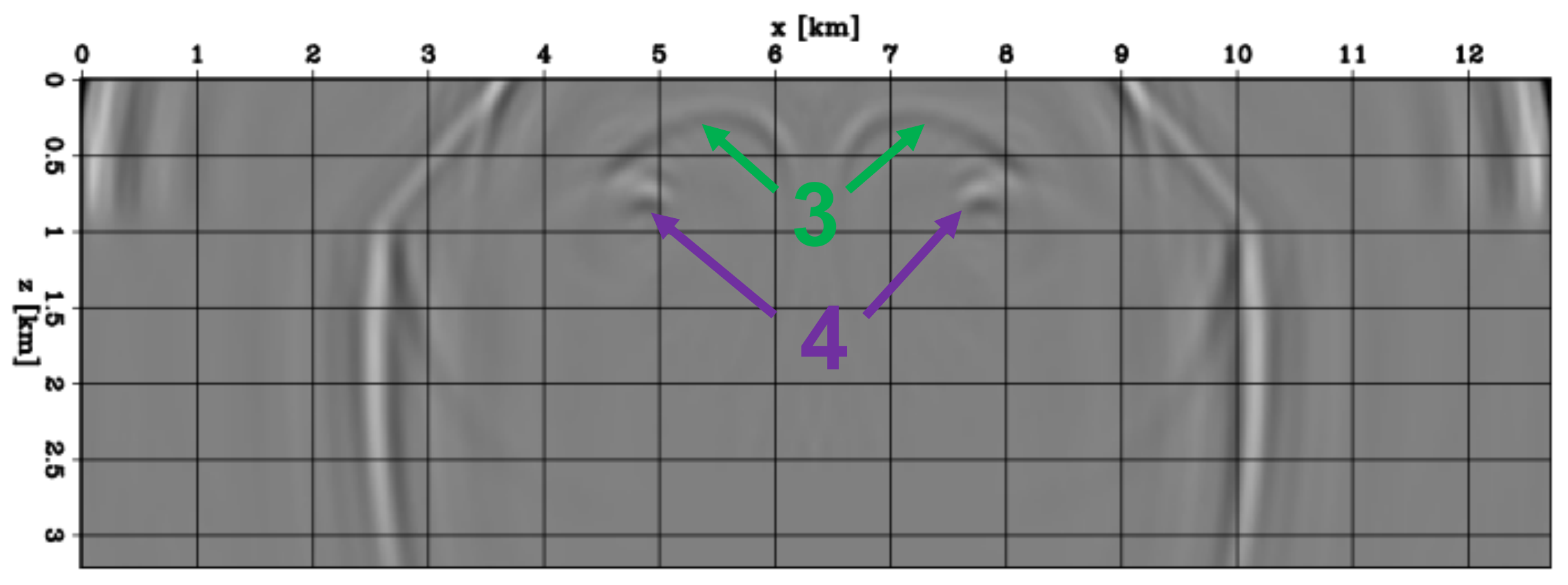}}\\    
    \caption{Wavefield snapshots generated by a source located at $x = 6.35$ km and containing energy within the 2-30 Hz frequency range, giving rise to the shot record shown in Figure~\ref{fig:Virieux_data_high_frequency}. Each panel shows the wavefield at a given time step. All panels are displayed with the same grayscale.}
    \label{fig:Virieux_wavefield}
\end{figure}

To better identify the various events in the data, we generate one shot gather using a seismic source placed at $x = 6.3$ km with a higher frequency content (Figure~\ref{fig:Virieux_data_shot1_high_frequency}). Figure~\ref{fig:Virieux_data_shot1_annotated} shows the same shot gather where five separate events have been identified and labeled with numbers. In order to better understand the nature of the complex recorded waveforms, Figure~\ref{fig:Virieux_wavefield} displays a sequence of snapshots extracted from the wavefield that gave rise to the shot gather shown in Figure~\ref{fig:Virieux_data_high_frequency}. 

\begin{itemize}
    \item The first event (event 0, white arrow in Figure~\ref{fig:Virieux_data_shot1_annotated}) displays a linear moveout and corresponds to the direct arrival of the incident wavefield.
    \item Event 1 (blue arrows in Figures~\ref{fig:Virieux_data_shot1_annotated} and~\ref{fig:Virieux_wave2}) corresponds to the back-scattering of the incident wavefield as it interacts with the sharp corners on the edges of the synclinal basin. The corners act as diffracting points and the recorded event shows a hyperbolic moveout observable in the shot gather.
    \item Event 2 (red arrow in Figures~\ref{fig:Virieux_data_shot1_annotated} and \ref{fig:Virieux_wavefield}d-g) is generated when the incident wavefield reflects from the bottom of the basin. The wavefield later refocuses at the focal point of the syncline (Figure~\ref{fig:Virieux_wave2}), which creates a wavefield similar to the one that would have been generated by a virtual point source located at the focal point (red arrow in Figure~\ref{fig:Virieux_data_shot1_annotated}), directly observable by examining the wavefield snapshots. 
    \item Event 3 (green arrows in Figure~\ref{fig:Virieux_data_shot1_annotated}) is generated by the overlapping of a reflection from the bottom of the synclinal basin with the ``tails" of the incident scattered wavefield after its interaction with the sharp corners (event 1). This event generates two up-going wavefields (event 3) observable in Figures~\ref{fig:Virieux_wavefield}f-k (green arrows).
    \item Event 4 (purple arrows in Figure~\ref{fig:Virieux_data_shot1_annotated}) stems from the diffraction of both up-going wavefields (i.e., event 3) by the corners of the synclinal basin. Their respective wave paths can be observed in Figures~\ref{fig:Virieux_wavefield}i and j.
\end{itemize}

\begin{figure}[t]
    \centering
    \subfigure[]{\label{fig:Virieux_data_true}\includegraphics[width=0.3\columnwidth]{Fig/Syncline/Virieux-data-dipole-s2.pdf}} \hspace{5mm}
    \subfigure[]{\label{fig:Virieux_data_init}\includegraphics[width=0.3\columnwidth]{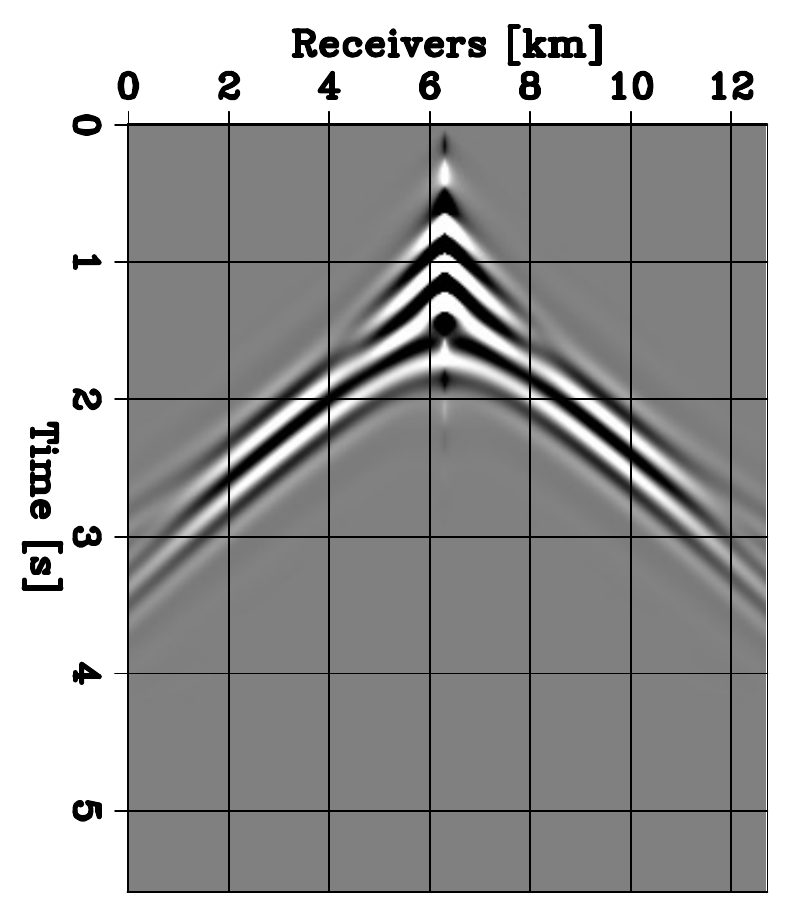}} \hspace{5mm}
    \subfigure[]{\label{fig:Virieux_data_init_diff}\includegraphics[width=0.3\columnwidth]{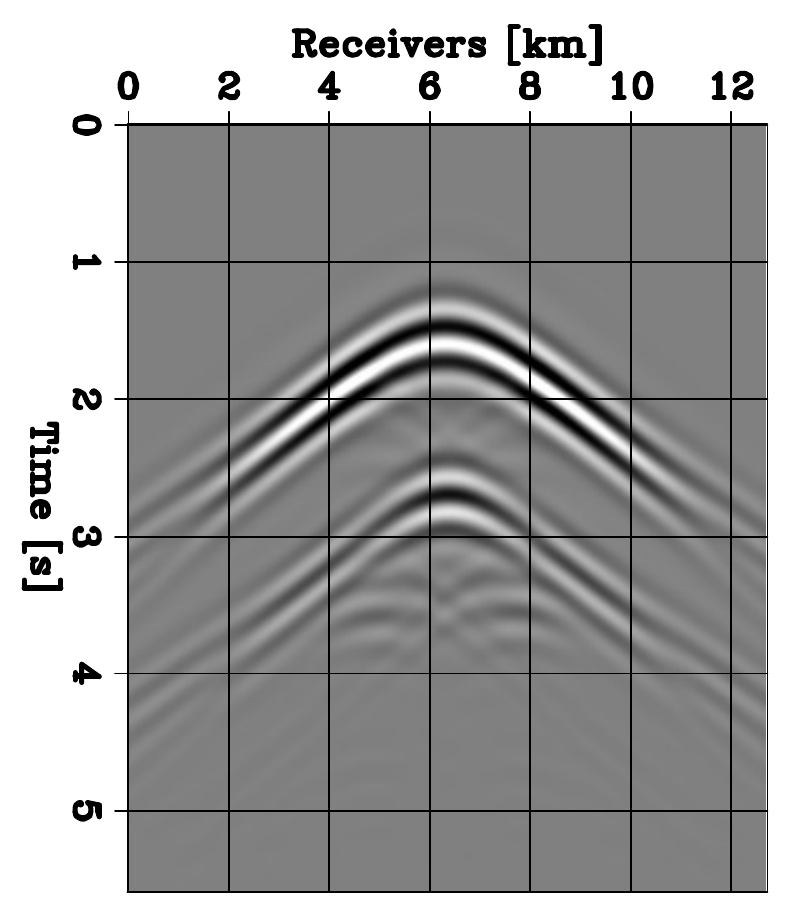}}
    \caption{Shot gather for a source located at $x = 6.3$ km containing energy within the 1.5-6 Hz frequency range. (a) Observed data, $\mathbf{d}^{obs}$. (b) Predicted data with the initial velocity model, $\mathbf{f}(\mathbf{m}_0)$. (c) Initial data difference, $\Delta \mathbf{d}(\mathbf{m}_0) = \mathbf{d}^{obs} - \mathbf{f}(\mathbf{m}_0)$. All panels are displayed with the same grayscale.}
    \label{fig:Virieux_data_init_recap}
\end{figure}

Finally, Figures~\ref{fig:Virieux_data_init_recap} shows the observed, predicted and data difference computed with the initial velocity model for a shot located at $x = 6.3$ km and maximum frequency of 6 Hz. Figures~\ref{fig:Virieux_data_init_diff} clearly illustrates the overlapping of true and wrongly predicted events (i.e., the reflection of the incident background wavefield from the misplaced horizontal interface present in the initial model).

\subsection{Conventional FWI}
\begin{figure}[t]
    \centering
    \subfigure[]{\label{fig:Marmousi_data_true}\includegraphics[width=0.3\columnwidth]{Fig/Syncline/Virieux-data-dipole-s2.pdf}} \hspace{5mm}
    \subfigure[]{\label{fig:Marmousi_data_fwi}\includegraphics[width=0.3\columnwidth]{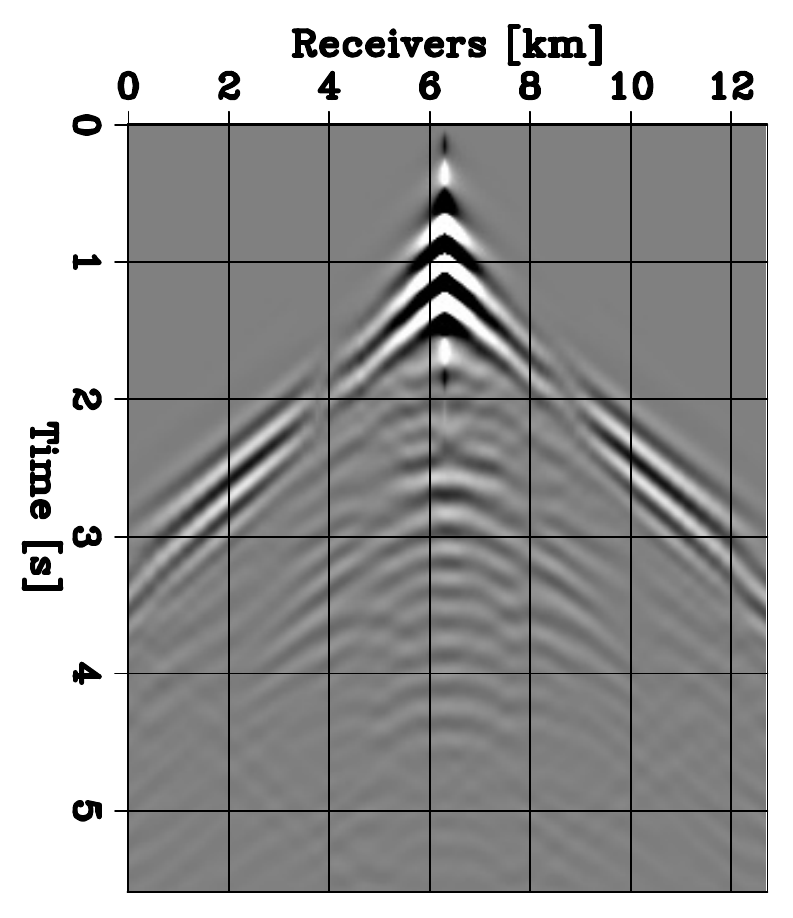}} \hspace{5mm}
    \subfigure[]{\label{fig:Marmousi_dataDiff_fwi}\includegraphics[width=0.3\columnwidth]{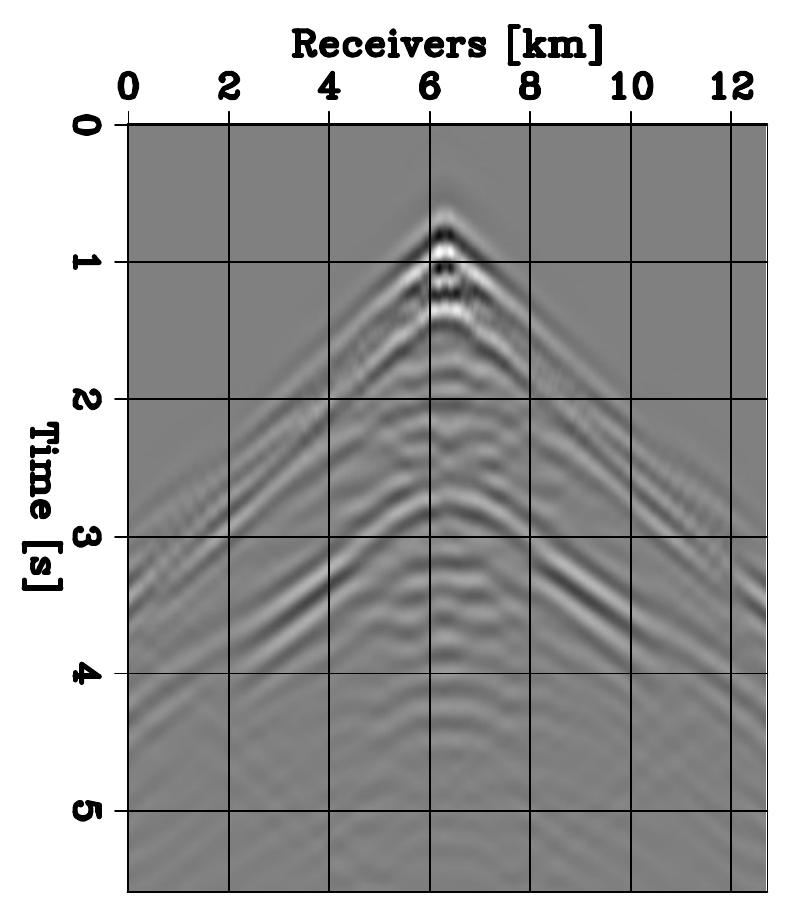}}
    \caption{Representative shot gather for a source placed at $x = 6.3$ km containing energy within the 1.5-6 Hz frequency range. (a) Observed data, $\mathbf{d}^{obs}$. (b) Predicted data with the inverted model using conventional FWI, $\mathbf{f}(\mathbf{m}_{FWI})$. (c) Data difference, $\Delta \mathbf{d}(\mathbf{m}_{FWI})=\mathbf{d}^{obs} - \mathbf{f}(\mathbf{m}_{FWI})$. All panels are displayed with the same grayscale.}
    \label{fig:Virieux_data_fwi}
\end{figure}

We first conduct conventional FWI to illustrate its inability to recover a useful model. Figures~\ref{fig:Virieux_fwi_mod} shows the inverted model after a total of 500 iterations of L-BFGS, confirming that FWI has converged to a local minimum. Moreover, Figures~\ref{fig:Virieux_data_fwi}b and \ref{fig:Virieux_data_fwi}c show the predicted data $\mathbf{f}(\mathbf{m}_{FWI})$ and data difference $\Delta \mathbf{d}(\mathbf{m}_{FWI}) = \mathbf{d}^{obs} - \mathbf{f}(\mathbf{m}_{FWI})$ computed with the final FWI inverted model. Clearly, the triplications generated by the synclinal feature are not accurately predicted. 

\subsection{FWIME} 
We apply FWIME by simultaneously inverting all data available within the 1.5-6 Hz frequency range. We use three spline grids throughout the inversion scheme. The first spline grid, $\mathbf{S}_0$, has a sampling of 500 m in both directions. However, in the vicinity of the initial horizontal interface between the two layers, we use a spacing of 50 m in the z-direction in order to preserve the sharpness present in the initial model. The second spline grid $\mathbf{S}_1$ halves the spacing of the first one (except in the vicinity of the horizontal interface) and the third grid $\mathbf{S}_2$ coincides with finite-difference grid (i.e., $\mathbf{S}_2= \mathbf{I}_d$). We use a horizontal subsurface-offset extension $h_x$ for $\tilde{\mathbf{p}}_{\epsilon}^{opt}$ which is adequate for a dataset dominated by reflection events \cite[]{biondi2014simultaneous}. Due to the large kinematic errors in the initial data-prediction, we employ 201 points of extension, which allows $h_x$ to range from -5 km to 5 km. Each variable projection step is conducted by minimizing objective function shown in equation~\ref{eqn:vp.obj} with 60 iterations of linear conjugate gradient. 

\begin{figure}[t]
    \centering
    \subfigure[]{\label{fig:Virieux_popt_cig4}\includegraphics[width=0.45\columnwidth]{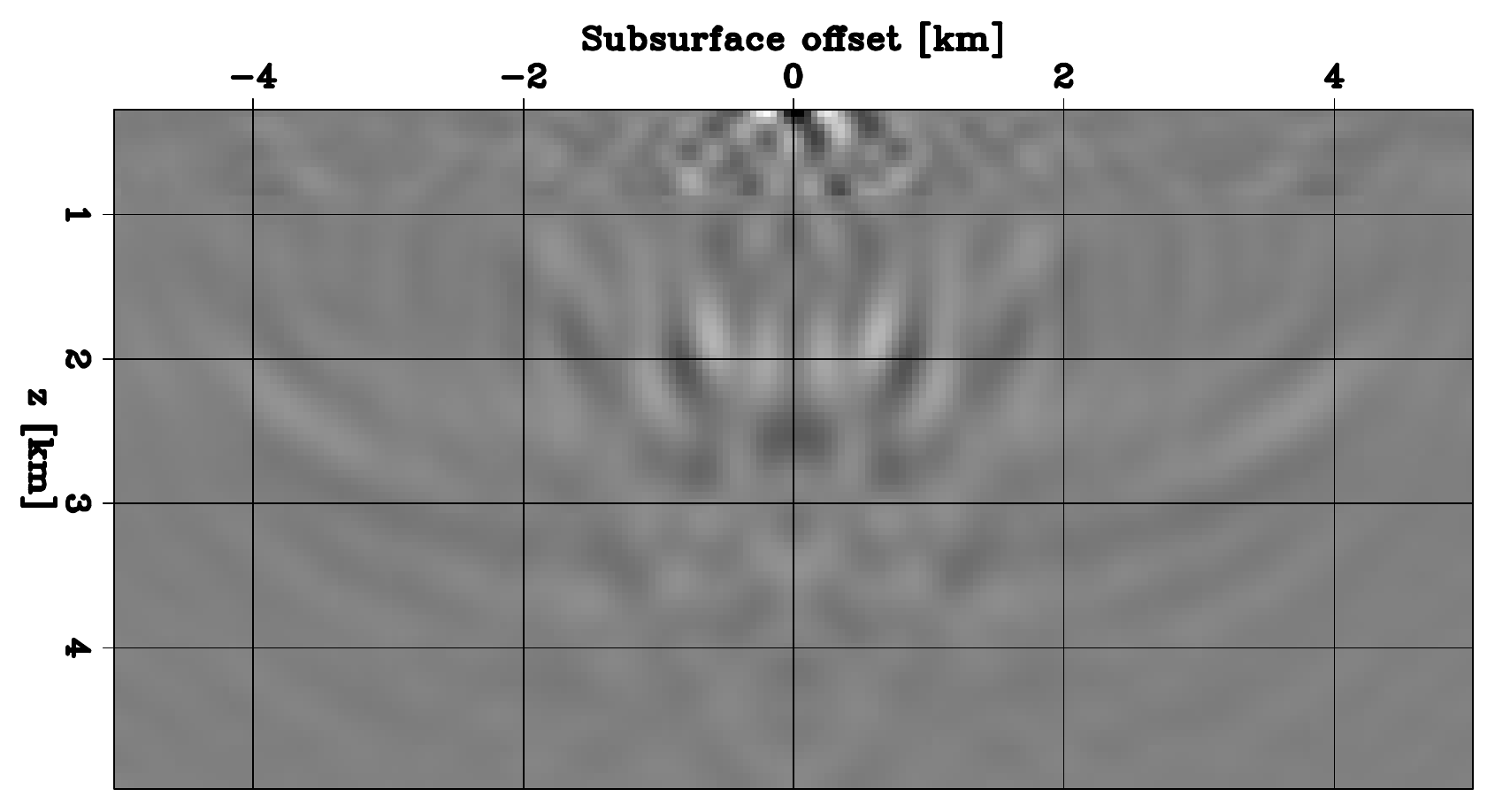}} \hspace{5mm}
    \subfigure[]{\label{fig:Virieux_popt_cig6}\includegraphics[width=0.45\columnwidth]{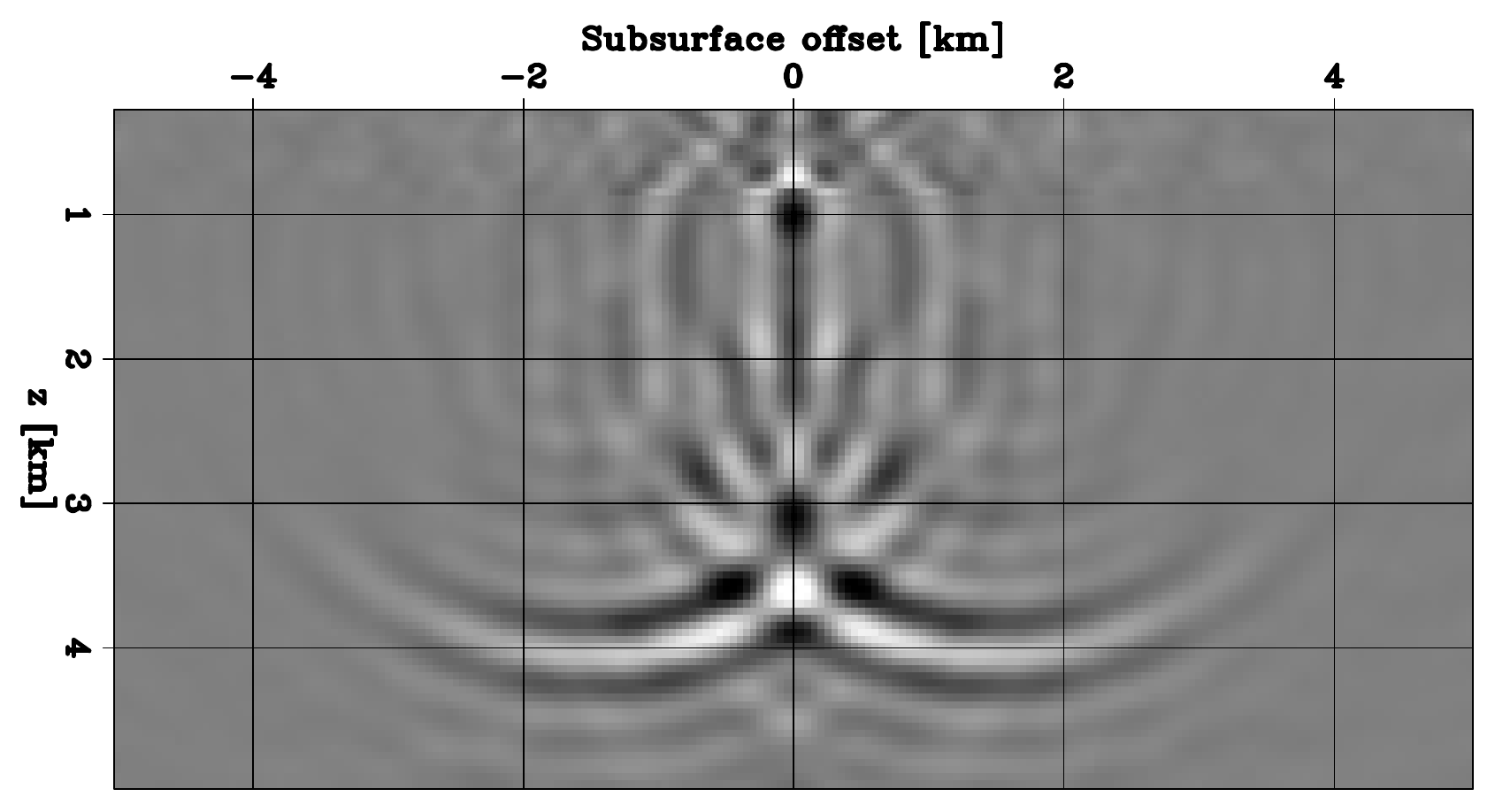}}
    \caption{Horizontal subsurface offset common image gathers (SOCIG) extracted from the extended optimal perturbation $\mathbf{\tilde{p}}_{\epsilon}^{opt}$ computed at the initial step and extracted at two horizontal positions. (a) $x=4.0$ km. (b) $x=6.0$ km. Both panels are displayed with the same grayscale.}
    \label{fig:Virieux_popt_cig}
\end{figure}

Figure~\ref{fig:Virieux_popt_cig} shows two horizontal subsurface-offset common image gathers (SOCIG) extracted from $\mathbf{\tilde{p}}_{\epsilon}^{opt}(\mathbf{S}_0\mathbf{m}_0)$ at $x=4.0$ km and $x=6.0$ km, respectively ($\mathbf{m}_0$ is the initial model parametrized on the spline grid). The SOCIG located at $x=4$ km contains weaker energy/events due to the fact that at that location, the initial model is already accurate, and thus $\mathbf{f}(\mathbf{S}_0\mathbf{m}_0) \approx \mathbf{d}^{obs}$. Consequently, less energy is mapped into $\mathbf{\tilde{p}}_{\epsilon}^{opt}$ during the minimization of equation~\ref{eqn:vp.obj}. Conversely, for the SOCIG closer to the basin (Figure~\ref{fig:Virieux_popt_cig6}), the initial data prediction is less accurate, and more energy is mapped into $\mathbf{\tilde{p}}_{\epsilon}^{opt}$. In fact, the initial kinematic error is so substantial that a maximum subsurface-offset span of 10 km is needed to fully capture the energy from the data misfit. 

\begin{figure}[t]
    \centering
    \subfigure[]{\label{fig:Virieux_gradient_no_spline_born}\includegraphics[width=0.45\columnwidth]{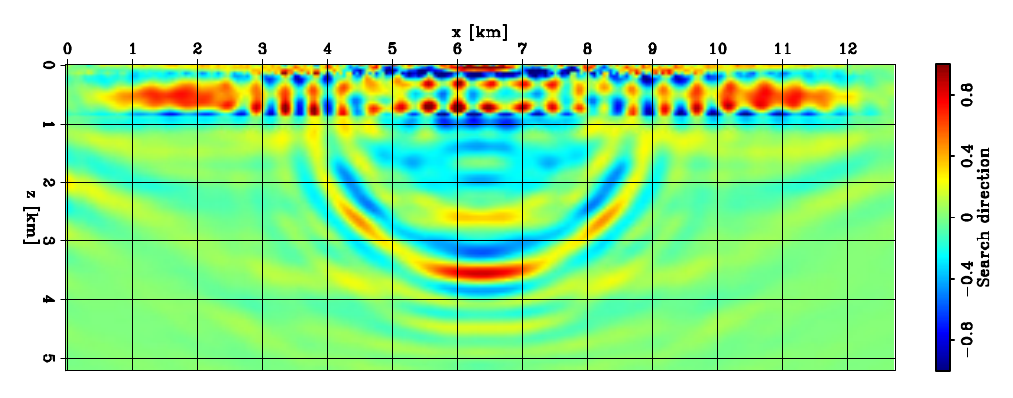}} \hspace{5mm}
    \subfigure[]{\label{fig:Virieux_gradient_no_spline_tomo}\includegraphics[width=0.45\columnwidth]{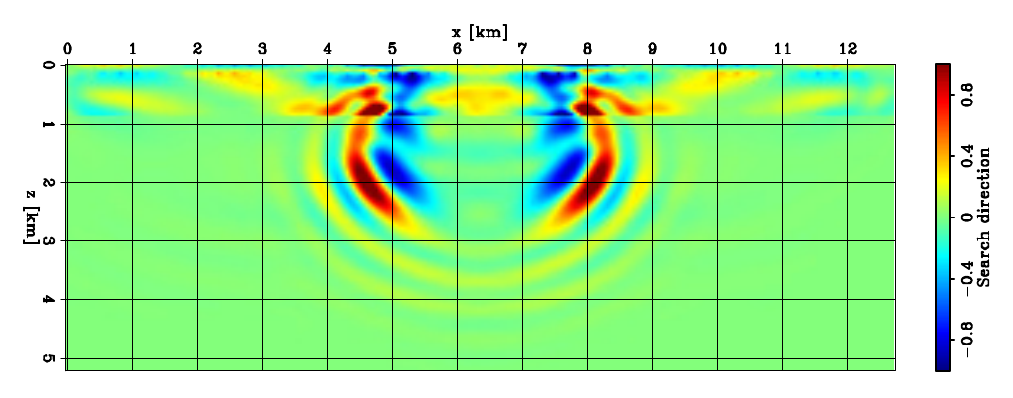}} \\
    \subfigure[]{\label{fig:Virieux_gradient_no_spline_total}\includegraphics[width=0.45\columnwidth]{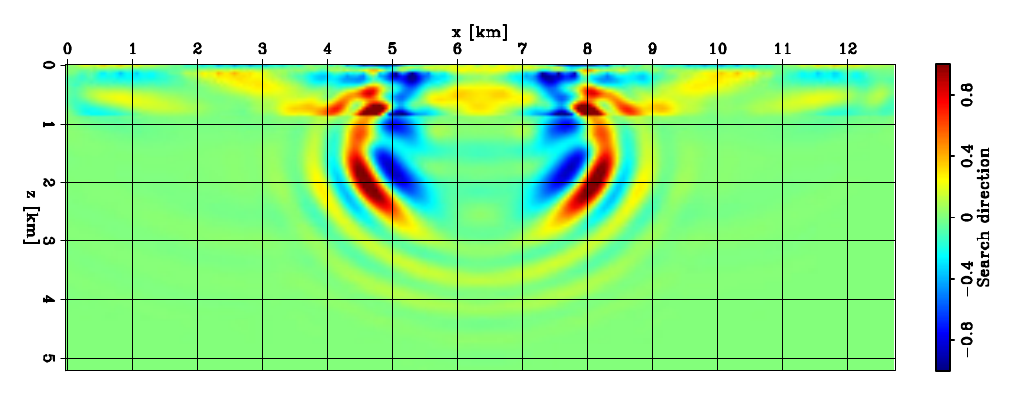}} \hspace{5mm}
    \subfigure[]{\label{fig:Virieux_gradient_no_spline_true}\includegraphics[width=0.45\columnwidth]{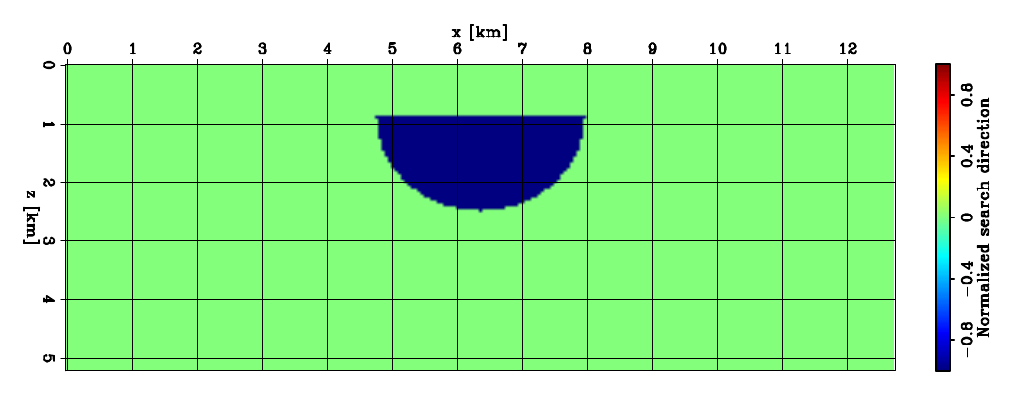}}        
    \caption{Scaled FWIME initial search directions on the finite-difference grid (before their mapping onto the initial spline grid). (a) Born search direction. (b) Tomographic search direction. (c) Total search direction (sum of panels (a) and (b)). (d) True search direction. Note that (a) is normalized by a different scaling factor than the one for (b) and (c) (for display purpose). The amplitude of panel (a) is approximately two orders of magnitude smaller than the one of panels (b) and (c).}
    \label{fig:Virieux_gradient_no_spline}
\end{figure}

\begin{figure}[t]
    \centering
    \subfigure[]{\label{fig:Virieux_gradient_spline_fwime_spline1}\includegraphics[width=0.45\columnwidth]{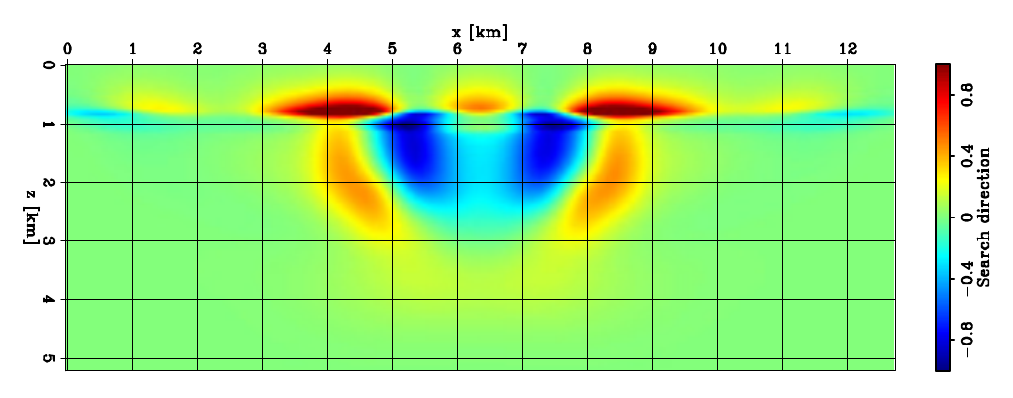}} \hspace{5mm}
    \subfigure[]{\label{fig:Virieux_gradient_spline_true}\includegraphics[width=0.45\columnwidth]{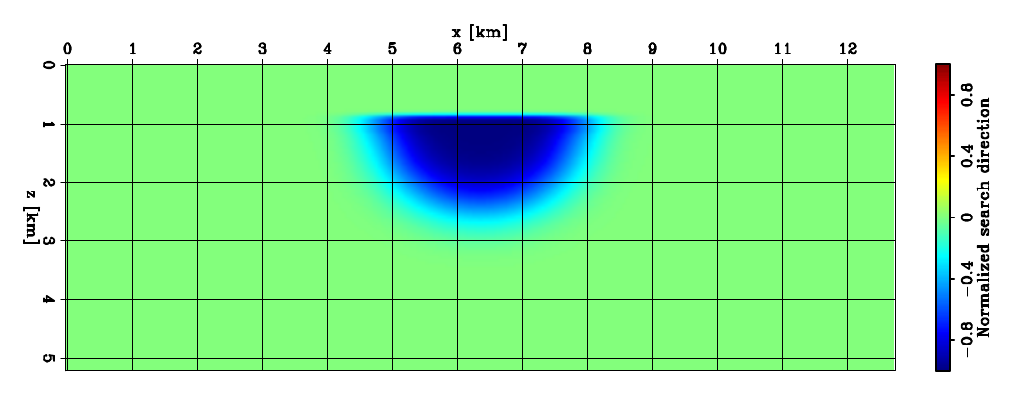}} 
    \caption{Scaled search directions. (a) FWIME initial search direction after applying $\mathbf{S}_0\mathbf{S}_0^*$ to panel \ref{fig:Virieux_gradient_no_spline_total}. (b) Search direction after applying $\mathbf{S}_0\mathbf{S}_0^*$ to panel~\ref{fig:Virieux_gradient_no_spline_true}.}
    \label{fig:Virieux_gradient_spline}
\end{figure}

Figures~\ref{fig:Virieux_gradient_no_spline}a and \ref{fig:Virieux_gradient_no_spline}b show the scaled Born and tomographic components of the first FWIME search direction, respectively. Note that Figures~\ref{fig:Virieux_gradient_no_spline}a and \ref{fig:Virieux_gradient_no_spline}b are not normalized with the same scaling factor: the Born component is approximately two orders of magnitude weaker than the tomographic component, even without the use of a spline parametrization. Indeed, this is expected (and desired) at early stages of the FWIME workflow when the model updates should primarily be guided by the tomographic gradient. Figure~\ref{fig:Virieux_gradient_no_spline_total} shows the total search direction (i.e., the sum of the two panels in Figures~\ref{fig:Virieux_gradient_no_spline}a and \ref{fig:Virieux_gradient_no_spline}b), which is almost identical to the tomographic component. Even though this search direction may seem promising (by comparing it to the ideal update in Figure~\ref{fig:Virieux_gradient_no_spline_true}), its mapping onto the first spline grid removes most of the high-wavenumber artifacts and provides a much more promising search direction (Figure~\ref{fig:Virieux_gradient_spline_fwime_spline1}). 

\begin{figure}[t]
    \centering
    \subfigure[]{\label{fig:Virieux_fwime_init}\includegraphics[width=0.45\columnwidth]{Fig/Syncline/Virieux-sep20-init-mod.pdf}}    
    \subfigure[]{\label{fig:Virieux_fwime_s1}\includegraphics[width=0.45\columnwidth]{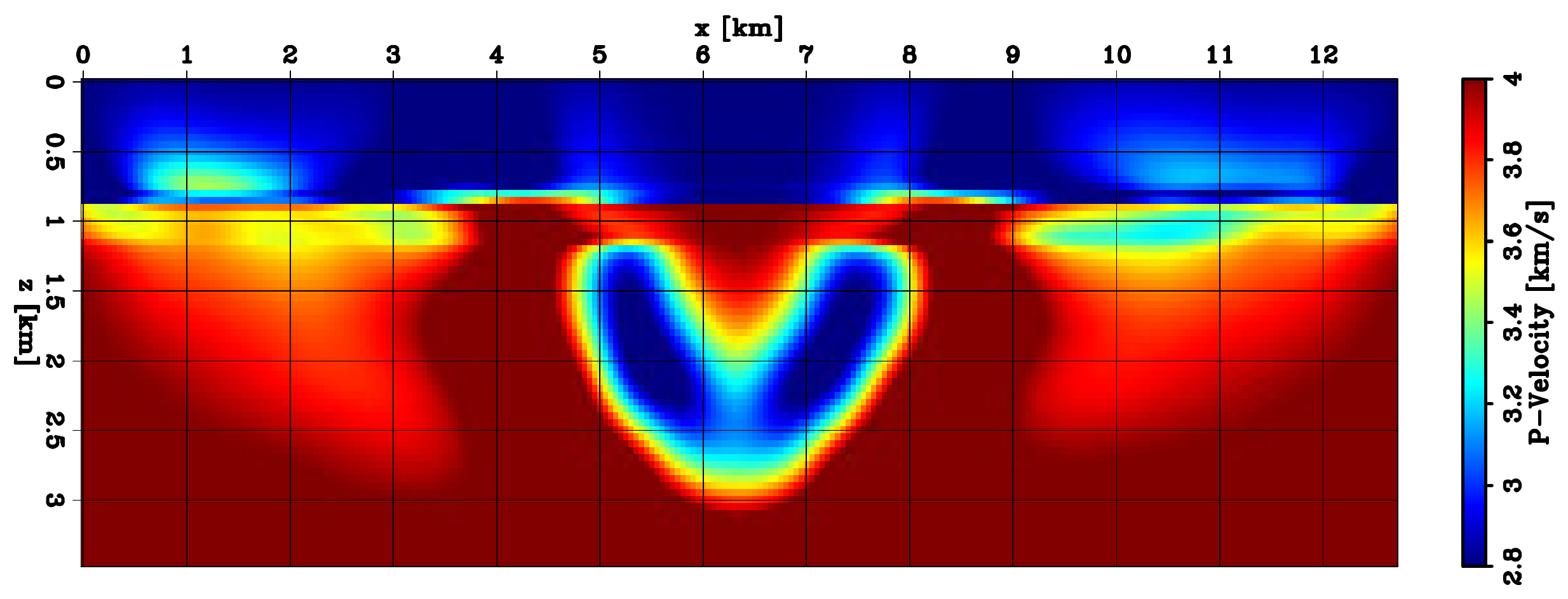}} \\
    \subfigure[]{\label{fig:Virieux_fwime_s1_no_spline}\includegraphics[width=0.45\columnwidth]{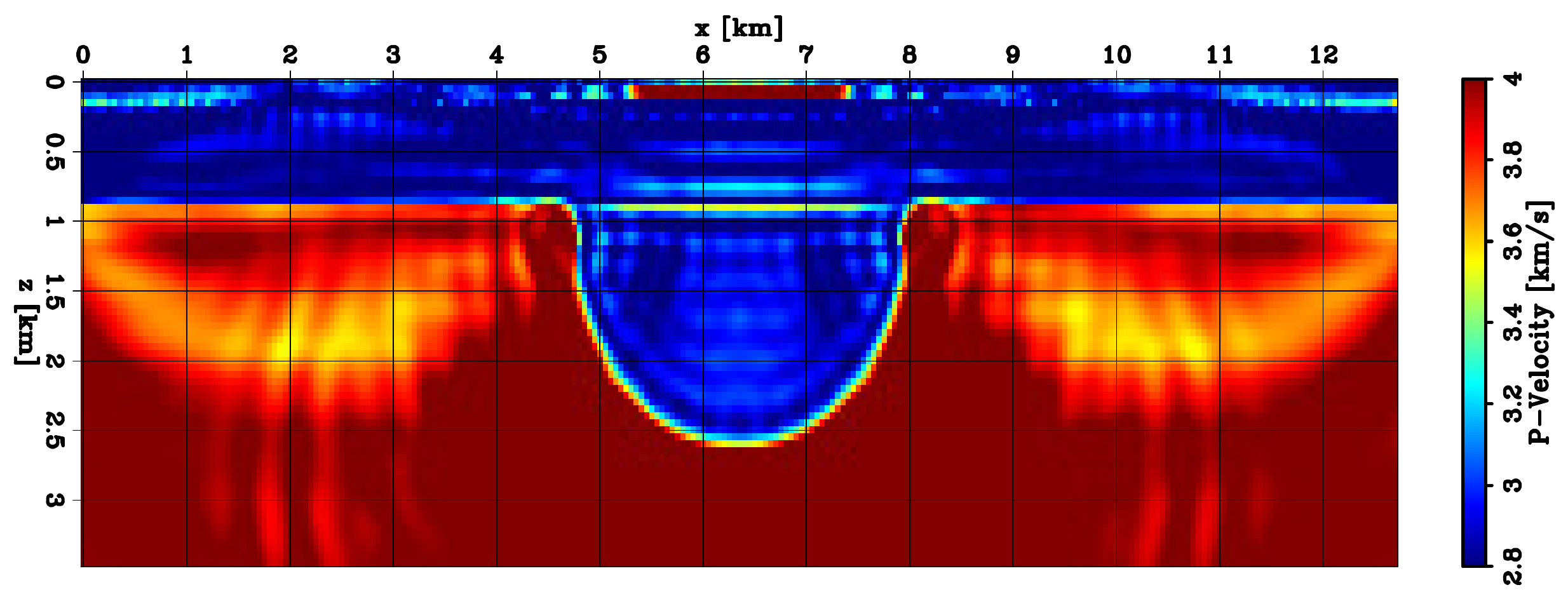}} 
    \subfigure[]{\label{fig:Virieux_fwime_s1_no_spline_fwi}\includegraphics[width=0.45\columnwidth]{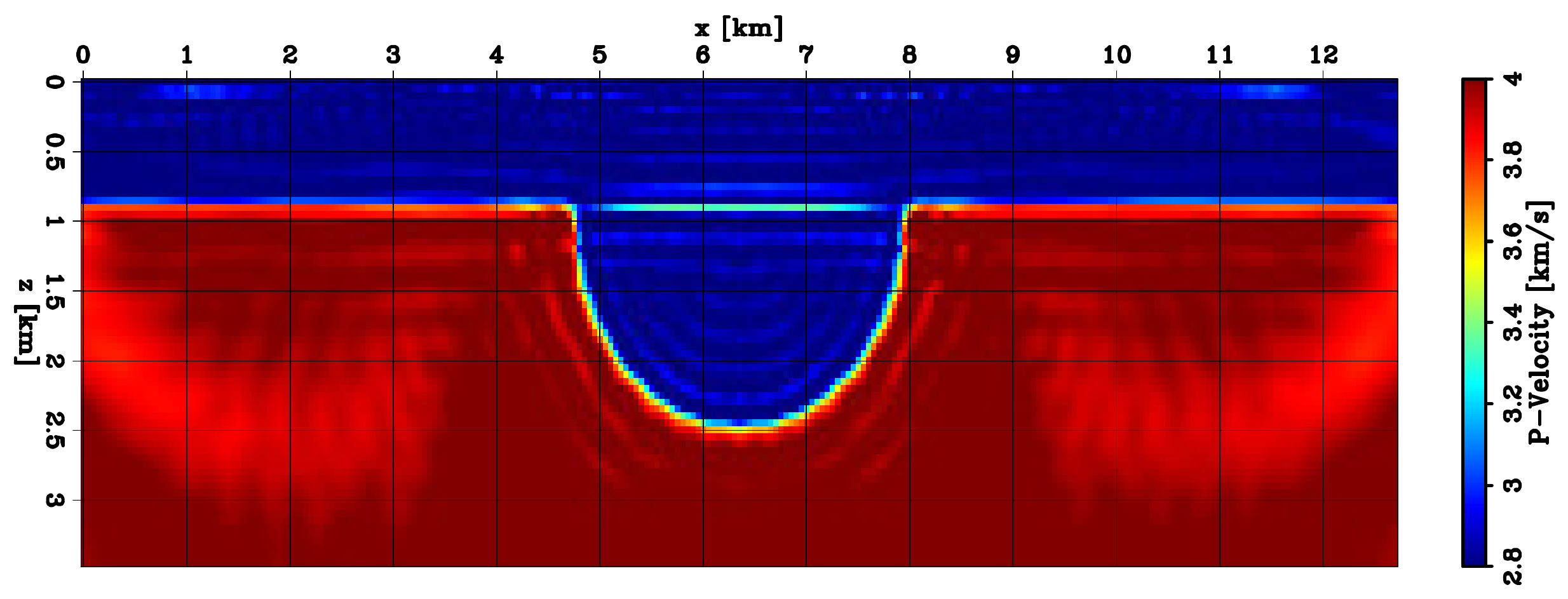}}\\
    \caption{Inverted models at various stages of the FWIME workflow, mapped onto the finite-difference grid. (a) Initial model. (b) Inverted model after 28 iterations of FWIME on $\mathbf{S}_0$. (c) Inverted model after 87 iterations of FWIME on $\mathbf{S}_1$ using (b) as initial guess. (d) Inverted model after applying 33 iterations of FWIME on the finite-difference grid using (c) as initial guess.}
    \label{fig:Virieux_mod_fwime}
\end{figure}

\begin{figure}[t]
    \centering
    \subfigure[]{\label{fig:Virieux_fwime_x6}\includegraphics[width=0.45\columnwidth]{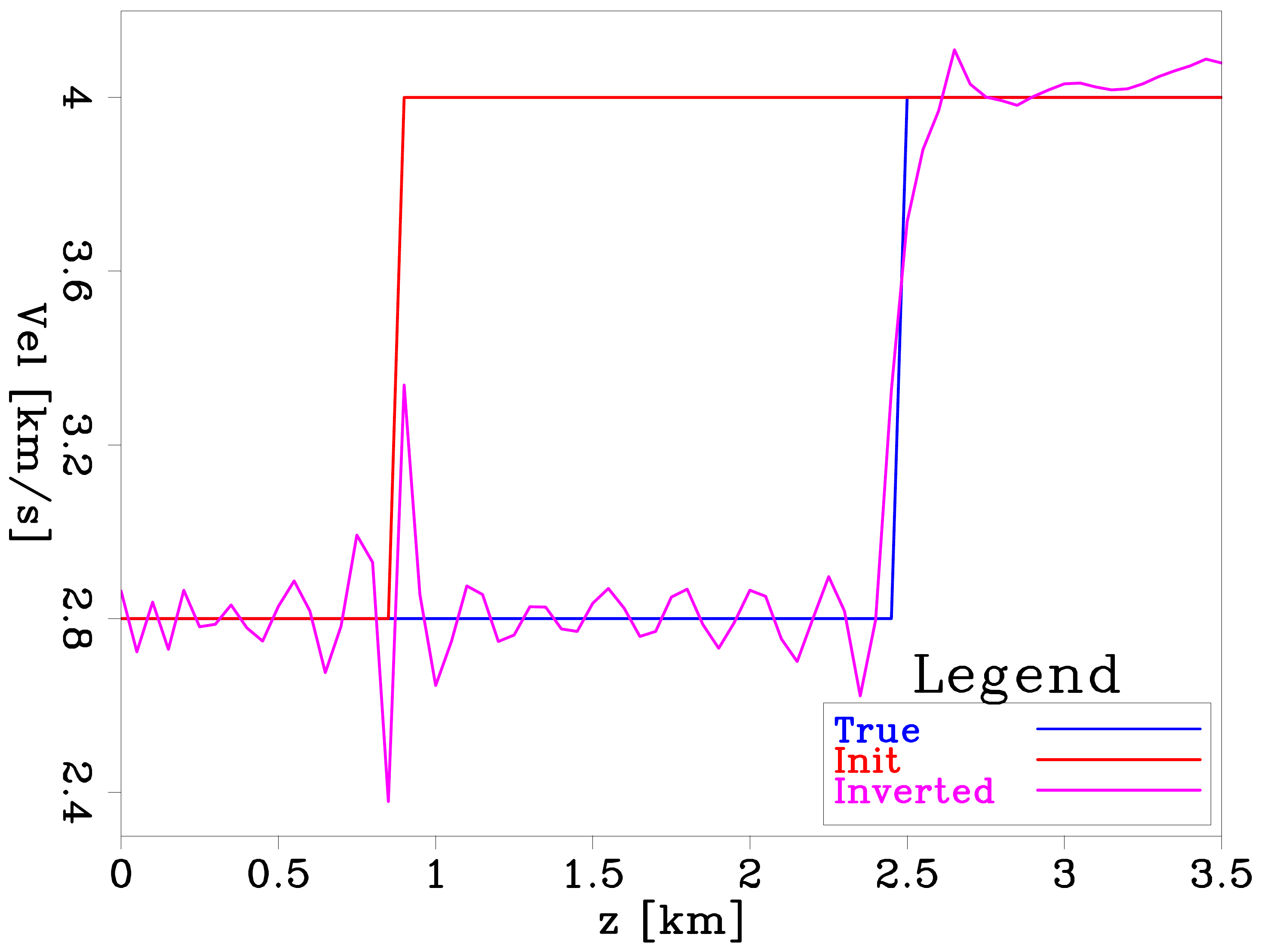}}\hspace{5mm}
    \subfigure[]{\label{fig:Virieux_fwime_z1.5}\includegraphics[width=0.45\columnwidth]{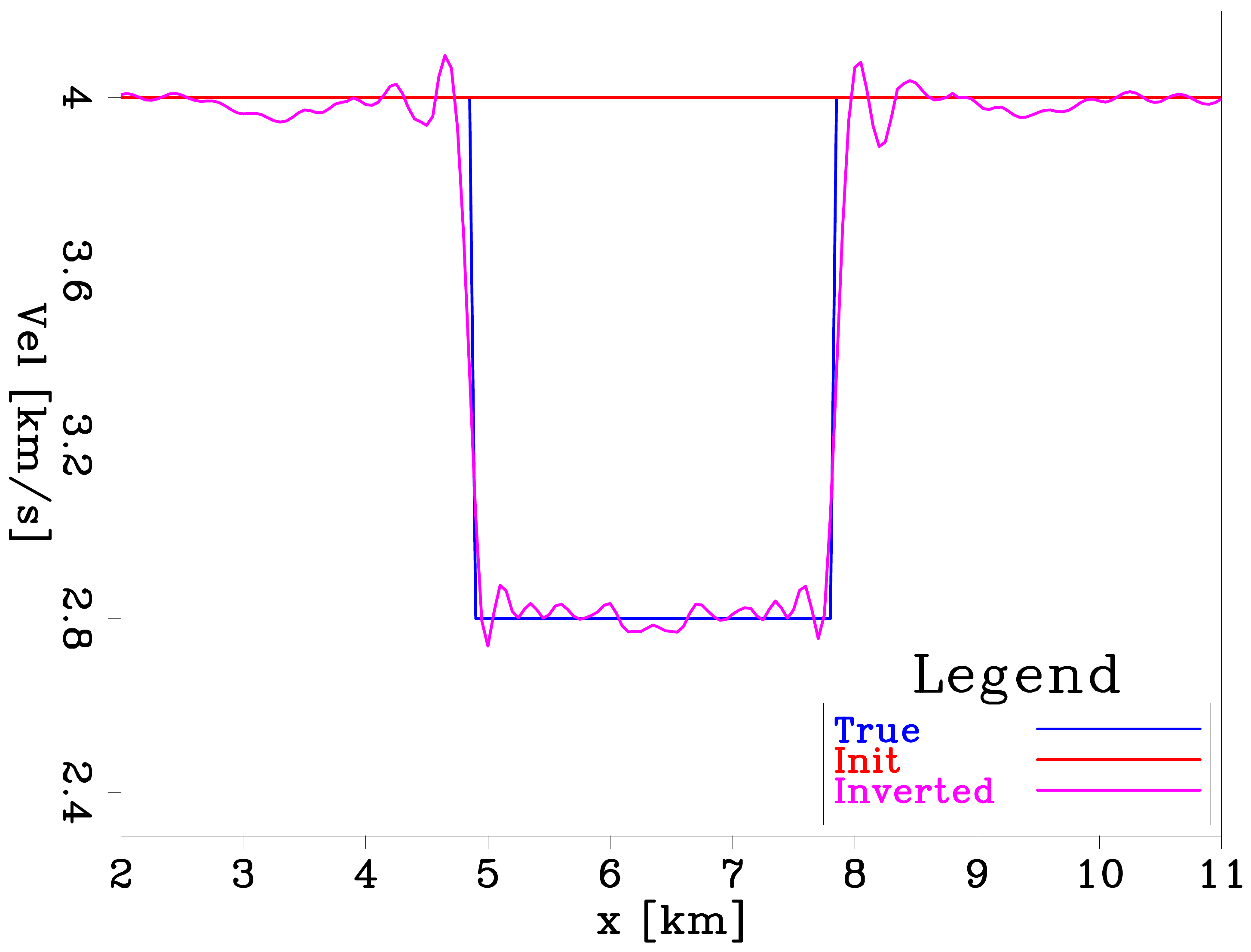}}    
    \caption{Velocity profiles of the initial model (red curve), true model (blue curve), and the final FWIME inverted model. (a) Vertical profile extracted at $x=6$ km. (b) Horizontal profile extracted at $z=1.5$ km.}
    \label{fig:Virieux_mod_fwime_1d}
\end{figure}

We minimize equation~\ref{eqn:fwime.obj} with 28 iterations of L-BFGS on the first spline grid, $\mathbf{S}_0$. The grid refinement is automatically triggered when the solver is unable to find a proper step length that decreases the objective function. The inverted model after the first spline grid, $\mathbf{m}_1$ is shown in Figure~\ref{fig:Virieux_fwime_s1} (for display purposes, this panel shows $\mathbf{S}_0 \mathbf{m}_1$ rather than $\mathbf{m}_1$). We use the inverted model on the first spline grid as the initial guess for the inversion on the second spline grid and we conduct 87 iterations of L-BFGS (Figure~\ref{fig:Virieux_fwime_s1_no_spline}). The final FWIME inverted model on the finite-difference grid is obtained after an additional 33 iterations of L-BFGS, and is shown in Figure~{\ref{fig:Virieux_fwime_s1_no_spline_fwi}}. Figures~\ref{fig:Virieux_mod_fwime_1d}a and \ref{fig:Virieux_mod_fwime_1d}b display vertical and horizontal velocity profiles of the initial (red curve), true (blue curve) and FWIME (pink curve) models at $x=6$ km and $z=1.5$ km, respectively. The final result is accurate but still contains a thin reflector present at an approximate depth of $z=0.8$ km (where the wrongly-positioned horizontal interface was initially located). However, this artifact would likely disappear if the frequency content of the data were to be increased. Even though the FWIME results shown in Figure~\ref{fig:Virieux_mod_fwime} are computed for a fixed $\epsilon$-value of $1.0 \times 10^{-5}$, we observe that different values (ranging from $\epsilon=0.5 \times 10^{-5}$ to $\epsilon=5.0 \times 10^{-5}$) lead to similar results.

\begin{figure}[t]
    \centering
    \subfigure[]{\label{fig:Virieux_data_fwime_true1}\includegraphics[width=0.22\columnwidth]{Fig/Syncline/Virieux-data-dipole-s2.pdf}} \hspace{3mm}
    \subfigure[]{\label{fig:Virieux_data_fwime_pred_init}\includegraphics[width=0.22\columnwidth]{Fig/Syncline/Virieux-data-init-dipole-s2.pdf}} \hspace{3mm}    
    \subfigure[]{\label{fig:Virieux_data_fwime_diff_init}\includegraphics[width=0.22\columnwidth]{Fig/Syncline/Virieux-data-init-diff-dipole-s2.pdf}} \\
    \subfigure[]{\label{fig:Virieux_data_fwime_true2}\includegraphics[width=0.22\columnwidth]{Fig/Syncline/Virieux-data-dipole-s2.pdf}} \hspace{3mm}    
    \subfigure[]{\label{fig:Marmousi_data_fwime_pred_spline1}\includegraphics[width=0.22\columnwidth]{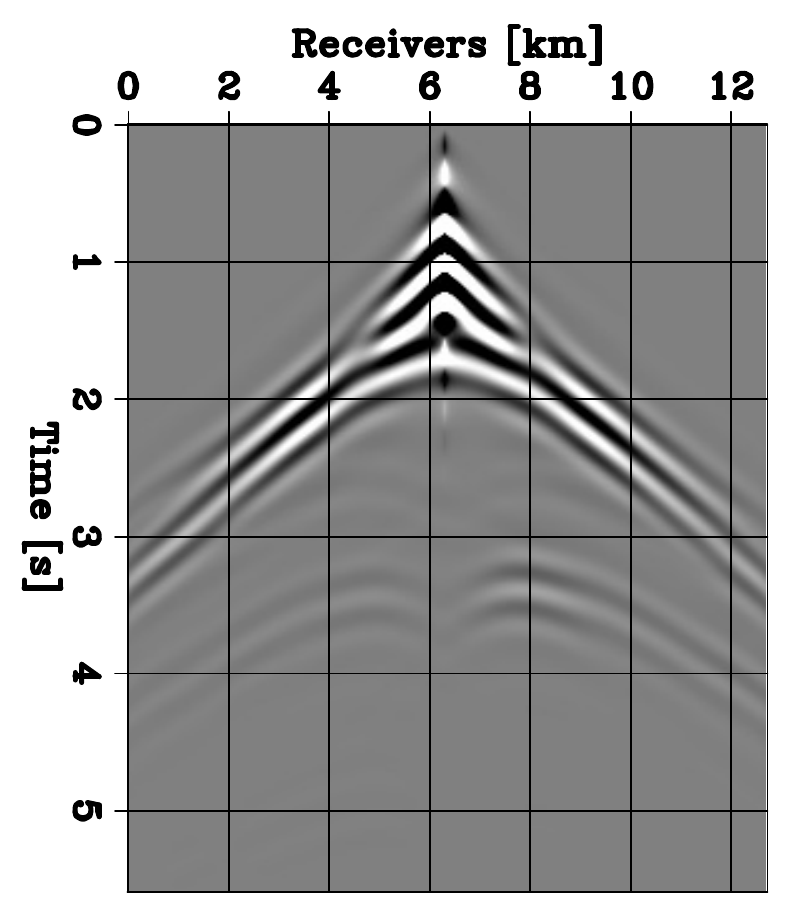}} \hspace{3mm} 
    \subfigure[]{\label{fig:Marmousi_data_fwime_diff_spline1}\includegraphics[width=0.22\columnwidth]{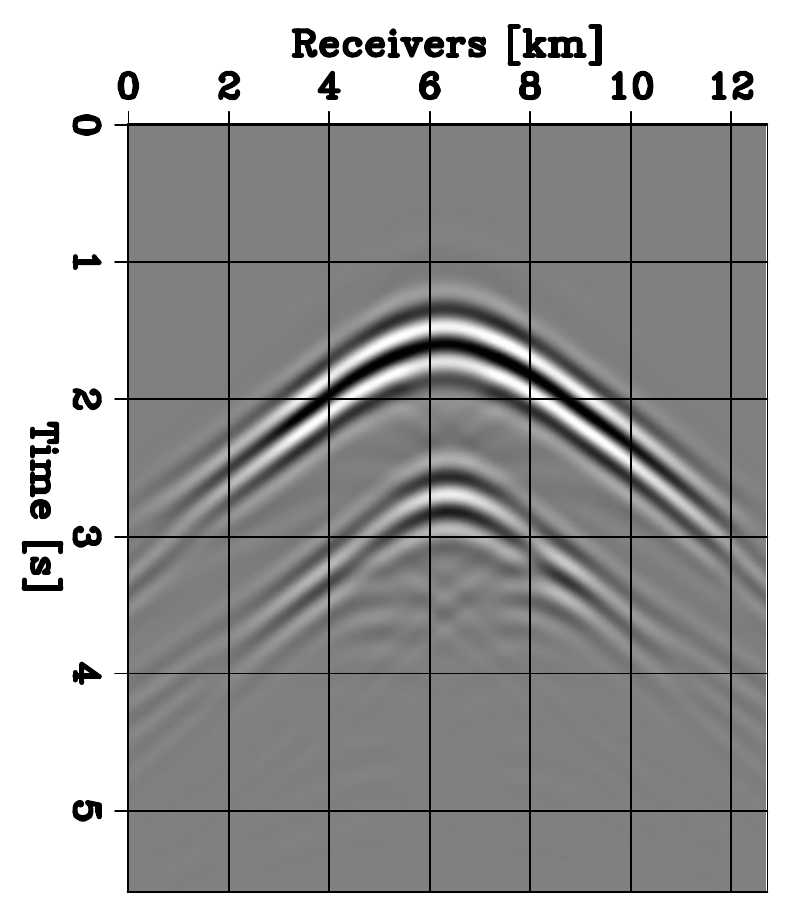}} \\
    \subfigure[]{\label{fig:Virieux_data_fwime_true3}\includegraphics[width=0.22\columnwidth]{Fig/Syncline/Virieux-data-dipole-s2.pdf}} \hspace{3mm}    
    \subfigure[]{\label{fig:Marmousi_data_fwime_pred_final}\includegraphics[width=0.22\columnwidth]{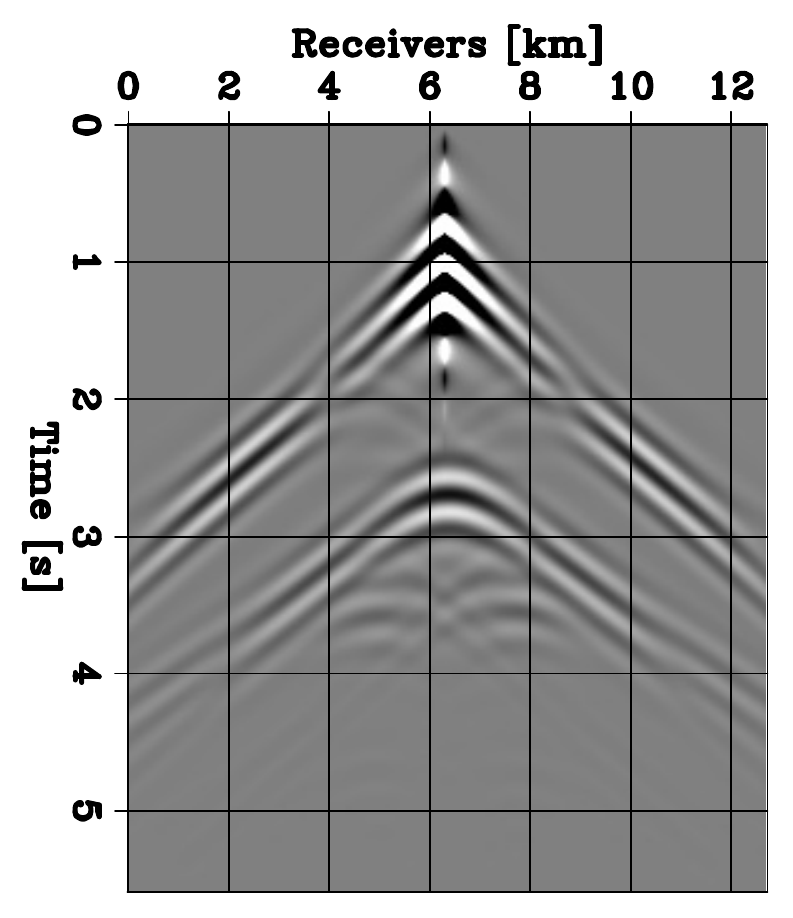}} \hspace{3mm} 
    \subfigure[]{\label{fig:Marmousi_data_fwime_diff_final}\includegraphics[width=0.22\columnwidth]{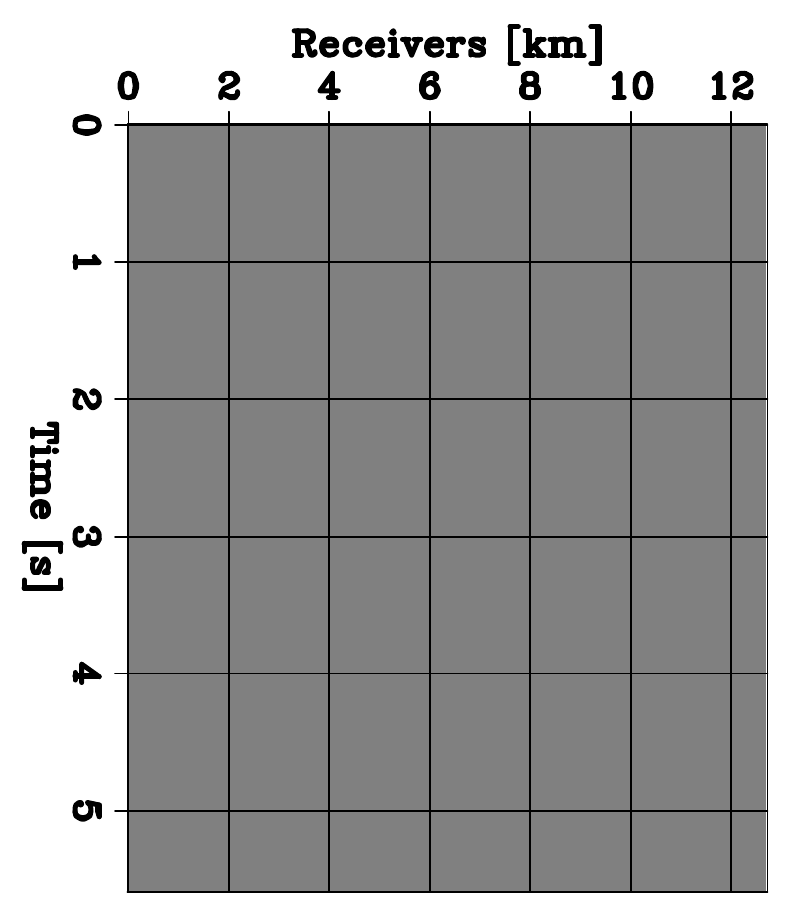}} \\        
    \caption{Observed data, $\mathbf{d}^{obs}$ (left column), predicted data, $\mathbf{f}(\mathbf{S}_i \mathbf{m}_i)$ (middle column), and data difference, $\Delta \mathbf{d}(\mathbf{m}_i) = \mathbf{d}^{obs} - \mathbf{f}(\mathbf{S}_i \mathbf{m}_i)$ (right column) for a shot located at $x = 6.3$ km computed with the FWIME inverted models at various stages of the FWIME workflow. Initial model (first row), inverted model on the first spline grid $\mathbf{S}_0$ (second row), and final inverted model (third row). All panels are displayed on the same grayscale.}
    \label{fig:Virieux_data_fwime}
\end{figure}

Figure~\ref{fig:Virieux_data_fwime} shows the evolution of the predicted data $\mathbf{f}(\mathbf{S}_i \mathbf{m}_i)$ (middle column) and the data-difference $\Delta \mathbf{d} (\mathbf{m}_i) = \mathbf{d}^{obs} - \mathbf{f}(\mathbf{S}_i \mathbf{m}_i)$ (right column) at various stages of the FWIME workflow (for a source placed at $x=6.3$ km). After the first spline grid (second row), the inversion recovers a model that is able to generate some of the triplications in the wavefield (Figure~\ref{fig:Marmousi_data_fwime_pred_spline1}) by removing a large portion of the incorrect high-velocity zone at the bottom of the basin (Figure~\ref{fig:Virieux_fwime_s1}). At the final stage (third row), most of the basin has been recovered, the reflection from the initially mis-positioned horizontal interface at $z \approx 0.8$ km has been removed from the predicted data (Figure~\ref{fig:Marmousi_data_fwime_pred_final}), and the data residuals have completely vanished (Figure~\ref{fig:Marmousi_data_fwime_diff_final}). 

\begin{figure}[t]
    \centering
    \subfigure[]{\label{fig:Virieux_fwime_zero_offset_init}\includegraphics[width=0.45\columnwidth]{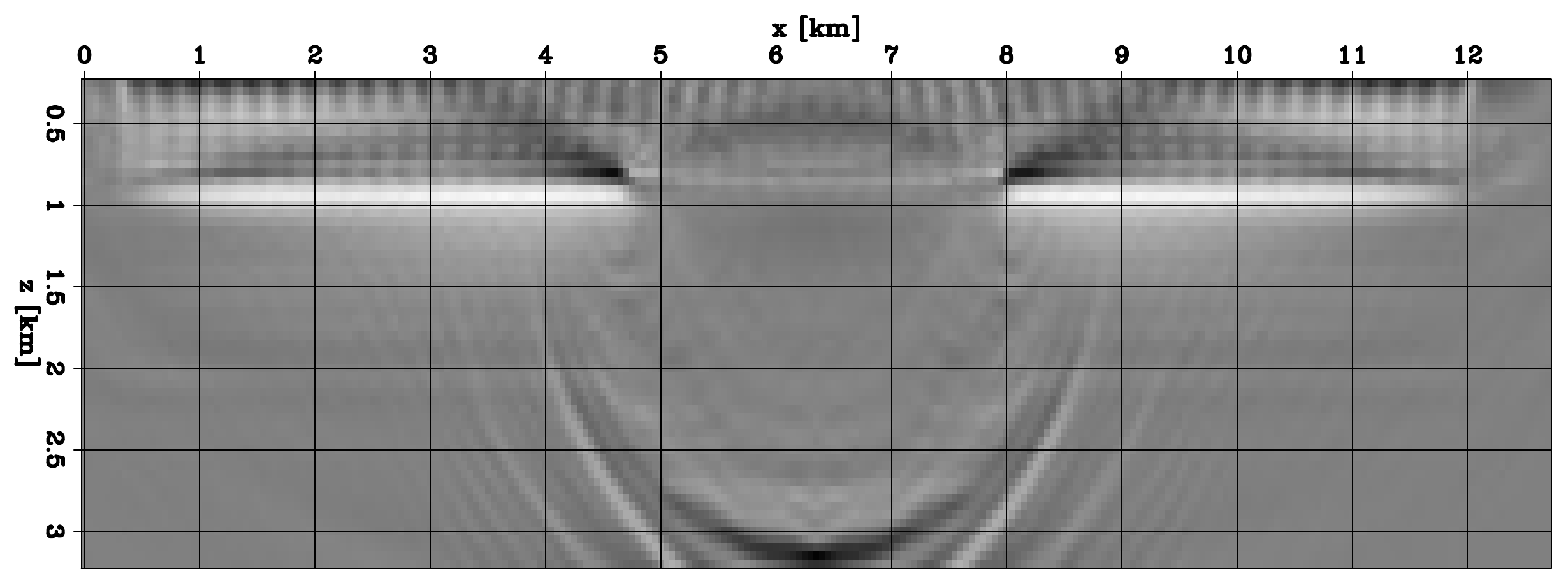}} \hspace{5mm} 
    \subfigure[]{\label{fig:Virieux_fwime_zero_offset_final}\includegraphics[width=0.45\columnwidth]{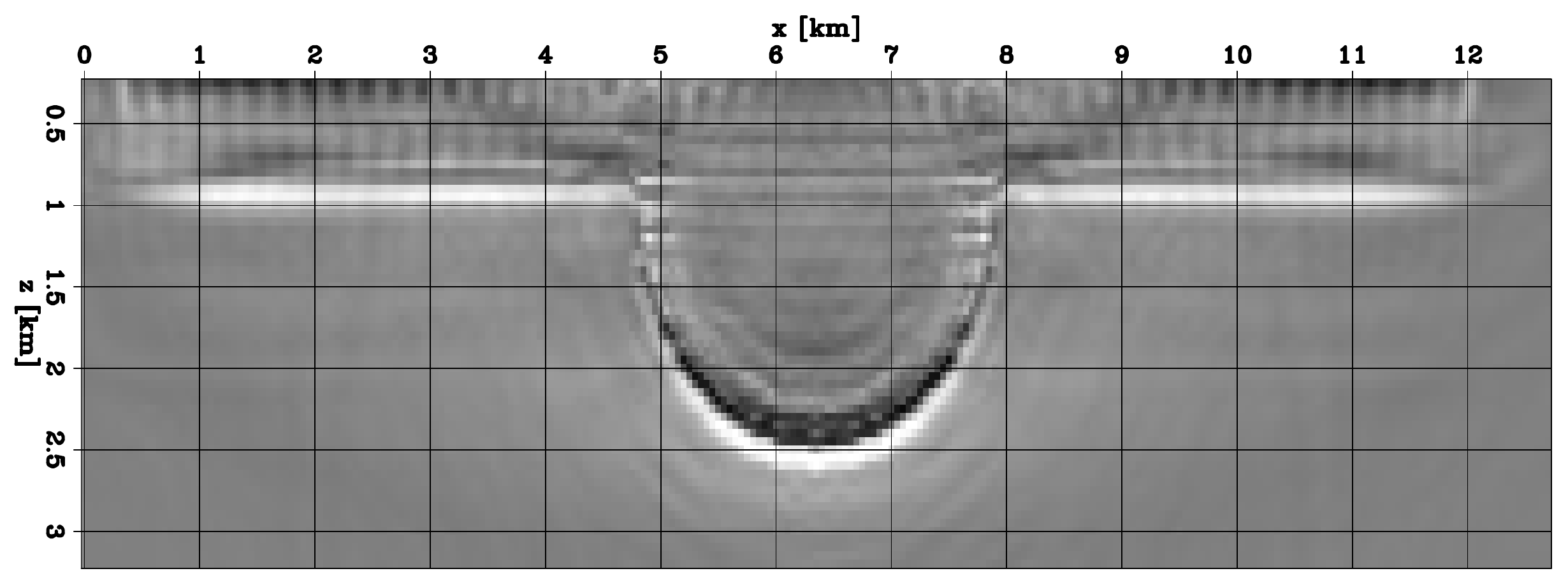}} \\
    \subfigure[]{\label{fig:Virieux_fwime_zero_offset_true}\includegraphics[width=0.45\columnwidth]{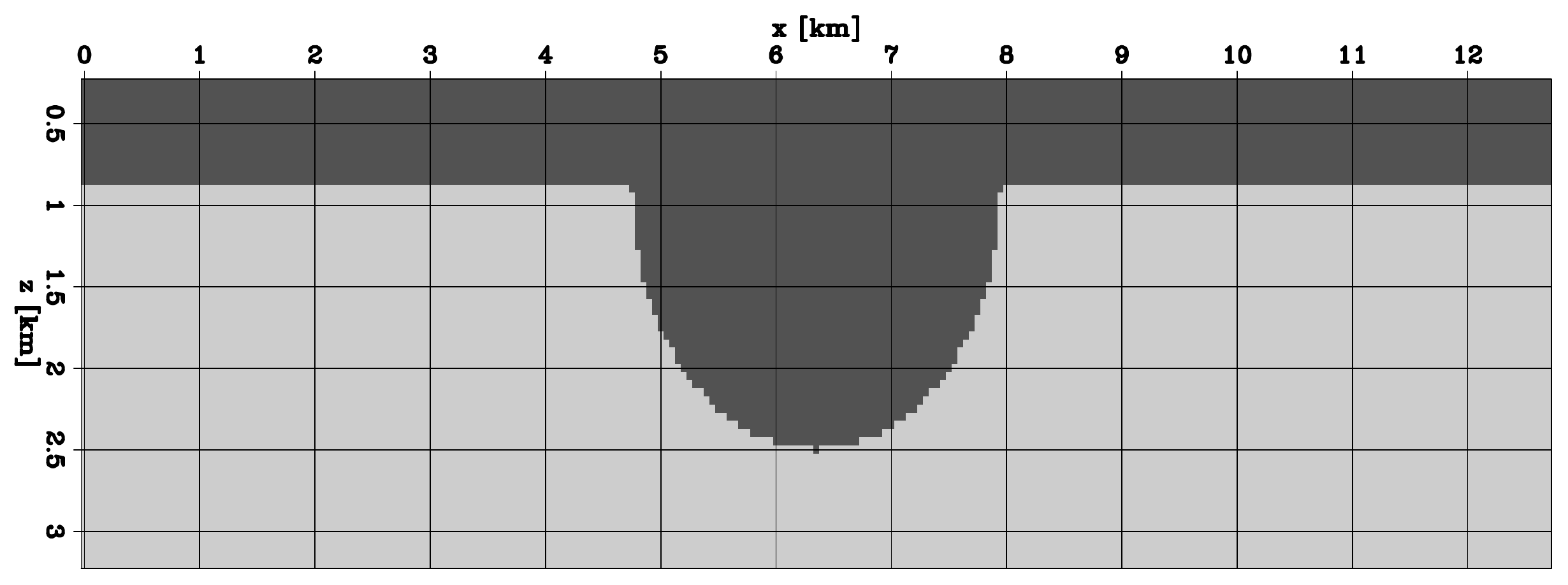}}
    \caption{RTM images computed with different velocity models. (a) Initial velocity model. (b) Final FWIME inverted model. (c) True velocity model (for reference). Panels (a) and (b) are displayed with the same grayscale.}
    \label{fig:Virieux_fwime_zero_offset}
\end{figure}

\begin{figure}[t]
    \centering
    \subfigure[]{\label{fig:Virieux_init_adcig3}\includegraphics[width=0.2\columnwidth]{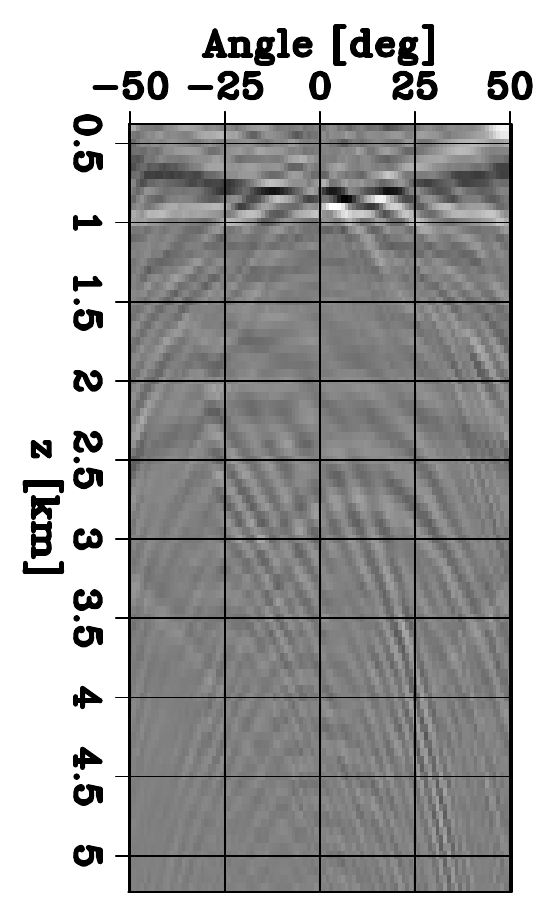}} \hspace{5mm} 
    \subfigure[]{\label{fig:Virieux_init_adcig4}\includegraphics[width=0.2\columnwidth]{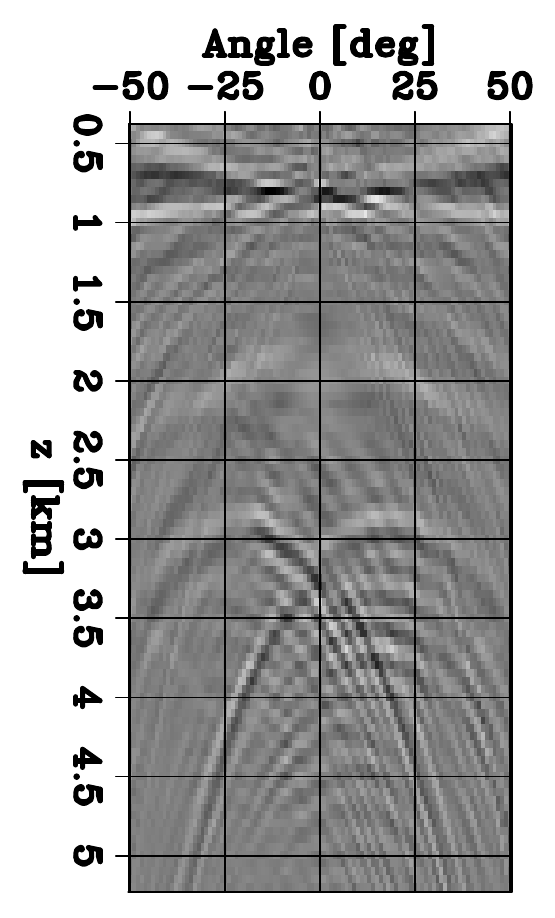}} \hspace{5mm} 
    \subfigure[]{\label{fig:Virieux_init_adcig5}\includegraphics[width=0.2\columnwidth]{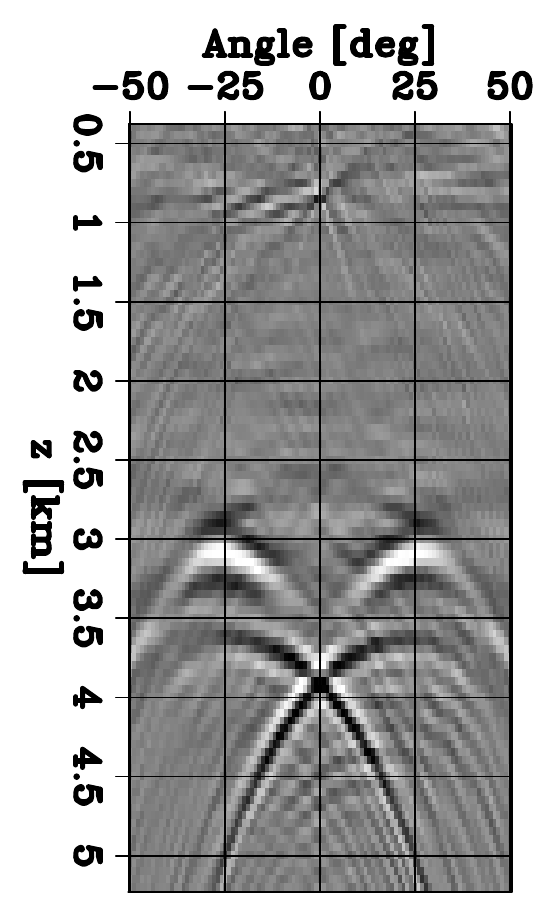}} \hspace{5mm} 
    \subfigure[]{\label{fig:Virieux_init_adcig6}\includegraphics[width=0.2\columnwidth]{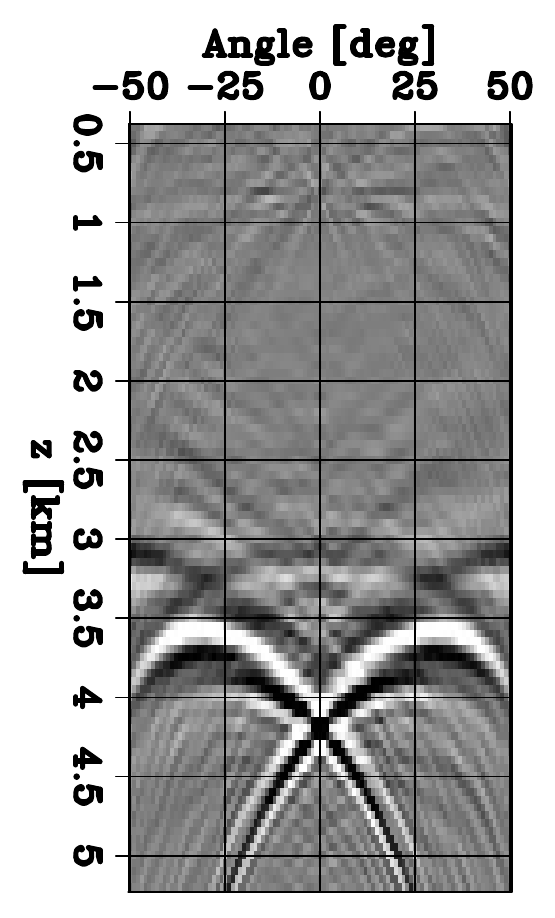}}\\
    \subfigure[]{\label{fig:Virieux_final_adcig3}\includegraphics[width=0.2\columnwidth]{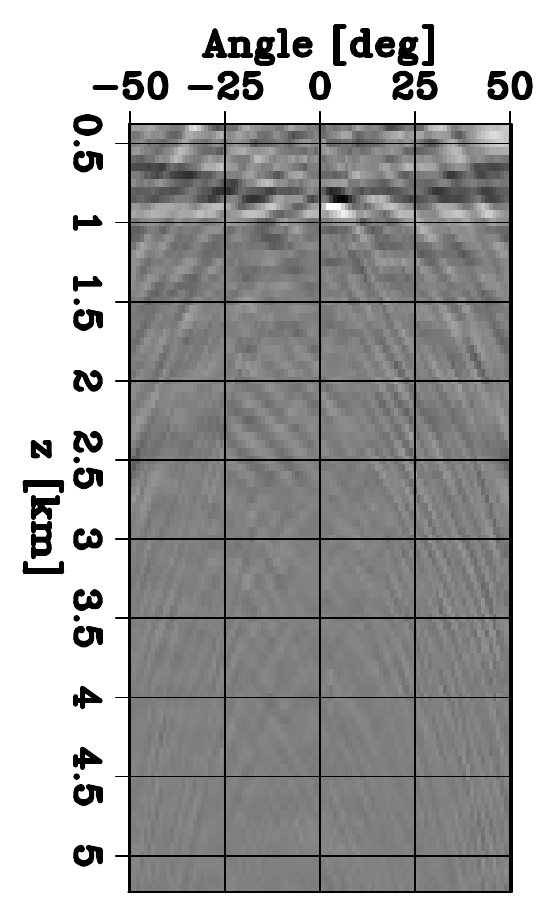}} \hspace{5mm} 
    \subfigure[]{\label{fig:Virieux_final_adcig4}\includegraphics[width=0.2\columnwidth]{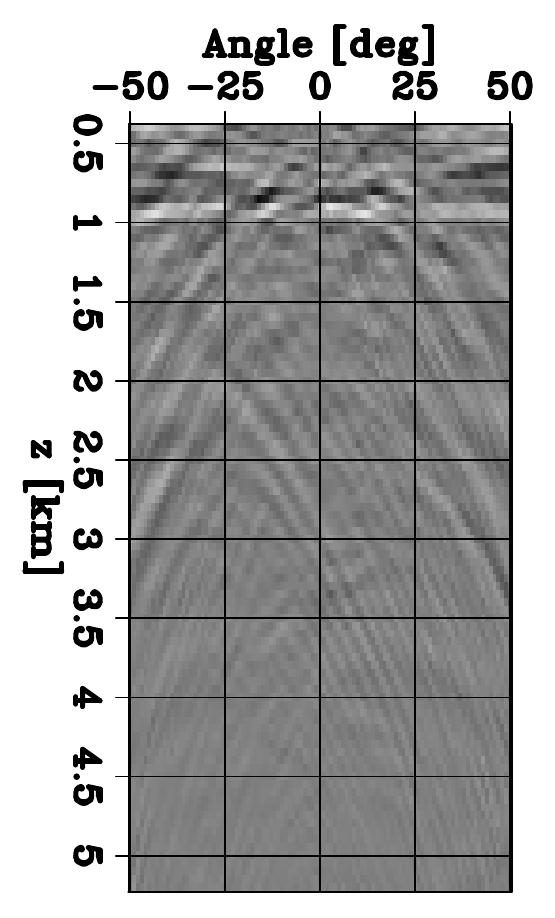}} \hspace{5mm} 
    \subfigure[]{\label{fig:Virieux_final_adcig5}\includegraphics[width=0.2\columnwidth]{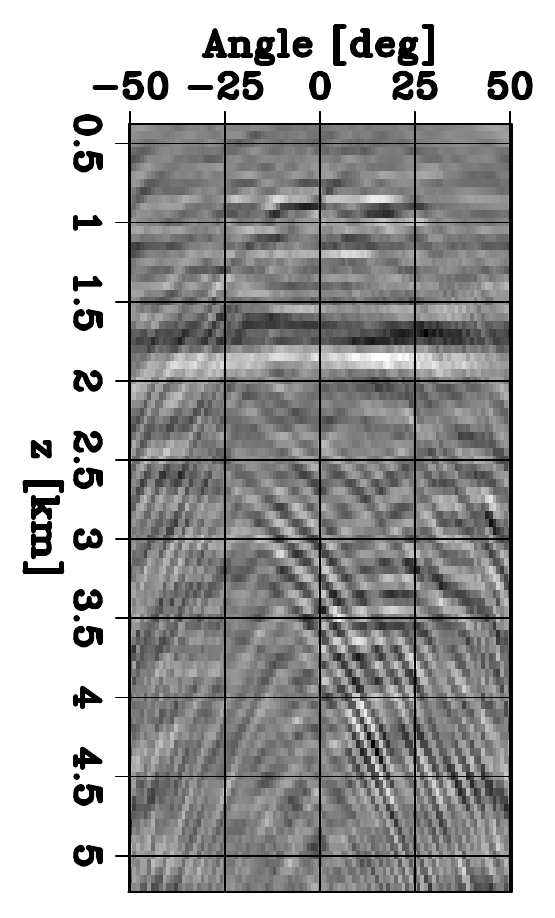}} \hspace{5mm} 
    \subfigure[]{\label{fig:Virieux_final_adcig6}\includegraphics[width=0.2\columnwidth]{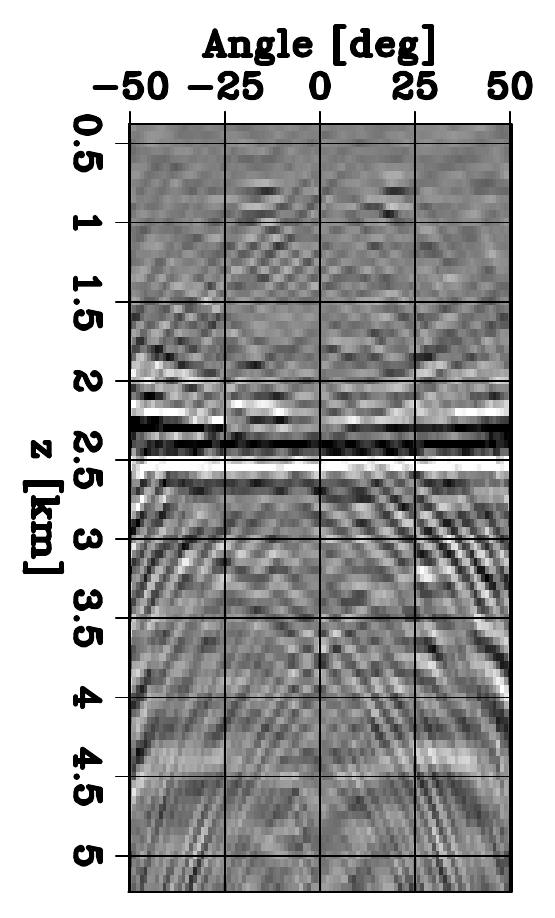}}
    \caption{Angle domain common image gathers (ADCIGs) computed with the initial model (top row) and the final FWIME inverted model (bottom row) at four different horizontal positions. First column is at $x = 3$ km, second column is at $x = 4$ km, third column is at $x = 5$ km, and fourth column is at $x = 6$ km. All panels are displayed on the same grayscale.}
    \label{fig:Virieux_adcig}
\end{figure}

For quality-control purposes, we migrate the observed data (after subtracting the direct arrival) using the initial velocity model (Figure~\ref{fig:Virieux_fwime_zero_offset_init}) and the final velocity model (Figure~\ref{fig:Virieux_fwime_zero_offset_final}). The migration is conducted with a RTM scheme. The second panel shows a clear improvement, which is confirmed by the ADCIGs extracted at various horizontal positions (Figure~\ref{fig:Virieux_adcig}). In the third and fourth columns of Figure~\ref{fig:Virieux_adcig}, the improvement of the flatness of the ADCIG confirms that the bottom of the basin is now accurately imaged. Furthermore, Figure~\ref{fig:Virieux_adcig} illustrates the ability of FWIME at leveraging any type of coherent information/moveout from the extended space of $\tilde{\mathbf{p}}_{\epsilon}^{opt}$ to accurately update the velocity model. Even though not tested here, we anticipate that more conventional techniques based on interpreting the curvature of the the moveouts within the ADCIGs \cite[]{zhang2015velocity} would not perform as well. 

In summary, model-space multi-scale FWIME performs well on this numerical example. First, it manages to automatically handle the phase identification issue occurring when complex waveforms are recorded in the data, as indicated by Seiscope. In addition, it successfully removes a large area filled with incorrect velocity values (with a strong velocity-contrast and sharp interfaces), which is a promising result for the next numerical example where we apply FWIME to recover a salt body from a pure sediment model.

%% file: bpSalt.tex
Following the recent deployment of ultra-long offset full-azimuth sparse-node acquisition surveys \cite[]{bate2021ultra}, we design a test to show that FWIME can leverage both refracted and reflected energy for velocity model building in complex subsalt regions. With optimal illumination, we show its potential at providing significant imaging uplifts without the need for low-frequency energy nor accurate initial model. 

\subsection{Ultra-long offset survey}
\begin{figure}[tbhp]
    \centering
    \subfigure[]{\label{fig:bpOz_true_mod}\includegraphics[width=0.45\columnwidth]{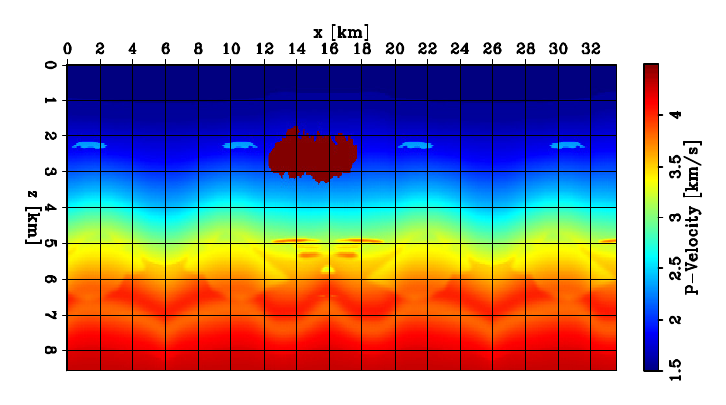}}
    \subfigure[]{\label{fig:bpOz_init_mod}\includegraphics[width=0.45\columnwidth]{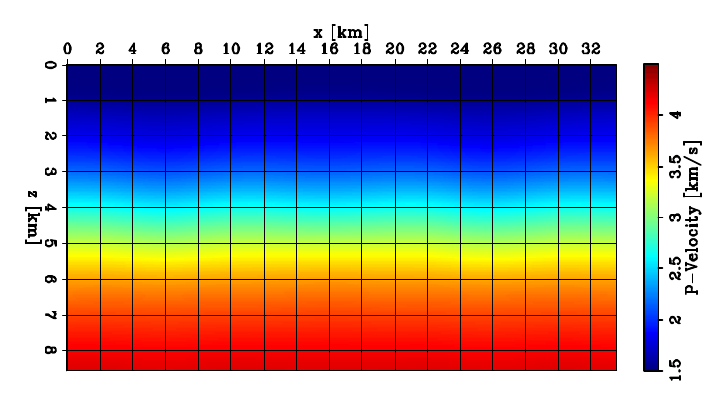}}    
    \caption{2D panels of velocity models. (a) True model. (b) Initial model.}
    \label{fig:bpOz_mod}
\end{figure}

\begin{figure}[tbhp]
    \centering
    \subfigure[]{\label{fig:bpOz_mod_1d_x14}\includegraphics[width=0.22\columnwidth]{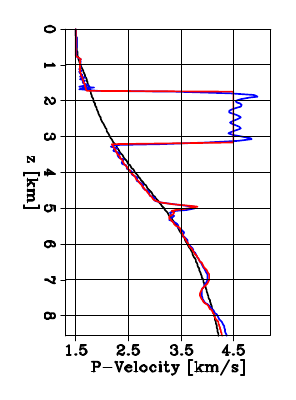}} \hspace{5mm}
    \subfigure[]{\label{fig:bpOz_mod_1d_x15}\includegraphics[width=0.22\columnwidth]{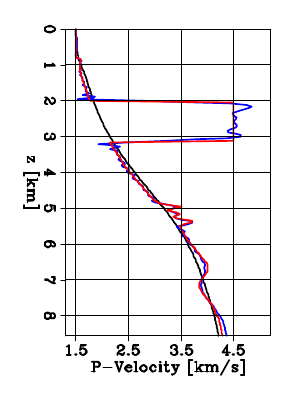}} 
    \subfigure[]{\label{fig:bpOz_mod_1d_x16}\includegraphics[width=0.22\columnwidth]{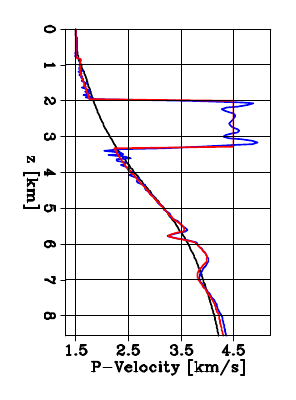}} \hspace{5mm}
    \subfigure[]{\label{fig:bpOz_mod_1d_x17}\includegraphics[width=0.22\columnwidth]{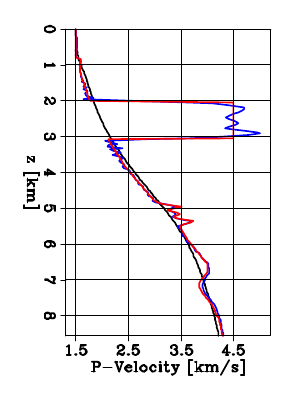}} 
    \caption{Depth velocity profiles extracted through the salt body at (a) $x=14$ km, (b) $x=15$ km, (c) $x=16$ km, and (d) $x=17$ km. The black curve represents the initial model, the red curve is the true model, and the blue curve is the final FWIME inverted model.}
    \label{fig:bpOz_mod_1d_x}
\end{figure}

The true model is 33 km wide and 8.5 km deep (Figure~\ref{fig:bpOz_true_mod}). It is composed of a sediment background (modified from \cite{billette20052004}) in which a 6 km-wide and 1 km-thick rugose salt body is embedded. The ultimate goal is to accurately image a set of four thin high-velocity layers and one low-velocity anomaly located underneath the salt, at an approximate depth of 5 km. The initial model is obtained by strongly smoothing the sediment background and does not contain any information about the presence of salt (Figure~\ref{fig:bpOz_init_mod} and black curves in Figures~\ref{fig:bpOz_mod_1d_x}). 

\begin{figure}[tbhp]
    \centering
    \subfigure[]{\label{fig:bpOz_obs_data_s0}\includegraphics[width=0.45\columnwidth]{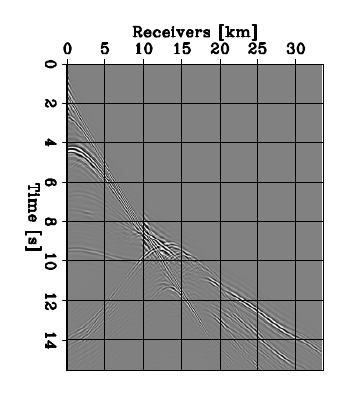}} \hspace{5mm}
    \subfigure[]{\label{fig:bpOz_obs_data_s70}\includegraphics[width=0.45\columnwidth]{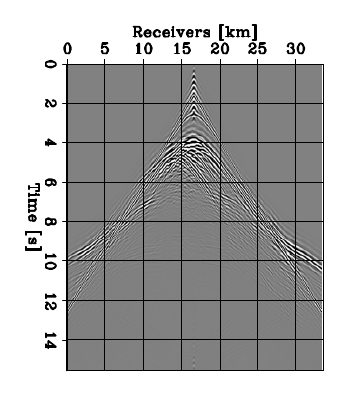}}
    \caption{Representative shot gathers of the observed data for sources placed at (a) $x = 0$ km and (b) $x=16$ km for the ultra-long offset acquisition. All panels are displayed with the same grayscale.}
    \label{fig:bpOz_obs_data}
\end{figure}

To simulate an ultra-long offset acquisition survey, we place 840 fixed receivers every 40 m and we generate noise-free pressure data with 139 sources. The distance between two consecutive sources is set to 240 m, and the maximum recorded offset for this test is 33 km. Both sources and receivers are placed at depth of 50 m below the surface. Figure~\ref{fig:bpOz_obs_data} shows two representative shot gathers for sources placed at $x=0$ km and $x=16$ km, which indicate the presence of refracted energy. For the FWIME workflow, the source wavelet contains energy limited to the 3-9 Hz range, and the data are modeled for 15 s with absorbing boundaries in all directions. This example proposes an ideal acquisition geometry (the salt body is well illuminated by diving waves propagating through it) and is conducted to show the usefulness of combining FWIME with novel long-offset node acquisition surveys to build accurate enough velocity models for FWI to succeed in the presence of complex overburdens. 

\begin{figure}[tbhp]
    \centering
    \subfigure[]{\label{fig:bpOz_wav_fwi_0_9}\includegraphics[width=0.3\columnwidth]{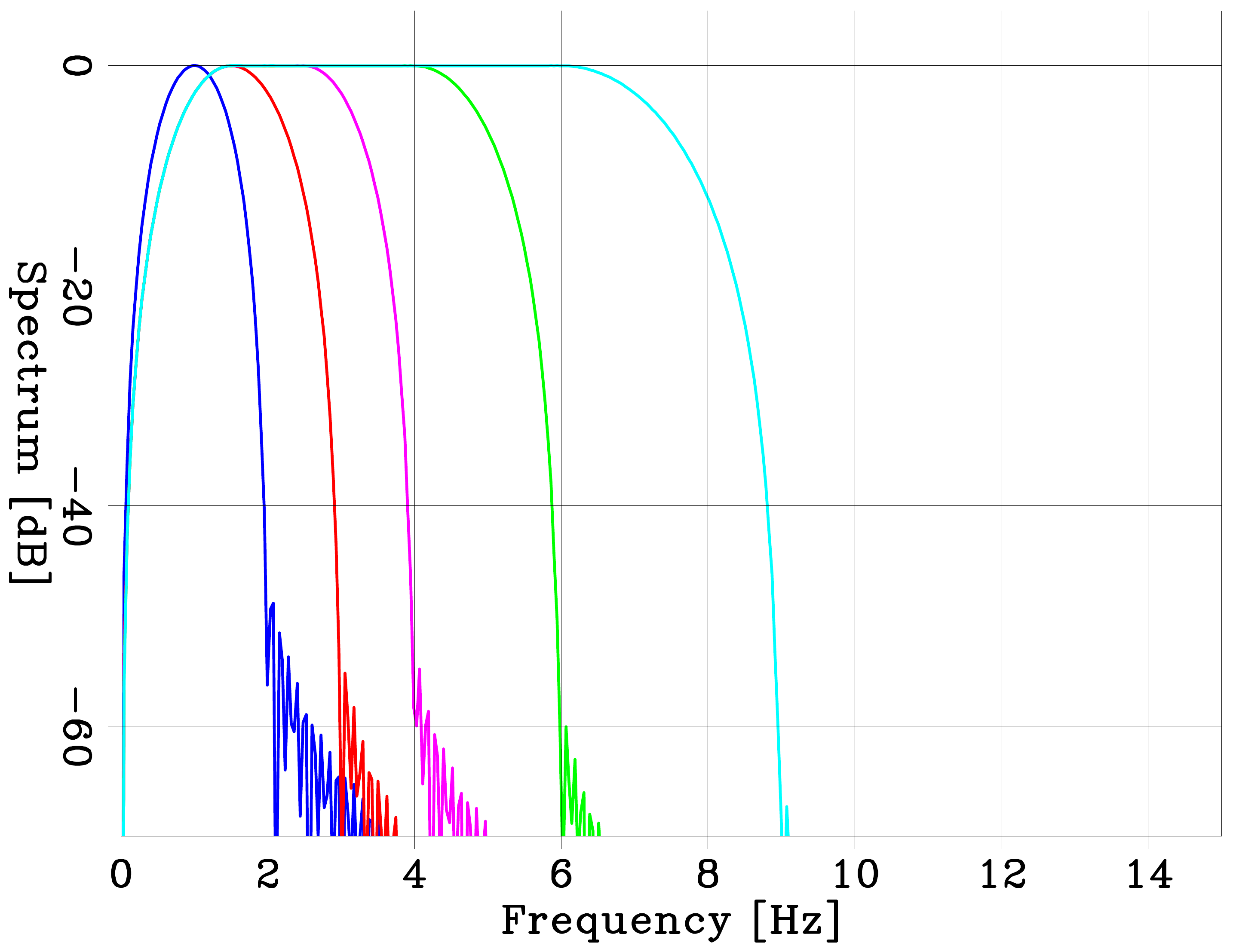}} \hspace{5mm}
    \subfigure[]{\label{fig:bpOz_wav_fwi_2_9}\includegraphics[width=0.3\columnwidth]{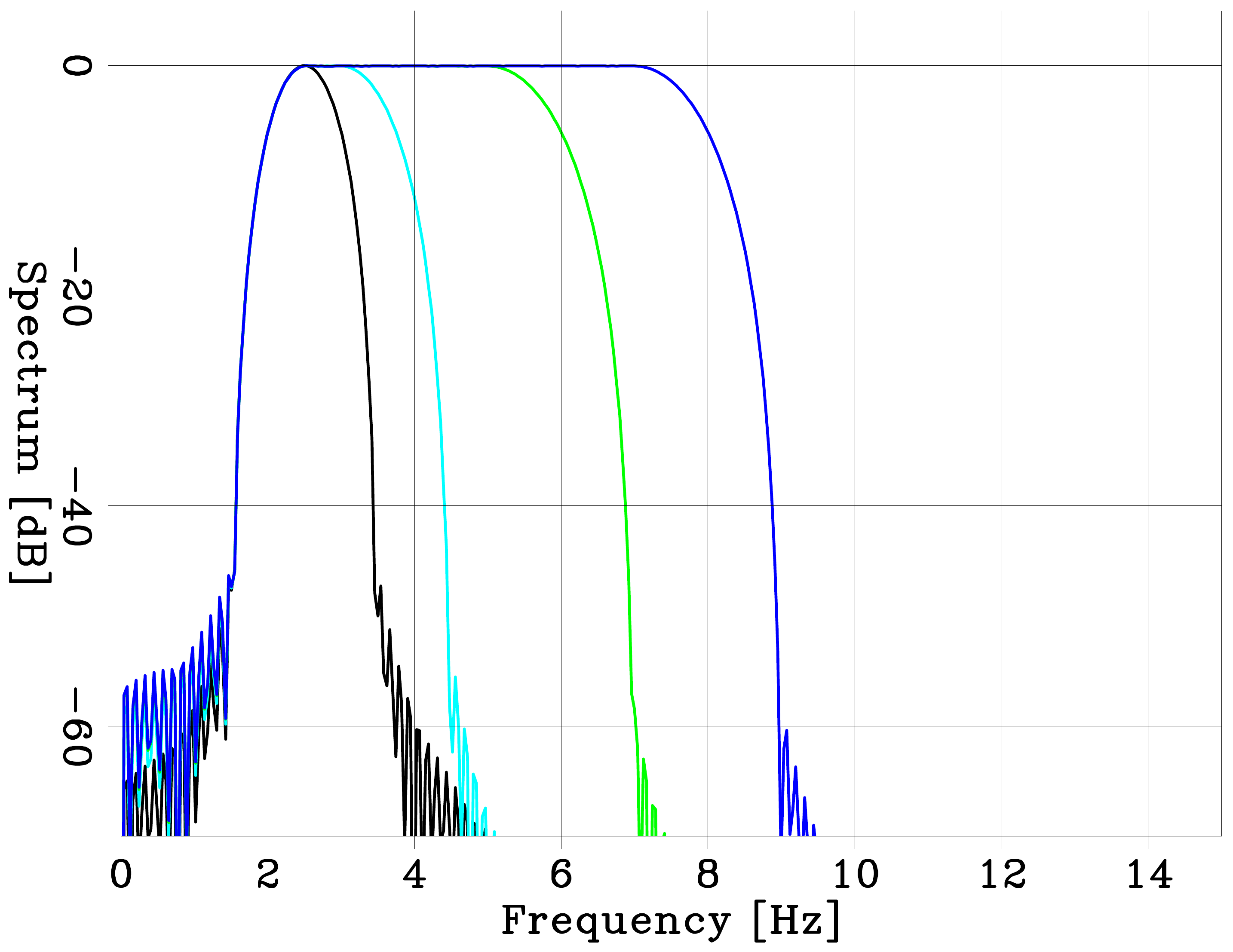}} \hspace{5mm}
    \subfigure[]{\label{fig:bpOz_wav_fwime}\includegraphics[width=0.3\columnwidth]{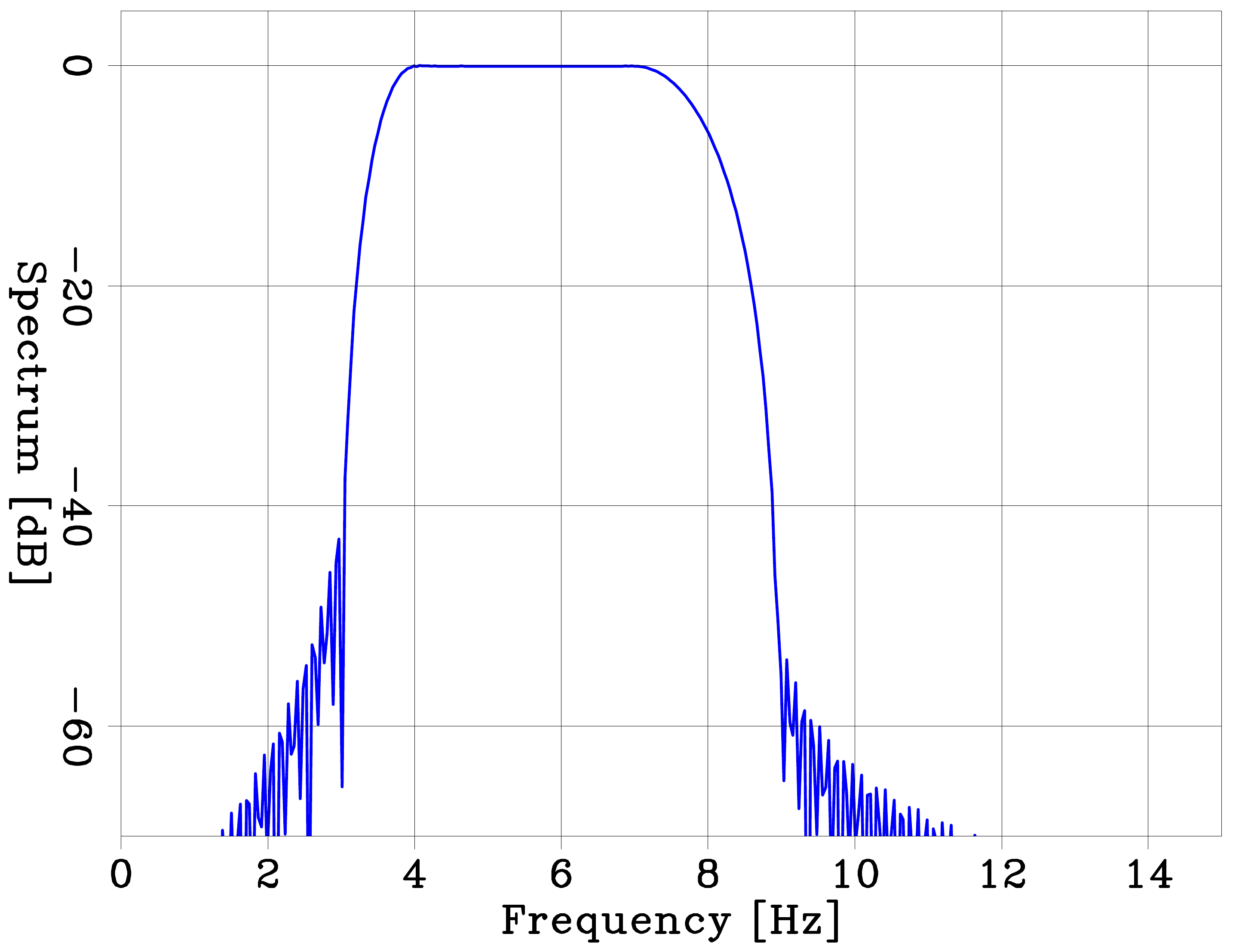}} \hspace{5mm}    
    \caption{Amplitude spectra of the seismic sources employed in this numerical example. (a) Sequence of sources used for the data-space multi-scale FWI workflow using unrealistic low-frequency energy. (b) Sequence of sources used for the data-space multi-scale FWI workflow using energy restricted to the 1.8-9 Hz bandwidth. (c) Source used for the FWIME workflow with energy restricted to 3-9 Hz.}
    \label{fig:bpOz_wav}
\end{figure}

\begin{figure}[tbhp]
    \centering
    \subfigure[]{\label{fig:bpOz_recap_init_mod}\includegraphics[width=0.45\columnwidth]{Fig/BP_Oz/bpSaltNew-init-mod.pdf}}
    \subfigure[]{\label{fig:bpOz_recap_fwi_0_9_mod}\includegraphics[width=0.45\columnwidth]{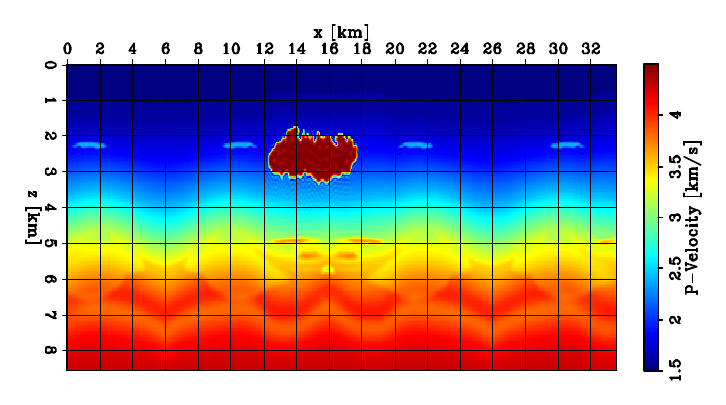}} \\
    \subfigure[]{\label{fig:bpOz_recap_fwi_2_9_mod}\includegraphics[width=0.45\columnwidth]{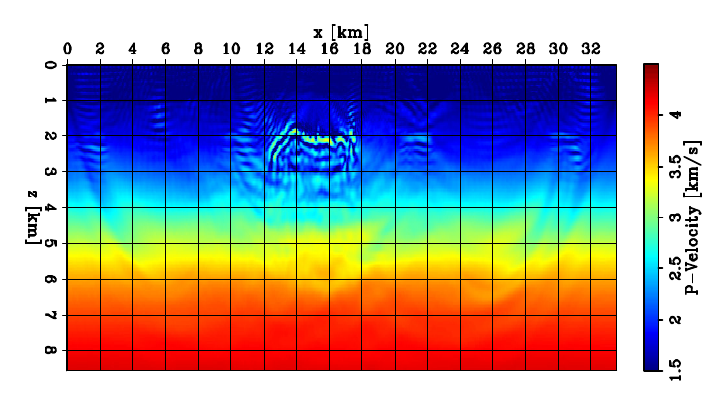}}   
    \subfigure[]{\label{fig:bpOz_recap_true_mod}\includegraphics[width=0.45\columnwidth]{Fig/BP_Oz/bpSaltNew-true-mod.pdf}}
    \caption{2D panels of velocity models. (a) Initial model used for both the FWI and FWIME schemes. (b) Inverted model after conventional multi-scale FWI using unrealistic low-frequency energy restricted to the 0-9 Hz range. (c) Inverted model after conventional multi-scale FWI using energy restricted to the 1.8-9 Hz range. (d) True model.}
    \label{fig:bpOz_fwi_mod}
\end{figure}

We first conduct multi-scale FWI using the initial model shown in Figure~\ref{fig:bpOz_init_mod} assuming the presence of unrealistic low-frequency energy in the recorded data, as shown by the five wavelet spectra in Figure~\ref{fig:bpOz_wav_fwi_0_9}. The final inverted model is extremely accurate (Figure~\ref{fig:bpOz_recap_fwi_0_9_mod}) and provides an estimate of the best recoverable solution using waveform-inversion schemes with this particular acquisition geometry. We apply a second data-space multi-scale FWI workflow (with the same initial model) assuming the presence of coherent energy as low as 1.8 Hz, which is still quite optimistic for field data. The sequence of wavelet spectra employed for this scheme is shown in Figure~\ref{fig:bpOz_wav_fwi_2_9}. In this case, FWI fails to converge to a physical solution (Figure~\ref{fig:bpOz_recap_fwi_2_9_mod}), which highlights the difficulty of retrieving the salt body from a pure sediment model. 

For the FWIME scheme, we invert data generated with a source wavelet whose spectrum is shown in Figure~\ref{fig:bpOz_wav_fwime}. To increase the difficulty of this test for FWIME, the lowest useful frequency is set to 3 Hz, which is even higher than the lowest available frequency from the first band used in conventional FWI (blue curve in Figure~\ref{fig:bpOz_wav_fwi_2_9}). We use a time-lag extension spanning the $[-1.2 \; \rm{s}, \; 1.2 \; \rm{s}]$ interval with 101 points sampled at $\Delta \tau=24$ ms and we set $\epsilon=2.5 \times 10^{-4}$. The parameters of the six spline grids employed are shown in Table~\ref{table:bpOz_spline}. In addition, we assume the bathymetry is approximately known and for the first, second, third, fourth, and fifth grids, we use a finer spatial sampling of 100 m in the vicinity of the water bottom to allow the recovered velocity models to correctly predict the strong reflection generated from this interface. 

\begin{table}[h!]
\centering
\begin{tabular}{ |c|c|c|c|  } 
\hline
\textbf{Grid number} & $\Delta z$ [km] & $\Delta x$ [km]\\
\hline
 0 & 0.5 & 1.2 \\
 1 & 0.3 & 0.8 \\
 2 & 0.25 & 0.6 \\
 3 & 0.12 & 0.4 \\
 4 & 0.06 & 0.12 \\
 5 & 0.04 & 0.04 \\
\hline
\end{tabular}
\caption{Parameters of the spline grid sequence used for the model-space multi-scale FWIME scheme applied to the salt model. Spline 5 coincides with the finite-difference grid. For spline grids 0, 1, 2, 3 and 4, we use a finer sampling in the z-direction in the vicinity of the water bottom (not shown in the table). }
\label{table:bpOz_spline}
\end{table}

\begin{figure}[tbhp]
    \centering
    \subfigure[]{\label{fig:bpOz_grad_Born}\includegraphics[width=0.45\columnwidth]{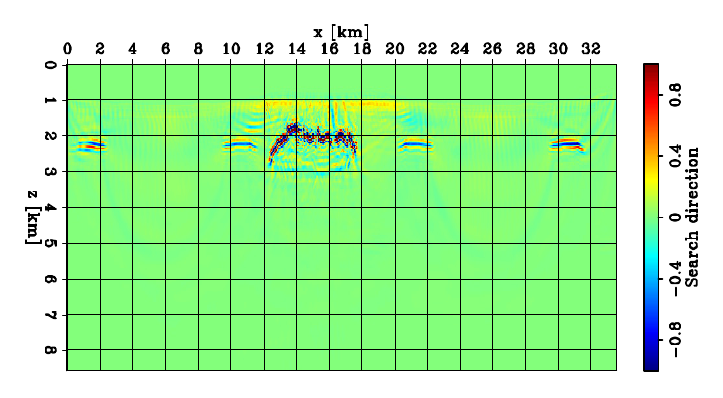}}
    \subfigure[]{\label{fig:bpOz_grad_tomo}\includegraphics[width=0.45\columnwidth]{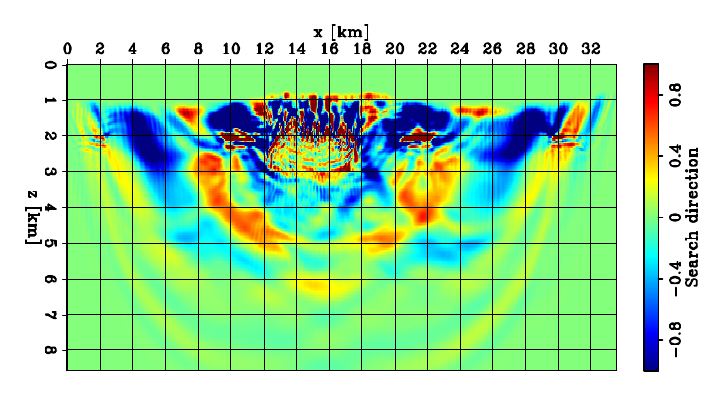}} \\
    \subfigure[]{\label{fig:bpOz_grad_total}\includegraphics[width=0.45\columnwidth]{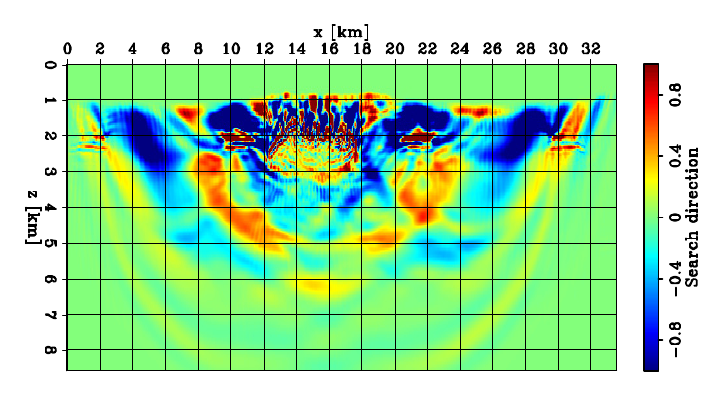}}
    \subfigure[]{\label{fig:bpOz_grad_true}\includegraphics[width=0.45\columnwidth]{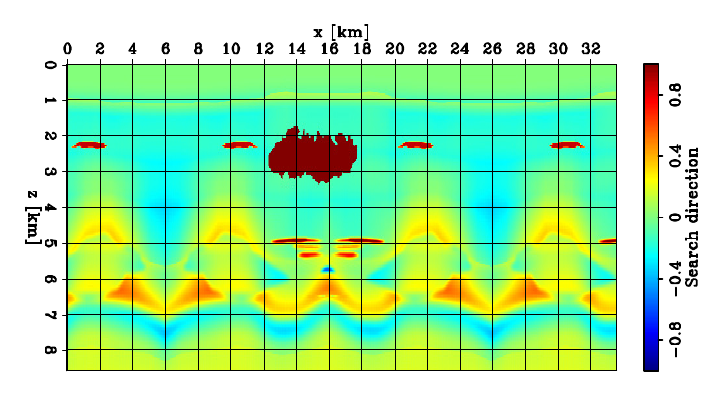}}
    \caption{Normalized initial search directions (before applying any spline parametrization). (a) Born component of the FWIME search direction. (b) Tomographic component of the search direction. (c) Total FWIME search direction (sum of panels (a) and (b)). (d) True search direction. Panels (a), (b), and (c) are normalized with the same value.}
    \label{fig:bpOz_grad}
\end{figure}

\begin{figure}[tbhp]
    \centering
    \subfigure[]{\label{fig:bpOz_grad_Born_spline}\includegraphics[width=0.45\columnwidth]{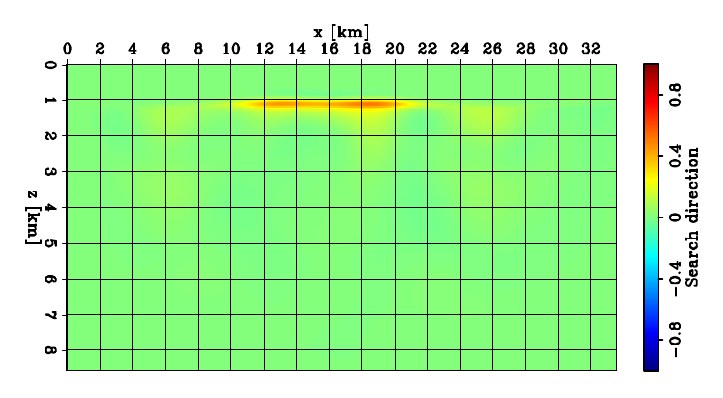}}
    \subfigure[]{\label{fig:bpOz_grad_tomo_spline}\includegraphics[width=0.45\columnwidth]{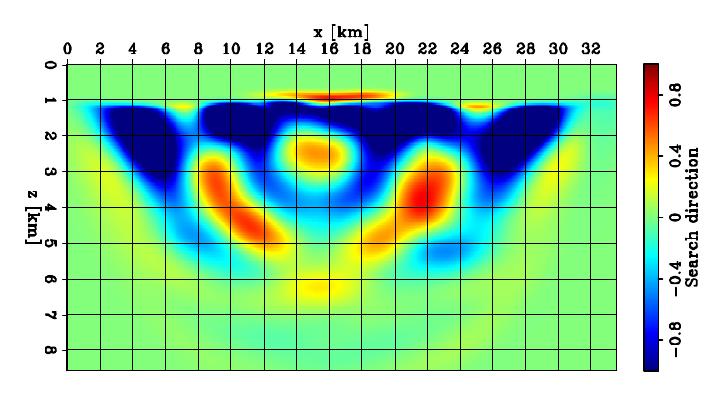}} \\
    \subfigure[]{\label{fig:bpOz_grad_total_spline}\includegraphics[width=0.45\columnwidth]{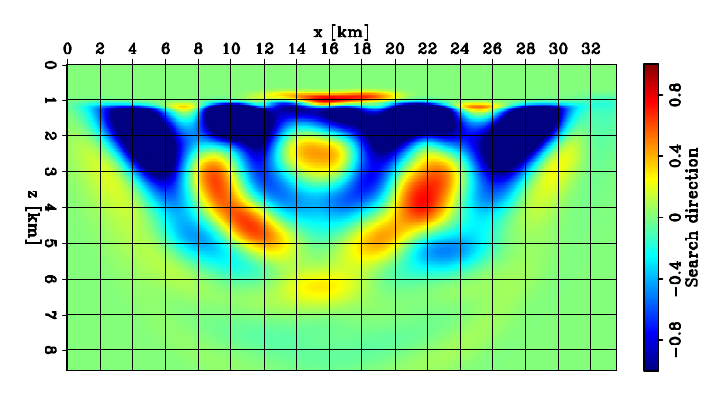}}
    \subfigure[]{\label{fig:bpOz_grad_true_spline}\includegraphics[width=0.45\columnwidth]{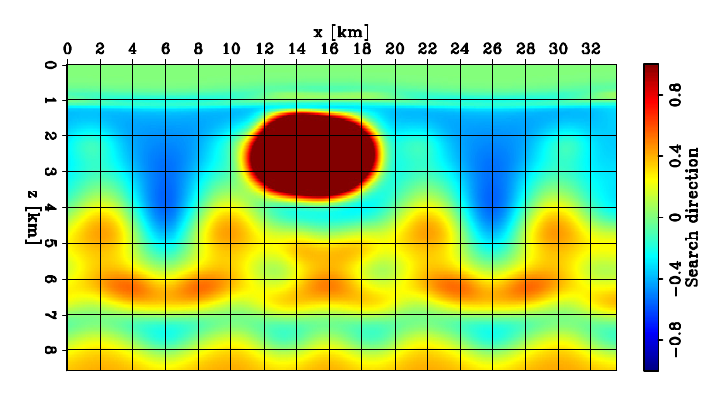}}
    \caption{Normalized initial search directions after being mapped on the first spline grid (obtained by applying $\mathbf{S}_0\mathbf{S}^*_0$ to the panels in Figure~\ref{fig:bpOz_grad}). (a) Born component of the FWIME search direction. (b) Tomographic component of the search direction. (c) Total FWIME search direction (sum of panels (a) and (b)). (d) True search direction. Panels (a), (b), and (c) are normalized with the same value.}
    \label{fig:bpOz_grad_spline}
\end{figure}

\begin{figure}[tbhp]
    \centering
    \subfigure[]{\label{fig:bpOz_grad_fwi_fd}\includegraphics[width=0.45\columnwidth]{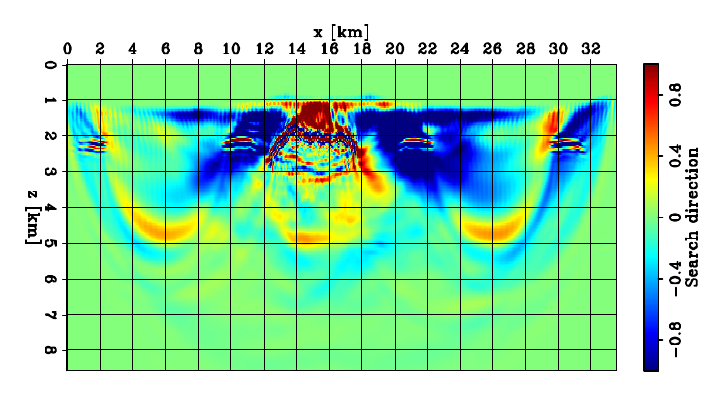}}
    \subfigure[]{\label{fig:bpOz_grad_fwi_spline}\includegraphics[width=0.45\columnwidth]{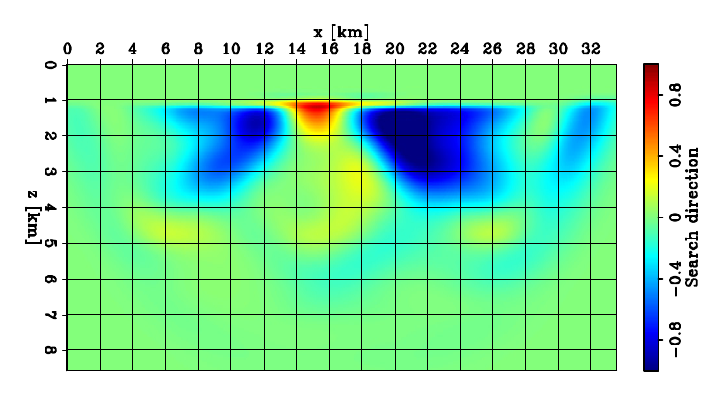}}
    \caption{Normalized initial search directions. (a) Conventional FWI. (b) Conventional FWI after applying $\mathbf{S}_0\mathbf{S}^*_0$ to panel (a).}
    \label{fig:bpOz_grad_fwi}
\end{figure}

We examine the FWIME Born, tomographic, and total search directions on the finite-difference grid shown in Figures~\ref{fig:bpOz_grad}a-c (normalized by the same value). The tomographic component dominates the Born and seems to guide the inversion towards the true solution (Figure~\ref{fig:bpOz_grad_true}). After our parametrization on the first spline grid, the FWIME initial update direction is further improved. The high-wavenumber artifacts located on top of the salt body have been removed, and the velocity values are increased at the exact position of the salt body. Underneath the salt, we also observe a good phase matching between Figures~\ref{fig:bpOz_grad_total_spline} and \ref{fig:bpOz_grad_true_spline}. This analysis confirms that the tomographic component plays an important role at early stages. In contrast, FWI's update direction is not able to capture the missing low-wavenumber components, even using a spline grid re-parametrization (Figure~\ref{fig:bpOz_grad_fwi}). 

\begin{figure}[tbhp]
    \centering
    \subfigure[]{\label{fig:bpOz_fwime_init_mod}\includegraphics[width=0.45\columnwidth]{Fig/BP_Oz/bpSaltNew-init-mod.pdf}}
    \subfigure[]{\label{fig:bpOz_fwime_s2b_mod}\includegraphics[width=0.45\columnwidth]{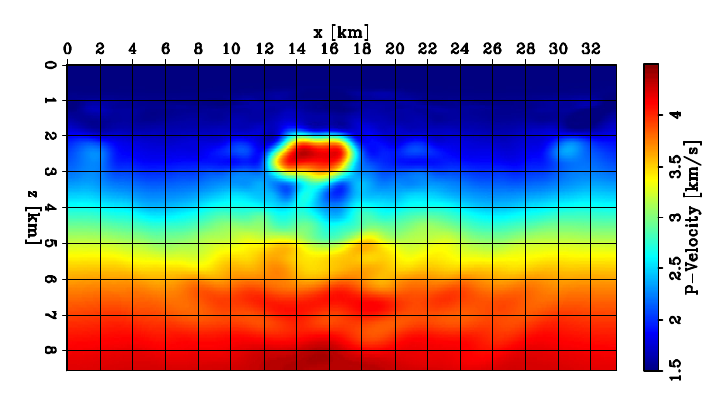}}  \\
    \subfigure[]{\label{fig:bpOz_fwime_s2_mod}\includegraphics[width=0.45\columnwidth]{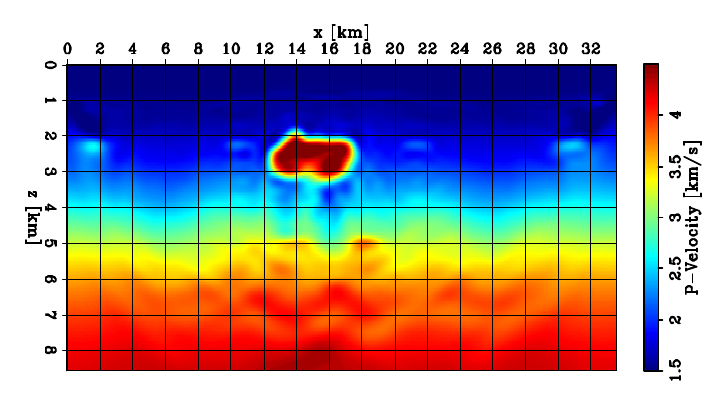}} 
    \subfigure[]{\label{fig:bpOz_fwime_s3_mod}\includegraphics[width=0.45\columnwidth]{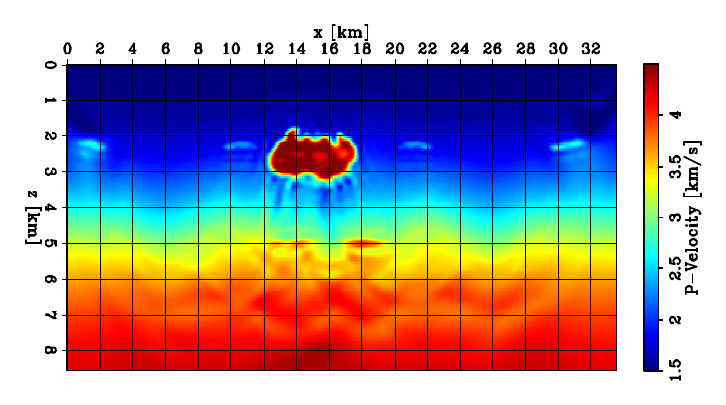}} \\
    \subfigure[]{\label{fig:bpOz_fwime_noSpline_mod}\includegraphics[width=0.45\columnwidth]{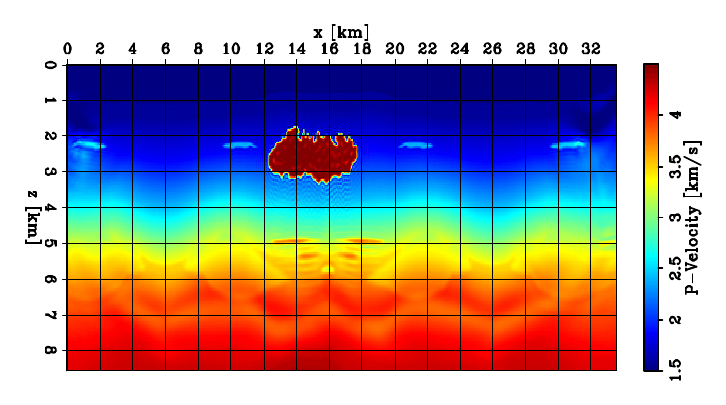}}
    \subfigure[]{\label{fig:bpOz_fwime_true}\includegraphics[width=0.45\columnwidth]{Fig/BP_Oz/bpSaltNew-true-mod.pdf}}
    \caption{2D panels of velocity models inverted on the various spline grids throughout the FWIME workflow: (a) Initial model (b) first grid, (c) second grid, (d) fourth grid, and (e) final inverted model after a total of 250 L-BFGS iterations. (f) True model.}
    \label{fig:bpOz_fwime_mod}
\end{figure}

The sequence of FWIME inverted models is shown in Figure~\ref{fig:bpOz_fwime_mod}. Each model is inverted on its respective spline grid and then mapped on the finite-difference grid for display purposes. The inversion on the first grid is crucial as it retrieves a low-resolution salt body (i.e., a geobody  with smooth contours) (Figure~\ref{fig:bpOz_fwime_s2b_mod}). Then, the contours of the salt are gradually sharpened as the spline grid is refined (Figures~\ref{fig:bpOz_fwime_mod}c and \ref{fig:bpOz_fwime_mod}d). Once the salt body is accurately reconstructed, the thin subsalt reflectors are then well image, and the inverted model is excellent (Figure~\ref{fig:bpOz_fwime_noSpline_mod}). 

\begin{figure}[tbhp]
    \centering
    \includegraphics[width=0.45\columnwidth]{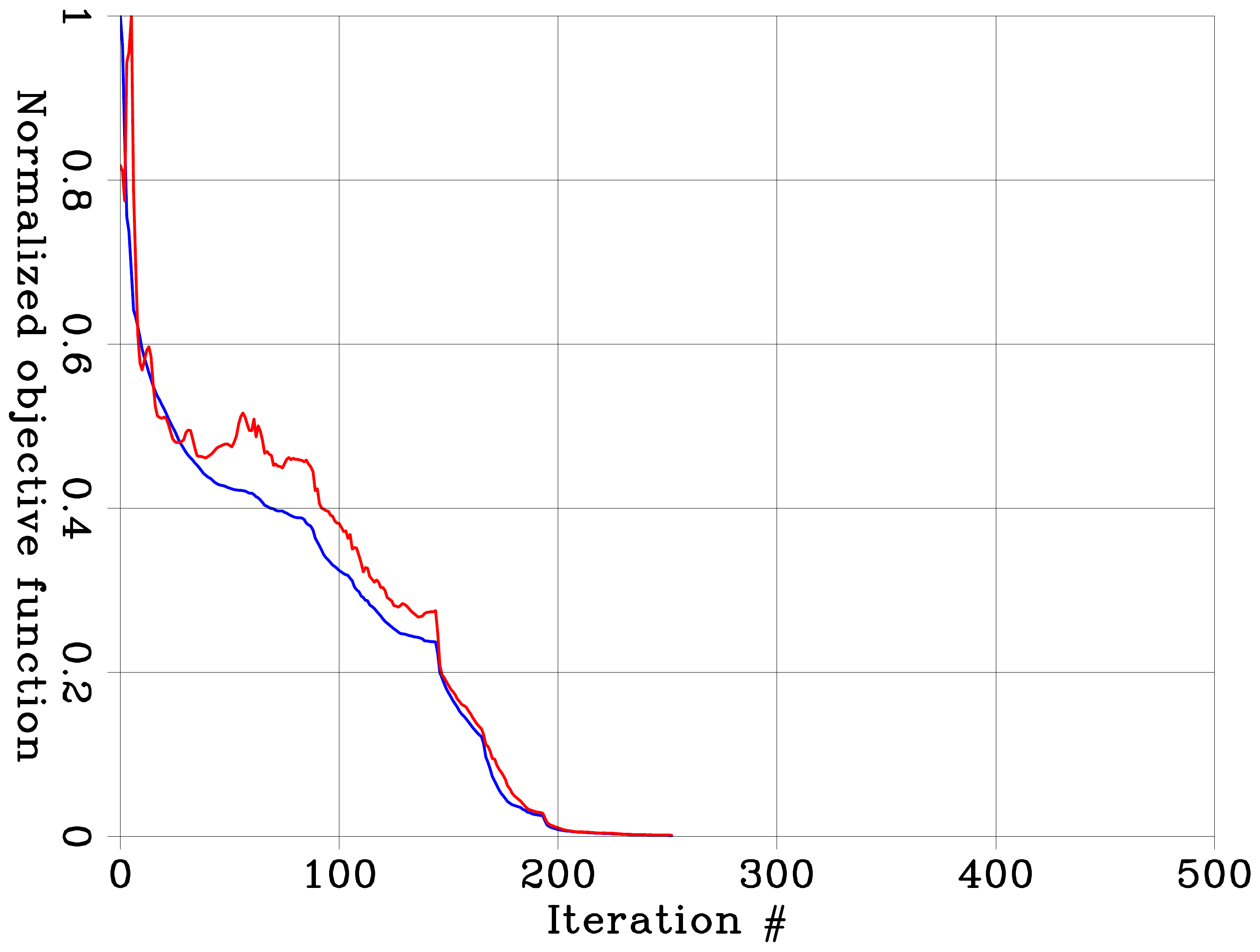}
    \caption{Convergence curves for the numerical test conducted on the wide salt model (Figure~\ref{fig:bpOz_true_mod}). Total normalized FWIME objective function (blue curve), and normalized FWI objective function evaluated at each FWIME inverted model (red curve). We obtain full convergence after 250 L-BFGS iterations.}
    \label{fig:bpSalt_obj}
\end{figure}

Figure~\ref{fig:bpSalt_obj} displays the normalized FWIME objective function (blue curve) along with the value of the conventional FWI objective function computed at each inverted model during the FWIME sequence (red curve). As expected, the red curve is not monotonically decreasing but behaves differently than the analogous curve obtained in the example based on the Marmousi2 model from the previous section (Figure~\ref{fig:Marmousi_obj_fwime_fwi}), where the inverted dataset is dominated by reflections. It is interesting to notice that the red curve becomes smoother and seems to contain less local minima than its counterpart from Figure~\ref{fig:Marmousi_obj_fwime_fwi}. This observation potentially corroborates the claim that transmitted energy makes seismic inversion problems better conditioned than pure reflections \cite[]{gauthier1986two}. 

\begin{figure}[tbhp]
    \centering
    \subfigure[]{\label{fig:bpOz_fwime_fail1_mod}\includegraphics[width=0.45\columnwidth]{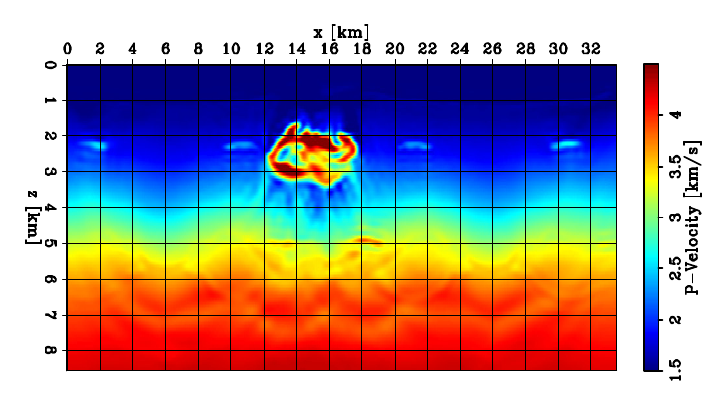}}
    \subfigure[]{\label{fig:bpOz_fwime_fail2_mod}\includegraphics[width=0.45\columnwidth]{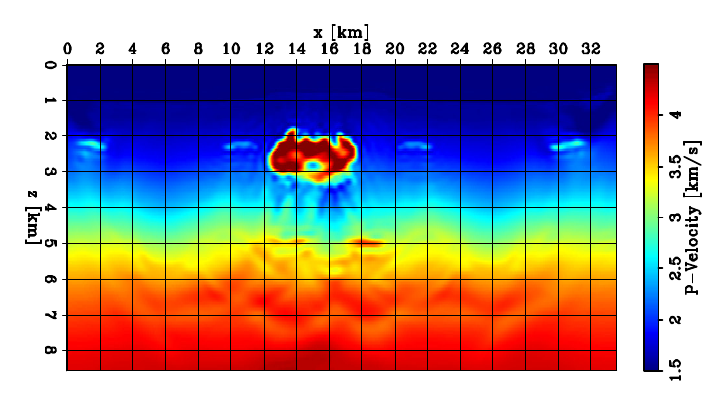}}    
    \caption{2D panels of FWIME models inverted with a different spline grid refinement schedule as the one shown in Table~\ref{table:bpOz_spline}. (a) Inverted model starting from spline 2. (b) Inverted model without the use of spline 2.}
    \label{fig:bpOz_fwime_fail_mod}
\end{figure}

We observe here that the quality of the inverted model is mostly affected by the selection of the spline grid sequence rather than by the trade-off parameter value $\epsilon$. Figure~\ref{fig:bpOz_fwime_fail_mod} shows two FWIME inverted models obtained with different spline grid sequences: Figure~\ref{fig:bpOz_fwime_fail1_mod} results from starting the inversion directly on the second grid, whereas Figure~\ref{fig:bpOz_fwime_fail2_mod} is computed by skipping the second grid. In both figures, the contours of salt body have been correctly reconstructed but the inversion is unable to automatically fill the salt with the correct velocity value. Moreover, by conducting additional tests (not shown here), we notice that the presence of complex overburdens (or strong velocity contrasts) may generally require a slower refinement rate in the spline grid sequence to ensure convergence to an accurate solution. 

\subsection{Reducing the acquisition offset}
\begin{figure}[tbhp]
    \centering
    \subfigure[]{\label{fig:bpNarrow_init_mod}\includegraphics[width=0.45\columnwidth]{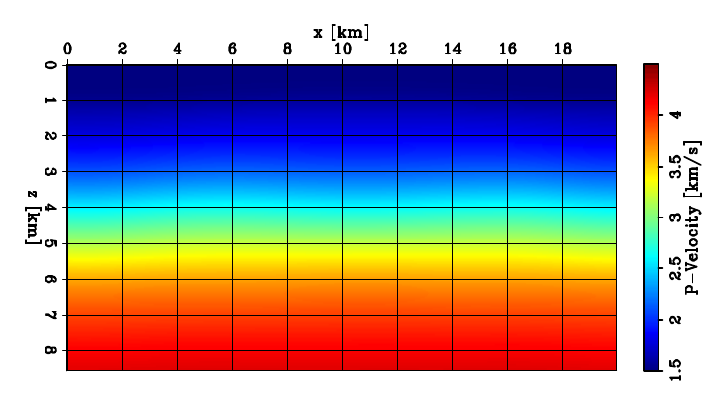}}
    \subfigure[]{\label{fig:bpOz_fwime_it250_mod}\includegraphics[width=0.45\columnwidth]{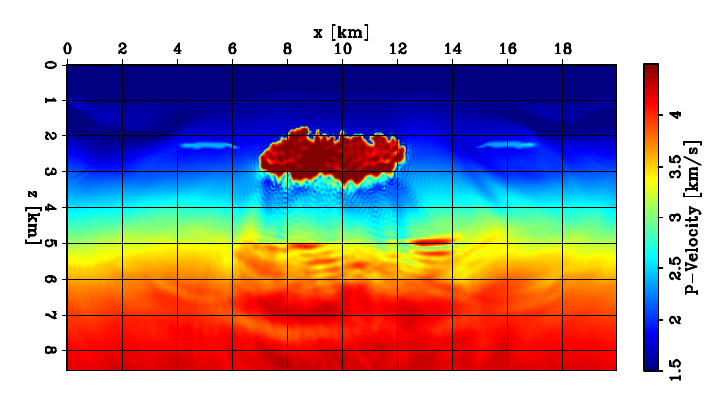}} \\
    \subfigure[]{\label{fig:bpOz_fwime_it500_mod}\includegraphics[width=0.45\columnwidth]{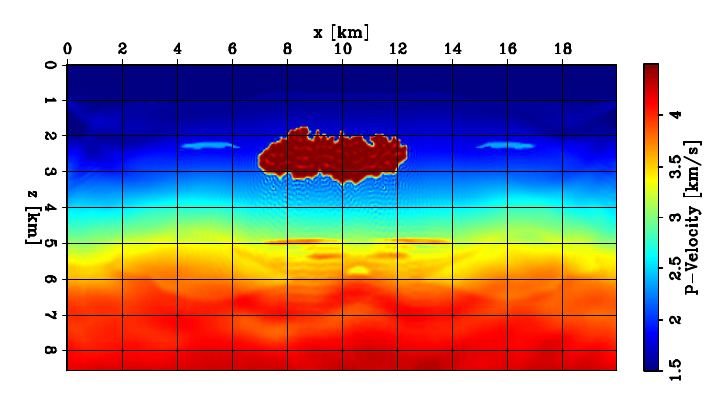}} 
    \subfigure[]{\label{fig:bpNarrow_true_mod}\includegraphics[width=0.45\columnwidth]{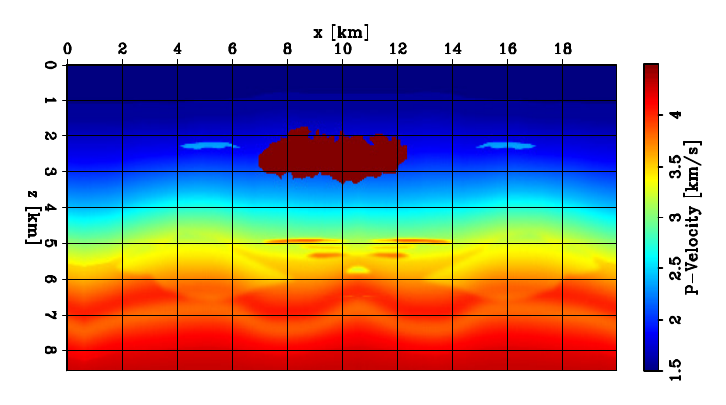}}    
    \caption{2D panels of velocity models inverted on the various spline grids throughout the FWIME workflow: (a) Initial narrow salt model, (b) FWIME model after 250 L-BFGS iterations, (c) FWIME inverted model after 500 L-BFGS iterations, and (d) true model.}
    \label{fig:bpNarrow_fwime_mod}
\end{figure}

We assess the effect of reducing the maximum acquisition offset on the quality of the FWIME results. The goal is to test whether FWIME can be used without deploying ultra-long offset (and costly) surveys to build accurate initial salt models. We design an analogous test as in the previous section but we decrease the width of our area of study. Figures~\ref{fig:bpNarrow_init_mod} and \ref{fig:bpNarrow_true_mod} show the initial and true models, which are 20 km wide instead of 34 km. The rest of the geological features are identical to the ones from Figure~\ref{fig:bpOz_mod}. We deploy 500 fixed receivers every 40 m and we generate noise-free pressure data with 83 sources. The frequency content is identical to the one used in the previous example (Figure~\ref{fig:bpOz_wav_fwime}). 

\begin{figure}[t]
    \centering
    \subfigure[]{\label{fig:}\includegraphics[width=0.45\columnwidth]{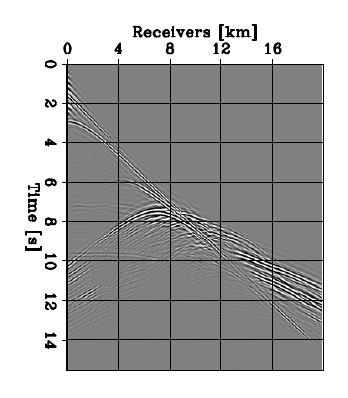}} \hspace{5mm}
    \subfigure[]{\label{fig:}\includegraphics[width=0.45\columnwidth]{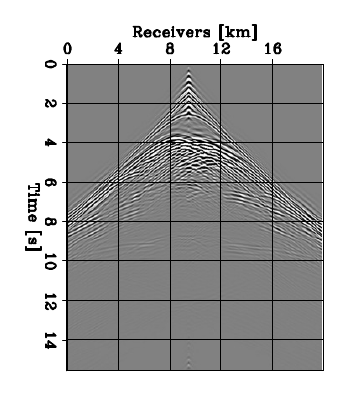}}
    \caption{Representative shot gathers of the observed data for sources placed at (a) $x = 0$ km and (b) $x=9$ km for the narrow salt model. All panels are displayed with the same grayscale. }
    \label{fig:bpNarrow_obs_data}
\end{figure}

Figure~\ref{fig:bpNarrow_obs_data} shows two shot gathers from this dataset, which indicate that less transmitted energy is present. For the FWIME workflow, we use the same spline grid refinement schedule, and we set $\epsilon=2.5 \times 10^{-4}$. As shown in Figure~\ref{fig:bpOz_fwime_it500_mod}, FWIME recovers a solution as accurate as the one from the previous section. However, the algorithm converged after approximately 500 L-BGFS instead of 250 L-BFGS. The inverted model after 250 L-BFGS iterations (Figure~\ref{fig:bpOz_fwime_it250_mod}) shows that more iterations are needed to correctly capture the edges of the salt body. At that point, the subsalt reflectors are poorly imaged. 

\begin{figure}[tbhp]
    \centering
    \includegraphics[width=0.45\columnwidth]{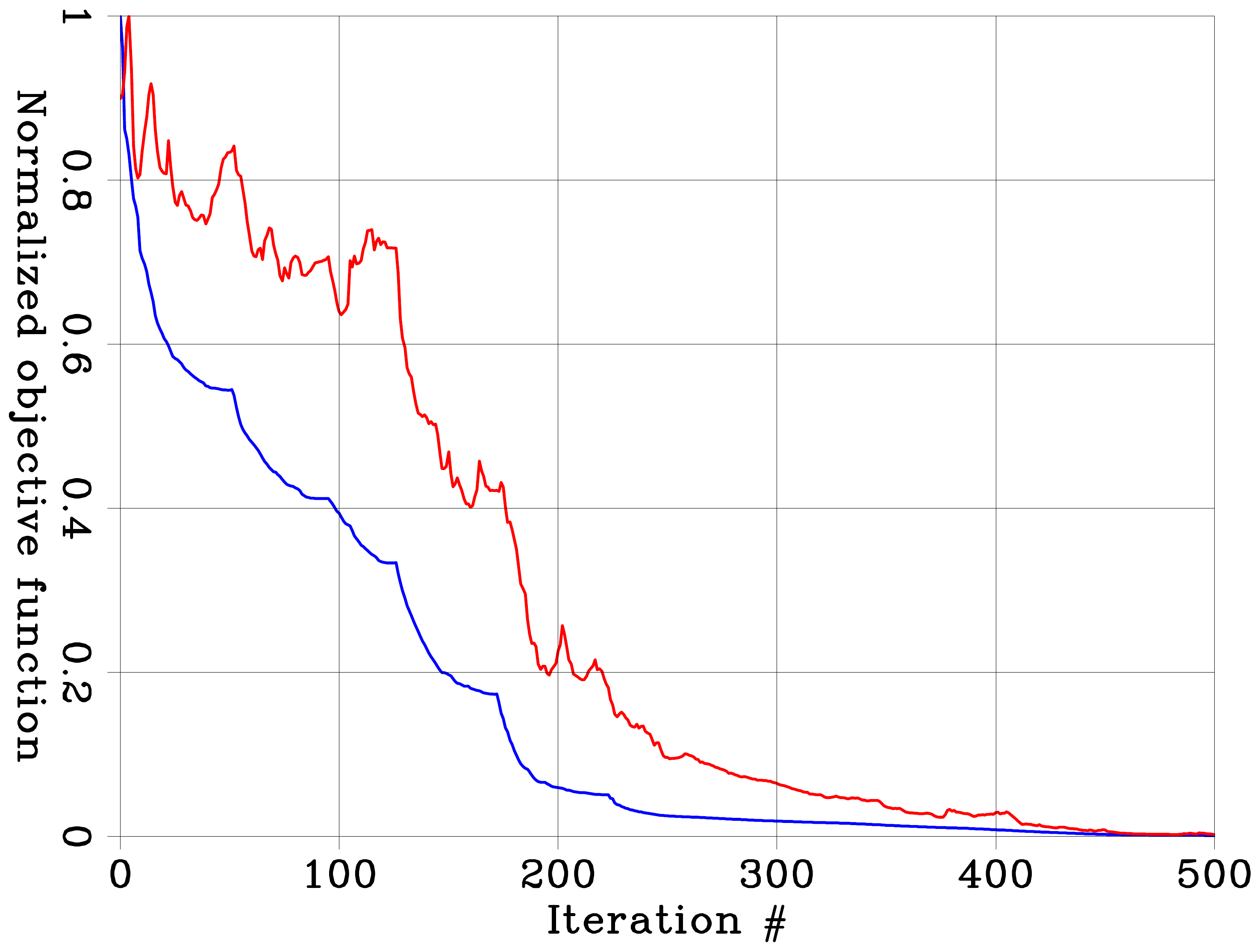}
    \caption{Convergence curves for the numerical test conducted on the narrow salt model (Figure~\ref{fig:bpNarrow_true_mod}). Total normalized FWIME objective function (blue curve), and normalized FWI objective function evaluated at each FWIME inverted model (red curve). We obtain full convergence after 500 L-BFGS iterations.}
    \label{fig:bpSaltNarrow_obj}
\end{figure}

Finally, Figure~\ref{fig:bpSaltNarrow_obj} displays the FWIME convergence curve (blue curve) along with the evaluated FWI objective function (red curve). By comparing it with Figure~\ref{fig:bpSalt_obj}, we can clearly observe that reducing the amount of transmitted energy has made the inversion problem more ill-posed, but FWIME is eventually able to mitigate the presence of local minima. 

%% file: bpSaltFeeder.tex
We design an experiment based on the BP 2004 benchmark model \cite[]{billette20052004} to assess the limitations of FWIME when imaging in the presence of thick and complex overburdens. Our study focuses on the central part of the original BP 2004 model, which is representative of the geological environments encountered in West Africa. We show that FWIME does not manage to recover an accurate solution. However, we anticipate that our method can be conveniently improved by designing and employing a more flexible coarse-model representation, such as radial basis functions (RBF) \cite[]{buhmann2000radial,martin2015seismic}. 

\begin{figure}[tbhp]
    \centering
    \subfigure[]{\label{fig:bpFeeder_init_mod}\includegraphics[width=0.45\columnwidth]{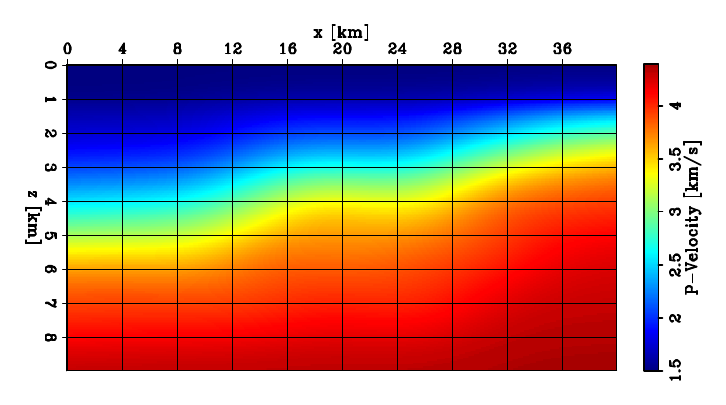}}
    \subfigure[]{\label{fig:bpFeeder_fwi_0_8_mod}\includegraphics[width=0.45\columnwidth]{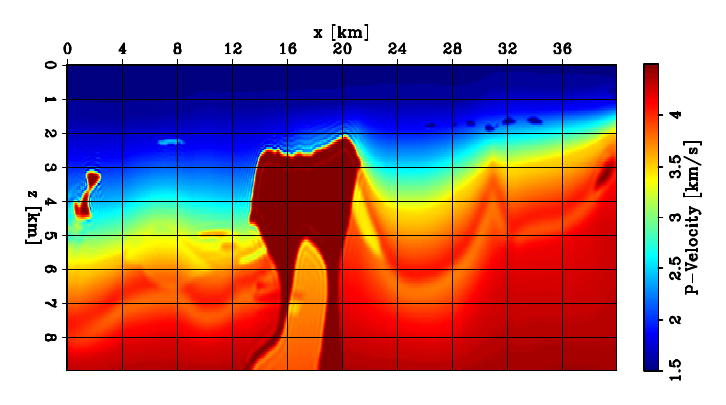}} \\
    \subfigure[]{\label{fig:bpFeeder_fwi_3_8_mod}\includegraphics[width=0.45\columnwidth]{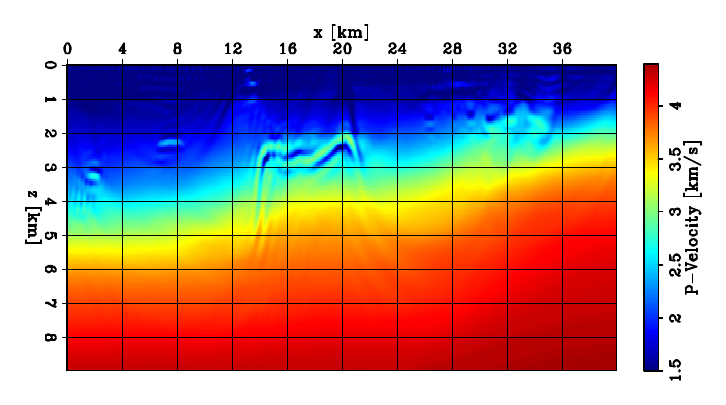}}
    \subfigure[]{\label{fig:bpFeeder_true_mod}\includegraphics[width=0.45\columnwidth]{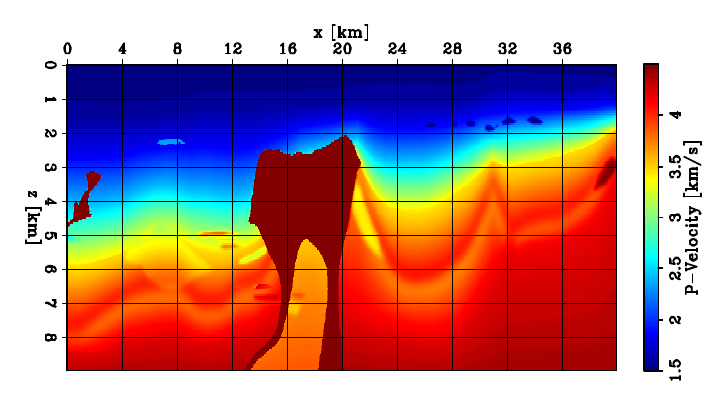}}    
    \caption{2D panels of velocity models. (a) Initial model. (b) Conventional data-space multi-scale FWI inverted model using a 0-8 Hz dataset. (c) Conventional data-space multi-scale FWI inverted model using a 3-8 Hz dataset. (d) True model.}
    \label{fig:bpFeeder_mod}
\end{figure}

The true velocity is 40 km wide and 9 km deep (Figure~\ref{fig:bpFeeder_true_mod}). The salt body is thick, deeply rooted, and presents steep flanks, which makes its boundaries challenging to delineate. In addition, a sharp interface between the base salt and a low-velocity zone is present underneath the main salt structure and ranges from an approximate depth of 5 km, down to the bottom of the model. Even with a long-offset acquisition geometry which generates diving waves, very little recorded energy can illuminate this region, making its recovery very difficult. 

\begin{figure}[tbhp]
    \centering
    \subfigure[]{\label{fig:bpFeeder_shot_gather_4km}\includegraphics[width=0.45\columnwidth]{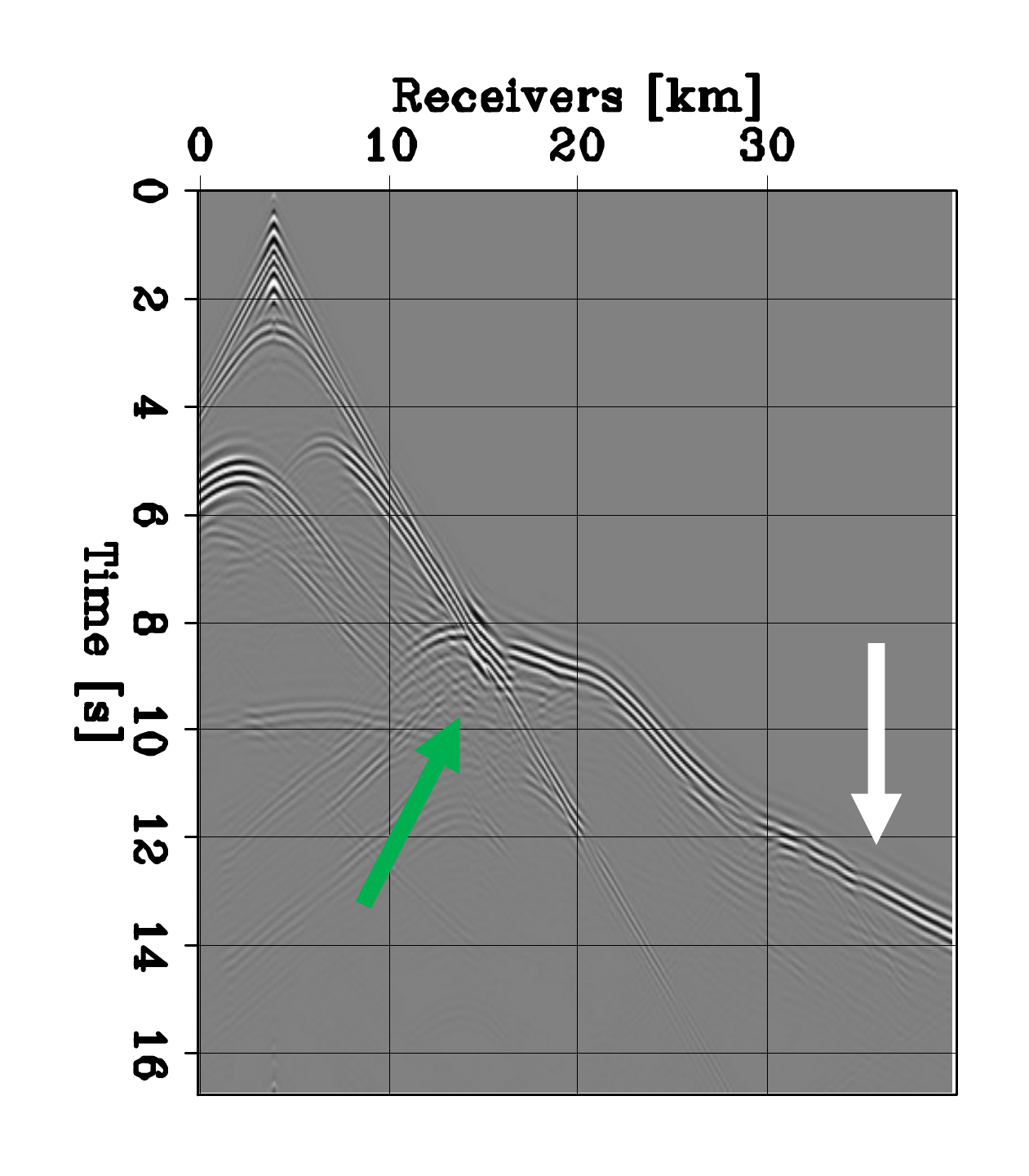}}
    \subfigure[]{\label{fig:bpFeeder_shot_gather_20km}\includegraphics[width=0.45\columnwidth]{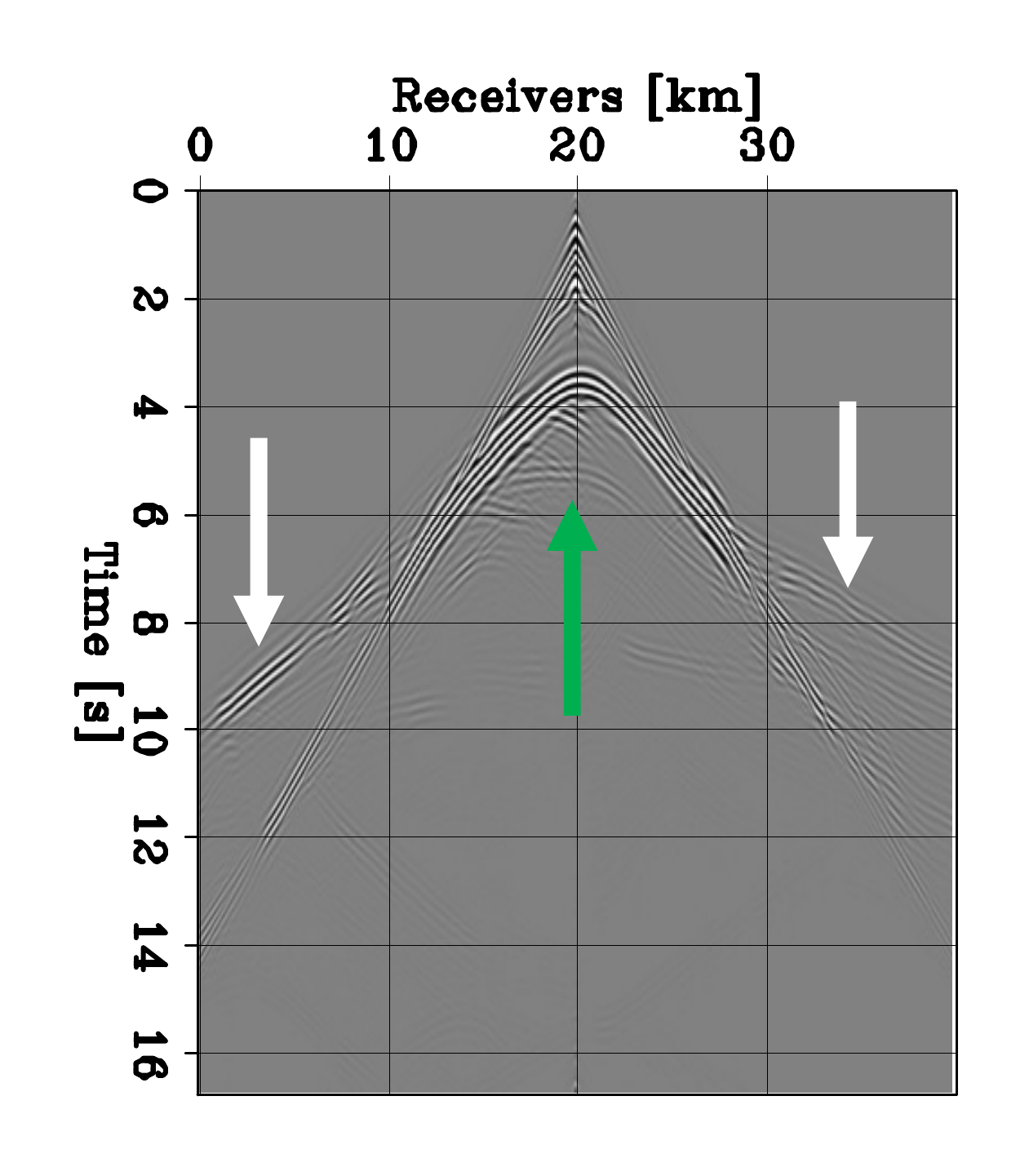}} 
    \caption{Representative shot gathers generated with a source wavelet contaning energy restricted to the 3-8 Hz range. (a) Source placed at $x=4$ km. (b) Source placed at $x=20$ km. All panels are displayed with the same grayscale. White arrows show the energy recorded from diving waves traveling through the low-velocity zone underneath the salt structure. Green arrows correspond to reflected energy from the base salt.}
    \label{fig:bpFeeder_shot_gather}
\end{figure}

We simulate a surface seismic acquisition with 160 sources placed every 250 m (50 m below the surface), and 800 fixed receivers every 50 m (also placed 50 m below the surface). The noise-free data are modeled and recorded for 17 s with an acoustic isotropic constant-density propagator and a spatial sampling of 50 m for both vertical and horizontal axes. In order to reduce the difficulty of this test, we use absorbing boundaries in all directions, and thus no free-surface related multiples are present within the dataset. The initial model is solely composed of sediments (Figure~\ref{fig:bpFeeder_init_mod}). Throughout this study, the maximum frequency used to generate the data is set to 8 Hz for both FWI and FWIME schemes. Figure~\ref{fig:bpFeeder_shot_gather} shows two representative shot gathers for sources placed at $x=4$ km and $x=20$ km, respectively. These panels confirm the difficulty to illuminate the low-velocity region underneath the salt. In fact, the only recorded signal interacting with this zone is carried by diving waves recorded at very long offsets, as shown by the white arrows in both panels. Moreover, for sources placed above the salt feature, the amplitude of the top-salt reflections dominate the ones generated by the interface between the base of the salt and the low-velocity zone (green arrows in Figures~\ref{fig:bpFeeder_shot_gather}a and \ref{fig:bpFeeder_shot_gather}b).

\begin{figure}[tbhp]
    \centering
    \subfigure[]{\label{fig:bpFeeder_wav_fwi_0_8}\includegraphics[width=0.3\columnwidth]{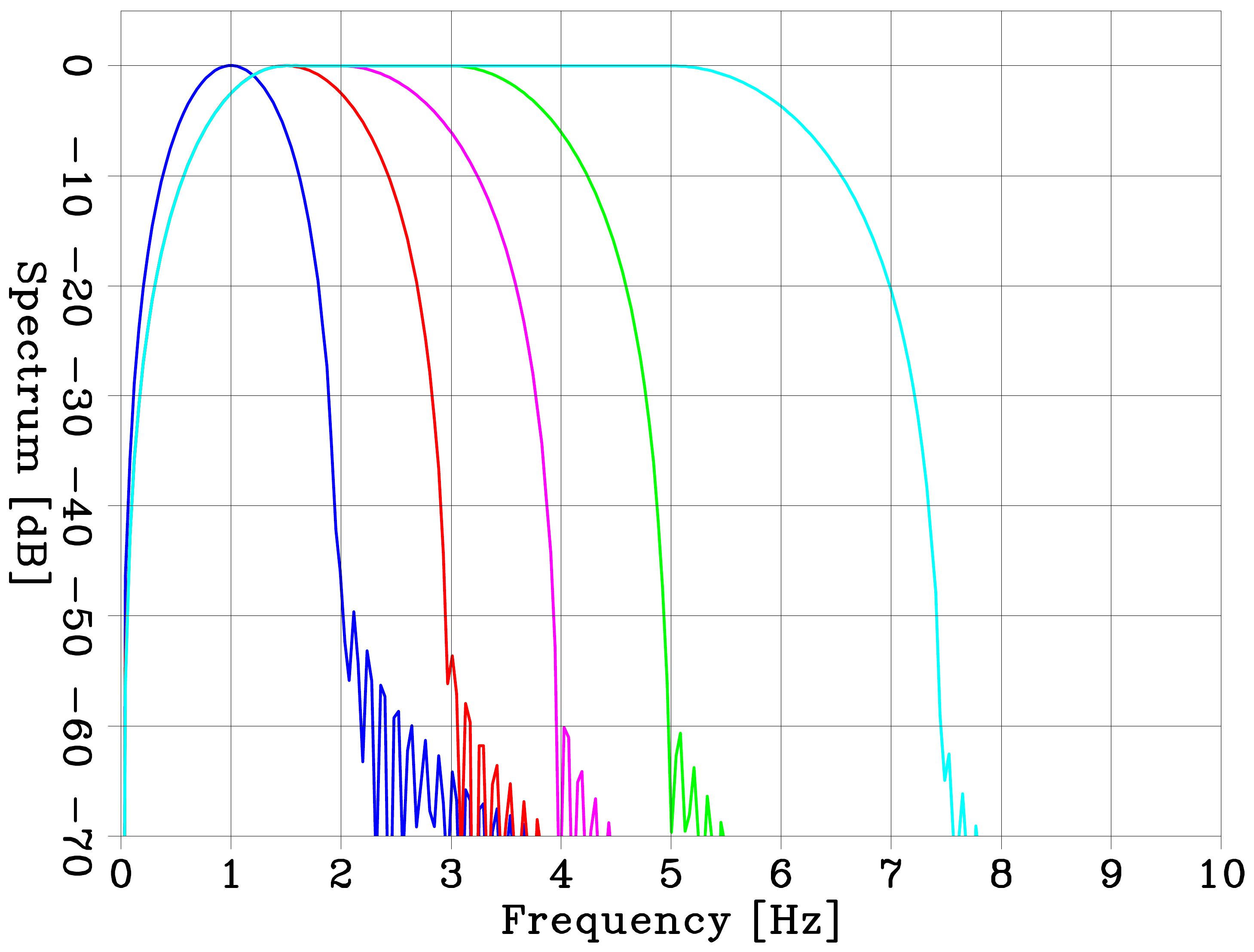}} \hspace{5mm}
    \subfigure[]{\label{fig:bpFeeder_wav_fwi_3_8}\includegraphics[width=0.3\columnwidth]{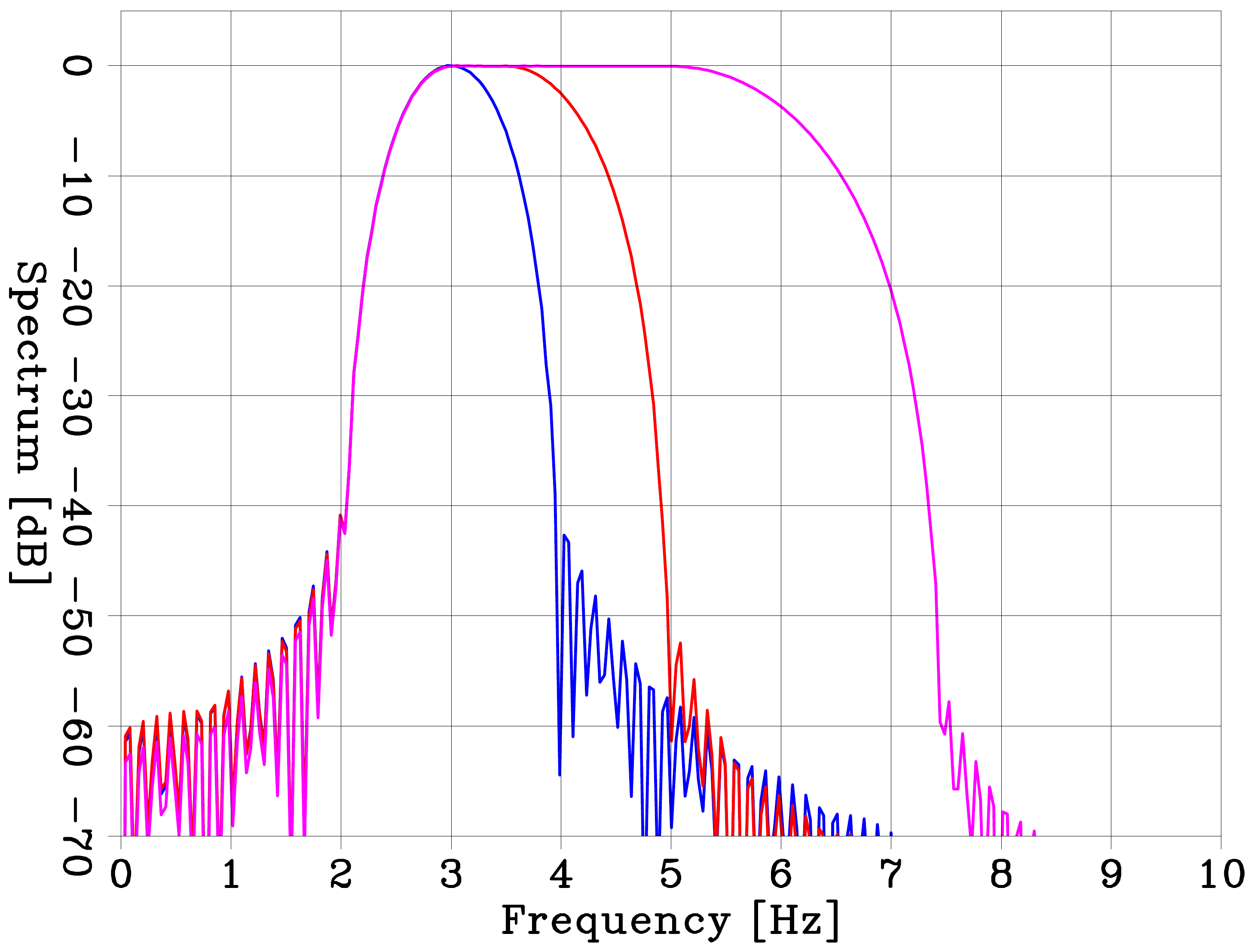}} \hspace{5mm}
    \subfigure[]{\label{fig:bpFeeder_wav_fwime_3_8}\includegraphics[width=0.3\columnwidth]{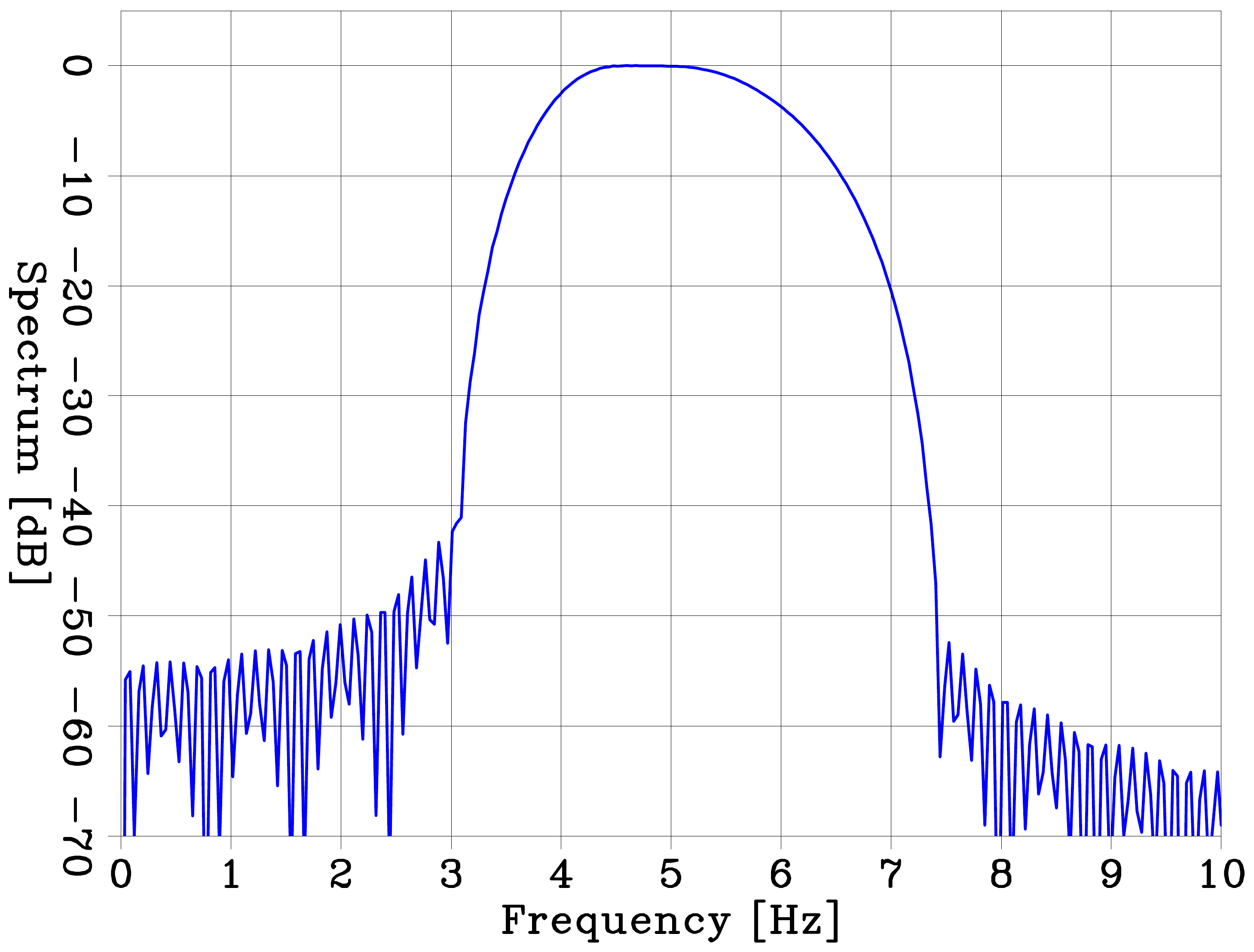}} \hspace{5mm}    
    \caption{Amplitude spectra of the seismic sources employed in this numerical example. (a) Sequence of sources used for the data-space multi-scale FWI workflow using unrealistic low-frequency energy. (b) Sequence of sources used for the data-space multi-scale FWI workflow using energy restricted to the 2-8 Hz bandwidth. (c) Source used for the FWIME workflow, containing energy within the 3-8 Hz bandwidth.}
    \label{fig:bpFeeder_wav}
\end{figure}

We first conduct FWI using unrealistic low-frequency energy ranging from 0.25 Hz to 8 Hz (Figure~\ref{fig:bpFeeder_wav_fwi_0_8}), which converges to an excellent solution (Figure~\ref{fig:bpFeeder_fwi_0_8_mod}). Then, we limit the available bandwidth to 2-8 Hz (Figure~\ref{fig:bpFeeder_wav_fwi_3_8}) and the final inverted model is unable to recover the salt body (Figure~\ref{fig:bpFeeder_fwi_3_8_mod}), even with the presence of coherent energy as low as 2 Hz. 

\begin{figure}[tbhp]
    \centering
    \subfigure[]{\label{fig:fwime_mod_init}\includegraphics[width=0.35\columnwidth]{Fig/BP_feeder/bpSaltFeeder-init-mod.pdf}}
    \subfigure[]{\label{fig:fwime_mod_s0}\includegraphics[width=0.35\columnwidth]{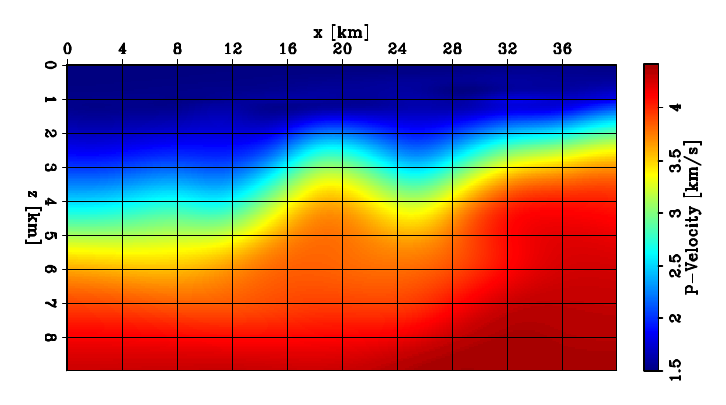}} \\
    \subfigure[]{\label{fig:fwime_mod_s1}\includegraphics[width=0.35\columnwidth]{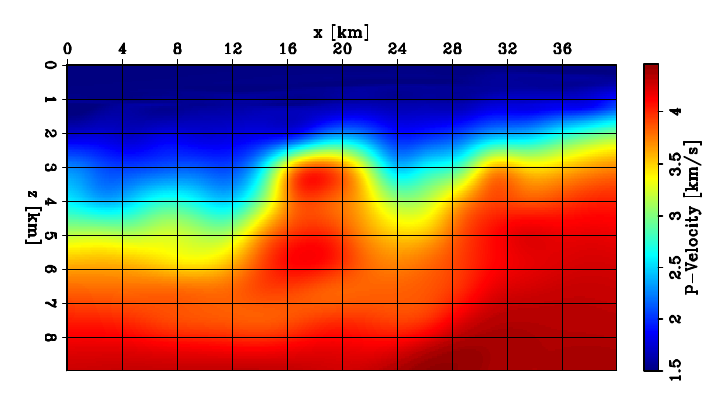}} 
    \subfigure[]{\label{fig:fwime_mod_s2}\includegraphics[width=0.35\columnwidth]{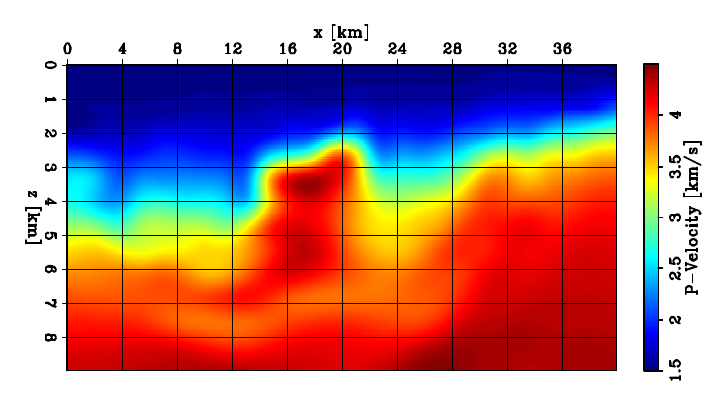}} \\    
    \subfigure[]{\label{fig:fwime_mod_s4}\includegraphics[width=0.35\columnwidth]{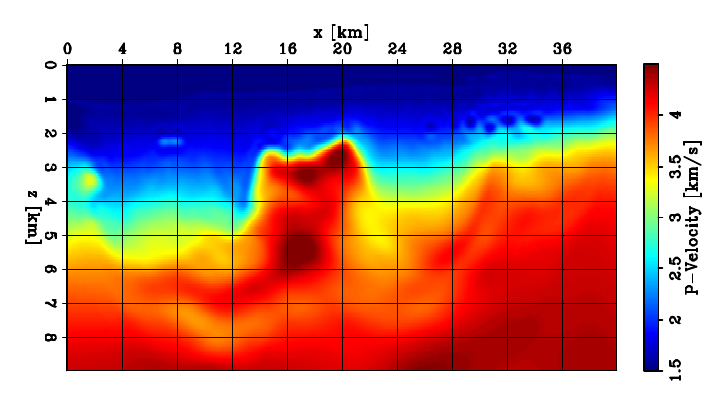}} 
    \subfigure[]{\label{fig:fwime_mod_s5}\includegraphics[width=0.35\columnwidth]{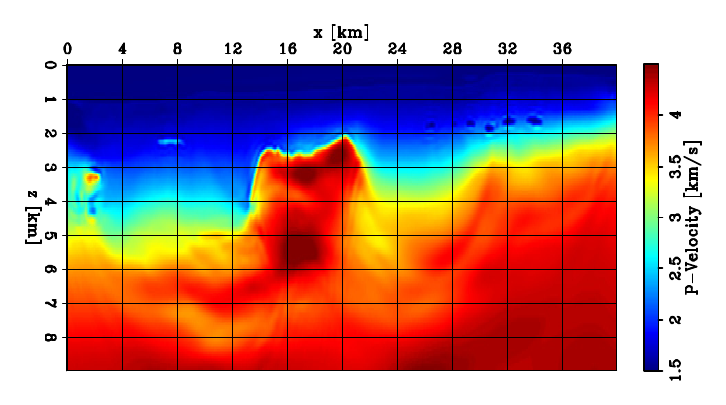}} \\
    \subfigure[]{\label{fig:fwime_mod_noSpline}\includegraphics[width=0.35\columnwidth]{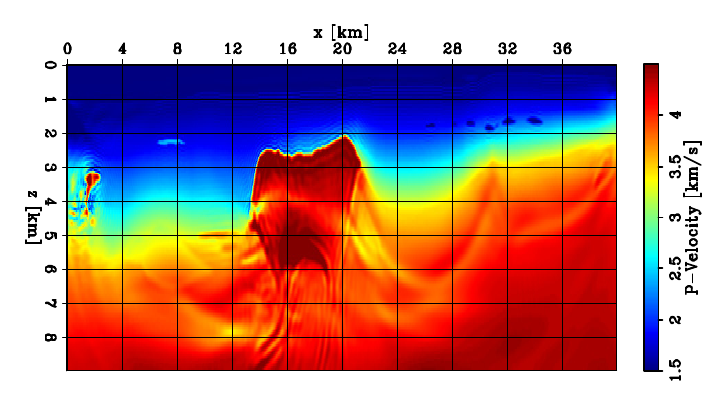}} 
    \subfigure[]{\label{fig:fwime_mod_true}\includegraphics[width=0.35\columnwidth]{Fig/BP_feeder/bpSaltFeeder-true-mod.pdf}}        
    \caption{2D panels of FWIME inverted model at various stages of the inversion process. (a) Initial step. (b) Spline 1. (c) Spline 2. (d) Spline 3. (e) Spline 4. (f) Spline 5. (g) Final FWIME inverted model on the last spline. (h) True model. The last spline grid coincides with the FD grid. Panel (g) is obtained after a total of 258 iteration of L-BFGS.}
    \label{fig:bpFeeder_mod_fwime}
\end{figure}

\begin{figure}[tbhp]
    \centering
    \includegraphics[width=0.5\columnwidth]{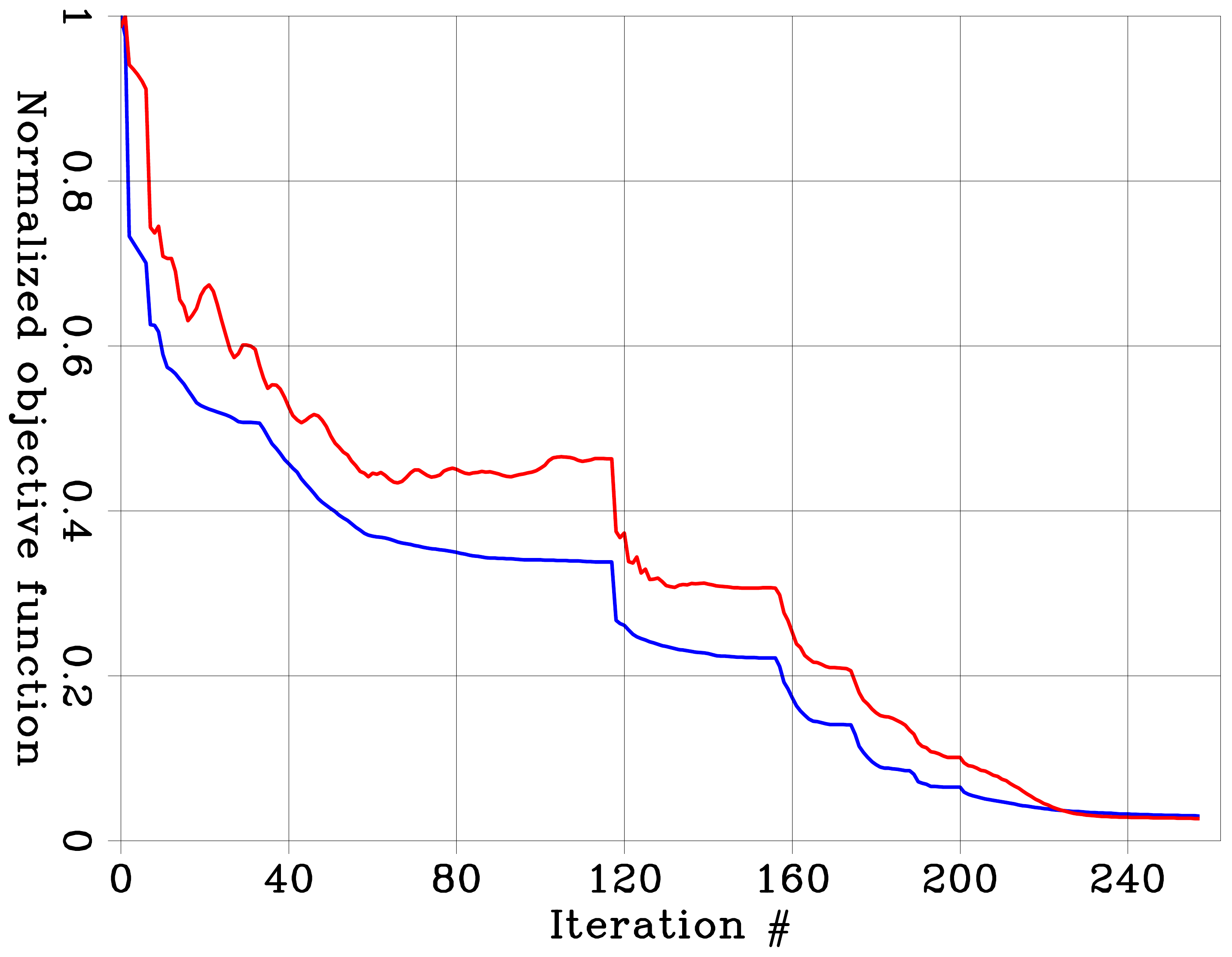}
    \caption{Total normalized FWIME objective function (blue curve), and normalized FWI objective function evaluated at each FWIME inverted model (red curve).}
    \label{fig:bpFeeder_obj}
\end{figure}

We apply FWIME with a time-lag extension and we simultaneously invert a dataset generated with a wavelet containing energy within the 3-8 Hz frequency band (Figure~\ref{fig:bpFeeder_wav_fwime_3_8}). Unfortunately, even with a thorough hyper-parameter search (i.e., by testing a multiple spline grid dispositions, refinement schedules, and a wide range of $\epsilon$-values), FWIME is not able to recover a model as accurate as the one obtained with FWI using unrealistic low frequencies (shown in Figure~\ref{fig:bpFeeder_fwi_0_8_mod}). Figures~\ref{fig:bpFeeder_mod_fwime} shows the set of the FWIME inverted models (displayed on the finite-difference grid) obtained by using the most optimal hyper-parameters (a sequence of six spline grids with $\epsilon=5.0 \times 10^{-4}$). The final inverted model is shown in Figure~\ref{fig:fwime_mod_noSpline} and fails to accurately capture the low-velocity region underneath the salt body, and the right flank is mis-positioned. Moreover, some low-velocity artifacts are present on the left flank of the salt. The blue curve in Figure~\ref{fig:bpFeeder_obj} shows the total FWIME objective function corresponding to the optimization sequence and does not converge to zero, which indicates that the algorithm converges to a local minimum. The red curve in Figure~\ref{fig:bpFeeder_obj}, which is the FWI objective function evaluated at each FWI inverted model during the optimization sequence, shows that FWIME was able to bypass some - but not all - local minima from the conventional FWI formulation. Nevertheless, FWIME performs better than conventional FWI, and the salt features in the shallower parts of the model are better imaged (for depths smaller than 4 km). 

\begin{figure}[tbhp]
    \centering
    \subfigure[]{\label{fig:bpFeeder_data_obs_4km}\includegraphics[width=0.3\columnwidth]{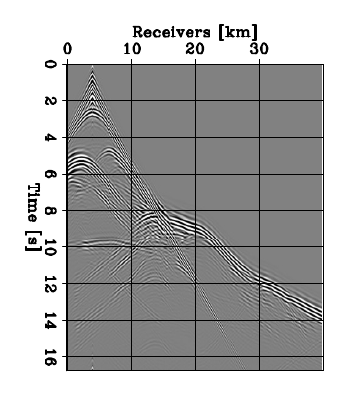}}
    \subfigure[]{\label{fig:bpFeeder_data_pred_4km}\includegraphics[width=0.3\columnwidth]{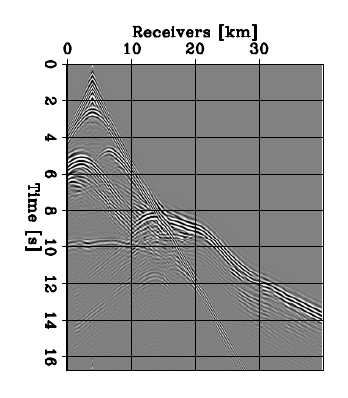}}
    \subfigure[]{\label{fig:bpFeeder_data_res_4km}\includegraphics[width=0.3\columnwidth]{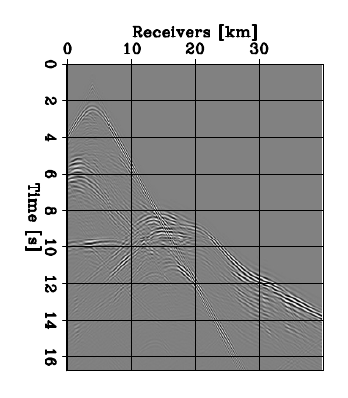}} \\
    \subfigure[]{\label{fig:bpFeeder_data_obs_20km}\includegraphics[width=0.3\columnwidth]{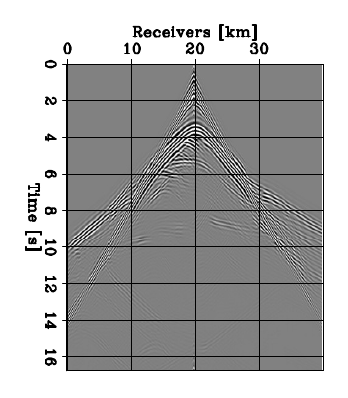}} 
    \subfigure[]{\label{fig:bpFeeder_data_pred_20km}\includegraphics[width=0.3\columnwidth]{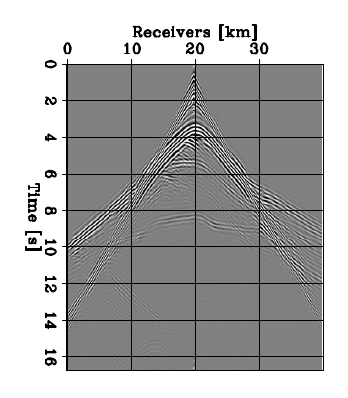}}
    \subfigure[]{\label{fig:bpFeeder_data_res_20km}\includegraphics[width=0.3\columnwidth]{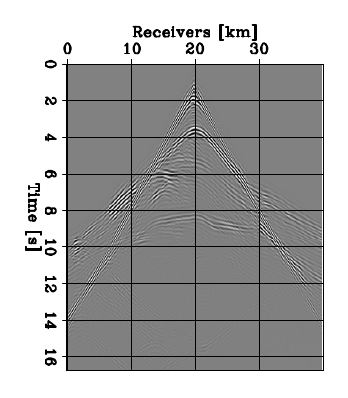}}    
    \caption{Representative shot gathers for sources placed at $x = 4$ km (first row) and $x=20$ km (second row). Observed data, $\mathbf{d}^{obs}$ (first column), Predicted data with the final FWIME model, $\mathbf{f}(\mathbf{m}_{FWIME})$ (second column), and data-difference, $\Delta \mathbf{d}(\mathbf{m}_{FWIME})=\mathbf{d}^{obs} - \mathbf{f}(\mathbf{m}_{FWIME})$ (third column). All panels are displayed with the same grayscale.}
    \label{fig:bpFeeder_data}
\end{figure}

Figure~\ref{fig:bpFeeder_data} shows the observed data, $\mathbf{d}^{obs}$ (first column), predicted data with the final FWIME inverted model $\mathbf{f}(\mathbf{m}_{FWIME})$ (second column), and the final data difference, $\mathbf{d}^{obs}-\mathbf{f}(\mathbf{m}_{FWIME})$ (third column). The first and second rows correspond to sources located at $x=4$ km, and $x=20$ km, respectively. As expected, a large data mismatch can be observed for events occurring at larger offsets (Figure~\ref{fig:bpFeeder_data_res_4km}) for receivers' positions greater than 25 km, which correspond to the recordings of diving waves traveling through the salt body. This prediction error highlights the inability of FWIME to accurately recover for the low-velocity zone underneath the salt. In addition, the last inverted model fails at predicting some reflected and scattered energy coming from the base salt, as shown in Figures~\ref{fig:bpFeeder_data_res_4km} and \ref{fig:bpFeeder_data_res_20km}, for near offsets. 

In the presence of complex geological structures and overburdens, we observe that the selection of the spline grid spatial sampling and the rate at which we adjust it throughout the optimization process strongly impacts the quality of inverted models (more than the $\epsilon$-value). More specifically, the inability to constrain the velocity values to vary less within salt bodies and more rapidly in the vicinity of its edges, seems to hamper the inversion process. Perhaps a more flexible model parametrization may improve the convergence properties of FWIME. When dealing with complex salt bodies, we intend to investigate in the near future the use of more adjustable basis functions for coarse model representation, such as the approach implemented by \cite{dahlke2020applied}. Our proposed method would consist in using migrated images to guide the spline nodes' spacing and spatial disposition. The grid density would be increased in regions containing salt edges, and would be reduced within the salt bodies to promote homogeneous velocity values.

%% file: computational_aspects.tex
\subsection{An efficient GPU implementation}
We design a 2D/3D fully-fledged GPU software package for FWI and FWIME. Our computational framework is composed of three abstract layers. (1) The first layer handles FWIME's most computationally intensive and parallelizable tasks (2D wavefield propagations and extended imaging/scattering operations), which are performed directly on the GPU device and are programmed in the parallel computing platform referred to as compute unified device architecture (CUDA). (2) The collection of CUDA functions (kernels) are then wrapped into a C++ framework which allows us to dispatch the various modeling tasks on multiple GPU devices available within a compute node. (3) A Python-based abstraction layer is then added to easily build, define, and solve large-scale complex numerical inverse problems on multi-node heterogeneous GPU clusters with the use of the objected-oriented optimization framework developed by \cite{biondi2021object}. Finally, the binding between C++ and Python user-defined classes is achieved with the pybind11 library \cite[]{pybind11}. 

\subsection{Computational cost analysis}
To compare the theoretical computational cost of FWIME, $C_{FWIME}$, with the cost of FWI, $C_{FWI}$, we evaluate the number of wavefield propagations required by each method to conduct one L-BFGS iteration. In FWI, one L-BFGS iteration typically requires four propagations: two propagations for the gradient computation, and two propagations for a parabolic line search \cite[]{gill2019practical}. Therefore, $C_{FWI} = 4$. 

In FWIME, the gradient computation is achieved with six propagations: two propagations for the Born component and four for the tomographic component. In addition, for each evaluation of the objective function defined in equation~\ref{eqn:fwime.obj}, one variable projection (VP) step needs to be conducted by minimizing the quadratic objective function defined in equation~\ref{eqn:vp.obj}. If we perform $N_{vp}$ linear conjugate gradient iterations for the VP step, the additional cost is given by 

\begin{eqnarray}
    \label{eqn:comp.cost.vp}
    C_{vp} = 4 \times N_{vp} (1 + C_{ext}),     
\end{eqnarray}

where $0 \le C_{ext} \le 1 $ is the percentage cost increase due to the fact that all migrations/demigrations are conducted on an extended space. The value of $C_{ext}$ depends on the length of the extended axis but is typically on the order of 40$\%$ for 3D applications with a time-lag extended length of approximately 50 points. However, for the 2D numerical examples shown in this paper, we do not try to optimize the length of the extended axis and $C_{ext}$ can be as high as 60$\%$. In equation~\ref{eqn:comp.cost.vp}, the factor 4 comes from the application of one forward and one adjoint extended migration per iteration, each corresponding to two propagations. To avoid evaluating the FWIME objective function (equation~\ref{eqn:fwime.obj}) multiple times during the step length estimation, we use the method proposed by \cite{more1994line} instead of a parabolic line search, which only requires one cost function evaluation in addition to the gradient computation. Finally, the total cost of one L-BFGS FWIME iteration is given by

\begin{eqnarray}
    \label{eqn:comp.cost}
    C_{FWIME} = 6 + 4 \times N_{vp} (1 + C_{ext}). 
\end{eqnarray}

Table~\ref{table:computational_cost_prop} summarizes this analysis assuming different values of $N_{vp}$. The sixth column corresponds to the ratio between the cost of the VP step and the total cost of one FWIME iteration, which illustrates that this step is our main computational bottleneck. The last column displays the ratio between the FWIME and the FWI cost. For 3D applications, we observe that setting $N_{vp}=30$ is sufficient for accurate convergence, which indicates that FWIME is approximately 40 to 70 times more costly than conventional FWI. 

\begin{table}[h!]
\centering
\begin{tabular}{ |c|c|c|c|c|c|c|c|c| } 
\hline
Method & $N_{vp}$ & $C_{ext}$ ($\%$) & $C_{vp}$ & $C_{total}$ & $\frac{C_{vp}}{C_{total}}$ ($\%$) & Ratio  \\
\hline
FWI & - & - &  -  & 4 & - & 1 \\
FWIME & 100 & 40 & 560 & 566 & 99 & 142 \\
FWIME & 50 & 40 & 280 & 286 & 98 & 72 \\
FWIME & 30 & 40 & 168 & 174 & 97 & 43 \\
FWIME & 20 & 40 & 112 & 118 & 95 & 30 \\
\hline
\end{tabular}
\caption{Table summarizing the computational cost comparison between FWIME and FWI. $N_{vp}$ is the number of linear conjugate gradient iterations for the VP step. $C_{ext}$ is the percentage cost increase due to the fact that all migrations/demigrations are conducted on an extended space. $C_{vp}$ is the number of propagations required for the VP step. $C_{total}$ is the total number of propagations for one L-BFGS iteration. The quantity $\frac{C_{vp}}{C_{total}}$ is the percentage cost of the VP step. The last column shows the ratio between the FWIME and the FWI computational costs.}
\label{table:computational_cost_prop}
\end{table}

\subsection{Computational time for 2D applications}
The numerical implementation for the 2D examples proposed in this paper are conducted on a GPU computer cluster, which includes four compute nodes, each containing four NVIDIA Tesla V100 GPU devices with 16 GB of global memory. Each node is equipped with 24 CPU cores and 512 GB of random access memory (RAM). Each example is solved using four GPU devices without any domain-decomposition strategy \cite[]{micikevicius20093d}: each GPU device simulates one shot throughout the full finite-difference domain of interest.

For reproducibility purposes, Table~\ref{table:computational_time} provides a breakdown of the FD parameters and an estimation of the computational time for the proposed numerical examples. $N_z$ and $N_x$ correspond to the number of spatial samples on the FD grid, which includes additional padding for the absorbing boundaries. $N_{ext}$ is the total number of samples on the extended axis for $\mathbf{\tilde{p}}_{\epsilon}^{opt}$. $T_{rec}$ is the total recording time for each shot, expressed in seconds. The quantities $t_{FWIME}$ and $t_{FWI}$ correspond to the approximate computational time taken for one L-BFGS iteration for FWIME and conventional FWI, respectively (expressed in minutes). The last column reports the ratio between the FWIME and the FWI computational times. 

For the 2D applications propsed in this paper, we do not focus on improving the computational efficiency of FWIME. We choose very conservative values for $N_{ext}$ and $N_{vp}$ to ensure accurate convergence for the VP step and assess the performance of our method in optimal conditions. For the Marmousi2 example, we investigate the effect of using a smaller number of linear iterations for the VP step. We notice that 15 iterations are sufficient to converge to a similar solution, thereby reducing FWIME's computational cost by a factor of three. 

\begin{table}[h!]
\centering
\begin{tabular}{ |c|c|c|c|c|c|c|c|c| } 
\hline
2D Model & $N_z$ & $N_x$ & $N_{ext}$ & $T_{rec}$ & $N_{vp}$ & $t_{FWIME}$ & $t_{FWI}$ & Ratio\\
\hline
Marmousi2 & 250 & 700 & 101 & 8 s & 60 & 50 min & 0.5 min & 100 \\
Marmousi2 & 250 & 700 & 101 & 8 s & 30 & 26 min & 0.5 min & 52 \\
Marmousi2 & 250 & 700 & 101 & 8 s & 15 & 14 min & 0.5 min & 28 \\
North Sea & 250 & 840 & 101 & 13 s & 60 & 80 min & 0.75 min & 106 \\
Seiscope & 220 & 470 & 201 & 6 s & 80 & 15 min & 0.2 min & 90 \\
Salt & 330 & 1000 & 101 & 15 s & 100 & 85 min & 0.75 min & 110  \\
West Africa & 330 & 1000 & 101 & 15 s & 100 & 85 min & 0.75 min & 110  \\
\hline
\end{tabular}
\caption{Table summarizing the FD modeling parameters and computational time for the five numerical examples proposed in this paper. Each example is conducted on four NVIDIA Tesla V100 GPU devices. $N_z$ and $N_x$ correspond to the number of samples for the FD grid. $N_{ext}$ is the total number of samples on the extended axis. $T_{rec}$ corresponds to the recording time for the simulated data. $N_{vp}$ is the number of linear conjugate-gradient iterations conducted for the VP step. $t_{FWIME}$ and $t_{FWI}$ correspond to the approximate computational time for one L-BFGS iteration of FWIME and FWI, respectively. The last column displays the ratio between $t_{FWIME}$ and $t_{FWI}$.}
\label{table:computational_time}
\end{table}

\subsection{Reproducibility}
To ensure the full reproducibility of our results from this paper, we create an online repository where our open-source software package can be freely accessed \cite[]{barnierFwimeGitHub2}. All the numerical examples proposed in this paper are implemented in Jupyter notebooks and can be replicated with the use of NVIDIA GPU devices. 

%% file: discussions.tex
To mitigate the computational cost of FWIME in a production workflow, we see two potential applications of our method. (1) FWIME can be conducted until the inverted model is accurate enough (which can be assessed by examining the optimal extended perturbation $\tilde{\mathbf{p}}_{\epsilon}^{opt}$) and used as an input for conventional FWI. (2) Alternatively, if we wish to better characterize a specific region (e.g., hydrocarbon reservoir), we do not need high accuracy/resolution everywhere in the subsurface. Instead of applying acoustic FWI on the entire domain, we can use the output of FWIME and feed it to target-oriented elastic FWI procedures \cite[]{wapenaar2014elastic,ravasi2017rayleigh,guo2019target,garg2020surface,biondi2021target}. Even though elastic FWI is more computationally intensive, it becomes tractable if applied to a smaller target volume. In addition, elastic FWI is better suited for retrieving elastic properties than conventional ray-based amplitude versus angle (AVA) approaches \cite[]{biondi2019amplitude,biondi2018target,biondi2021target}.

%% file: conclusions.tex
The main benefit of FWIME is its ability to converge to useful acoustic Earth models by inverting any type of waves without the need for coherent low-frequency signal or accurate initial guesses. We do not provide a mathematical proof of global convergence for our method, but we provide plenty of numerical evidence that support this claim. Our method leverages the robustness of WEMVA with the high-resolution nature and accuracy of FWI by judiciously pairing them into one workflow which is more efficient than applying these two techniques separately/sequentially. The automatic coupling between WEMVA and FWI is achieved with the variable projection method, which substantially reduces the number of hyper-parameters to two: the trade-off variable $\epsilon$ and the spline grid schedule. Our algorithm is applied with the same mechanism regardless of the type of waves being inverted, which makes it simple to use. Thus, the need to design user-intensive ad hoc inversion strategies is mitigated. 

We apply FWIME on five realistic synthetic tests that replicate some of the most challenging environments encountered by energy companies: inaccurate initial model, lack of low-frequency data, lack of illumination, and presence of complex overburdens. In each example, we guide the reader step by step through the hyper-parameter tuning process and we show that our method can retrieve excellent solutions, whereas conventional methods converge to unsatisfactory models. FWIME is more computationally demanding than conventional FWI. However, from energy companies' perspective, the potential upside (in terms of safety and efficiency) gained from the image quality uplift in certain regions may offset this cost. 

We identify some current limitations and potential opportunities to improve the convergence properties of FWIME. They include: (1) employing a numerical modeling scheme that simulates more realistic physics (anisotropy, elastic effects, attenuation) and thus handle multi-parameter inversions, (2) applying a surface-related multiple elimination pre-processing step to the data for offshore acquisitions, and (3) designing a more flexible sparse model representation to efficiently delineate and recover complex geobodies. 

In an online repository, we provide a fully-reproducible open-source 2D GPU software suite (along with Jupyter notebooks) to easily replicate the results presented in this paper. In a third paper, we successfully apply FWIME to a 3D OBN field dataset from the Gulf of Mexico and we make the 3D FWIME implementation freely accessible. 